\documentclass[fleqn,usenatbib]{mnras}

\usepackage{newtxtext,newtxmath}

\usepackage[T1]{fontenc}

\DeclareRobustCommand{\VAN}[3]{#2}
\let\VANthebibliography\thebibliography
\def\thebibliography{\DeclareRobustCommand{\VAN}[3]{##3}\VANthebibliography}

\usepackage{graphicx}	
\usepackage{amsmath}	
\usepackage{tikz}
\usepackage{multirow}   
\usepackage{footnote}
\usepackage{supertabular,booktabs}
\usepackage{afterpage}

\newcommand{\h}{$^{\text{h}}$}
\newcommand{\m}{$^{\text{m}}$}
\newcommand{\s}{$^{\text{s}}$}
\renewcommand{\d}{$^{\circ}$}
\newcommand{\M}{$^{\prime\hspace{0.1mm}}$}
\renewcommand{\S}{$^{\prime\prime}$}

\newcommand{\figscale}{0.32}
\newcommand{\imgheight}{200pt}

\newcommand{\figxi}{2.7}
\newcommand{\figxj}{8.6}
\newcommand{\figxk}{14.5}
\newcommand{\figyi}{20.15}
\newcommand{\figyj}{14.25}
\newcommand{\figyk}{8.35}
\newcommand{\figyl}{2.45}
\newcommand{\figyii}{-4.0}
\newcommand{\figyjj}{-9.85}
\newcommand{\figykk}{-15.7}
\newcommand{\figyll}{-21.55}

\title[Relation between accretion rate and jet power]{The relation between accretion rate and jet power in early-type galaxies with thermally unstable hot atmospheres}

\author[T. Pl\v{s}ek et al.]{
T. Pl\v{s}ek,$^{1}$\thanks{E-mail: plsek@physics.muni.cz}
N. Werner,$^{1}$
R. Grossová,$^{1,2}$
M. Topinka,$^{1,3}$
A. Simionescu$^{4,5,6}$ and
S. W. Allen$^{7,8,9}$\\
$^{1}$ Department of Theoretical physics and astrophysics, Faculty of Science, Kotlá\v{r}ská 2, Masaryk University, Brno, 611 37, Czech Republic
\\
$^{2}$ Astronomical Institute of the Czech Academy of Sciences, Bo\v{c}ní II 1401, 141 00, Prague, Czech Republic\\
$^{3}$ INAF-Istituto di Astrofisica Spaziale e Fisica Cosmica, Via A. Corti 12, I-20133 Milano, Italy
\\
$^{4}$ SRON Netherlands Institute for Space Research, Niels Bohrweg 4, 2333 CA Leiden, The Netherlands
\\
$^{5}$ Leiden Observatory, Leiden University, PO Box 9513, 2300 RA Leiden, The Netherlands\\
$^{6}$ Kavli Institute for the Physics and Mathematics of the Universe (WPI), The University of Tokyo, Kashiwa, Chiba 277-8583, Japan\\
$^{7}$ Department of Physics, Stanford University, 382 Via Pueblo Mall, Stanford CA 94305\\
$^{8}$ SLAC National Accelerator Laboratory, 2575 Sand Hill Road, Menlo Park, CA 94025\\
$^{9}$ Kavli Institute for Astrophysics and Cosmology, Stanford University, 452 Lomita Mall, Stanford, CA 94305\\
}

\date{Accepted XXX. Received YYY; in original form ZZZ}

\pubyear{2022}

\begin{document}
\label{firstpage}
\pagerange{\pageref{firstpage}--\pageref{lastpage}}
\maketitle

\begin{abstract}
We use {\it Chandra} X-ray data and Very Large Array radio observations for a sample of 20 nearby, massive, X-ray bright, early-type galaxies to investigate the relation between the Bondi accretion rates and the mechanical jet powers. We find a strong correlation ($\rho = 0.96^{+0.03}_{-0.09}$; BF$_{10} > 100$) between the Bondi accretion power, $P_{\text{Bondi}}$, and the mechanical jet power, $P_{\text{jet}}$, for a subsample of 14 galaxies, which also host cool H$\alpha$+[N\textsc{ii}] line emitting gas and thus likely have thermally unstable atmospheres. The relation between the Bondi accretion power and the mechanical jet power for this subsample is well described by a power-law model $\log \frac{P_{\mathrm{Bondi}}}{{10^{43} \, \mathrm{erg \, s^{-1}}}}  = \alpha + \beta \log \frac{P_{\mathrm{jet}}}{{10^{43} \, \mathrm{erg \, s^{-1}}}}$, where $\alpha = 1.10 \pm 0.25$ and $\beta = 1.10 \pm 0.24$ with an intrinsic scatter $\sigma = 0.08^{+0.14}_{-0.06}$ dex. The results indicate that in all galaxies with thermally unstable atmospheres the cooling atmospheric gas feeds the central black holes at a similar jet-to-Bondi power ratio. For the full sample of 20 galaxies, the correlation is weaker and in a subset of galaxies with no signs of H$\alpha$+[N\textsc{ii}] emission, we see a hint for a systematically lower jet-to-Bondi power ratio. We also investigate the dependence of jet power on individual quantities in the Bondi formula such as the supermassive black hole mass ($M_{\bullet}$) and the specific entropy of the gas ($K$) at the Bondi radius. For the subsample of H$\alpha$+[N\textsc{ii}] emitting galaxies, we find a very tight correlation of $P_{\text{jet}}$ with $M_{\bullet}$ ($\rho = 0.91^{+0.06}_{-0.11}$; BF$_{10} > 100$) and, although poorly constrained, a hint of an anti-correlation for $P_{\text{jet}}$ and $K$ ($\rho = -0.47^{+0.60}_{-0.37}$; BF$_{10} = 1.1$).
\end{abstract}

\begin{keywords}
accretion, accretion discs -- galaxies: active -- galaxies: jets -- galaxies: nuclei -- radio continuum: galaxies -- X-ray: galaxies
\end{keywords}



\section{Introduction}

All massive early-type galaxies harbour supermassive black holes (SMBH) in their centre and are permeated by extended, hot, X-ray emitting atmospheres. Although the central black holes represent a small fraction in terms of total mass and size compared to the proportions of the host galaxy, they are often outperforming its energetic output and thus playing a substantial role in the energetics of the whole galactic atmosphere.

As the supermassive black holes accrete the surrounding material, a big part of its rest mass (up to 40 \% for fast rotating black holes) may be turned into energy. Part of the energy is then expelled in the form of electromagnetic radiation or outflows of relativistic particles (jets), typically depending on the type and rate of the accretion flow. However, the vast majority of low-redshift active galactic nuclei (AGNs) in early-type galaxies operate in the radio-mechanical (kinetic) mode, which is observed in the form of radio lobes and X-ray cavities \citep[e.g.][]{Fabian2012,Shin2016,Grossova2021}.

A detailed description of the accretion of supermassive black holes is nontrivial. Assuming a matter-to-radiation conversion efficiency of $\epsilon=10$ per cent, a spherical Bondi accretion \citep{Bondi1952} from the hot atmospheres of early-type galaxies would result in luminosities that are several orders of magnitude above the observed values. However, it is likely that most of the accretion power in early-type galaxies is converted into jets and the estimates of mechanical jet powers from the observed X-ray cavities are within an order of magnitude comparable to powers predicted from the Bondi formula \citep{bohringer2002,churazov2002,dimatteo2003,Allen2006}.

\citet{Allen2006} studied the relation between Bondi accretion powers and mechanical jet powers in nine nearby giant ellipticals using a combination of {\it Chandra} X-ray and VLA radio observations finding a tight power-law correlation. Their result indicates that $2.2^{+1.0}_{-0.7}$ per cent of the rest mass energy of the accreted material is converted into jet power. A relation with a much larger scatter has been observed by \citet{Russell2013}; however, these authors derived their estimates of jet powers using X-ray data only.

Detailed studies of the gas distribution within the Bondi radii were performed using very deep {\it Chandra} observations of NGC 3115 \citep{wong2011,wong2014} and M87 \citep{dimatteo2003,Russell2015,Russell2018}. These observations reveal shallow density profiles, which are consistent with the presence of significant outflows and multi-phase gas at the Bondi radii of central black holes. The clumpy cool gas within the hot multi-phase flow could reach the core episodically, triggering larger outbursts than the continuous activity driven by a more steady hot inflow \citep{werner2019}. 

Here, we study how the presence of thermal instabilities and multi-phase gas affects the relationship between the Bondi accretion power and the mechanical jet power in 20 nearby early-type galaxies.
Ten galaxies in our sample harbour extended H$\alpha$+[N\textsc{ii}] nebulae, four systems display nuclear optical line emission, and six galaxies show no indications for the presence of cool gas. The Bondi accretion rates are calculated from density and temperature profiles of the hot atmospheric gas in the centre of the galaxies determined using {\it Chandra} data and using black hole mass measurements from the literature (see Sect. \ref{section:sample}). The jet powers are estimated from the energies and time scales needed to inflate cavities in the X-ray emitting gas, where the cavity volumes are estimated based on the extent of the most recent generation of radio lobes. The relation between Bondi accretion powers and jet powers is studied for both the full sample and separately for galaxies which also show nebular line emission and their atmospheres are thus presumably prone to thermally unstable cooling \citep[e.g.][]{Lakhchaura2018}. We also investigate the correlations of the jet power, separately, with the mass of the central SMBH and with the gas entropy within the Bondi accretion radius and examine whether the estimated jet power offsets the radiative cooling within radii where the cooling time is shorter than 1 Gyr.

\subsection{Sample selection}
\label{section:sample}

We selected a sample of galaxies with high-quality archival \textit{Chandra} observations that allow us to determine the properties of their X-ray emitting atmospheres within an order of magnitude of the Bondi radii of their central supermassive black holes (SMBH). Importantly, all of the selected galaxies host radio lobes, which appear to have been inflated recently or still undergoing inflation. The sample galaxies and their basic parameters are listed in Table \ref{tab:basic}. SMBH masses were taken either from direct measurements reported by \cite{Kormendy2013} and \cite{Saglia2016} or derived from the $M_{\bullet}-\sigma_v$ scaling relation (see Table \ref{tab:basic})
\begin{equation}
\log \left( \frac{M_{\bullet}}{10^9 \, M_{\sun}} \right) = \alpha + \beta \, \log \left( \frac{\sigma_v}{200 \, \mathrm{km/s}} \right),
\end{equation}
where $\alpha = 8.49 \pm 0.05$ and $\beta = 4.38 \pm 0.29$ \citep{Kormendy2013} and velocity dispersions were taken from the \textit{HyperLeda}\footnote{\href{http://leda.univ-lyon1.fr/}{http://leda.univ-lyon1.fr/}} database \citep{Makarov2014}. Galactic redshifts were taken from \textit{Nasa Extragalactic Database} (NED). The distances were derived from measurements of Surface Brightness Fluctuations \citep{Blakeslee2001,Jensen2003,Cantiello2005,Blakeslee2009} except for NGC$\,$507 and NGC$\,$6166 \citep[Fundamental Plane method;][]{Tully2013}, and NGC$\,$1600 and NGC$\,$4778 \citep[Tully-Fisher relation;][]{Theureau2007}.

\setlength{\tabcolsep}{4.05pt}
\renewcommand{\arraystretch}{1.4}
\begin{table*}
\caption{Initial parameters for the sample of galaxies: right ascension and declination coordinates of the SMBH position, morphology of cool gas tracer H$\alpha$+[N\textsc{ii}] emission (adopted from \protect\citealp{Lakhchaura2018}; extended emission extends beyond central 2 kpc), total ACIS exposure time, VLA configuration used for radio lobe size-estimation, distance $d$, redshift $z$ (\textit{Nasa Extragalactic Database}), hydrogen column density $n_{\text{H}}$ \citep{HI4PI2016}, velocity dispersion $\sigma_v$ \citep[\textit{HyperLeda} database; ][]{Makarov2014}, supermassive black hole masses $M_{\bullet}$ and their references.}
\begin{tabular}{lccccccccccc}
 	\toprule \vspace{-0.7mm}
    \multirow{2}{3.1em}{Galaxy} & \multirow{2}{3.1em}{RA$\,^a$} & \multirow{2}{3.1em}{DEC$\,^a$} & H$\alpha$+[N\textsc{ii}] & Exptime & VLA & $d$ & \multirow{2}{1.0em}{$z$} & $n_{\text{H}}$ & $\sigma_v$ & $M_{\bullet}\,^c$ & \multirow{2}{1.35em}{Ref.}\\
     & & & morphology & (ks) & conf.$^b$ & (Mpc) & & ($10^{20}$ cm$^{-2}$) & (km s$^{-1}$) & ($10^9 \, M_{\sun}$) & \\
    \bottomrule
    IC$\,$4296&13\h36\m39.0325\s$\:$&-33\d57\M57.072\S$\:$&Extended&28.5&D&49.0&0.0125&3.96&$327\pm5$&$1.30\,^{+0.24}_{-0.20}$ & [1]\\ 
    NGC$\,$507&01\h23\m39.9409\s$\:^*$&33\d15\M21.858\S$\:^*$&No&62.2&C&64.6&0.0165&5.24&$292\pm6$&$2.14\,^{+0.58\:\ddagger}_{-0.46}$ & -\\ 
    NGC$\,$708&01\h52\m46.458\s$\:$&36\d09\M06.485\S$\:$&Extended&139.4&A$^{\dagger}$&62.8&0.0162&6.87&$222\pm8$&$0.49\,^{+0.07\:\ddagger}_{-0.07}$ & -\\ 
    NGC$\,$1316&03\h22\m41.7052\s$\:$&-37\d12\M28.557\S$\:$&Extended&228.4&A&22.7&0.0059&2.11&$223\pm3$&$0.17\,^{+0.03}_{-0.03}$ & [2]\\ 
    NGC$\,$1399&03\h38\m29.0170\s$\:$&-35\d27\M01.507\S$\:$&No&204.5&A (B)&20.9&0.0048&1.39&$332\pm5$&$1.26\,^{+0.52}_{-0.63}$ & [2]\\
    NGC$\,$1407&03\h40\m11.79\s$\:^*$&-18\d34\M48.9\S$\:^*$&No&44.5&A&25.4&0.0059&4.94&$266\pm5$&$4.65\,^{+0.73}_{-0.41}$ & [2]\\ 
    NGC$\,$1600&04\h31\m39.8510\s$\:^*$&-05\d05\M10.476\S$\:^*$&No&243.6&A&63.7&0.0156&3.19&$331\pm7$&$17.0\,^{+1.5}_{-1.5}$ & [3]\\ 
    NGC$\,$4261&12\h19\m23.2357\s$\:$&05\d49\M29.650\S$\:$&Nuclear&135.3&C&32.4&0.0074&1.65&$297\pm4$&$1.67\,^{+0.39}_{-0.24}$ & [4]\\ 
    NGC$\,$4374&12\h25\m3.743\s$\:$&12\d53\M13.139\S$\:$&Nuclear&882.4&B&18.5&0.0034&2.85&$278\pm2$&$0.93\,^{+0.10}_{-0.09}$ & [2]\\ 
    NGC$\,$4472&12\h29\m46.7619\s$\:$&08\d00\M01.713\S$\:$&No&367.0&A (C)&16.7&0.0033&1.56&$282\pm3$&$2.54\,^{+0.58}_{-0.10}$ & [2]\\ 
    NGC$\,$4486&12\h30\m49.4234\s$\:$&12\d23\M28.044\S$\:$&Extended&370.7$/$135.4$\,^e$&A&16.7&0.0043&1.23&$323\pm4$&$6.5\,^{+0.7}_{-0.7}$ & [5]\\ 
    NGC$\,$4552&12\h35\m39.8141\s$\:$&12\d33\M22.732\S$\:$&Extended$\,^{d}$&201.4&C&16.0&0.0011&2.72&$250\pm3$&$0.50\,^{+0.06}_{-0.06}$ & [1]\\ 
    NGC$\,$4636&12\h42\m49.827\s$\:$&02\d41\M15.99\S$\:$&Nuclear&197.4&A&14.7&0.0031&1.82&$200\pm3$&$0.33\,^{+0.04\:\ddagger}_{-0.04}$ & -\\ 
    NGC$\,$4649&12\h43\m39.986\s$\:$&11\d33\M09.86\S$\:$&No&293.8&B&16.5&0.0037&2.02&$330\pm5$&$4.72\,^{+1.04}_{-1.05}$ & [2]\\ 
    NGC$\,$4696&12\h48\m49.2762\s$\:^*$&-41\d18\M39.532\S$\:^*$&Extended&711.1&A&42.5&0.0099&7.78&$243\pm6$&$0.89\,^{+0.18\:\ddagger}_{-0.15}$ & -\\ 
    NGC$\,$4778&12\h53\m05.7003\s$\:^*$&-09\d12\M14.676\S$\:^*$&Nuclear&167.0&B/A&66.2&0.0147&3.86&$251\pm21$&$0.84\,^{+0.17\:\ddagger}_{-0.14}$ & -\\ 
    NGC$\,$5044&13\h15\m23.9727\s$\:^*$&-16\d23\M07.779\S$\:^*$&Extended&563.7&A&32.2&0.0093&4.92&$225\pm9$&$0.22\,^{+0.12}_{-0.07}$ & [6]\\ 
    NGC$\,$5813&15\h01\m11.2345\s$\:^*$&01\d42\M07.244\S$\:^*$&Extended&638.2&B&32.2&0.0065&4.29&$236\pm3$&$0.71\,^{+0.10}_{-0.09}$ & [1]\\ 
    NGC$\,$5846&15\h06\m29.284\s$\:^*$&01\d36\M20.25\S$\:^*$&Extended&113.4&A&24.9&0.0057&4.31&$237\pm4$&$1.10\,^{+0.16}_{-0.14}$ & [1]\\
    NGC$\,$6166&16\h28\m38.245\s$\:$&39\d33\M04.234\S$\:$&Extended&158.3&A$^{\dagger}$&125.0&0.0304&0.79&$301\pm6$&$2.11\,^{+0.57\:\ddagger}_{-0.45}$ & -\\
	\bottomrule
\end{tabular}\\
\vspace{-0.8mm}
\begin{flushleft}
\footnotesize{$^a$ Coordinates marked with $^*$ represent SMBH positions derived from the hard X-ray ($3 - 7\:$keV) central peak, for the rest of the galaxies the positions are based on radio observations and they were either derived from VLBI data or taken from the literature (NGC$\,$4636, NGC$\,$5846; ALMA; \citealp{Temi2018}).}\\ \vspace{0.5mm}
\footnotesize{$^b$ VLA array configurations marked with $^{\dagger}$ were taken from the NRAO VLA Archive Survey (NVAS). Array configurations in parentheses represent shorter baselines capturing more extended radio emission.}\\ \vspace{0.5mm}
\footnotesize{$^c$ SMBH masses marked with $^{\ddagger}$ were calculated from the velocity dispersion $\sigma_v$ using the $M_{\bullet}- \sigma_v$ scaling relation. Masses for all other galaxies were taken from direct measurements of gas or star kinematics from within the sphere of influence of the SMBH.}\\ \vspace{0.5mm}
\footnotesize{$^d$ In the case of NGC\,4552, the H$\alpha$+[N\textsc{ii}] morphology was taken from \cite{Boselli2021}.}\\ \vspace{0.5mm}
\footnotesize{$^e$ The first number represents the total exposure time for 1/8th subarray observations (OBSIDs 18232-21458) and the second number is the total exposure time for full-array observations, which are strongly affected by pile-up effects (OBSIDs 352, 2707).}\\ \vspace{1.5mm}
\footnotesize{References:\, [1] \cite{Saglia2016}\: [2] \cite{Kormendy2013}\: [3] \cite{Thomas2016}\: [4] \cite{Boizelle2021}\: [5] \cite{EHT2019}\:  [6]~\cite{Schellenberger2021}}
\end{flushleft}
\label{tab:basic}
\end{table*}

\section{Data reduction and analysis}

The parameters of Bondi accretion were determined using observations from the \textit{Chandra X-ray Observatory}. The mechanical powers of relativistic jets emanating from AGNs were approximated by the work performed by the jet to inflate the radio lobes, where the volumes were estimated from the extended radio emission observed by the Karl G. Jansky Very Large Array (VLA) telescope and the pressure was inferred from \textit{Chandra} data. The exact positions of SMBHs were either determined as the centre of hard X-ray emission ($3-7\;$keV) observed with the \textit{Chandra X-ray Observatory} or small scale radio emission observed with Very-long-baseline interferometry (VLBI), or they were taken from Atacama Large Millimeter Array (ALMA) measurements (NGC$\,$4636, NGC$\,$5846;  \citealp{Temi2018}) (see Table \ref{tab:basic}).

\subsection{\textit{Chandra} data}

Throughout this analysis, we used archival \textit{Chandra} observations of 20 nearby early-type galaxies. The \textit{Chandra} data were reprocessed using standard \textsc{ciao 4.14} procedures \citep{Fruscione2006} and current calibration files (\textsc{caldb 4.9.6}). For most objects, the observations were obtained by the ACIS-S chip in the VFAINT mode, but for some galaxies (NGC$\,$507, NGC$\,$4636, NGC$\,$5846 and NGC$\,$6166) we also included ACIS-I observations (see Table \hyperref[tab:obsids]{D1}). 

Point sources, as well as regions of strong nonthermal emission emanating from relativistic jets (e.g. NGC4261, NGC4486), were found using the \texttt{wavdetect} procedure both in the hard ($3-7$ keV) and broad ($0.5-7$ keV) band, visually inspected and excluded from further analysis. The light curves, extracted in the $0.5-7.0\:$keV band, were deflared using the \texttt{lc\_clean} algorithm within the \texttt{deflare} routine and good time intervals were obtained.

For objects with multiple observations, the individual OBSIDs were reprojected onto one of the observations via the \texttt{reproject\_obs} script. For background subtraction, we used blanksky background files, which were reprojected onto the observations, filtered for VFAINT events, and scaled to match the particle background of observations in the $9-12\;$keV band.

\subsubsection{Spectral analysis}
\label{section:spectra}

Spectral files were produced for each observation separately using the \texttt{specextract} script in the $0.5 - 7.0\;$keV energy band\footnote{For low-energy limit of 0.6 keV, consistent results were obtained.}. The spectra were extracted from concentric annuli with increasing radii centred at the SMBH positions stated in Table~\ref{tab:basic}.
The radii of individual annuli were chosen to maximize the central spatial resolution while still obtaining a reasonable number of counts required to constrain the electron density with a relative uncertainty smaller than 25 per cent.

In the case of the galaxy NGC$\,$4486 (M87), the bright central AGN and jet emission leads to very strong pile-up effects within the central few arcsec of \textit{Chandra} ACIS-S observations obtained in the classical full-array 3.2$\:$s exptime mode. Following the analysis of \cite{Wilson2002}, \cite{dimatteo2003} and \cite{Russell2015}, we also utilized archival observations in the 1/8th subarray 0.4~s exptime mode. Spectral files for the short frame-time observations were extracted from circular annuli starting at 2 arcsec and up to 30 arcsec from the centre, whereas for observations in the full-frame 3.2 s exptime mode the spectra were extracted in the $15-230$ arcsec range.

For each galaxy, spectra for all annuli and OBSIDs were fitted simultaneously using PyXspec v2.0.5 \citep[\textsc{Xspec} v12.12.0;][]{Arnaud1996} and \textsc{Atomdb} v3.0.9 \citep{Foster2012}. The deprojection was performed within PyXspec using the \textit{projct} model, which was applied only to the thermal component (\textit{apec}) that describes the extended emission of the galactic atmosphere. For the deprojection, spherical symmetry and constant temperatures and electron number densities within individual annuli are assumed.

To account for the single-phase collisionally ionised diffuse gas, we used a single \textit{apec} model component, which describes the state of the gas by its temperature $kT$, metallicity $Z$, redshift $z$ and normalization $Y$, which is directly proportional to the emission measure of the X-ray emitting gas
\begin{equation}
Y = \frac{10^{-14} \int n_{\text{e}} n_{\text{i}} \text{d}V}{4 \pi D^2_{\text{A}} (1 + z)^2},
\end{equation}
where $D_{\text{A}}$ is the angular diameter distance and $n_{\text{e}}$ and $n_{\text{p}}$ are electron and ion concentrations, respectively, where for fully ionised medium with Solar abundances $n_{\text{e}} = 1.18 \, n_{\text{p}}$. The redshift for all objects was fixed to values stated in Table \ref{tab:basic}. Temperatures and abundances were allowed to vary during the fitting, however, for most galaxies, these were tied for two or more neighbouring shells in order to constrain these parameters with a relative uncertainty smaller than 10 and 25 per cent, respectively. The abundances were measured with respect to the proto-solar abundances reported by \cite{Lodders2003}.

For some of the galaxies, we also included an absorbed power-law component \textit{zphabs
(powerlaw)} in the innermost annulus to account for the obscured emission of the active galactic nucleus. The redshifted absorbing hydrogen column density $n_{\mathrm{H},z}$, as well as the photon index $\Gamma$, were freed during fitting. However, for galaxies for which the fitted values were not significant, we eventually fixed the parameters to $n_{\mathrm{H},z} = 0$ and $\Gamma = 1.9$ \citep[following][]{Russell2013}, where $\Gamma = 1.9$ represents a mean value of the photon index for nearby AGNs \citep{Gilli2007}. For galaxies, for which even the normalization of the non-thermal power-law component was insignificant, this component was eventually excluded from the model.

For the remaining outer annuli, we added a bremsstrahlung component (\textit{bremss}) with the temperature fixed to 7.3$\:$keV \citep{Irwin2003}, which should well describe the hard X-ray contribution of unresolved low-mass X-ray binaries, cataclysmic variables, and coronally active binaries.

The Galactic absorption was modelled using the \textit{phabs} model with the \textit{bcmc} cross-section \citep{Bcmc1992}. The hydrogen column densities $n_{\text{H}}$ were fixed to the values obtained from the \textit{HEASARC}\,\footnote{\href{https://heasarc.gsfc.nasa.gov/cgi-bin/Tools/w3nh/w3nh.pl}{https://heasarc.gsfc.nasa.gov/cgi-bin/Tools/w3nh/w3nh.pl}} database which uses the values reported by \cite{HI4PI2016}.

The best-fit parameters were found using the Levenberg-Marquardt minimization method and cash statistics \citep{Cash1979}. The final values of parameters and their uncertainties were obtained using Markov Chain Monte Carlo (MCMC) simulations, where we used Gaussian priors centred at the best-fit values. Unless stated otherwise, all uncertainties are expressed in the $1 \, \sigma$ credible interval (for asymmetric distributions, this corresponds to distances of $15.9\,\%$ and $84.1\,\%$  quantiles from the median value). During the fitting, we assumed the standard $\Lambda$CDM cosmology with $H_0 = 70\;$km$\,$s$^{-1}\,$Mpc$^{-1}$, $\Lambda_0 = 0.73$ and $q_0 = 0$. 

In addition to the spectral analysis used for determining electron number densities, the fitting was also performed using a \textit{cflux} model component and the fluxes of thermal \textit{apec} components for individual annuli were derived. Each \textit{apec} component was multiplied by a convolutional \textit{cflux} component with an energy range of $0.01 - 100\;$keV (bolometric X-ray flux). Total X-ray luminosities were calculated from the derived fluxes and distances stated in Table \ref{tab:basic}.

\subsection{VLA observations}

We used VLA observations in A, B, C, or D configurations centred at around 1.4~GHz, which were calibrated and `imaged' using the {\sc NRAO} Common Astronomy Software Applications pipeline \citep[\textsc{casa},][]{McMullin2007} version 4.7.2 and 5.6.1. Two categories of data were analysed depending on the year of their observation including both `historical' data observed before the major upgrade in 2011 \citep{Perley2009,Perley2011} and `Karl G. Jansky VLA/EVLA' data obtained after this upgrade. Three galaxies, NGC$\,$4552, NGC$\,$4636 and NGC$\,$4649, were observed by the upgraded Karl G. Jansky VLA and calibrated using the automatic \textsc{casa} pipeline version 1.3.11. Reduction and `imaging' follows standard procedures described in \cite{Grossova2019,Grossova2021}. The historical observations were manually calibrated using the NRAO pre-upgrade calibration methods\footnote{\href{https://casaguides.nrao.edu/index.php/Jupiter:\_continuum\_polarization\_calibration}{https://casaguides.nrao.edu/index.php/Jupiter}}. For galaxies NGC$\,$708 and NGC$\,$6166, reduced images were obtained from the NRAO VLA Archive Survey (NVAS)\footnote{\href{https://archive.nrao.edu/nvas/}{https://archive.nrao.edu/nvas/}}.

The mechanical jet powers were estimated from the sizes of radio lobes that were identified using processed VLA images. For observations showing extended radio emission, we produced radio contours and estimated the sizes of radio lobes manually by overlaying the radio contours with ellipse regions using the \textit{SAOImageDS9} software \citep{Joye2003}. The minimum level of radio contours was set to be 5 times the root mean square error (RMSE) of the surrounding background.

In order to probe the most recent radio-mechanical AGN activity, we used L-band ($1-2$ GHz) images with the best available angular resolution, which for the A configuration data is comparable to our \textit{Chandra} images. However, for some objects in our sample, no A configuration data were available or they did not capture the extended emission fully. For these objects, we used the more compact array configurations (B, C, or D), always aiming for the best possible spatial resolution (see Table~\ref{tab:basic}). For NGC\,1399 and NGC\,4472 it is not entirely clear whether the smaller scale radio emission seen in the A-configuration data, which also appears to be associated with surface brightness depressions in the \textit{Chandra} residual images, corresponds to the most recent radio lobes or to a channel feeding the more extended lobes. For these galaxies, we show the jet power estimates and the resulting correlations based on both the more compact and the larger scale structures.

\subsection{VLBI observations}

Observations from the VLBI were retrieved from the \textit{Astrogeo VLBI FITS image database}\footnote{\href{http://astrogeo.org/vlbi\_images/}{http://astrogeo.org/vlbi\_images/}}. All utilized observations were observed in the X-band ($8-8.8$ GHz) under the \textit{VLBI 2MASS Survey} (V2M) or \textit{Wide-Field VLBA Calibration Survey} (WFCS) and were all analysed by Leonid Petrov (\citealp{Condon2013,Petrov2021}; Petrov \& Kovalev in prep.). The exact positions of SMBHs were determined from the peak of the small scale VLBI radio emission.

\section{Results}
\label{section:results}

The deprojected radial profiles of temperature, $kT$, metallicity, $Z$, and electron number density, $n_{\mathrm{e}}$, were determined from spatially resolved spectral analysis of galaxies. During the spectral fitting, the metallicities were derived using a single-temperature \textit{apec} component, but due to the fact that the central parts of early-type galaxies are often multiphase, we checked the metallicity estimates also using a multi-temperature model. Nevertheless, fixing the central abundances to those obtained from the multi-temperature model introduced only minor changes to the final values of thermodynamic properties and the derived Bondi powers.

For some of the galaxies, especially those for which the temperatures and abundances of neighbouring annuli were tied, we observed strong degeneracy (mostly two-peak) in the posterior distributions of some of these parameters.  However, for all galaxies with strong temperature and abundance degeneracy, the posterior peaks were relatively nearby and the differences were maximally of the order of $0.05\;\text{keV}$ and $0.1\;\textit{Z}_{\sun}$, respectively.

For most of the galaxies, the spatial resolution of \textit{Chandra} was not sufficient to constrain the thermodynamic properties at the Bondi radius and, therefore, extrapolations were necessary. The central temperature and metallicity were assumed to be constant within the innermost radial bin, but the electron number density had to be extrapolated (see Figs. \ref{fig:density}, \ref{fig:temperature} and \ref{fig:abund}). For the extrapolation, we used 3 different profiles (similarly as \citealp{Russell2013}): a~power-law model, a $\beta$-model \citep{Cavaliere1976} and a Sersic profile \citep{Sersic1963}. The mean of these three profiles was then taken as the final value of the electron number density and the scatter was accounted in the uncertainty\footnote{We also tried extrapolating the electron number densities only by the power-law model, which resulted in systematically higher densities. The final conclusions were, however, consistent with the results obtained by averaging.}. In the case of the power-law model, we only used the inner 3 to 4 radial bins in order to properly fit the innermost substructure. When fitting the $\beta$-model and the Sersic profile, we included more radial bins up to the point where the profile slope changes significantly or starts flattening again. To account for uncertainties in both axes, the profile fitting was carried out using the Orthogonal Distance Regression \citep[Scipy v1.4.1;][]{ODR}.

The inferred temperatures and electron number densities were used to calculate the values of central specific entropy $K = kT n_{\text{e}}^{-2/3}$ and also profiles of other thermodynamic quantities such as thermal pressure $p = n kT$, free fall time $t_{\text{ff}}$, and cooling time $t_{\text{cool}}$ of the X-ray emitting gas (see Table \ref{tab:derived}). The cooling time is defined as the timescale needed for a gas of certain density, temperature and metallicity to thermally emit all of its energy via bremsstrahlung, recombination and line emission
\begin{equation}\label{eq:tcool}
t_{\text{cool}} = \frac{3}{2} \frac{n kT}{n_{\text{e}} n_{\text{i}} \Lambda (T, Z)},
\end{equation}
where $n$ is the total particle density $n = n_{\text{e}} + n_{\text{i}}$ and $\Lambda (T, Z)$ is the temperature and abundance-dependent cooling function, values of which were taken from \cite{Schure2009}. The free-fall time resembles a dynamical timescale required for a condensed clump with zero momentum to fall into the centre of the galaxy and is given by $t_{\mathrm{ff}} = \sqrt{2r/g}$, where the local gravitational acceleration $g$ was calculated from the velocity dispersion $\sigma$ and the galactocentric distance $r$ using the assumption of an isothermal sphere $g = 2 \sigma^2 /r$ \citep{Binney1987}.

To probe how susceptible the galactic atmospheres are to thermal instabilities, we calculated the profiles of cooling time to free-fall time ratios. Formerly, the atmospheres of galaxies were expected both from analytical \citep{Nulsen1986} and numerical computations \citep{McCourt2012} to become thermally unstable when the cooling time falls below the free-fall time. However, the most up-to-date observations \citep[e.g.][]{Voit2015,Hogan2017} and simulations \citep{Sharma2012,Gaspari2012} have shown that atmospheres of realistic galaxies may become thermally unstable even when the cooling time to free-fall time ratio falls below $t_{\mathrm{cool}} / t_{\mathrm{ff}} \lessapprox 10$, which is often referred to as the \textit{precipitation limit} (we note that it resembles an approximate division line rather than a strict limit). The susceptibility of galactic atmospheres to thermal instabilities was, therefore, assessed from the minimal values in the $t_{\text{cool}} / t_{\text{ff}}$ profiles (Fig. \ref{fig:tcooltff}).

Total power outputs of hot galactic atmospheres were approximated by the total X-ray luminosities within a defined radius such as the cooling radius (\textit{cooling luminosity}). The cooling radius is defined as the radius at which the cooling time profile (Equation \ref{eq:tcool}) reaches a certain value. Commonly used are the values of $3$ Gyr \citep{Panagoulia2014} and $7.7$ Gyr \citep[lookback time of $z = 1$;][]{Rafferty2006,Nulsen2009}. However, since most galaxies in our sample are relatively nearby, such radii would often be outside of ACIS chip's edges. Instead, we used the radius where the cooling time profile reaches the value of 1 Gyr (see Fig. \ref{fig:tcool}). The energetic balance of galactic atmospheres was then probed by comparing the cooling luminosities to the mechanical powers of the jets (Table \ref{tab:cooling}).

\setlength{\tabcolsep}{3.35pt}
\renewcommand{\arraystretch}{1.4}
\begin{table*}
\centering
\caption{Summary of parameters derived from \textit{Chandra} observations. Listed quantities: temperature $kT$, electron number density $n_{\text{e}}$, metallicity $Z$, and specific entropy $K$ at the Bondi radius, minimum of the cooling time over free fall time ratio $t_{\text{cool}}\,/\,t_{\text{ff}}$, sound speed $c_{\text{sound}}$, Bondi radius $r_{\text{Bondi}}$, Bondi accretion rate $\dot{m}_{\text{Bondi}}$, Bondi accretion power $P_{\text{Bondi}}$, and mechanical jet power $P_{\text{jet}}$. 
}
\begin{tabular}{lcccccccccc}
\toprule \vspace{-0.5mm}
\multirow{2}{3.2em}{Galaxy$\,^a$} & $kT$ & $n_{\text{e}}$ & $Z$ & $K$ & min & $c_{\mathrm{sound}}$ & $r_{\text{Bondi}}$ & $\dot{m}_{\text{Bondi}}$ & $P_{\text{Bondi}}$ & $P_{\text{jet}}$\\
 & (keV) & (cm$^{-3}$) & ($Z_{\sun}$) & (keV cm$^{-2/3}$) & $t_{\text{cool}}\,/\,t_{\text{ff}}$ & (km\:s$^{-1}$) & (pc) & ($ 10^{-3} M_{\sun} \, \text{yr}^{-1}$) & ($10^{43}$ erg$\,$s$^{-1}$) & ($10^{43}$ erg$\,$s$^{-1}$)\\
\bottomrule
IC\,4296&$0.69_{-0.06}^{+0.05}$&$1.56_{-0.24}^{+0.28}$&$0.87_{-0.11}^{+0.15}$&$0.47_{-0.14}^{+0.39}$&$5.4\pm0.5$&$421_{-17}^{+15}$&$63_{-11}^{+13}$&$48_{-30}^{+44}$&$27_{-17}^{+25}$&$4.2_{-1.5}^{+3.0}$\\ 
NGC\,507$\,^*$&$0.82_{-0.05}^{+0.04}$&$0.72_{-0.13}^{+0.17}$&$0.65_{-0.15}^{+0.16}$&$1.0_{-0.3}^{+0.8}$&$8.7\pm0.8$&$460_{-15}^{+12}$&$87_{-20}^{+24}$&$44_{-28}^{+48}$&$25_{-16}^{+27}$&$6.8_{-2.3}^{+4.6}$\\ 
NGC\,708&$1.03\pm0.04$&$0.081 \pm 0.006$&$0.88_{-0.05}^{+0.06}$&$5.0_{-1.5}^{+4.4}$&$6.1_{-0.2}^{+0.3}$&$514\pm11$&$16_{-2}^{+3}$&$0.4_{-0.25}^{+0.36}$&$0.23_{-0.14}^{+0.20}$&$0.06_{-0.021}^{+0.041}$\\ 
NGC\,1316&$0.77\pm0.02$&$0.74_{-0.07}^{+0.09}$&$0.42\pm0.02$&$0.85_{-0.25}^{+0.73}$&$6.9\pm0.3$&$444_{-7}^{+6}$&$7.4_{-1.3}^{+1.2}$&$0.32_{-0.20}^{+0.31}$&$0.18_{-0.11}^{+0.17}$&$0.012_{-0.004}^{+0.008}$\\ 
NGC\,1399$\,^*$&$1.16\pm0.05$&$0.84_{-0.13}^{+0.32}$&$1.07\pm0.06$&$1.2_{-0.3}^{+1.0}$&$10.8\pm0.3$&$545_{-11}^{+12}$&$37_{-17}^{+15}$&$9.1_{-7.2}^{+15.2}$&$5.1_{-4.1}^{+8.6}$&$0.037_{-0.013}^{+0.026}$ $^b$\\ 
NGC\,1407&$0.94\pm0.04$&$0.13 \pm 0.01$&$1.36_{-0.16}^{+0.20}$&$3.3_{-1.0}^{+2.8}$&$19.9_{-0.9}^{+1.2}$&$493_{-10}^{+9}$&$170_{-20}^{+30}$&$35_{-21}^{+29}$&$20_{-12}^{+17}$&$0.12_{-0.04}^{+0.08}$\\ 
NGC\,1600&$0.96\pm0.02$&$0.059_{-0.006}^{+0.007}$&$1.25\pm0.10$&$5.8_{-1.8}^{+5.0}$&$18.8\pm0.7$&$498\pm6$&$590 \pm 50$&$190_{-120}^{+150}$&$110_{-70}^{+90}$&$0.13_{-0.05}^{+0.10}$\\ 
NGC\,4261&$0.7_{-0.03}^{+0.02}$&$1.17_{-0.19}^{+0.18}$&$0.57_{-0.04}^{+0.06}$&$0.58_{-0.18}^{+0.48}$&$5.9\pm0.3$&$424_{-8}^{+7}$&$80_{-12}^{+19}$&$60_{-37}^{+58}$&$34_{-21}^{+33}$&$1.2_{-0.41}^{+0.81}$\\ 
NGC\,4374&$0.73\pm0.02$&$0.16 \pm 0.02$&$0.86_{-0.09}^{+0.11}$&$2.2_{-0.7}^{+1.9}$&$6.0\pm0.3$&$434\pm6$&$42 \pm 4$&$2.4_{-1.5}^{+1.9}$&$1.35_{-0.82}^{+1.06}$&$0.2_{-0.07}^{+0.15}$\\ 
NGC\,4472&$0.85_{-0.03}^{+0.04}$&$0.37_{-0.03}^{+0.02}$&$0.81\pm0.02$&$1.5_{-0.5}^{+1.3}$&$10.4\pm0.2$&$469_{-9}^{+10}$&$100_{-10}^{+18}$&$36_{-22}^{+32}$&$20_{-13}^{+18}$&$0.05_{-0.017}^{+0.034}$ $^b$\\ 
NGC\,4486&$0.79\pm0.05$&$0.21_{-0.02}^{+0.03}$&$0.64\pm0.05$&$2.1_{-0.6}^{+1.8}$&$8.4\pm0.3$&$451\pm13$&$270 \pm 30$&$130_{-80}^{+110}$&$75_{-46}^{+60}$&$1.31_{-0.49}^{+0.97}$\\  NGC\,4552&$1.12_{-0.12}^{+0.14}$&$2.01_{-0.31}^{+0.34}$&$0.47_{-0.03}^{+0.04}$&$0.65_{-0.20}^{+0.56}$&$7.3_{-0.7}^{+0.8}$&$537_{-29}^{+34}$&$15_{-2}^{+3}$&$4.4_{-2.7}^{+3.7}$&$2.5_{-1.5}^{+2.1}$&$0.11_{-0.04}^{+0.07}$\\ 
NGC\,4636&$0.3\pm0.02$&$0.14 \pm 0.04$&$0.91_{-0.03}^{+0.04}$&$1.1_{-0.3}^{+0.8}$&$4.9_{-0.4}^{+0.5}$&$278\pm10$&$37 \pm 5$&$0.93_{-0.55}^{+0.73}$&$0.53_{-0.31}^{+0.41}$&$0.016_{-0.006}^{+0.011}$\\ 
NGC\,4649$\,^*$&$1.46\pm0.07$&$0.43_{-0.05}^{+0.06}$&$1.21_{-0.07}^{+0.05}$&$2.3_{-0.7}^{+2.0}$&$13.9\pm0.3$&$613_{-16}^{+14}$&$111 \pm 19$&$54_{-35}^{+58}$&$31_{-20}^{+33}$&$0.024_{-0.009}^{+0.018}$\\ 
NGC\,4696&$0.88\pm0.02$&$0.54 \pm 0.06$&$0.51\pm0.01$&$1.2_{-0.4}^{+1.0}$&$3.5\pm0.1$&$475\pm5$&$34_{-6}^{+7}$&$5.5_{-3.5}^{+5.3}$&$3.1_{-2.0}^{+3.0}$&$0.42_{-0.14}^{+0.29}$\\ 
NGC\,4778$\,^*$&$0.76\pm0.03$&$0.26 \pm 0.02$&$0.99_{-0.06}^{+0.08}$&$1.7_{-0.5}^{+1.5}$&$7.3\pm0.6$&$443\pm10$&$37_{-6}^{+8}$&$2.9_{-1.8}^{+2.8}$&$1.6_{-1.0}^{+1.6}$&$0.071_{-0.024}^{+0.049}$\\ 
NGC\,5044&$0.75_{-0.03}^{+0.02}$&$0.10 \pm 0.05$&$0.43_{-0.02}^{+0.03}$&$3.3_{-0.8}^{+2.0}$&$4.7\pm0.2$&$440_{-8}^{+6}$&$9.9_{-3.2}^{+5.3}$&$0.073_{-0.049}^{+0.117}$&$0.041_{-0.028}^{+0.066}$&$0.016_{-0.006}^{+0.011}$\\ 
NGC\,5813&$0.8\pm0.01$&$0.12 \pm 0.01$&$0.82\pm0.02$&$3.0_{-0.9}^{+2.5}$&$5.5\pm0.2$&$454\pm2$&$30 \pm 4$&$0.92_{-0.56}^{+0.78}$&$0.52_{-0.32}^{+0.44}$&$0.065_{-0.023}^{+0.046}$\\ 
NGC\,5846&$0.77\pm0.02$&$0.20 \pm 0.01$&$0.58_{-0.03}^{+0.04}$&$2.0_{-0.6}^{+1.8}$&$7.0\pm0.3$&$445\pm6$&$48_{-6}^{+7}$&$3.9_{-2.4}^{+3.4}$&$2.2_{-1.4}^{+1.9}$&$0.14_{-0.05}^{+0.10}$\\ 
NGC\,6166&$1.6_{-0.17}^{+0.20}$&$0.37_{-0.05}^{+0.06}$&$1.53_{-0.09}^{+0.12}$&$2.9_{-0.9}^{+2.5}$&$11.5\pm0.4$&$641_{-34}^{+39}$&$44_{-10}^{+13}$&$8.3_{-5.4}^{+9.5}$&$4.7_{-3.1}^{+5.4}$&$0.68_{-0.23}^{+0.47}$\\ 
	\bottomrule
\end{tabular}\\
\begin{flushleft}
\footnotesize{$^a$ For galaxies marked with $^*$, the absorbed power-law component describing the non-thermal central AGN emission was not significant and therefore it was not included in the final fit.}\\ \vspace{0.5mm}
\footnotesize{$^b$ For NGC\,1399 and NGC\,4472, the jet power estimates based on the potentially older, larger scale lobes are $0.29_{-0.09}^{+0.18} \times 10^{43}$ erg$\,$s$^{-1}$ and $0.27_{-0.10}^{+0.21} \times 10^{43}$ erg$\,$s$^{-1}$, respectively.}
\end{flushleft}
\label{tab:derived}
\end{table*}

\subsection{Bondi power}
\label{bondipowers}

The total power input that the supermassive black hole acquires by accreting the surrounding material was estimated under the assumption of the spherical Bondi accretion model \citep{Bondi1952}. Bondi accretion assumes steady spherical accretion from the circumnuclear medium, which most likely rarely happens in realistic galactic nuclei due to the presence of angular momentum, outflows, and magnetic fields. However, since the precise geometry of the accretion flow is typically not known, it still provides an order-of-magnitude estimate of the total accretion power \citep{churazov2002,Allen2006} and is parameterizable only by the temperature and density of the ambient gas and the mass of the SMBH.

The radius for which the gravitational influence of the SMBH is dominant over the thermal energy of the gas is the Bondi radius $r_{\mathrm{Bondi}} = G M_{\bullet} c^{-2}_{\mathrm{s}}$, where $M_{\bullet}$ is the SMBH mass and $c_{\mathrm{s}} = \sqrt{\gamma kT / \mu m_{\mathrm{p}}}$ is the speed of sound in the surrounding medium, where $\mu \approx 0.62$ is the mean atomic weight for fully ionised gas and $\gamma = 5/3$ is the adiabatic index of the X-ray emitting plasma. The accretion rate can be expressed as a flux of matter through a spherical shell of Bondi radius 
\begin{equation} \label{eq:mBondi}
\dot{m}_{\mathrm{Bondi}} = 4 \pi \lambda \rho (G M_{\bullet})^2 c_{\mathrm{s}}^{-3} = \pi \lambda \rho c_{\mathrm{s}} r_{\mathrm{Bondi}}^2,
\end{equation}
with a numerical coefficient for adiabatic gas $\lambda = 0.25$. The accreted matter is then with an efficiency $\eta$ turned into energy $P_{\mathrm{Bondi}} = \eta \dot{m}_{\mathrm{Bondi}} c^2$, which can be expelled in the form of relativistic jets. Throughout this analysis, we assume an efficiency of 10 per cent.

The parameters of the Bondi accretion were derived from the spectral properties of the X-ray emitting gas surrounding the supermassive black hole (see Table \ref{tab:basic} for the black hole masses). The results are listed in Table \ref{tab:derived}.

\begin{figure}
\centering
\includegraphics[width=\linewidth]{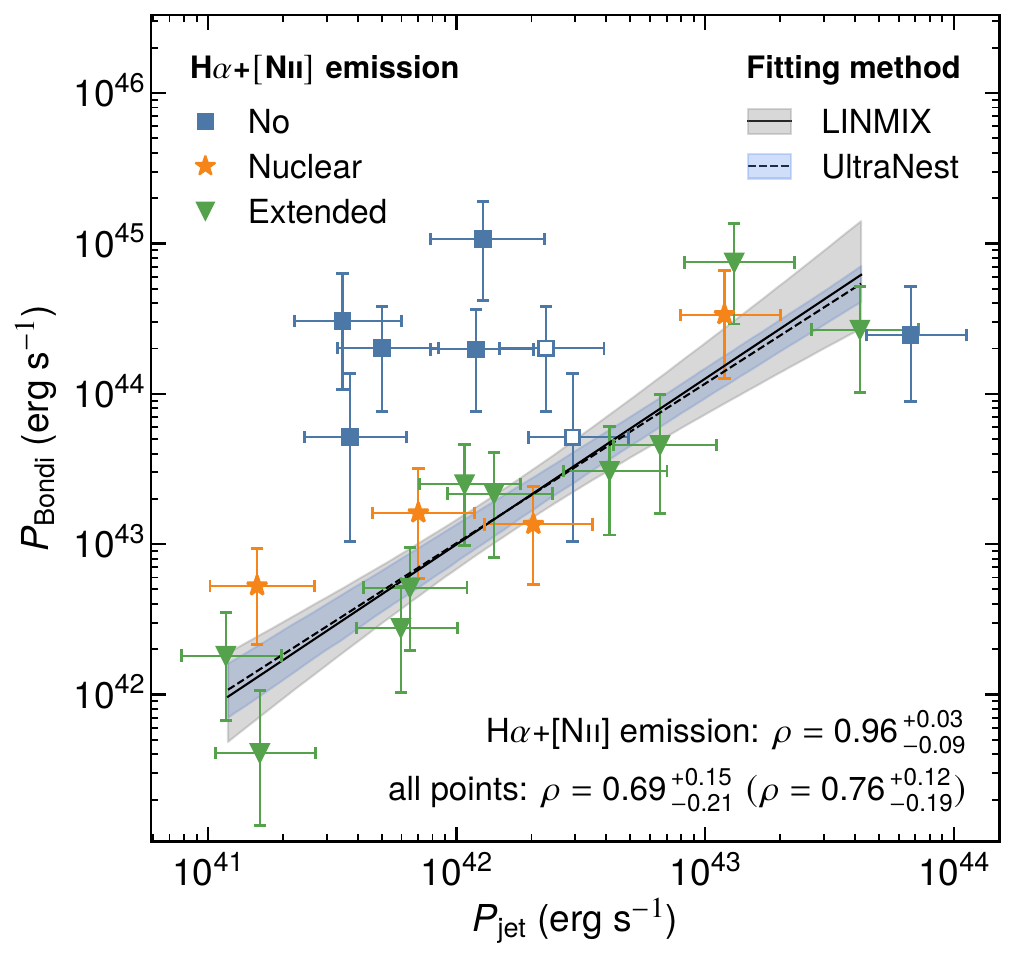}
\caption{Correlation between Bondi accretion power and mechanical jet power. The solid line is the LINMIX power-law fit for the subsample of galaxies containing cold gas (either extended or nuclear H$\alpha$+[N\textsc{ii}] emission), while the grey area represents its 1$\: \sigma$ confidence band. For comparison, we show also UltraNest fit (dashed line) with its confidence band (blue area). Correlation coefficients for the H$\alpha$+[N\textsc{ii}] emitting sub-sample and the full sample are shown in the lower right corner. The larger scale radio lobes for NGC\,1399 and NGC\,4472 (blue empty squares) and the corresponding correlation coefficient (stated in parentheses) are also shown.}
\label{fig:Pjet_Pbondi}
\end{figure}

\subsection{Mechanical jet power}

The mechanical power of a relativistic jet emanating from the vicinity of a supermassive black hole can be estimated either from the size of the extended radio emission (radio lobes) or from the corresponding X-ray cavities inflated in the ambient medium. Throughout this analysis, we estimated the jet powers from the approximate volumes of radio lobes assuming that the same volume of thermal plasma was displaced by the jet producing X-ray cavities.

The total kinetic energy needed to inflate a cavity of volume $V$ into the circumgalactic medium with pressure $p$ is equal to its total enthalpy $H$, which is the sum of the internal energy of the relativistic plasma and the work done on the ambient gas
\begin{equation}
H = \frac{1}{\gamma - 1} p V + pV,
\end{equation}
where $\gamma$ is the adiabatic index (ratio of specific heats) of plasma filling the cavity. The value of the adiabatic index depends on whether the gas pressure support is caused by nonrelativistic ($\gamma=5/3$) or relativistic particles ($\gamma = 4/3$). Modern observations of massive clusters of galaxies via the Sunyaev-Zeldovich effect \citep[][]{Abdulla2019} have indicated that the X-ray cavities are filled mainly with relativistic gas ($\gamma = 4/3$) and the total enthalpy is therefore given by $H = 4 pV$, which is also in a good agreement with magnetohydrodynamical simulations \citep{Mendygral2011,McNamara2012}.

The total enthalpy was numerically integrated on a linearly scaled Cartesian 3D grid of pressure, which was interpolated from the pressure profile using linear interpolation in log-log space (see Fig.~\ref{fig:pressure}) and under the assumption of spherical symmetry. In the case of IC$\,$4296, the thermal pressure of the gas at the position of the cavity was taken from \cite{Grossova2019}, which is based on \textit{XMM-Newton} measurements. Compared to the commonly used method based on estimating the $4pV$ work by using only the exact value of the pressure in the centre of the cavity and its total volume, this approach introduces for most prolate cavities and galaxies with a non-smooth pressure profile relative changes of up to 25 per cent.

The geometry of cavities was assumed to be prolate or oblate ellipsoids with rotational symmetry along the semi-axis closer to the direction towards the centre of the galaxy. The volume of the ellipsoid is in general given by $V = 4 \pi / 3 \, r_{\text{l}} \, r_{\text{w}} \, r_{\text{d}}$, where $r_{\text{w}}$ (width) is the semi-axis of the ellipsoid which is perpendicular to the direction of the jet, $r_{\text{l}}$ (length) is the semi-axis along that direction and $r_{\text{d}}$ is the unknown depth of the cavity. For the calculations, we assumed the depth of the cavity to be equal to its width. To account for uncertainties caused by the unknown depths of cavities, by the potential irregularities of their shapes and also by projection effects, we introduced an additional factor of 2 uncertainty for their volumes.

The total mechanical power required for inflating a cavity with internal energy $E$ is given by $P_{\mathrm{jet}} = \frac{E}{t_{\mathrm{age}}},$ where $t_{\mathrm{age}}$ is the age of the cavity. The ages of individual cavities were calculated from their galactocentric distances, $R$, and their inflation velocities, $v$, as $t_{\mathrm{age}} = R / v$. The velocity of the inflation was assumed to be equal to the average speed of sound in the hot medium, $c_{\mathrm{s}}$, along the path of the cavity. When integrating the surrounding pressure and estimating the age of the cavities, the galactocentric distances of cavities were calculated based on their angular distances.

The derived properties of each radio lobe, such as its volume, galactocentric distance, approximate age, and derived mechanical jet power can be found in the Appendix (Table \ref{tab:cavities}), while sums of the mechanical powers for both radio lobes are stated in Table~\ref{tab:derived}.

\setlength{\tabcolsep}{4.5pt}
\renewcommand{\arraystretch}{1.4}
\begin{table}
    \caption{Derived quantities describing the power output of galactic atmospheres: cooling radius $r_{\text{cool}}$ in units of kpc and arcsec (radius at which the cooling time profile reaches the value of 1 Gyr), cooling luminosity $L_{\text{cool}}$ in the $0.01-100\;$keV range (luminosity of thermal emission within the cooling radius), and the ratio of cooling luminosity to mechanical jet power $L_{\text{cool}} / P_{\text{jet}}$.}
    \centering
    \begin{tabular}{l c c c c}
    \toprule \vspace{-0.7mm}
    \multirow{2}{4.0em}{Galaxy} & $r_{\text{cool}}$ & $r_{\text{cool}}$ & $L_{\text{cool}}$ & \multirow{2}{4.0em}{$L_{\text{cool}} / P_{\text{jet}}$}\\
    & $(\text{kpc})$ & $(\text{arcsec})$ & $(\text{erg s}^{-1})$ & \\
    \bottomrule
IC\,4296&$5.4$&$22.1$&$1.7_{-0.2}^{+0.1} \times 10^{41}$&$0.0041_{-0.0017}^{+0.0022}$\\ 
NGC\,507&$8.4$&$26.8$&$2.02_{-0.07}^{+0.08} \times 10^{41}$&$0.003_{-0.0012}^{+0.0014}$\\ 
NGC\,708&$13.0$&$45.9$&$1.427_{-0.013}^{+0.014} \times 10^{42}$&$2.38_{-0.95}^{+1.22}$\\ 
NGC\,1316&$8.7$&$85.4$&$5.3_{-0.3}^{+0.1} \times 10^{40}$&$0.43_{-0.16}^{+0.21}$\\ 
NGC\,1399&$5.2$&$51.2$&$1.24_{-0.03}^{+0.02} \times 10^{41}$&$0.33_{-0.13}^{+0.16}$\\ 
NGC\,1407&$5.8$&$47.5$&$5.5_{-0.5}^{+0.2} \times 10^{40}$&$0.044_{-0.017}^{+0.024}$\\ 
NGC\,1600&$5.5$&$18.0$&$1.216_{-0.012}^{+0.016} \times 10^{41}$&$0.094_{-0.04}^{+0.06}$\\ 
NGC\,4261&$5.5$&$35.3$&$7.4_{-0.2}^{+0.3} \times 10^{40}$&$0.0061_{-0.0025}^{+0.0034}$\\ 
NGC\,4374&$5.4$&$60.4$&$4.035_{-0.002}^{+0.003} \times 10^{40}$&$0.0199_{-0.0089}^{+0.0095}$\\ 
NGC\,4472&$6.1$&$75.5$&$9.72_{-0.11}^{+0.15} \times 10^{40}$&$0.189_{-0.074}^{+0.101}$\\ 
NGC\,4486&$12.0$&$148.7$&$2.35_{-0.02}^{+0.01} \times 10^{41}$&$0.0178_{-0.0075}^{+0.0103}$\\ 
NGC\,4552&$6.3$&$81.7$&$2.47_{-0.02}^{+0.01} \times 10^{40}$&$0.0213_{-0.0083}^{+0.0125}$\\ 
NGC\,4636&$9.3$&$131.0$&$1.539_{-0.016}^{+0.021} \times 10^{41}$&$0.98_{-0.4}^{+0.46}$\\ 
NGC\,4649&$5.9$&$73.5$&$8.5_{-0.2}^{+0.3} \times 10^{40}$&$0.35_{-0.16}^{+0.21}$\\ 
NGC\,4696&$11.7$&$60.4$&$3.13_{-0.02}^{+0.03} \times 10^{42}$&$0.76_{-0.31}^{+0.42}$\\ 
NGC\,4778&$15.8$&$52.2$&$9.4_{-0.3}^{+0.2} \times 10^{41}$&$1.32_{-0.52}^{+0.72}$\\ 
NGC\,5044&$23.0$&$147.6$&$2.40_{-0.10}^{+0.09} \times 10^{42}$&$14.43_{-5.61}^{+8.05}$\\ 
NGC\,5813&$17.6$&$112.9$&$6.675_{-0.017}^{+0.015} \times 10^{41}$&$1.04_{-0.43}^{+0.58}$\\ 
NGC\,5846&$14.2$&$117.6$&$2.76_{-0.05}^{+0.04} \times 10^{41}$&$0.193_{-0.077}^{+0.107}$\\ 
NGC\,6166&$17.9$&$29.5$&$8.72_{-0.09}^{+0.12} \times 10^{42}$&$1.29_{-0.51}^{+0.68}$\\ 
    \bottomrule
    \end{tabular}
    \label{tab:cooling}
\end{table}

\subsection{Correlations}

The relationships between mechanical jet power, Bondi accretion power, SMBH mass, and specific entropy were fitted using a power-law model and the corresponding correlations between the quantity pairs were probed. Since the fit was performed using a linearized power-law model, the uncertainties of all the fitted quantities were recomputed either based on laws of error propagation or using random sampling. For fitting purposes, we approximated asymmetric distributions of all quantities by a normal distribution.

The fitting was performed using a hierarchical Bayesian approach to linear regression included in the LINMIX package\footnote{\href{https://github.com/jmeyers314/linmix}{https://github.com/jmeyers314/linmix}} \citep{Kelly2007}. For verification purposes, we performed the linear fit also using MLFriends algorithm \citep{Buchner2016,Buchner2019} within the UltraNest 3.3.2 package\footnote{\href{https://github.com/JohannesBuchner/UltraNest}{https://github.com/JohannesBuchner/UltraNest}} \citep{Buchner2021}, which enables incorporating asymmetric and non-gaussian uncertainties of individual data points by randomly sampling from a given distribution.

The correlations were probed via the LINMIX linear correlation coefficient between the latent variables ($\rho$) and the significance of the correlation was verified by the Bayes factor for the two-sided correlation test $\text{BF}_{10}(\rho \neq 0)$ \citep[Savage-Dickey density ratio\footnote{The Bayes factor ($\text{BF}_{10}$), as a Bayesian alternative to classical hypothesis test, expresses the ratio of marginal likelihoods of the alternative and null hypothesis. For the alternative hypothesis, we assume that the data are correlated ($\rho \neq 0$), while the null hypothesis assumes no correlation ($\rho = 0$). For equality constrained models, one can obtain the Bayes factor by directly comparing posterior and prior probabilities evaluated at a given point ($\rho = 0$) (Savage-Dickey density ratio).};][]{Dickey1970}, where for $\rho$ we assumed a flat prior. For comparison, we also state the classical Pearson product-moment correlation coefficient \citep{fisher1944} between the observed variables calculated without accounting for measurement errors, and the corresponding significance test (two-tailed $p$-value).

\begin{figure}
\centering
\includegraphics[width=\linewidth]{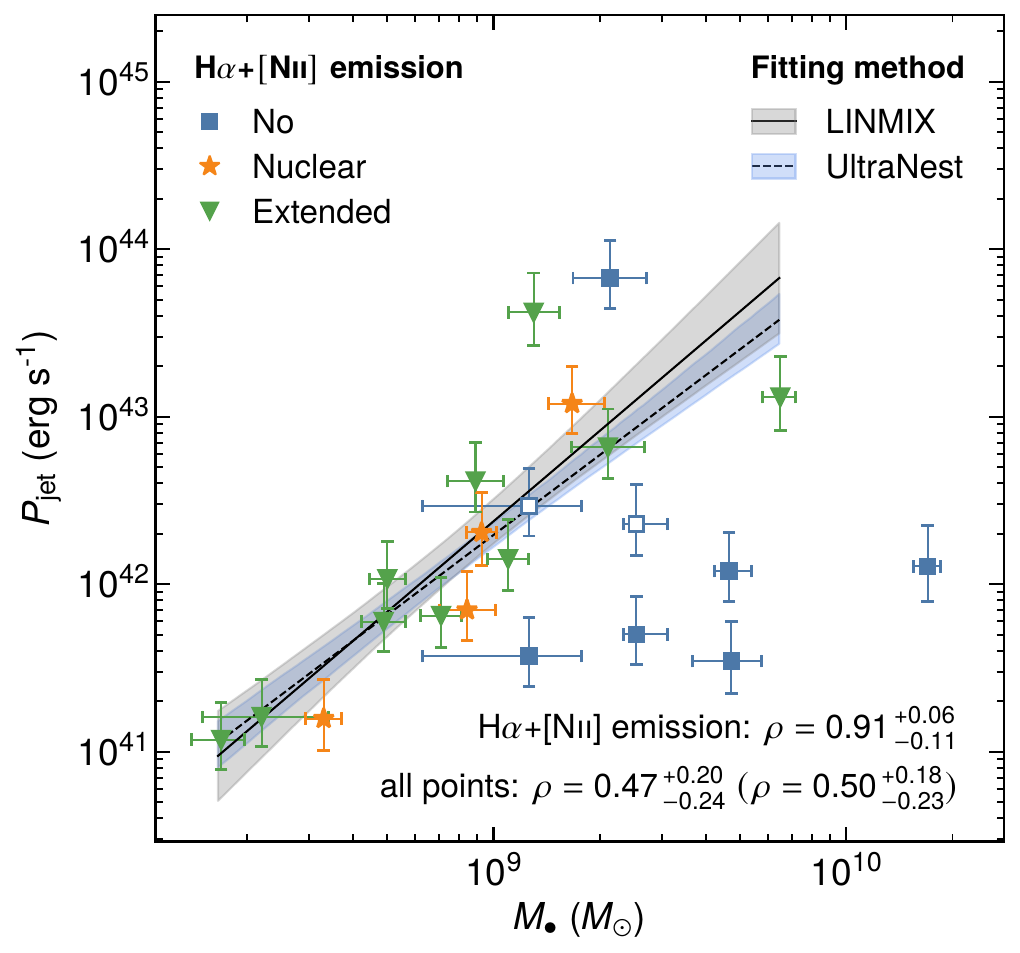}
\caption{Mechanical jet power versus the mass of SMBH.  The solid line is the LINMIX power-law fit for the subsample of galaxies containing cold gas (either extended or nuclear H$\alpha$+[N\textsc{ii}] emission), while the grey area represents its 1$\: \sigma$ confidence band. For comparison, we also show the UltraNest fit (dashed line) with its confidence band (blue area). Correlation coefficients for the H$\alpha$+[N\textsc{ii}] emitting sub-sample and for the full sample are shown in the lower right corner. The larger scale radio lobes for NGC\,1399 and NGC\,4472 (blue empty squares) and the corresponding correlation coefficient (stated in parentheses) are also shown.}
\label{fig:Pjet_MBH}
\end{figure}

During the fitting, we distinguished between galaxies that do contain signs of cool gas tracers (extended or nuclear H$\alpha$+[N\textsc{ii}] emission) and galaxies that do not. We, therefore, fitted the whole sample of galaxies and the H$\alpha$+[N\textsc{ii}] subsample separately and compared the result. Obtained linear correlation coefficients for individual pairs of quantities and subsamples can be found in Table~\ref{tab:correl}. For significantly correlated quantity pairs ($P_{\text{Bondi}} - P_{\text{jet}}$, and $P_{\text{jet}} - M_{\bullet}$), we also state the fitted intercept and slope parameters and we show their best-fit power-law models together with 1$\,\sigma$ confidence bands (see Fig.~\ref{fig:Pjet_Pbondi} and Fig. \ref{fig:Pjet_MBH}). For the full sample, we also provide the correlation coefficients when for NGC\,1399 and NGC\,4472 the jet powers were determined from larger scale radio lobes.

\setlength{\tabcolsep}{4.2pt}
\renewcommand{\arraystretch}{1.48}
\begin{table*}
\centering
\caption{Parameters obtained by fitting the linearized power-law model $(\log y = \alpha + \beta \log x)$: fitted quantities, fitted subsample, Pearson correlation coefficient $r$ and the corresponding $p-$value; LINMIX parameters: intercept $\alpha$, slope $\beta$, intrinsic scatter $\sigma$, correlation coefficient $\rho$, Bayes factor BF$_{10}(\rho \neq 0)$; and the UltraNest parameters: intercept $\alpha$ and slope $\beta$. During the fitting, the Bondi and jet powers were expressed in units of $10^{43} \, \text{erg s}^{-1}$, black hole masses in units of $10^9 \, M_{\sun}$, specific entropy in units of $\text{keV} \: \text{cm}^{2}$ and cooling luminosities in units of $10^{43} \, \text{erg s}^{-1}$. In the case of uncorrelated or non-significantly correlated pairs of quantities the fitted slope and intercept parameters were not stated. All stated uncertainties correspond to 68.3 per cent credible intervals and were either symmetrized or are expressed asymmetrically.}
\begin{tabular}{c c c c c c c c c c c c c c}
  & & & \multicolumn{2}{c}{Pearson} & & \multicolumn{5}{c}{LINMIX} & & \multicolumn{2}{c}{UltraNest} \\
\toprule
\phantom{aa} Quantities$^a$ \phantom{aaa} & Subsample & & $r$ & $p-$value & & $\alpha$ & $\beta$ & $\sigma$ & $\rho$ & BF$_{10}$ & & $\alpha$ & $\beta$\\
\bottomrule \vspace{-4.5mm}
\\
\multirow{3}{5.5em}{\hbox{$P_{\mathrm{Bondi}} - P_{\mathrm{jet}}$}} & all & & 0.61 & 0.004 & & $1.17 \pm 0.29$ & $0.80 \pm 0.28$ & $0.46^{+0.32}_{-0.19}$ & $0.69^{+0.15}_{-0.21}$ & 13 & & $1.18 \pm 0.28$ & $0.65 \pm 0.22$\\
 & all (larger) & & 0.67 & 0.0013 & & $1.19 \pm 0.27$ & $0.89 \pm 0.27$ & $0.38^{+0.28}_{-0.18}$ & $0.76^{+0.12}_{-0.19}$ & 25 & & $1.19 \pm 0.27$ & $0.64 \pm 0.22$\\
 & H$\alpha$+[N\textsc{ii}] & & 0.91 & $< 0.001$ & & $1.10 \pm 0.25$ & $1.10 \pm 0.24$ & $0.08^{+0.14}_{-0.06}$ & $0.96^{+0.03}_{-0.09}$ & $> 100$ & & $1.08 \pm 0.23$ & $1.05 \pm 0.20$\\
 \hline
 \multirow{3}{4.3em}{\hbox{$P_{\mathrm{jet}} - M_{\bullet}$}} & all & & 0.45 & 0.046 & & - & - & - & $0.47^{+0.20}_{-0.24}$ & 1.8 & - & - & - \\
 & all (larger) & & 0.48 & 0.030 & & - & - & - & $0.50^{+0.18}_{-0.23}$ & 2.5 & - & - & - \\
 & H$\alpha$+[N\textsc{ii}] & & 0.88 & $< 0.001$ & & $-0.62 \pm 0.14$ & $1.79 \pm 0.36$ & $0.15^{+0.14}_{-0.07}$ & $0.91^{+0.06}_{-0.11}$ & $> 100$ & & $-0.71 \pm 0.07$ & $1.57 \pm 0.15$\\
 \hline
 \multirow{3}{4.3em}{\hbox{$P_{\mathrm{jet}} - K$}} & all & & -0.32 & 0.17 & & - & - & - & $-0.49^{+0.57}_{-0.34}$ & 1.2 & & - & - \\
  & all (larger) & & -0.36 & 0.12 & & - & - & - & $-0.57^{+0.52}_{-0.31}$ & 1.7 & & - & - \\
 & H$\alpha$+[N\textsc{ii}] & & -0.31 & 0.29 & & - & - & - & $-0.47^{+0.60}_{-0.37}$ & 1.1 & & - & - \\
 \hline
 \multirow{3}{5.3em}{\hbox{$P_{\mathrm{jet}} - L_{\text{cool}}$}} & all & & 0.07 & 0.76 & & - & - & - & $0.07^{+0.25}_{-0.26}$ & 0.33 & & - & - \\
 & all (larger) & & 0.02 & 0.93 & & - & - & - & $0.02^{+0.25}_{-0.26}$ & 0.32 & & - & - \\ \vspace{0.5mm}
 & H$\alpha$+[N\textsc{ii}] & & 0.01 & 0.96 & & - & - & - & $0.01^{+0.32}_{-0.32}$ & 0.38 & & - & -\\
\bottomrule
\end{tabular}\\
\label{tab:correl}
\end{table*}

\begin{figure}
\centering
\includegraphics[width=\linewidth]{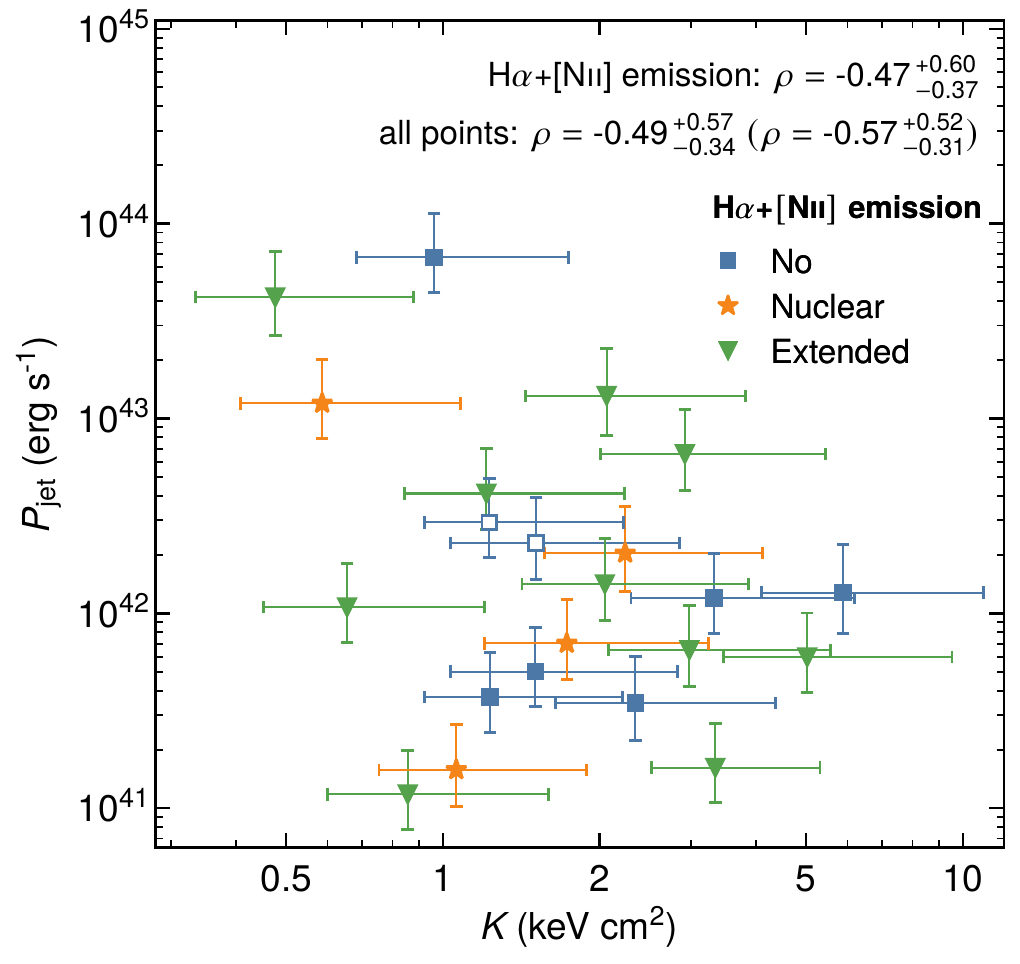}
\caption{Mechanical jet power versus central specific entropy of the X-ray emitting gas. Correlation coefficients for the H$\alpha$+[NII] emitting sub-sample and the full sample are shown in the upper right corner. The larger scale radio lobes for NGC\,1399 and NGC\,4472 (blue empty squares) and the corresponding correlation coefficient (stated in parentheses) are also shown.}
\label{fig:Pjet_entropy}
\end{figure}

\section{Discussion}

In our sample of 20 nearby massive early-type galaxies, we have found a correlation between the Bondi accretion rate and mechanical jet power (Fig. \ref{fig:Pjet_Pbondi}). The correlation for the whole sample shows moderate significance ($\rho = 0.69^{+0.15}_{-0.21}$, BF$_{10}$ = 13). However, a strong correlation with extreme evidence ($\rho = 0.96^{+0.03}_{-0.09}$, BF$_{10} > 100$) is detected for galaxies with thermally unstable atmospheres containing signs of cool gas tracers (H$\alpha$+[N\textsc{ii}] line emission).

The obtained relation between the Bondi accretion power and the mechanical jet power is for the H$\alpha$+[N\textsc{ii}] subsample well described by a power-law model
\begin{equation}
\log \frac{P_{\mathrm{Bondi}}}{{10^{43} \, \mathrm{erg \, s^{-1}}}}  = \alpha + \beta \log \frac{P_{\mathrm{jet}}}{{10^{43} \, \mathrm{erg \, s^{-1}}}},
\end{equation}
where $\alpha = 1.10 \pm 0.25$ and $\beta = 1.10 \pm 0.24$ with an intrinsic scatter $\sigma = 0.08^{+0.14}_{-0.06}$ dex. We note that the relation is remarkably close to linear, which yields the jet-to-Bondi power ratio to be practically constant ($11_{-3}^{+6}$ per cent) with increasing jet power. Assuming that the accretion is Bondi-like and accounting for the previously presumed ten per cent efficiency, for galaxies with H$\alpha$+[N\textsc{ii}] emission, we obtain a $1.1_{-0.3}^{+0.6}$ per cent efficiency of converting the rest mass of the accreted material into the energy of relativistic outflows.

Although the accretion, which is also affected by angular momentum of the infalling gas, outflows and magnetic fields, is most probably not spherical nor Bondi-like, the Bondi formula appears to provide a useful parametrization for estimating the total accretion power.

As already discussed by \cite{Allen2006}, the positive correlation between $P_{\mathrm{jet}}$ and $P_{\mathrm{Bondi}}$ is not a result of the Bondi accretion rates and mechanical jet powers being a function of galactic distances. Firstly, measured distances of all galaxies (except for NGC$\,$6166) are within an order of magnitude similar. And more importantly, while the Bondi accretion rates depend on distance approximately as $P_{\text{Bondi}} \propto d^{-0.5}$ (without accounting for the distance dependence of the SMBH mass estimates), the mechanical jet powers are proportional to distance as $P_{\text{jet}} \propto d^{1.5}$.

Instead, based on our results, we suggest that the observed $P_{\mathrm{jet}} - P_{\mathrm{Bondi}}$ correlation originates mainly from the positive correlation between the mechanical jet power and SMBH mass (Fig. \ref{fig:Pjet_MBH}). The Bondi accretion rate is by definition proportional to the square of SMBH mass ($P_{\text{Bondi}} \propto M_{\bullet}^2$), while, based on our measurements, the mechanical jet power depends on SMBH mass as $P_{\mathrm{jet}} \propto M_{\bullet}^{1.79 \pm 0.36}$. The Bondi accretion rate is also proportional to thermodynamic properties of gas inside the Bondi radius (as $\propto K^{-3/2}$), but the central specific entropy appears to correlate with mechanical jet power only mildly and with a much larger scatter (see Fig. \ref{fig:Pjet_entropy}), although the data are rather uninformative (BF$_{10}$ = 1.1). The thermodynamic properties at the Bondi radius were fairly resolved (innermost radius $\lesssim 2 \, r_{\text{Bondi}}$) only for 8 out of 20 galaxies, 4 of which contain no H$\alpha$+[N\textsc{ii}] emission. For the rest of the galaxies, the central parts were only poorly resolved and the extrapolations do not necessarily reflect the real thermodynamic state of the gas at the Bondi radius. A possible strong correlation between mechanical jet power and specific entropy could have therefore been lost in the scatter.

Similarly as for the Bondi power, a tight correlation between SMBH mass and jet power ($\rho = 0.91^{+0.06}_{-0.11}$; BF$_{10} > 100$) is only observed for the H$\alpha$+[N\textsc{ii}] emitting subsample, while for the whole sample the correlation is relatively weak and poorly constrained ($\rho = 0.47^{+0.20}_{-0.24}$; BF$_{10} = 1.8$).
Based on our results, when the galactic atmospheres are thermally unstable, the jet power will scale with the SMBH mass as described by the power-law model 
\begin{equation}\label{eq:mbh}
\log \frac{P_{\text{jet}}}{10^{43} \; \text{erg} / \text{s}} = \alpha + \beta \log \frac{M_{\bullet}}{10^9 \; M_{\sun}},
\end{equation}
where $\alpha = -0.62 \pm 0.14$ and $\beta = 1.79 \pm 0.36$ with an intrinsic scatter of $\sigma = 0.15^{+0.14}_{-0.07}$, while for thermally stable atmospheres the jet power will be unaffected by SMBH mass and remain approximately constant ($10^{41}-10^{42}$ erg$\,$s$^{-1}$). A similar relation between SMBH mass and jet power, although with a much larger scatter, was reported by \cite{McNamara2011} and also by \cite{McDonald2021} using data from \cite{Russell2013}. 

For one of the most powerful jets ever recorded with jet power of $1.7^{+0.6}_{-0.5} \times 10^{46}\:$erg\,s$^{-1}$ \citep[MS$\,$0735.6+7421;][]{Vantyghem2014}, the extrapolation of our relation (Eq. \ref{eq:mbh}) yields a SMBH mass of $8.0^{+14.1}_{-3.9} \times 10^{10}\:\text{M}_{\sun}$, which is actually consistent with the most up-to-date mass estimate obtained using the break radius scaling relation \citep[$5.1 \times 10^{10}\:\text{M}_{\sun}$;][]{Dullo2021}.

\subsection{Comparison to previous studies}

The relation between Bondi accretion power and mechanical jet power has been previously studied for 9 early-type galaxies by \cite{Allen2006} and for 13 sources by \cite{Russell2013}. In this work, we analysed data for 20 galaxies, including all sources used in these previous studies and adding 7 new systems. Despite obtaining comparable results for most sources, in several cases order-of-magnitude discrepancies are observed. 

One of the most significant sources of potential discrepancies are the assumed black hole masses, which have a significant influence on the Bondi accretion power estimation. In this work, for most objects, we were able to collect recent, direct SMBH mass estimates from within their sphere of influence (Table~\ref{tab:basic}). We only had to utilise scaling relations in 6 systems (see Section~\ref{section:sample}).

Compared to \cite{Allen2006}, in this work, we estimated the jet power of NGC\,4472 to be approximately 16 times lower because we used a newer generation of radio lobes (with corresponding X-ray cavities present). Nevertheless, in all our results, we also report the jet powers determined using the older generation of radio lobes, which are comparable to \cite{Allen2006}. With respect to the results of \cite{Russell2013}, most discrepant results for both Bondi power and jet power were obtained for NGC\,1316 and NGC\,5044. For the jet power, we also see large discrepancies in NGC\,4778. Mechanical jet powers estimated in our work are almost an order-of-magnitude lower, which is in all of these cases most probably due to the use of younger generations of lobes and cavities. The discrepancy in Bondi power is in both NGC\,1316 and NGC\,5044 caused by different electron densities at the Bondi radius. In this work, a combination of increased spatial resolution and a different extrapolation method, based on fitting the innermost substructure, resulted in almost an order of magnitude lower electron density estimates.

Significant differences between our measurements and the results of \cite{Allen2006} and \cite{Russell2013} were also obtained for the central densities in NGC\,4374 and NGC\,4486. In the case of NGC\,4374, we analysed significantly deeper \textit{Chandra} observations ($\sim 890$ ks) than in the previous studies ($\sim 120$ ks). In the latter case, the discrepancy can be caused by several reasons, however, we speculate that the main cause is the use of additional short frametime observations observed in 2016, which allowed us to avoid pile-up effects and thus better study the gas properties at smaller radii.

Except for the most distinct cases, modest differences in mechanical jet powers compared to the results of \cite{Allen2006} are most probably a result of using VLA radio observations with slightly different S/N ratio (e.g. NGC\,4374). The resulting parameters of the jet-to-Bondi power correlation are, however, consistent within uncertainties with those reported by \cite{Allen2006} ($\alpha = 0.65 \pm 0.16$ and $\beta = 0.77 \pm 0.20$). The differences between our mechanical jet powers and those of \cite{Russell2013} are primarily due to differently defined cavity volumes, with our study using combined X-ray and radio imaging data.

\subsection{Feeding from thermally unstable atmospheres}

\begin{figure}
\centering
\includegraphics[width=\linewidth]{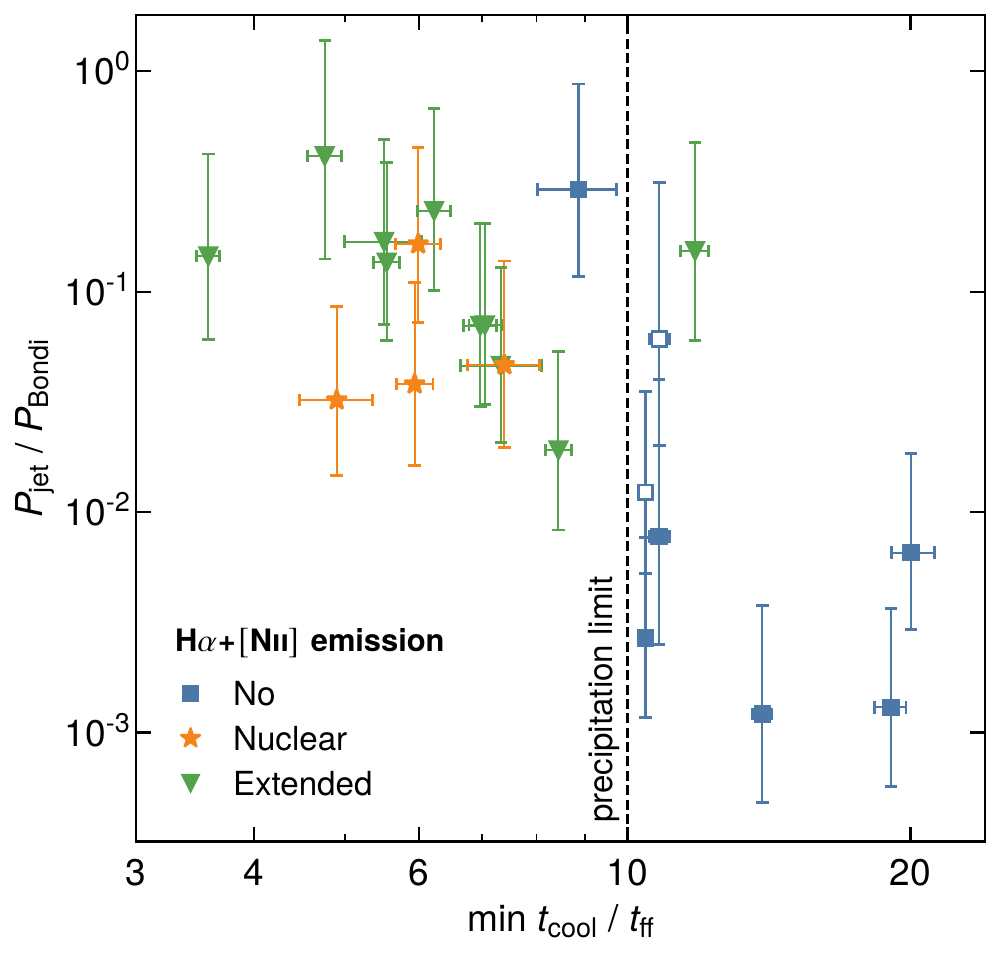}
\caption{Dependence of jet power to Bondi power efficiency on the susceptibility to thermal instabilities expressed by the minimum of the cooling time to free-fall time ratio. The vertical dashed line represents the precipitation limit $t_{\text{cool}} / t_{\mathrm{ff}} \approx 10$. The larger scale radio lobes for NGC\,1399 and NGC\,4472 (blue empty squares) are also shown.}
\label{fig:eff_ratio}
\end{figure}

The thermal stability of atmospheres was, besides the presence or absence of H$\alpha$+[N\textsc{ii}] line emission, also probed using the minimum value of the cooling time to free-fall time ratio. Fig.~\ref{fig:eff_ratio} shows that, for galaxies with $\text{min} (t_{\text{cool}} / t_{\text{ff}})$ lower than or close to the precipitation limit, the jet-to-Bondi power efficiency is approximately constant ($\approx 10$ per cent), while for higher values of $\text{min} (t_{\text{cool}} / t_{\text{ff}})$ the efficiency appears to decrease. Interestingly, the presence of the H$\alpha$+[N\textsc{ii}] emission correlates well with the minimal value of $t_{\text{cool}} / t_{\text{ff}}$ ratio, which supports the idea of galaxies with $\text{min} (t_{\text{cool}} / t_{\text{ff}}) \lesssim 10$ hosting thermally unstable atmospheres.

The fact that we only see a strong correlation for thermally unstable atmospheres with lower $\min(t_{\mathrm{cool}}/t_{\mathrm{ff}}$) suggests that the black holes producing the jets and lobes are fed by thermally unstable gas from the galactic atmospheres. The ratios of jet powers and Bondi powers inferred for the given black hole masses are within an order of magnitude similar for all these systems (of the order of $10^{-1}$ of the Bondi power) and the scatter in the inferred ratio is remarkably small. 
It appears that once the atmosphere becomes thermally unstable, the cooling gas feeds the black hole in the centres of all galaxies at a similar jet-to-Bondi power ratio, possibly indicating a key universal property of black hole accretion in early-type galaxies. The fact that all 14 thermally unstable atmospheres have similar jet powers relative to the inferred Bondi powers, indicates that the accretion in early-type galaxies is stable and the accreted cooling gas is relatively uniform and not particularly clumpy. 

Interestingly, for 5/6 galaxies with no detected H$\alpha$+[N\textsc{ii}] emission and thus most likely hosting thermally stable atmospheres, the jet power is of the order $10^{41}-10^{42}$ erg$\,$s$^{-1}$ and does not appear to trace the black hole mass. All these galaxies have within the order of magnitude similar stellar populations, and similar stellar masses. Given our interpretation, we expect that the hot atmosphere of NGC507 is thermally unstable and future, more sensitive observations should detect multi-phase gas in this system. We speculate that in these thermally stable atmospheres the stellar mass loss material provides a similar amount of fuel in all these systems \citep{Matthews2007,Voit2011}, resulting in a similar `floor' jet power. But once the atmospheres become thermally unstable, the amount of additional fuel that reaches the black hole and results in jet production will scale with the black hole mass.

\subsection{Lack of `true AGN feedback'?}

\begin{figure}
\centering
\includegraphics[width=\linewidth]{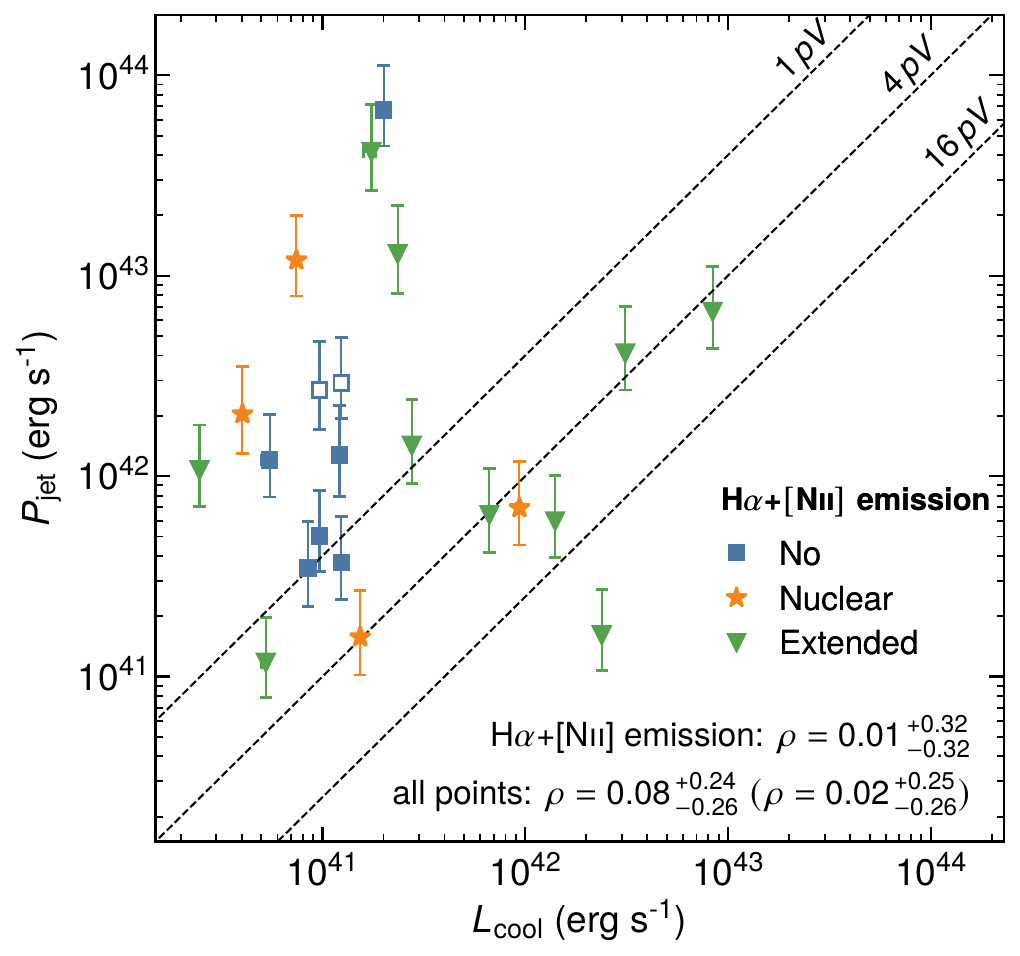}
\caption{Relation between mechanical jet power and bolometric X-ray luminosity of the galactic atmosphere from within the cooling radius (\textit{cooling luminosity}), which is the radius where the cooling time profile reaches the value of 1 Gyr. The diagonal dashed lines represent power input-output equalities for various values of cavity enthalpy (1pV, 4pV and 16pV). Correlation coefficients for the H$\alpha$+[NII] emitting sub-sample and the full sample are shown in the upper right corner. The larger scale radio lobes for NGC\,1399 and NGC\,4472 (blue empty squares) and the corresponding correlation coefficient (stated in parentheses) are also shown.}
\label{fig:Pjet_Lcool}
\end{figure}

The obtained ratios of cooling luminosities to jet powers are lower than in studies, which include more massive galaxy clusters \citep{Rafferty2006,Panagoulia2014,Hlavacek2015}. On the other hand, our results are consistent with the findings of \cite{Nulsen2009}, that for lower luminosity early-type galaxies the atmospheres are in general not precisely energetically balanced. The energy balance of our galactic atmospheres appears systematically inclined in favour of radio-mechanical heating. In other words, most of the galactic atmospheres seem to acquire more energy than they emit (Fig. \ref{fig:Pjet_Lcool}). Partially, it may be due to our choice of a lower cooling time threshold of 1 Gyr (compared to $3$ or $7.7$ Gyr) resulting in smaller cooling radii. 

Interestingly, we observe no correlation between the cooling luminosity and the mechanical jet power (Table \ref{tab:correl}, see Fig. \ref{fig:Pjet_Lcool}), which is most likely a result of the relatively small range of cooling luminosities and jet powers in our study. Similarly, for X-ray luminosities extracted from within a predefined fixed radius (e.g. 1 or 10~kpc), we observed at most weak correlations. At first glance, these results indicate a lack of fine-tuning between the jet power and the thermodynamic global state of hot atmospheres.

Instead, the atmospheric properties seem to provide an `on/off switch' - they determine whether the atmosphere will be thermally stable or not. The maximal possible jet power will be primarily set by the black hole mass. For thermally unstable galaxies, the jet power will trace this maximal limit (Eq. \ref{eq:mbh}), while for stable atmospheres the jet power will be orders of magnitude smaller.

Heavier dark matter haloes with more massive atmospheres will also host larger black holes (e.g. see \citealp{Bogdan2015}; \citealp{Lakhchaura2019}; \citealp{Truong2021}). For galaxies with thermally unstable atmospheres (central cooling times shorter than 1 Gyr), heavier haloes will also have more powerful AGNs \citep{Main2017}. These more powerful AGNs powered by heavier black holes will provide more jet heating, compensating for the larger cooling luminosities of more massive systems. We propose that the correlation between the $P_{\text{jet}}$ and $L_{\text{cool}}$ observed over a relatively large dynamic range from galaxies to clusters of galaxies (e.g. see \citealp{Hlavacek2015}) is primarily due to the underlying correlation between the black hole mass and halo mass.

\cite{Lakhchaura2019} showed that the average atmospheric gas temperature in early-type ellipticals correlates with the mass of the central black hole \citep[see also][]{Gaspari2019}. By comparing observations and state-of-the-art numerical simulations (Illustris TNG), \cite{Truong2021} shows that this is primarily due to an underlying correlation between the halo mass and the mass of the central supermassive black hole and jet heating will have a secondary effect. Our results show that more massive black holes will provide more heat to galactic atmospheres \citep[see also][]{Martin-Navarro2020}, which could also contribute to the correlation between black hole mass and atmospheric gas temperature.

\section*{Conclusions}

We have confirmed the presence of a correlation between the Bondi accretion power and the mechanical jet power in early-type galaxies previously reported by \cite{Allen2006}. We show that a particularly strong correlation holds for galaxies with thermally unstable atmospheres, as indicated by the presence of cool gas traced by H$\alpha$+[N\textsc{ii}] emission and with $\min (t_{\text{cool}} / t_{\text{ff}})\lesssim10$, while for the whole sample of galaxies the correlation is weaker.

Interestingly, according to the power-law fit for the H$\alpha$+[N\textsc{ii}] subsample, the Bondi power scales with jet power as $\propto \, {P_{\text{jet}}}^{1.10 \pm 0.25}$ with correlation coefficient of $\rho = 0.96^{+0.03}_{-0.09}$. We note that the exponent is remarkably close to unity, which yields a constant jet-to-Bondi power efficiency ($11_{-3}^{+6}$ per cent). 

Importantly, we find a strong correlation between the mechanical jet power ($P_{\mathrm{jet}}$) and the mass of the central supermassive black hole ($M_{\bullet}$) and, although poorly constrained, a hint of an anti-correlation with the specific entropy ($K$) of the ambient gas inside the Bondi radius. The mechanical jet power for the galaxies with H$\alpha$+[N\textsc{ii}] emission scales with the supermassive black holes mass as 
\begin{equation}
\log \frac{P_{\text{jet}}}{10^{43} \, \text{erg} \, \text{s}^{-1}} = \alpha + \beta \log \frac{M_{\bullet}}{10^9 \, M_{\sun}},
\end{equation}
where $\alpha = -0.62 \pm 0.14$ and $\beta = 1.79 \pm 0.36$ with a correlation coefficient of $\rho = 0.91^{+0.06}_{-0.11}$, while for the full sample the correlation is weaker.


The results indicate that at least for thermally unstable systems, the jet power is set primarily by the supermassive black hole mass. Since the central black hole mass of X-ray luminous early-type galaxies correlates with the total mass of the host halo \citep[see][]{Lakhchaura2019,Truong2021}, more massive systems undergoing thermally unstable cooling will naturally have larger jet powers.

\section*{Acknowledgements}

This research was supported by the GACR grant 21-13491X. 
A.S. is supported by the Women In Science Excel (WISE) programme of the Netherlands Organisation for Scientific Research (NWO), and acknowledges the World Premier Research Center Initiative (WPI) and the Kavli IPMU for the continued hospitality. SRON Netherlands Institute for Space Research is supported financially by NWO. We acknowledge support from the U.S. Department of Energy under contract number DE-AC02-76SF00515. This research has made use of the NASA/IPAC Extragalactic Database (NED), which is funded by the National Aeronautics and Space Administration and operated by the California Institute of Technology. We acknowledge the usage of the HyperLeda database (\href{http://leda.univ-lyon1.fr}{http://leda.univ-lyon1.fr}). The use of Very Long Baseline Array (VLBA) data is acknowledged. VLBA is operated by the National Radio Astronomy Observatory, which is a facility of the National Science Foundation, and operated under cooperative agreement by Associated Universities, Inc. 

\section*{Data availability}

The data in this article are available on request to the corresponding author.



\bibliographystyle{mnras}
\bibliography{main} 




\clearpage

\appendix

\section{Radial profiles}
\label{appendix:profiles}


\begin{figure}
\begin{tikzpicture}[overlay]
\draw (\figxi, \figyii) node {\includegraphics[scale=\figscale]{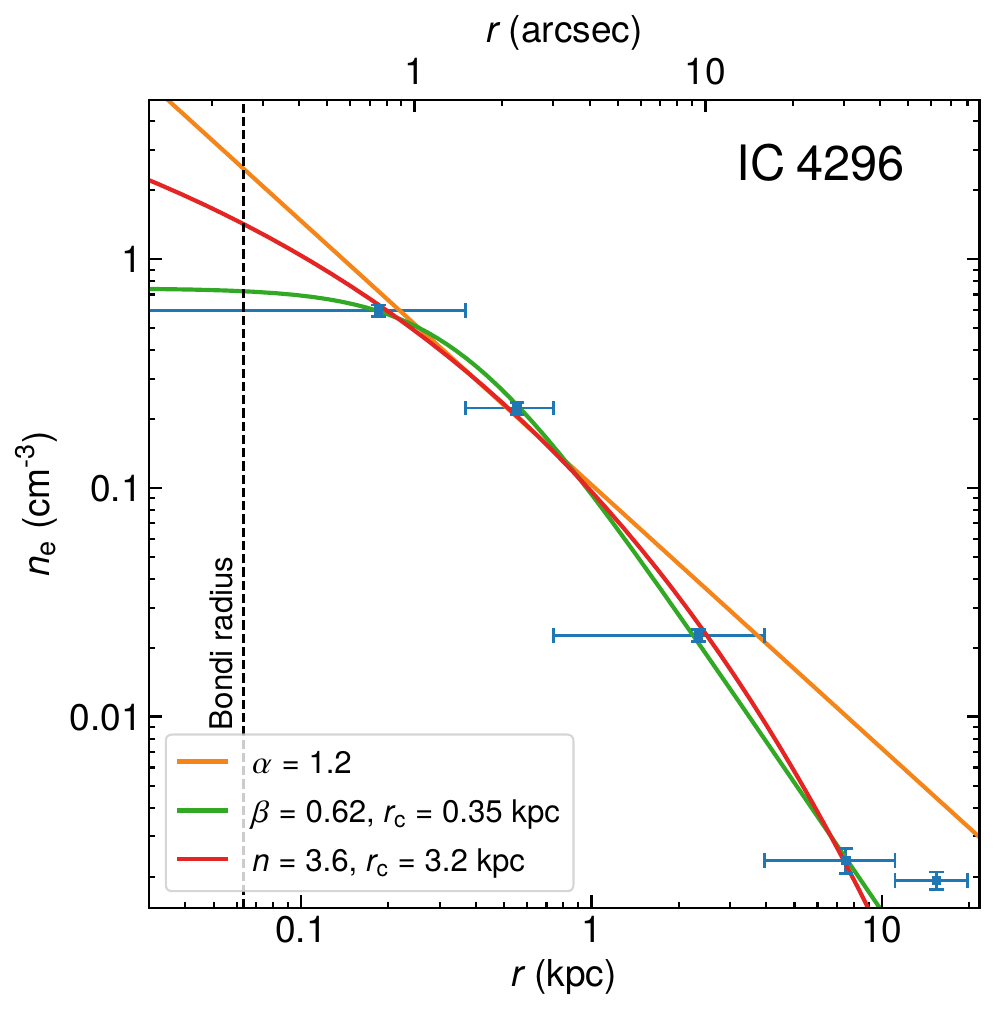}};
\draw (\figxj, \figyii) node {\includegraphics[scale=\figscale]{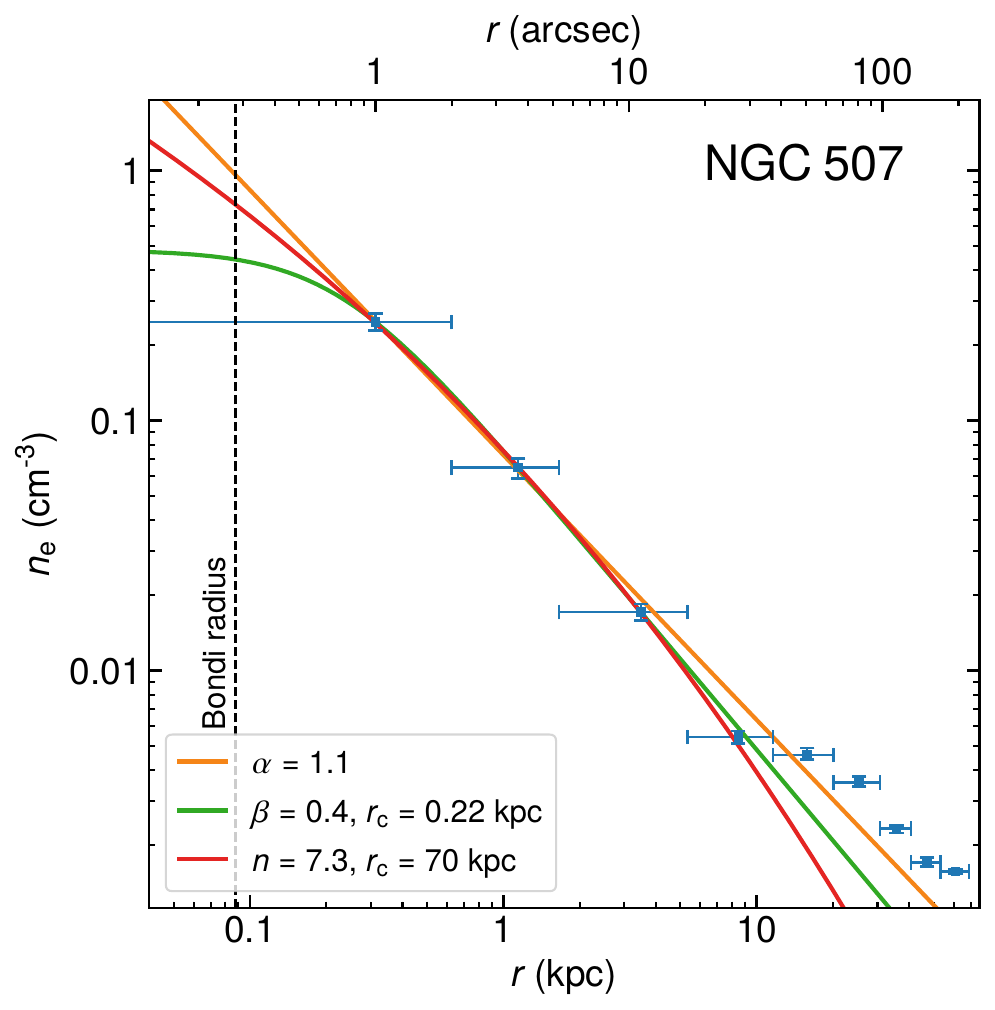}};
\draw (\figxk, \figyii) node {\includegraphics[scale=\figscale]{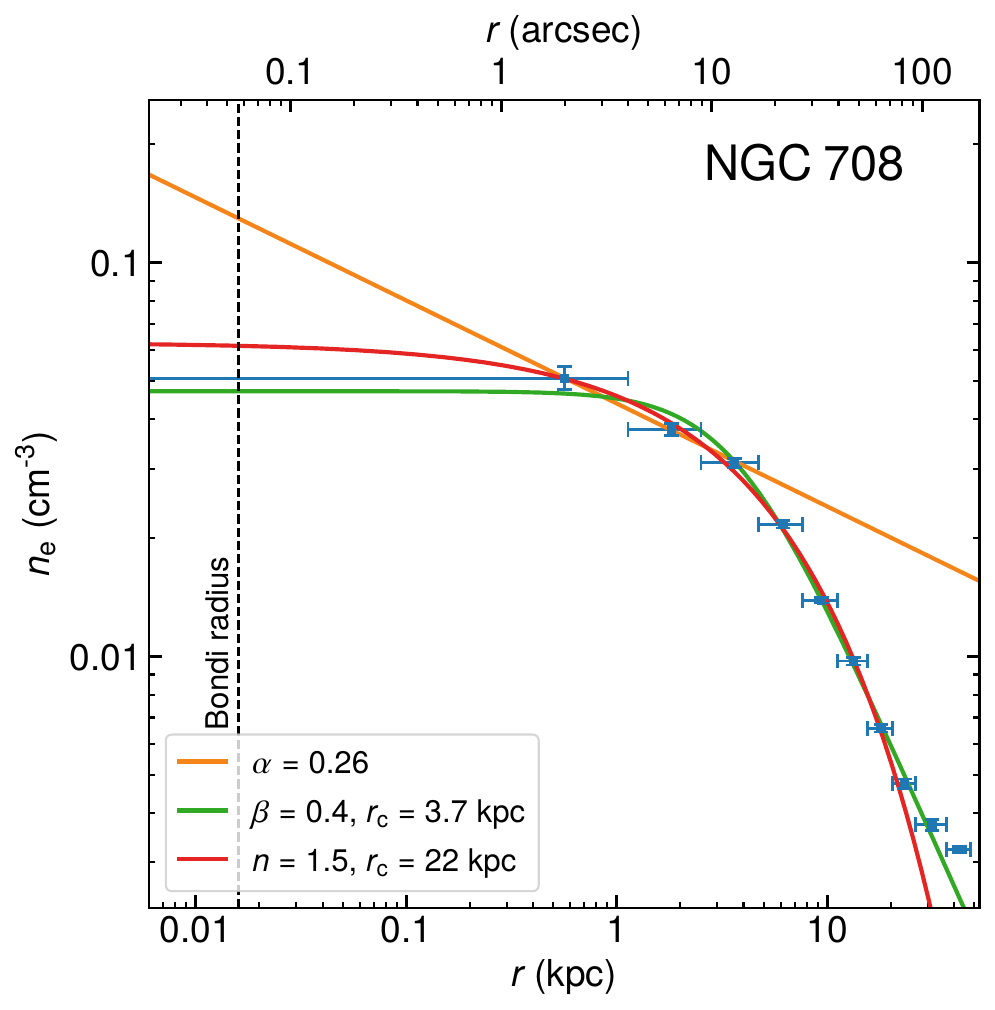}};
\draw (\figxi, \figyjj) node {\includegraphics[scale=\figscale]{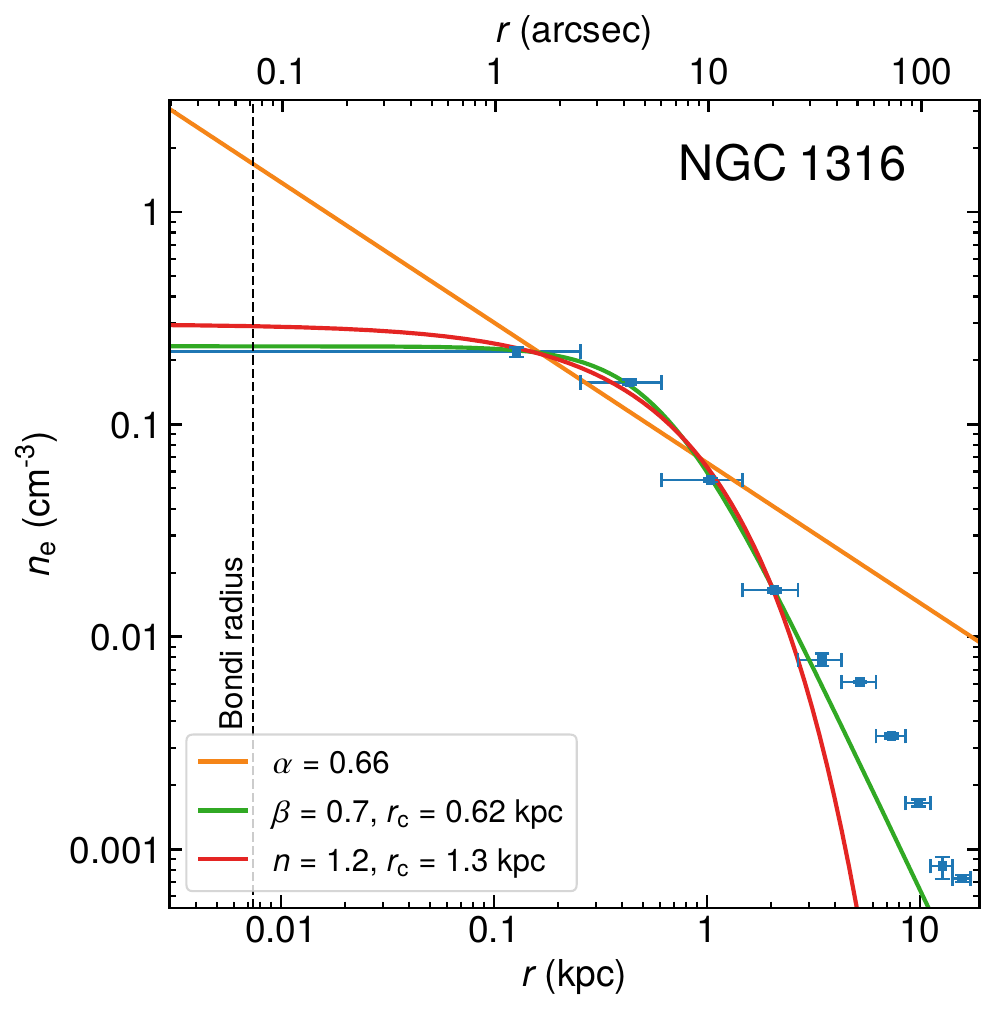}};
\draw (\figxj, \figyjj) node {\includegraphics[scale=\figscale]{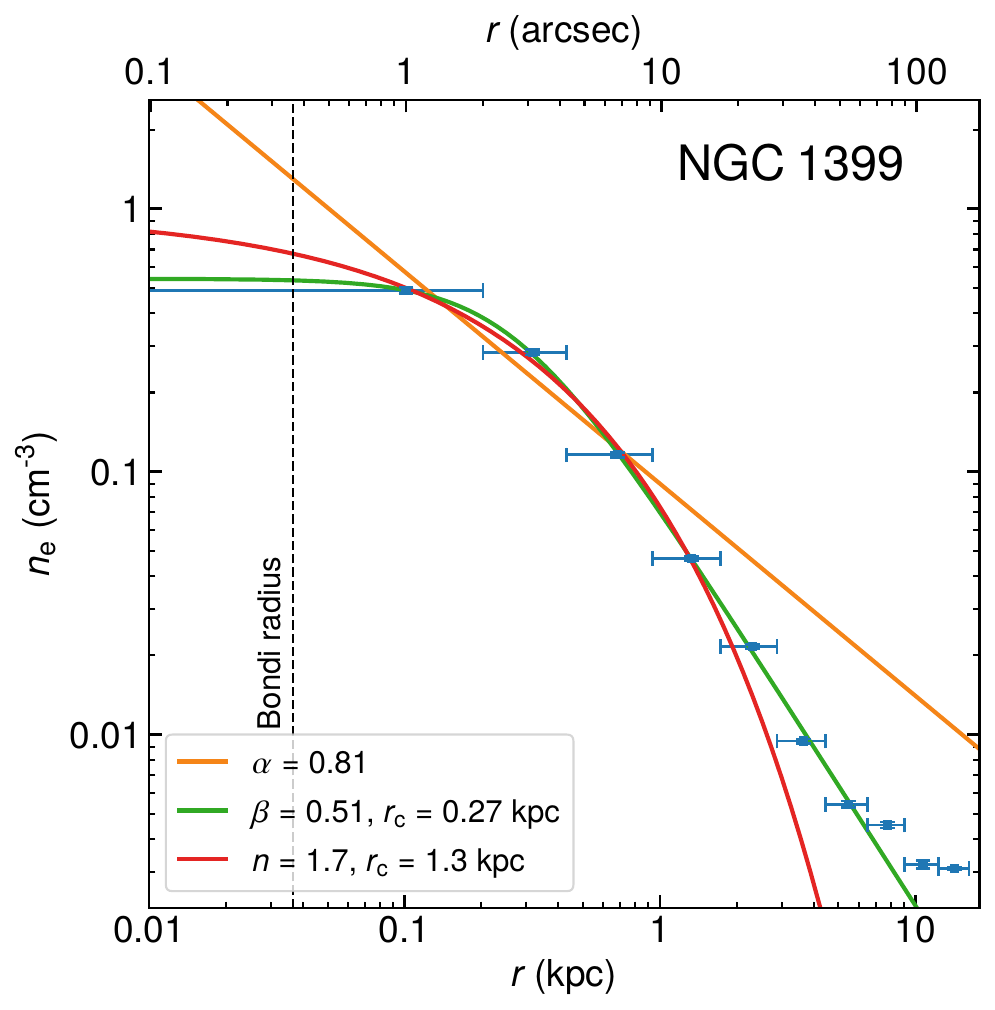}};
\draw (\figxk, \figyjj) node {\includegraphics[scale=\figscale]{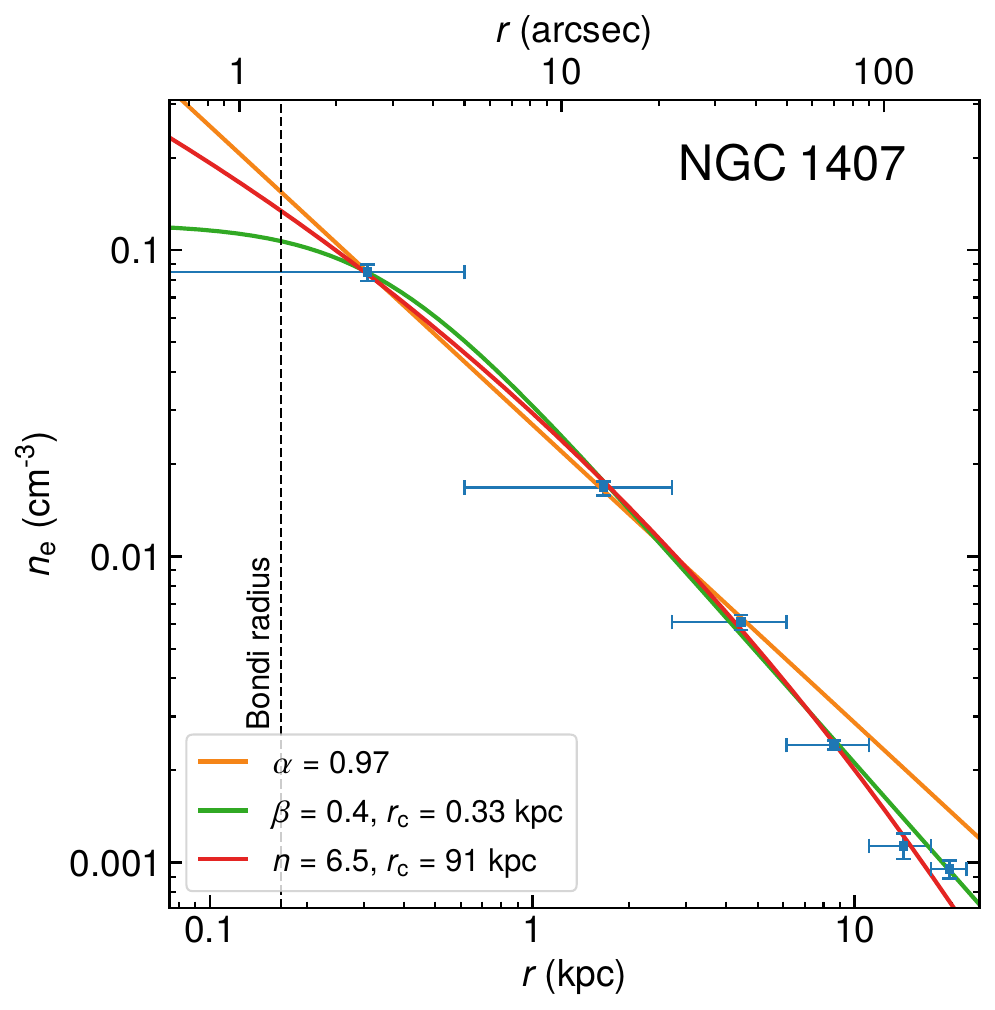}};
\draw (\figxi, \figykk) node {\includegraphics[scale=\figscale]{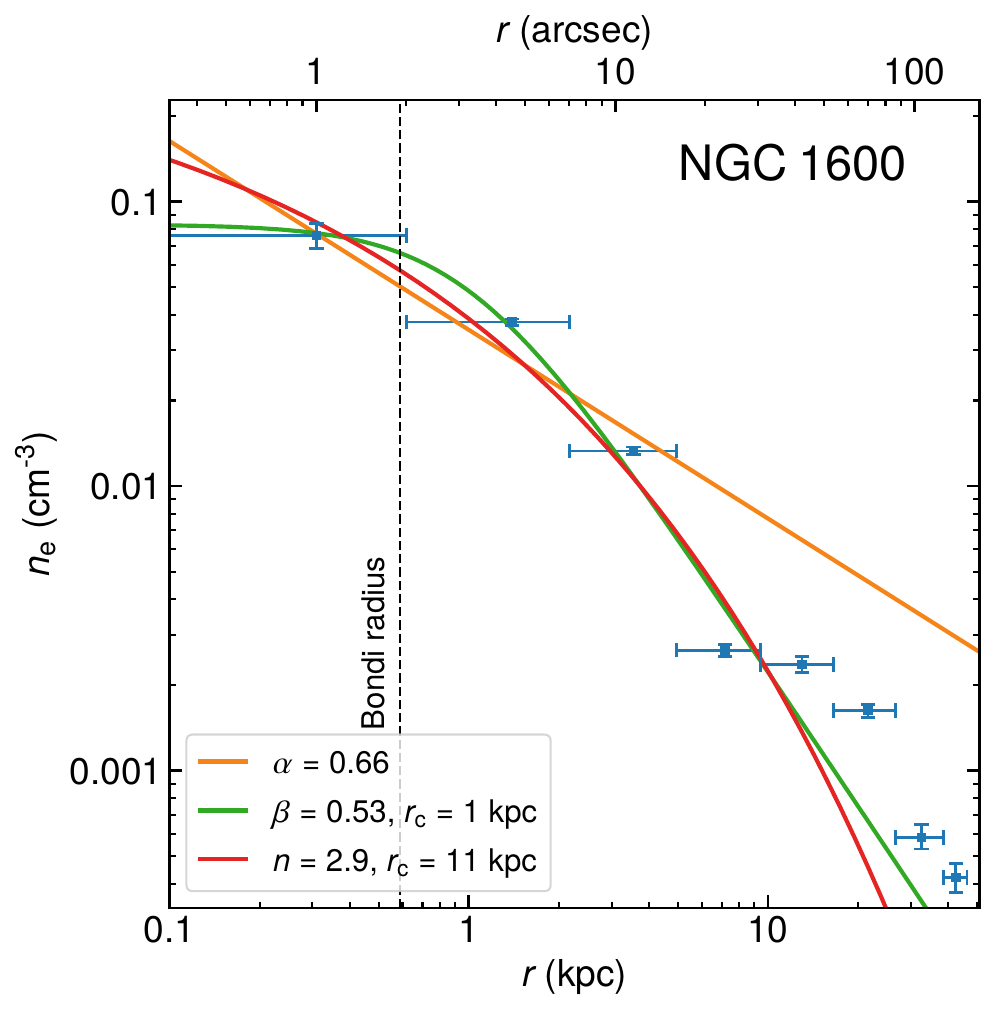}};
\draw (\figxj, \figykk) node {\includegraphics[scale=\figscale]{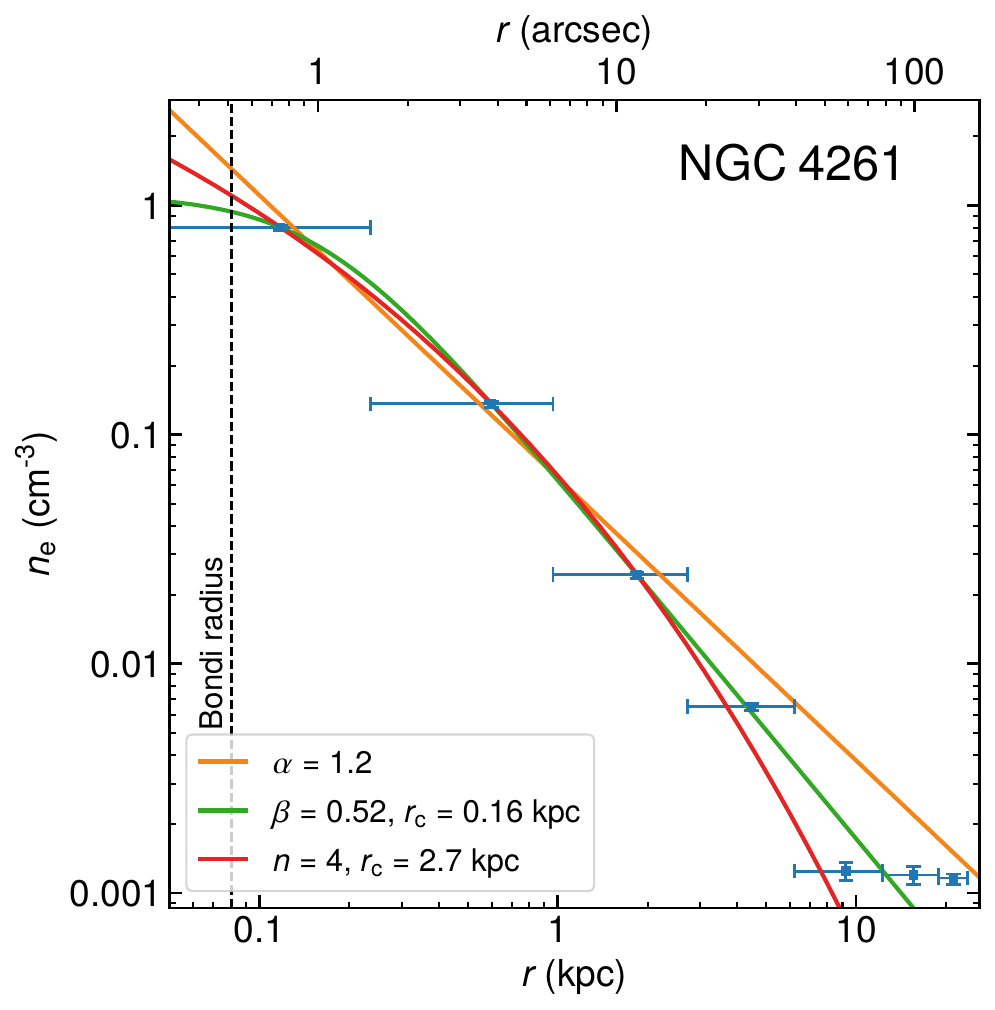}};
\draw (\figxk, \figykk) node {\includegraphics[scale=\figscale]{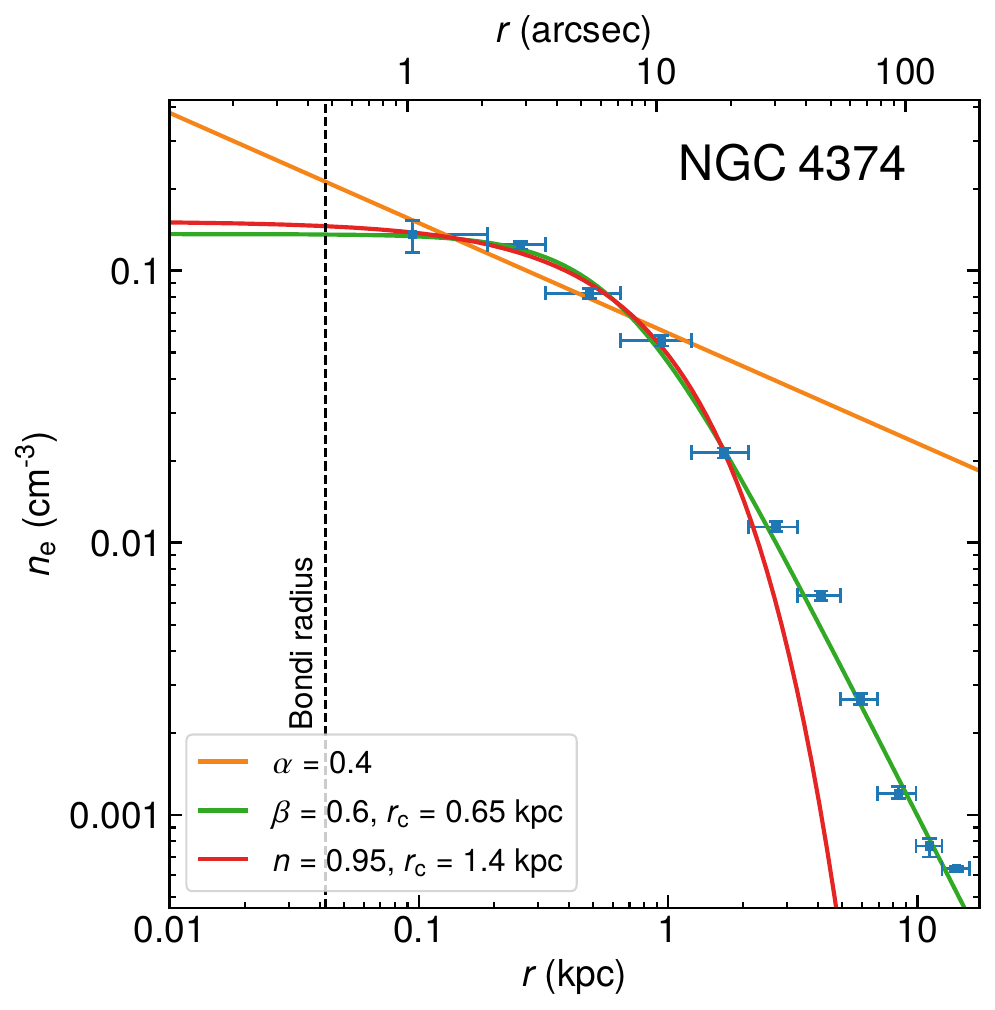}};
\draw (\figxi, \figyll) node {\includegraphics[scale=\figscale]{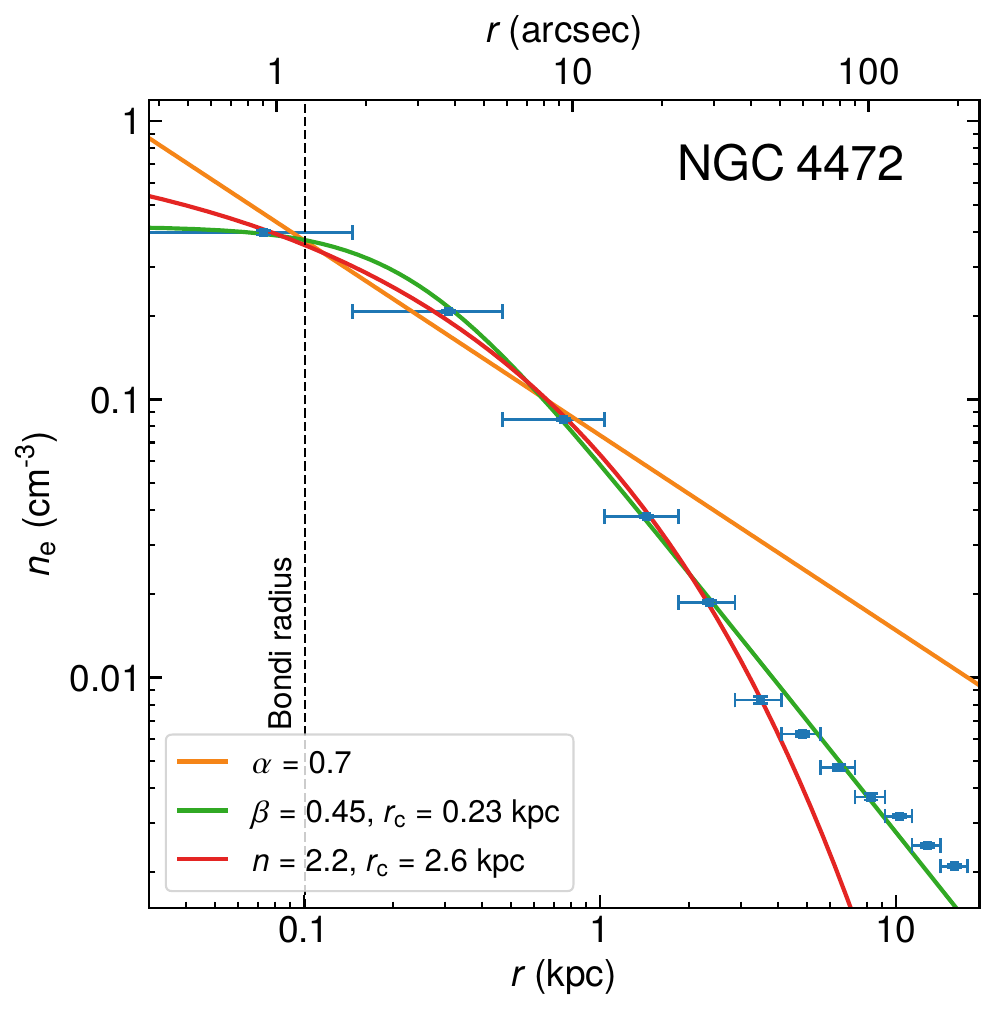}};
\draw (\figxj, \figyll) node {\includegraphics[scale=\figscale]{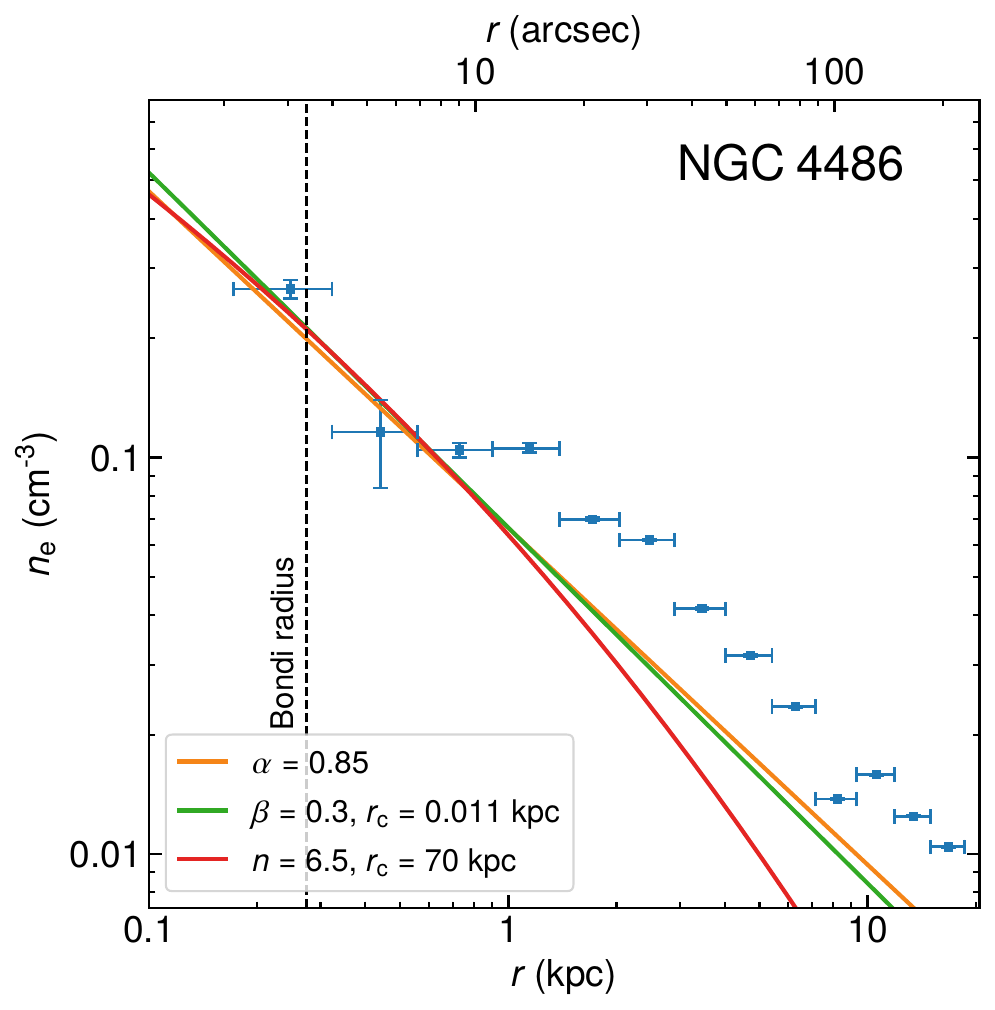}};
\draw (\figxk, \figyll) node {\includegraphics[scale=\figscale]{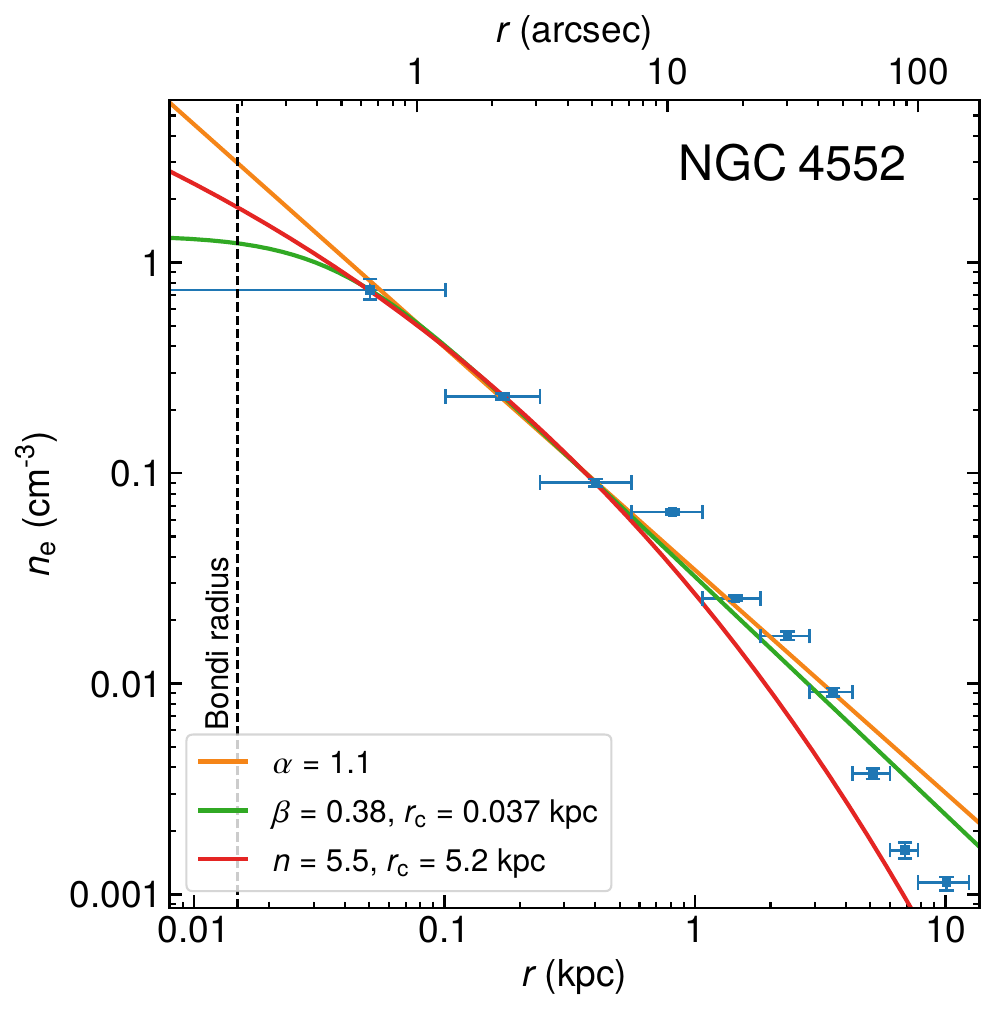}};
\end{tikzpicture}
\end{figure}

\begin{figure*}
\begin{tikzpicture}
\draw (\figxi, \figyi) node {\includegraphics[scale=\figscale]{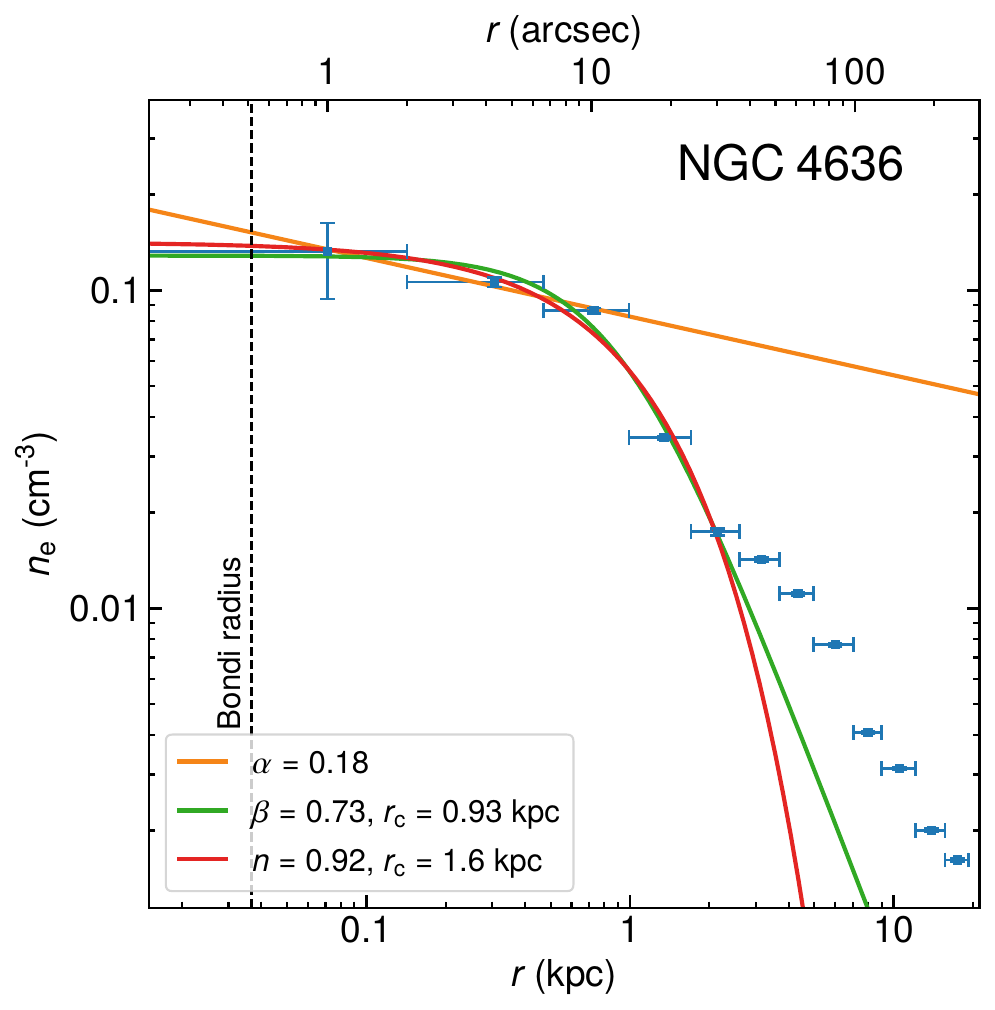}};
\draw (\figxj, \figyi) node {\includegraphics[scale=\figscale]{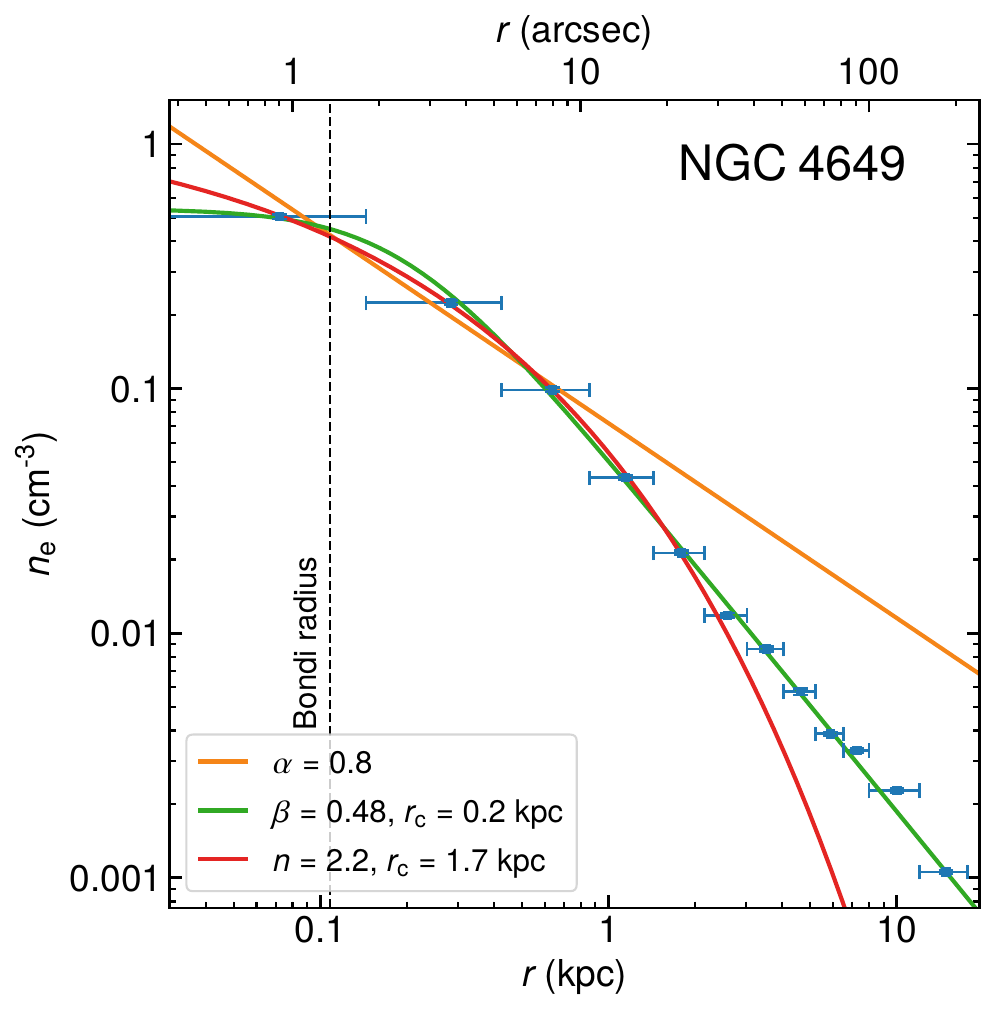}};
\draw (\figxk, \figyi) node {\includegraphics[scale=\figscale]{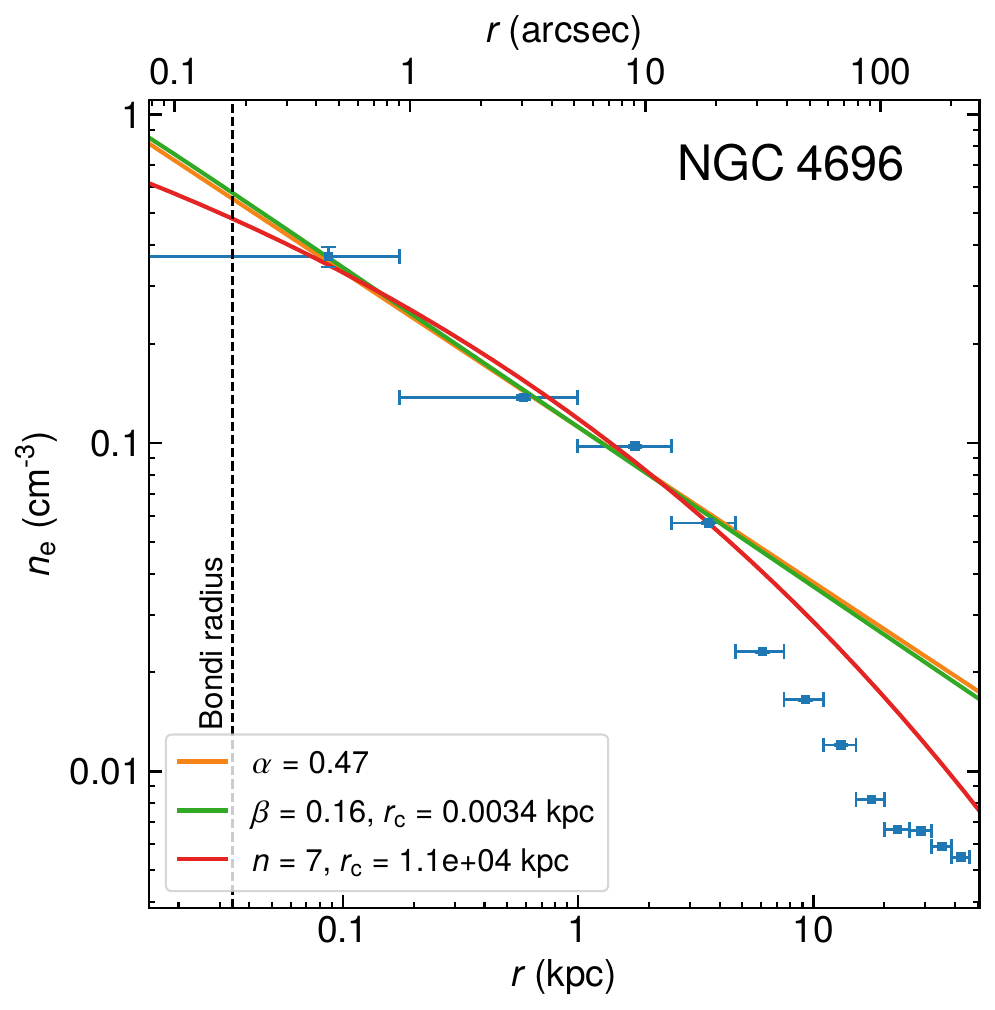}};
\draw (\figxi, \figyj) node {\includegraphics[scale=\figscale]{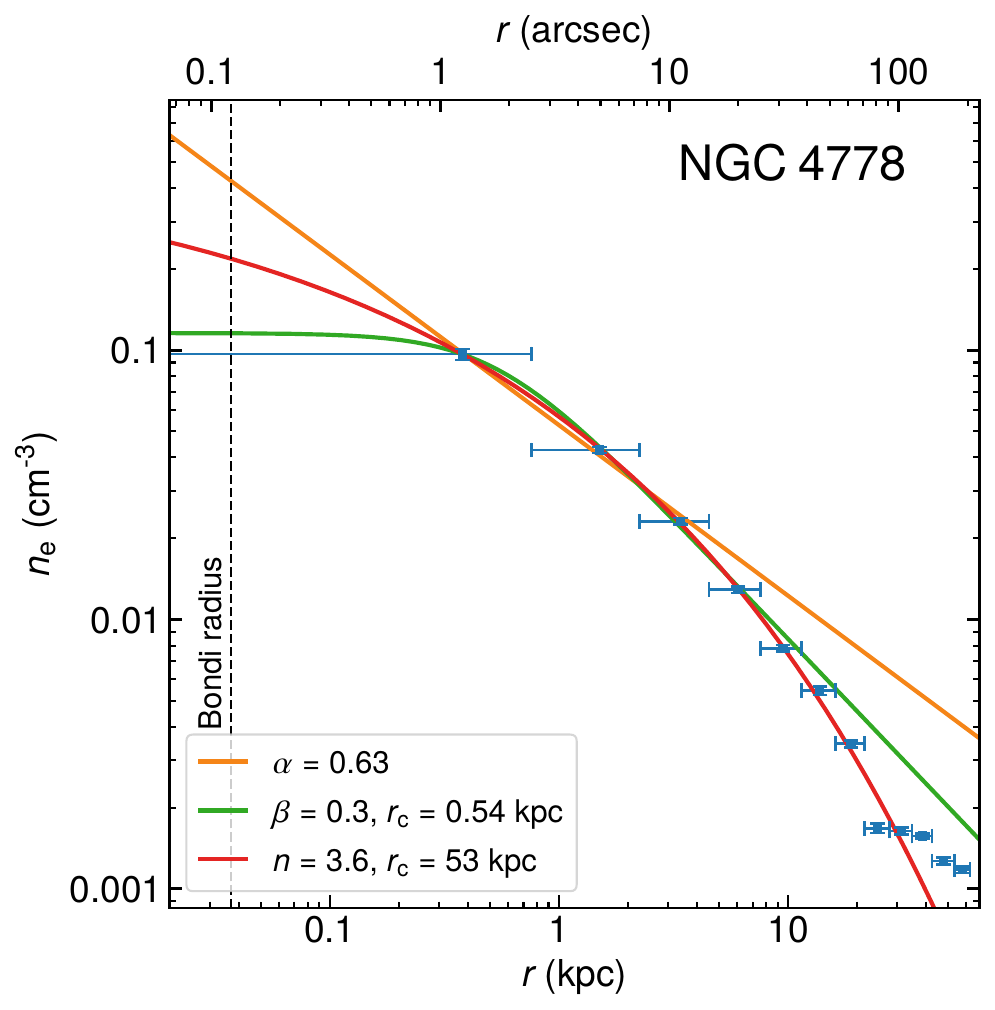}};
\draw (\figxj, \figyj) node {\includegraphics[scale=\figscale]{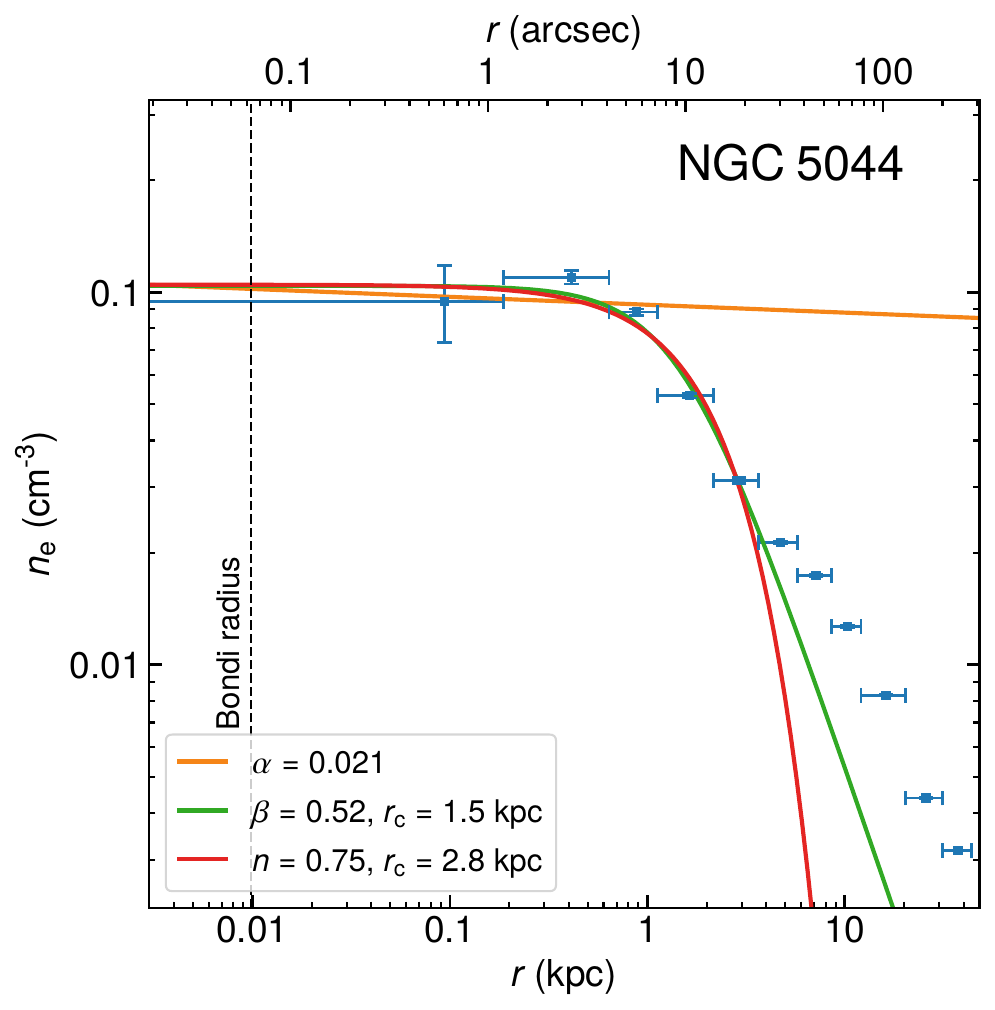}};
\draw (\figxk, \figyj) node {\includegraphics[scale=\figscale]{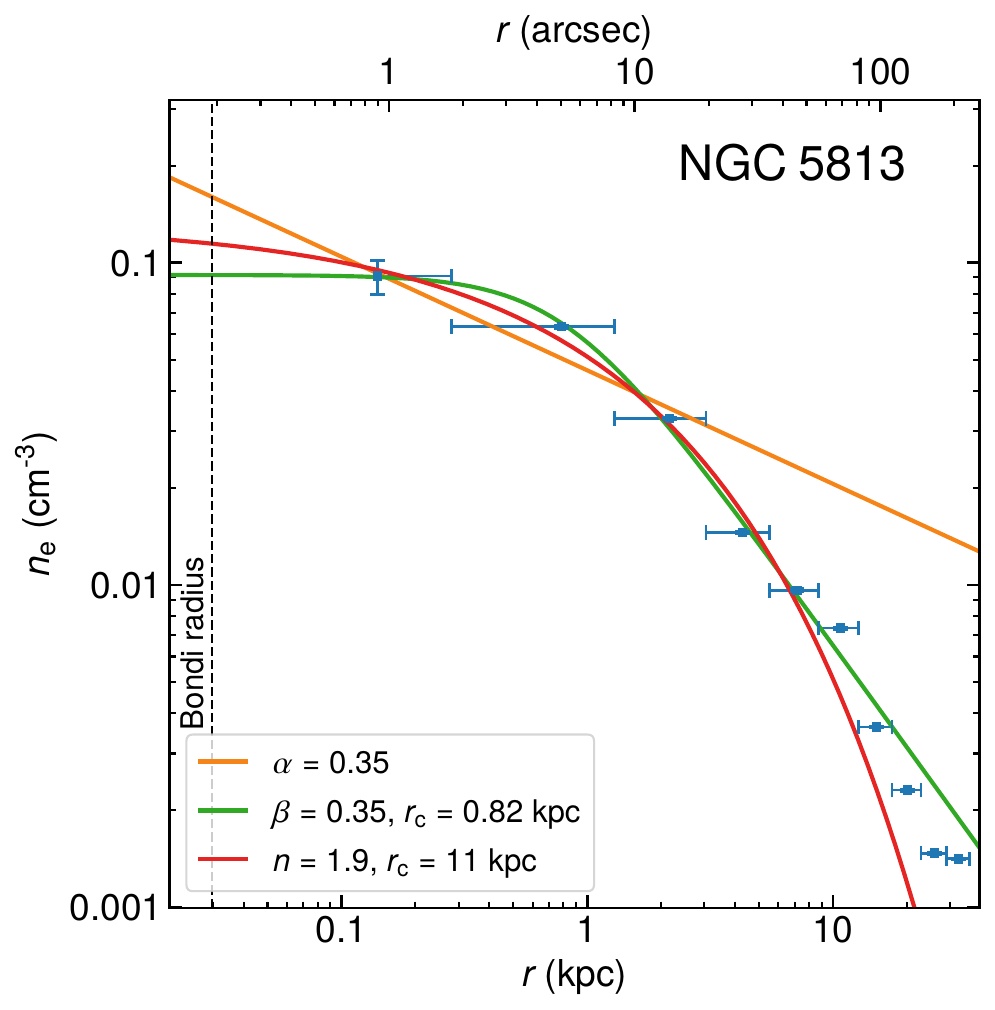}};
\draw (\figxi, \figyk) node {\includegraphics[scale=\figscale]{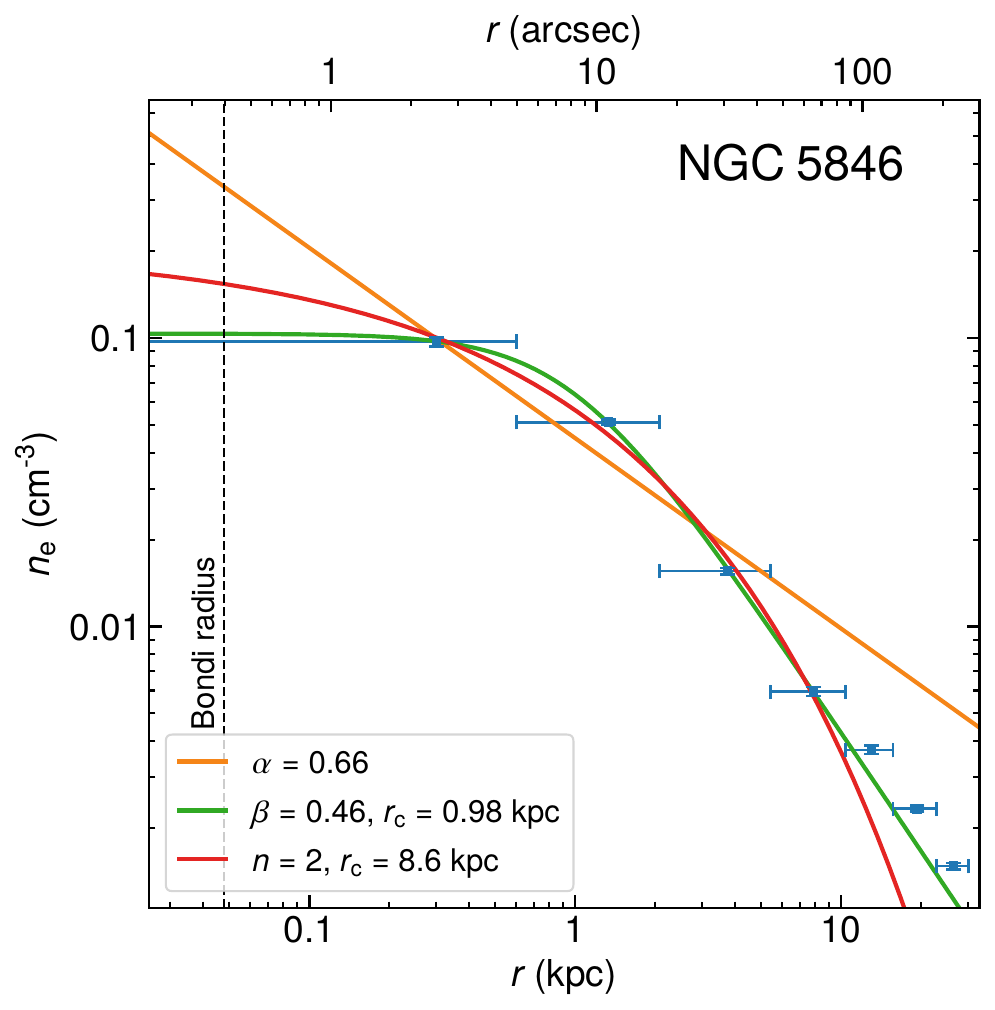}};
\draw (\figxj, \figyk) node {\includegraphics[scale=\figscale]{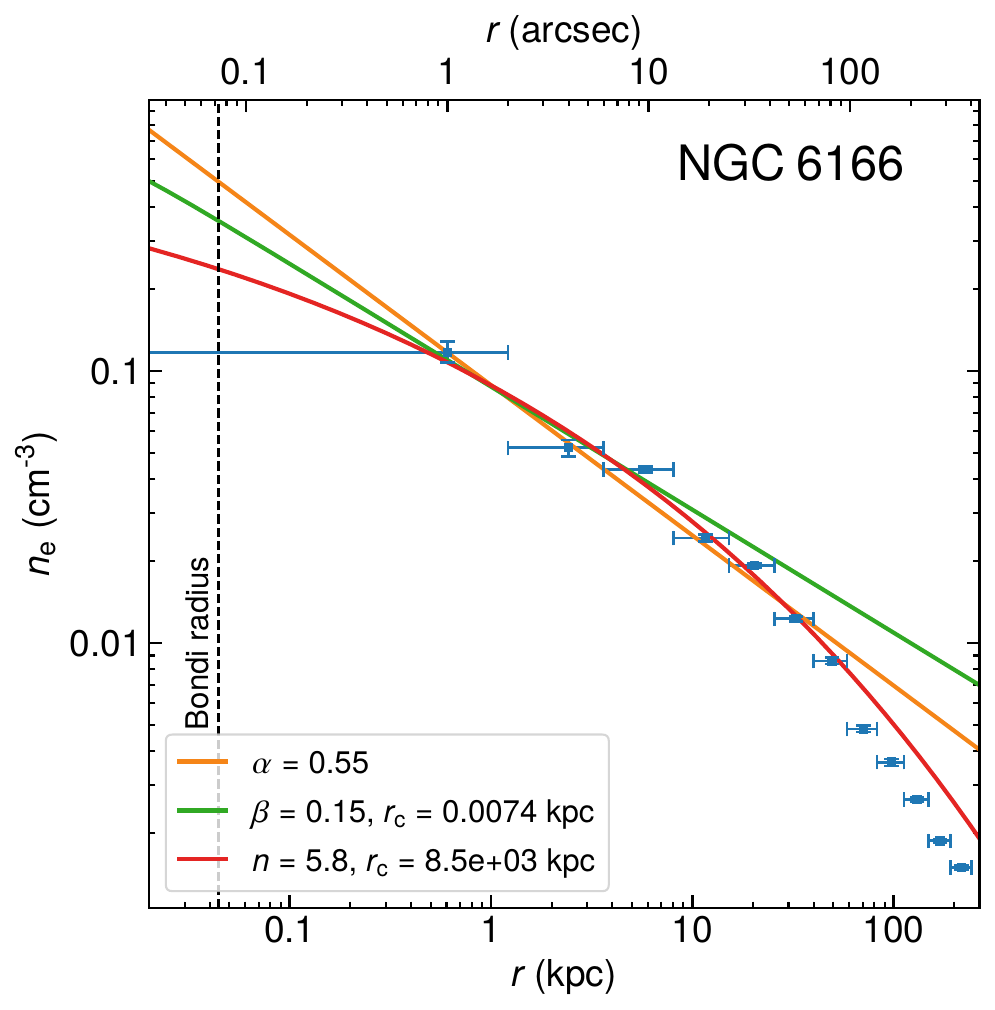}};
\end{tikzpicture}
\caption{Azimuthally averaged radial profiles of electron number densities $n_{\text{e}}$. The vertical dashed lines represent the Bondi radius $r_{\text{Bondi}}$ to which the electron number densities were extrapolated. For the extrapolation, we used three different profiles: power-law model (\textit{orange}), $\beta$-model (\textit{green}) and sersic profile with freed parameter $n$ (\textit{red}). The final value of electron number density at the Bondi radius was calculated as a mean from these three profiles and the scatter was accounted in the uncertainty.}
\label{fig:density}
\end{figure*}


\begin{figure}
\begin{tikzpicture}
\draw (\figxi, \figyi) node {\includegraphics[scale=\figscale]{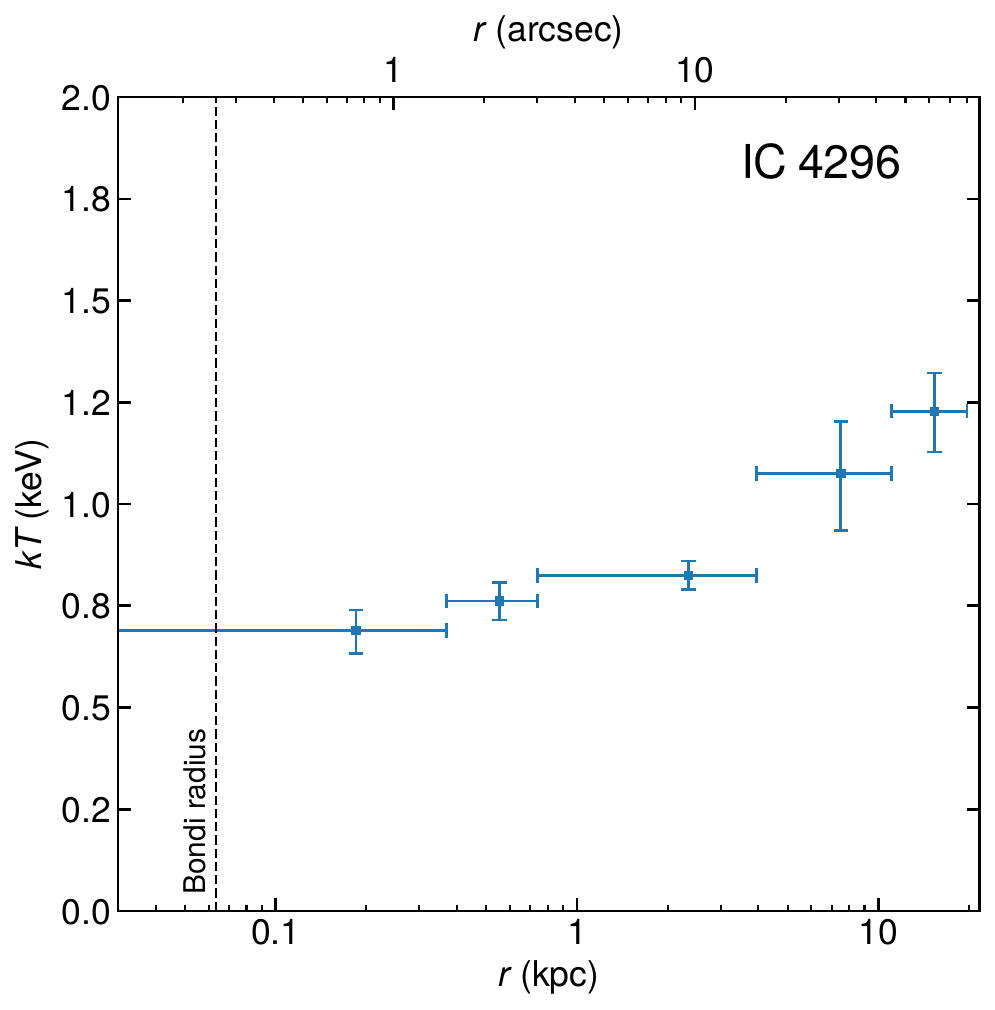}};
\draw (\figxj, \figyi) node {\includegraphics[scale=\figscale]{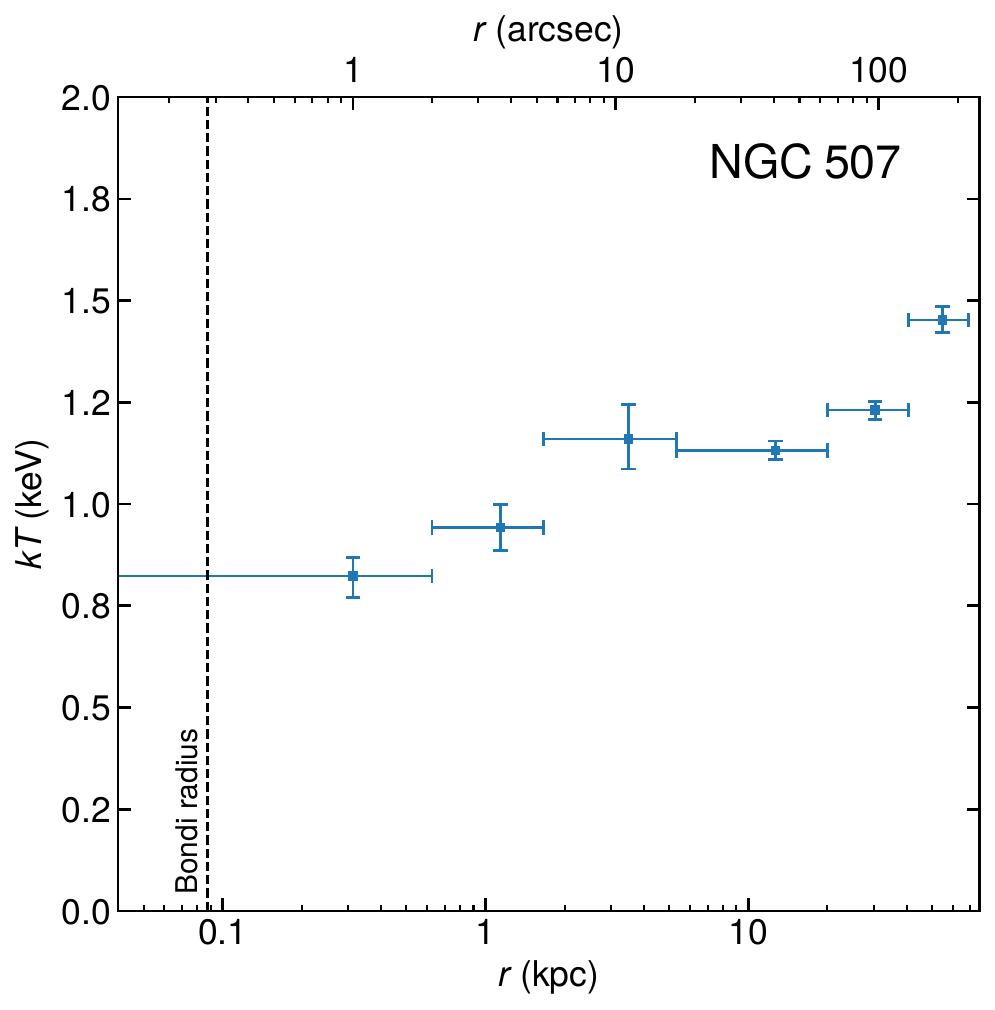}};
\draw (\figxk, \figyi) node {\includegraphics[scale=\figscale]{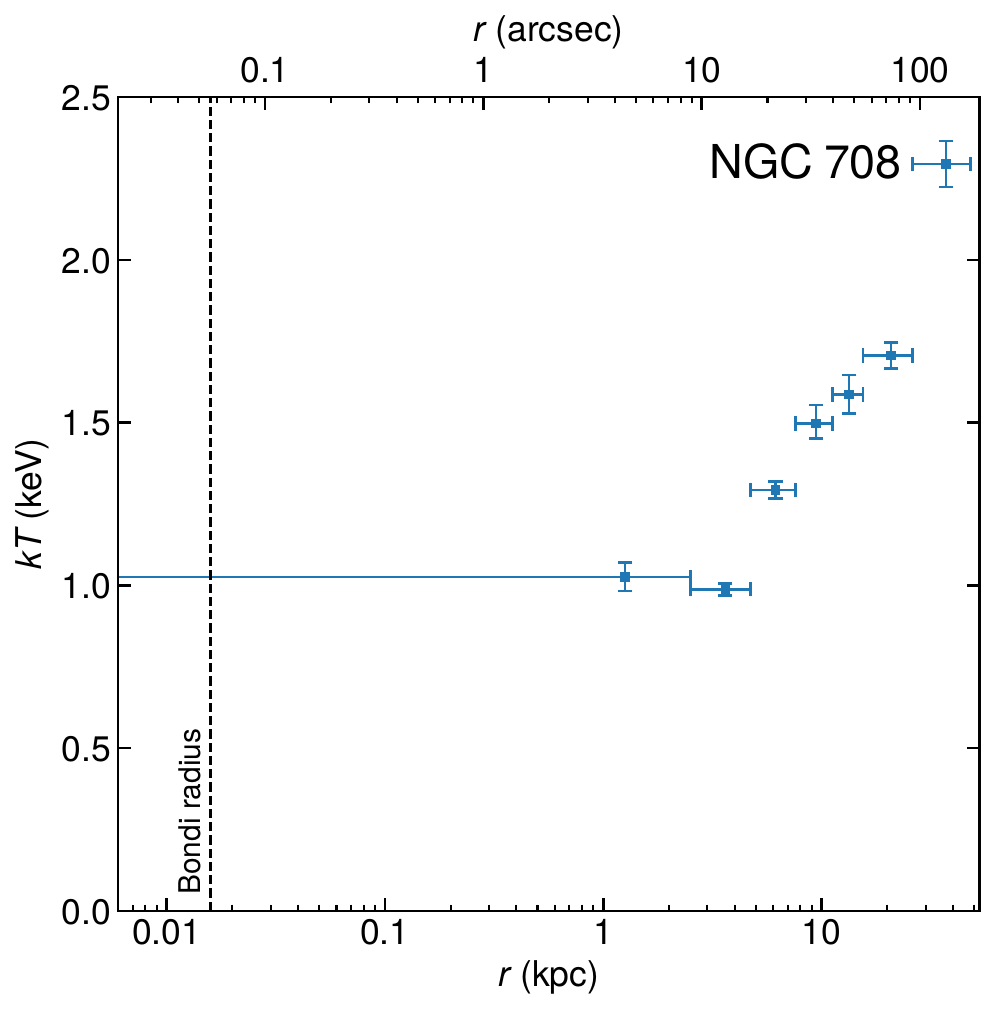}};
\draw (\figxi, \figyj) node {\includegraphics[scale=\figscale]{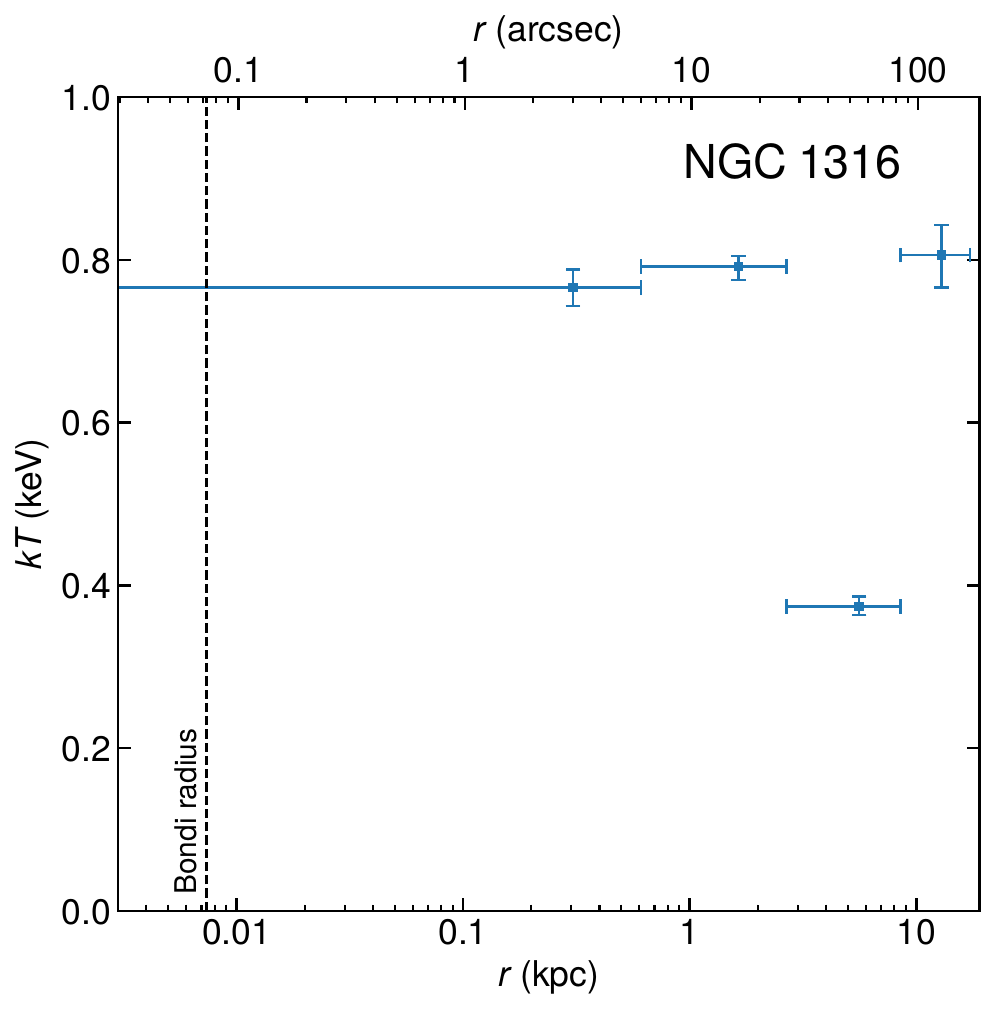}};
\draw (\figxj, \figyj) node {\includegraphics[scale=\figscale]{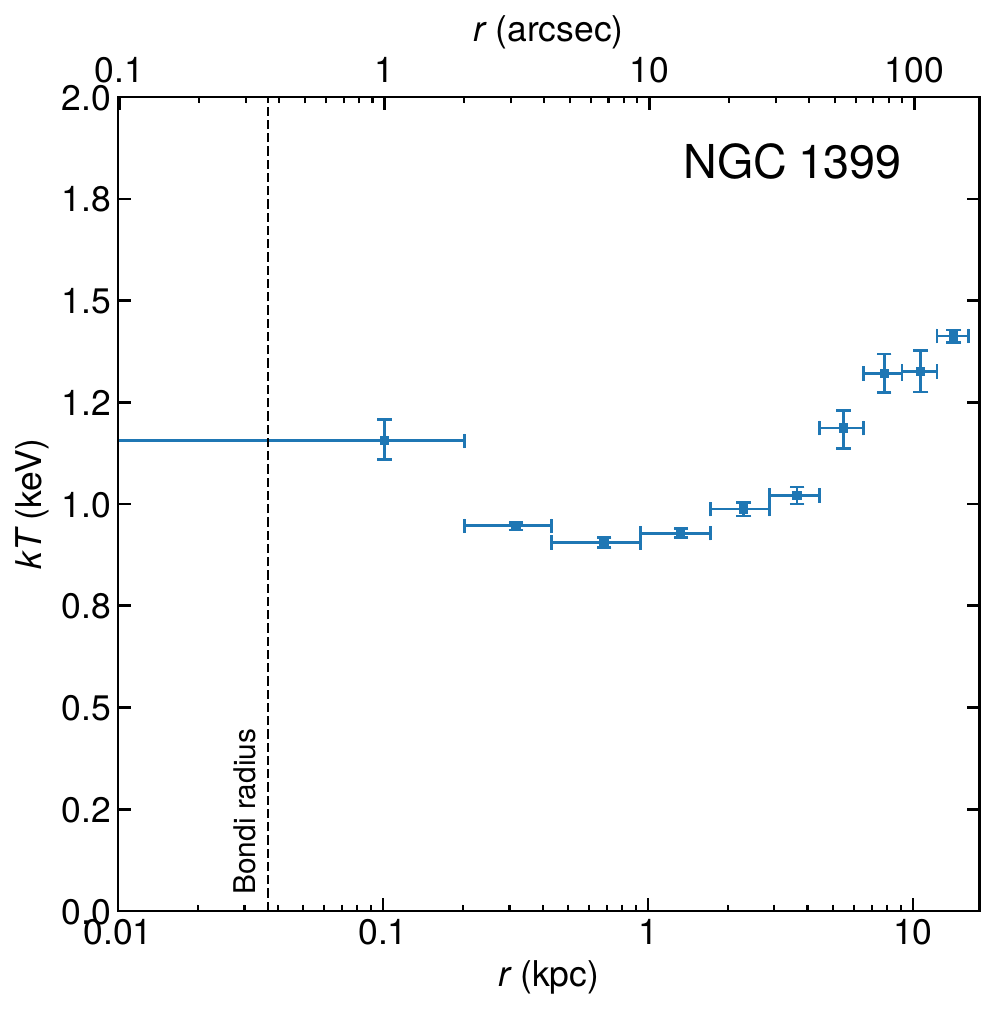}};
\draw (\figxk, \figyj) node {\includegraphics[scale=\figscale]{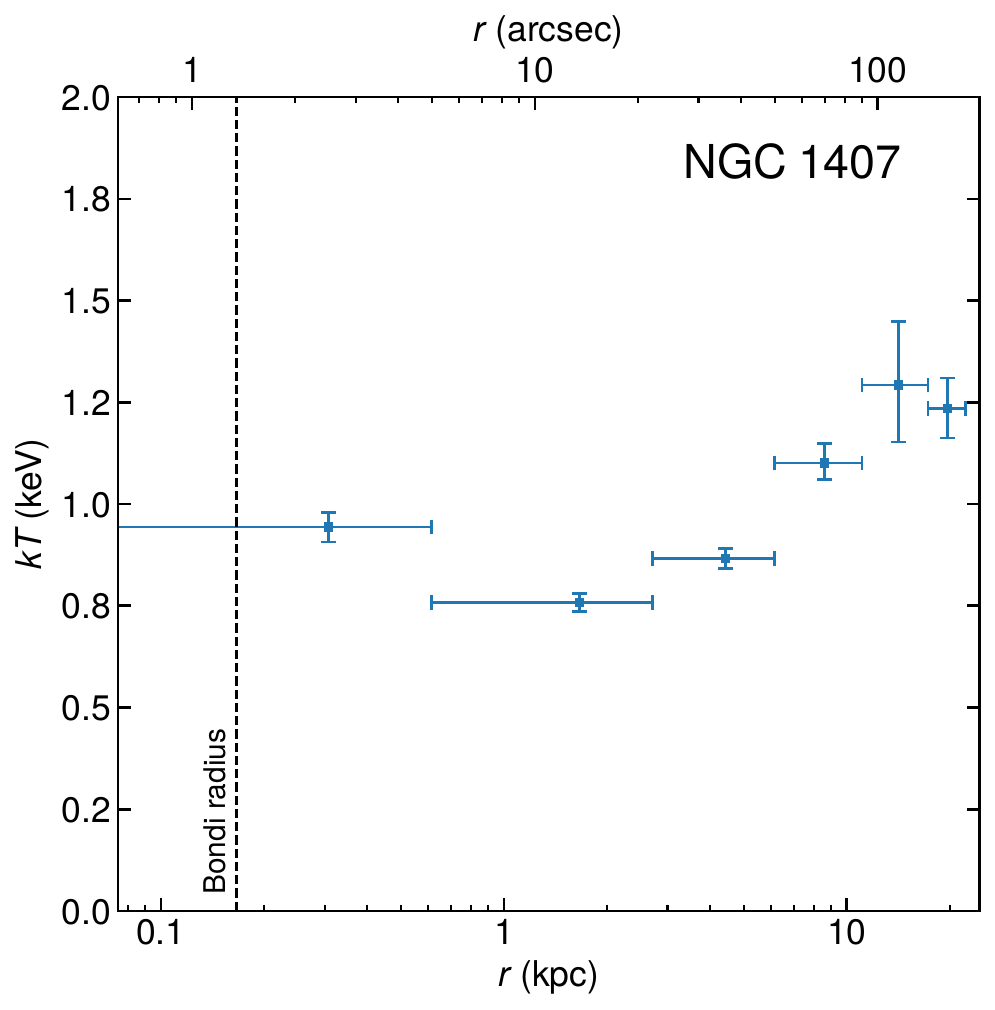}};
\draw (\figxi, \figyk) node {\includegraphics[scale=\figscale]{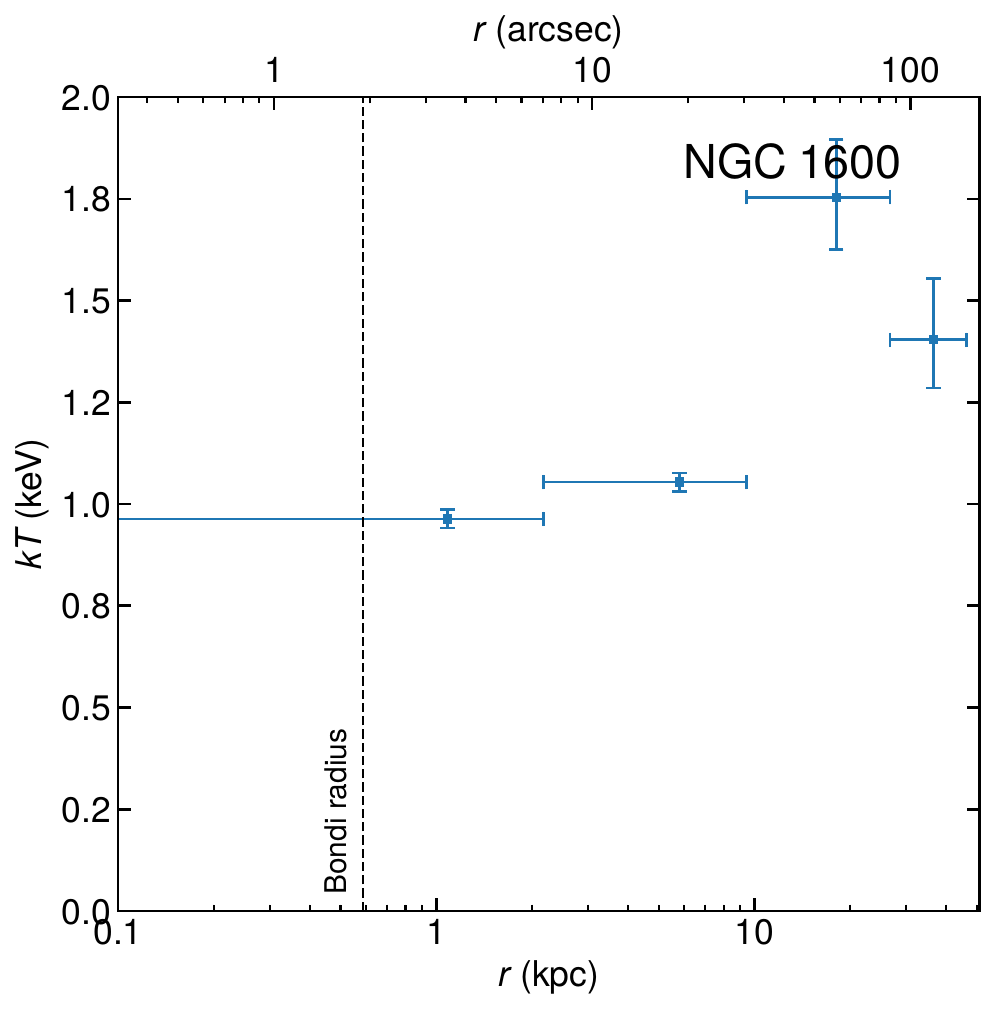}};
\draw (\figxj, \figyk) node {\includegraphics[scale=\figscale]{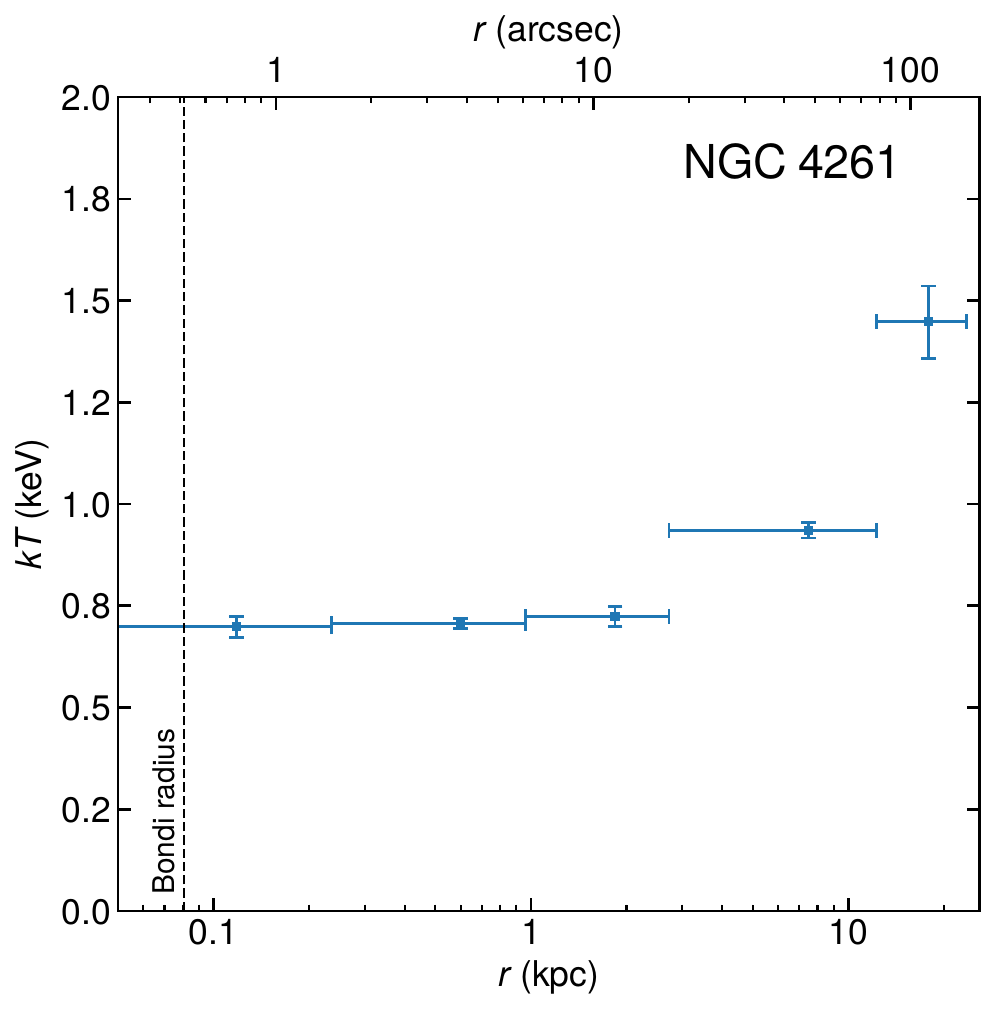}};
\draw (\figxk, \figyk) node {\includegraphics[scale=\figscale]{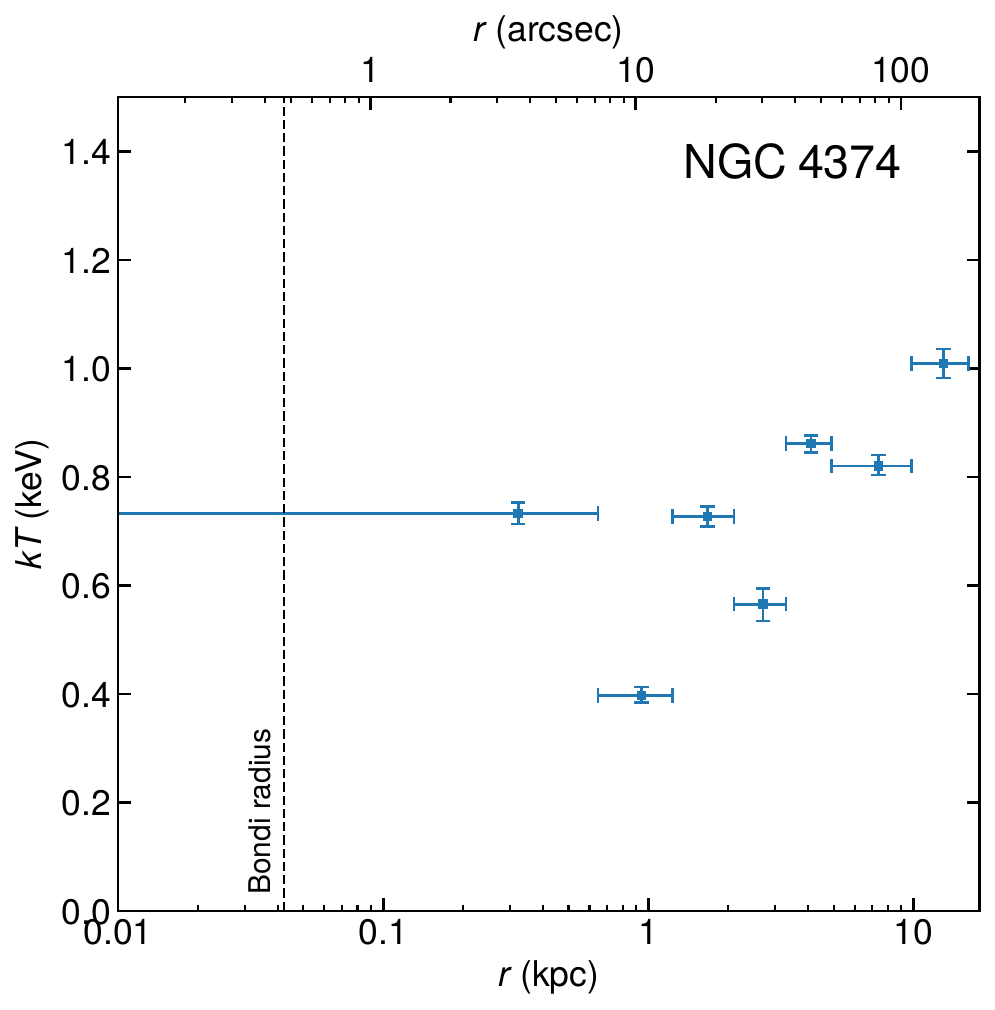}};
\draw (\figxi, \figyl) node {\includegraphics[scale=\figscale]{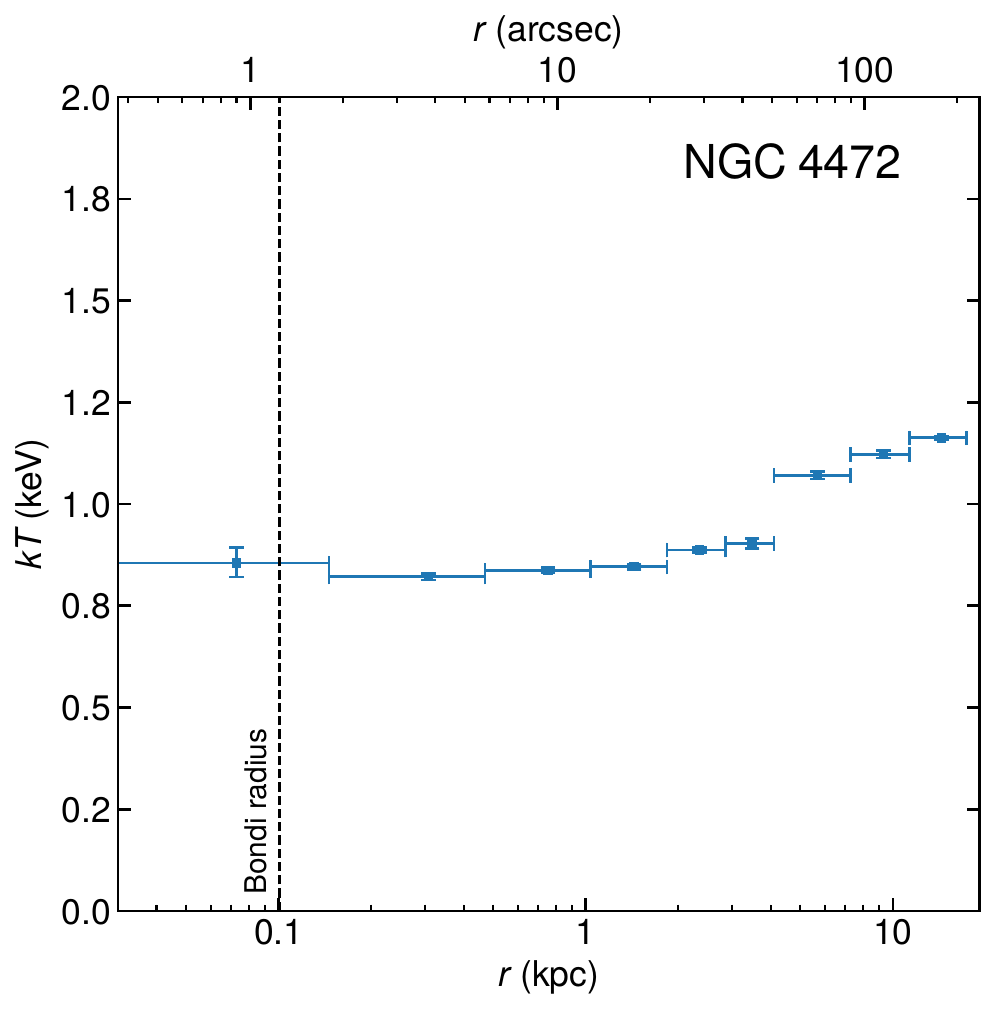}};
\draw (\figxj, \figyl) node {\includegraphics[scale=\figscale]{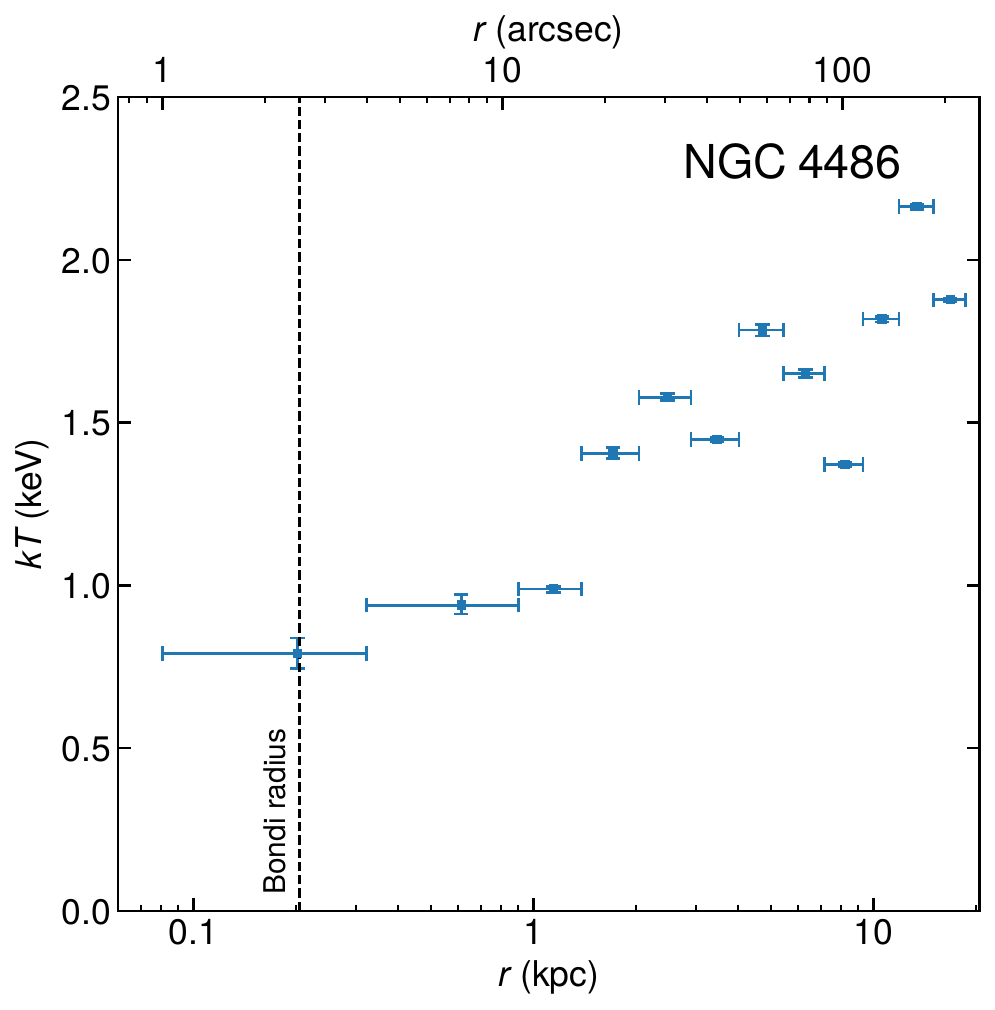}};
\draw (\figxk, \figyl) node {\includegraphics[scale=\figscale]{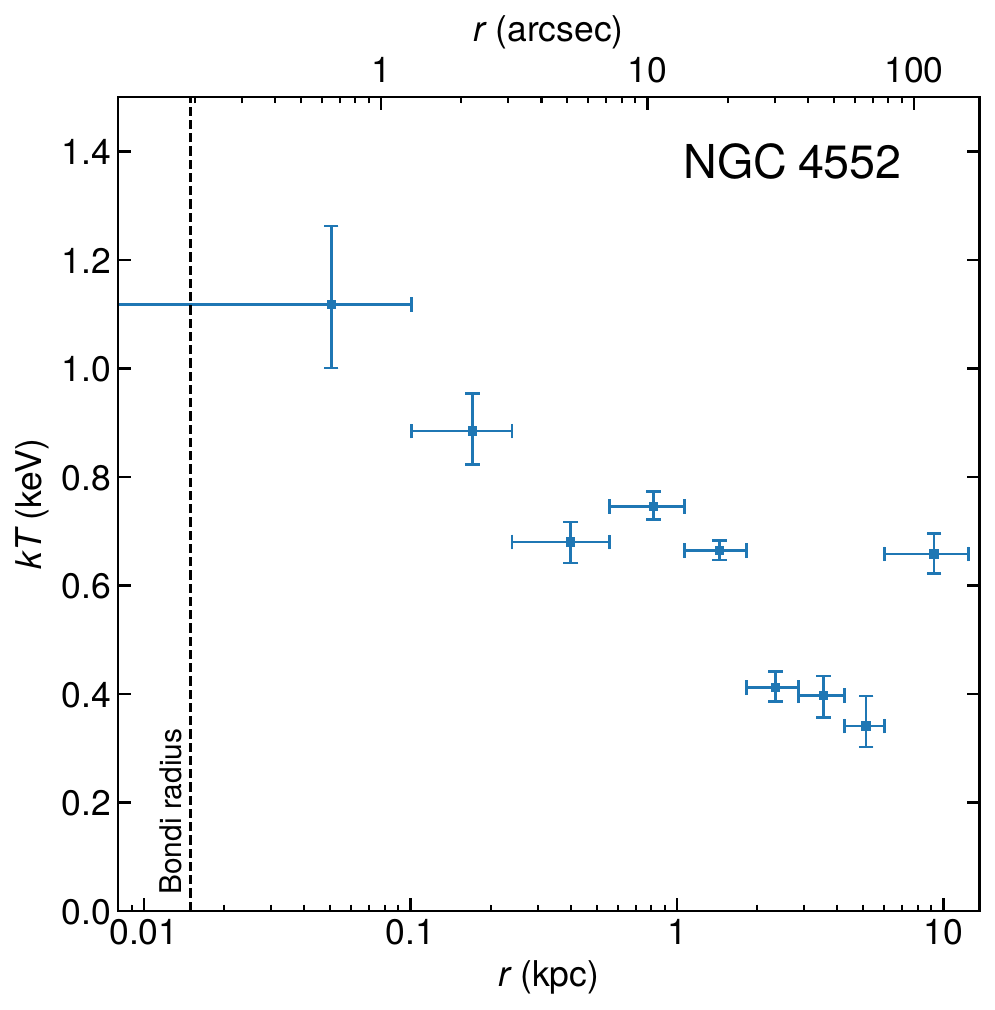}};
\end{tikzpicture}
\end{figure}

\begin{figure*}
\begin{tikzpicture}
\draw (\figxi, \figyi) node {\includegraphics[scale=\figscale]{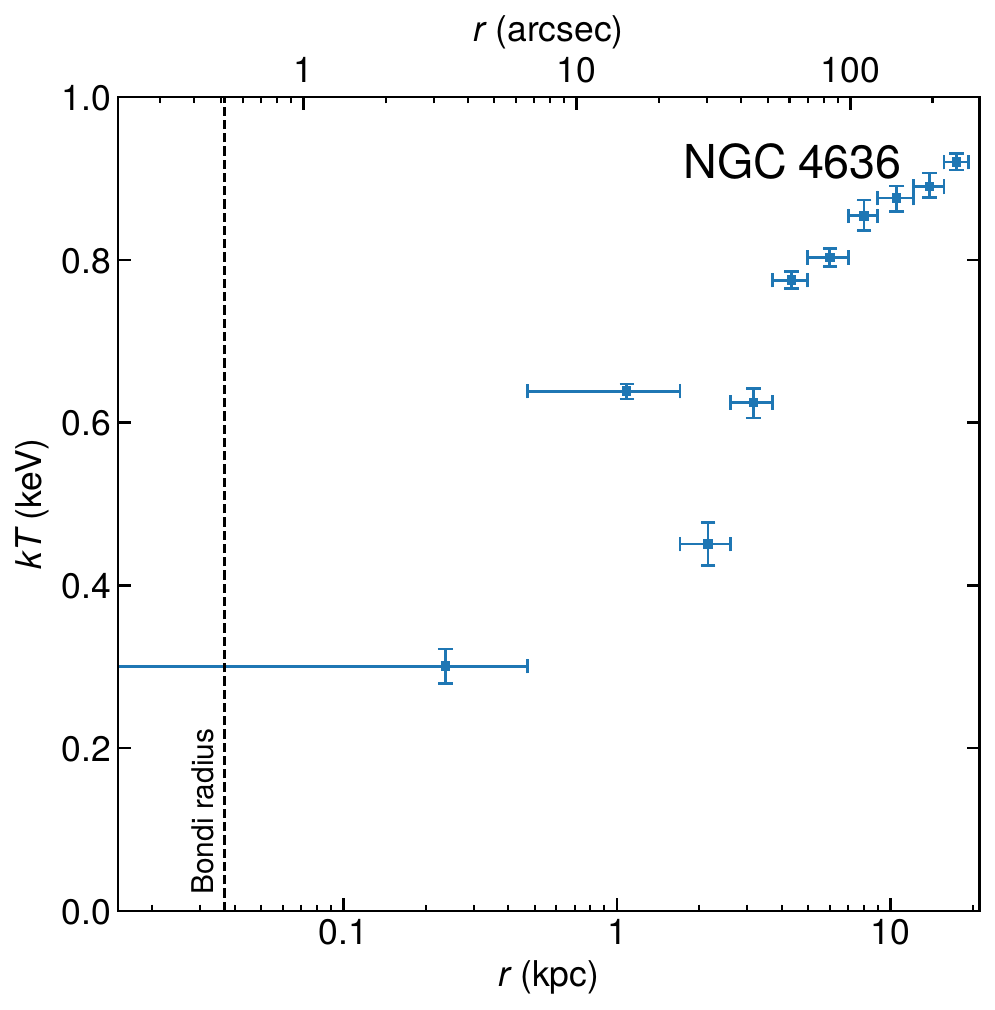}};
\draw (\figxj, \figyi) node {\includegraphics[scale=\figscale]{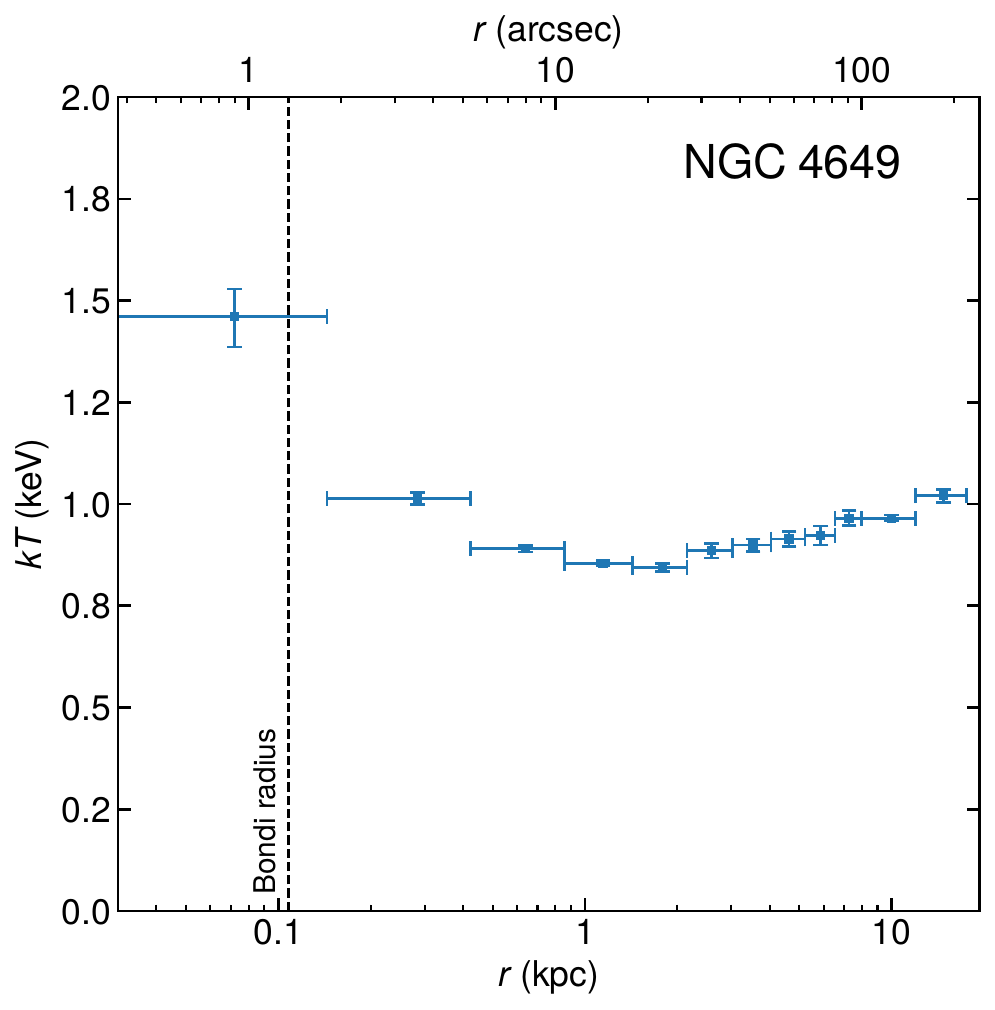}};
\draw (\figxk, \figyi) node {\includegraphics[scale=\figscale]{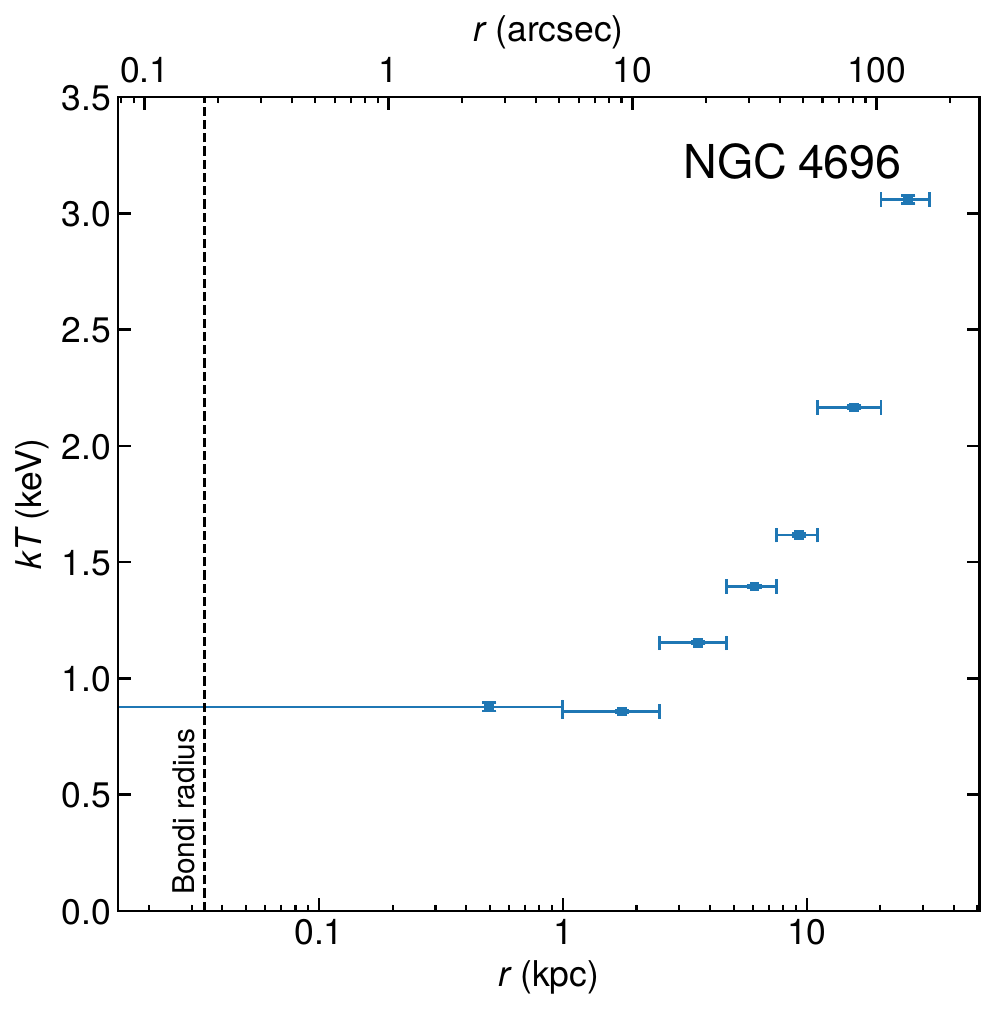}};
\draw (\figxi, \figyj) node {\includegraphics[scale=\figscale]{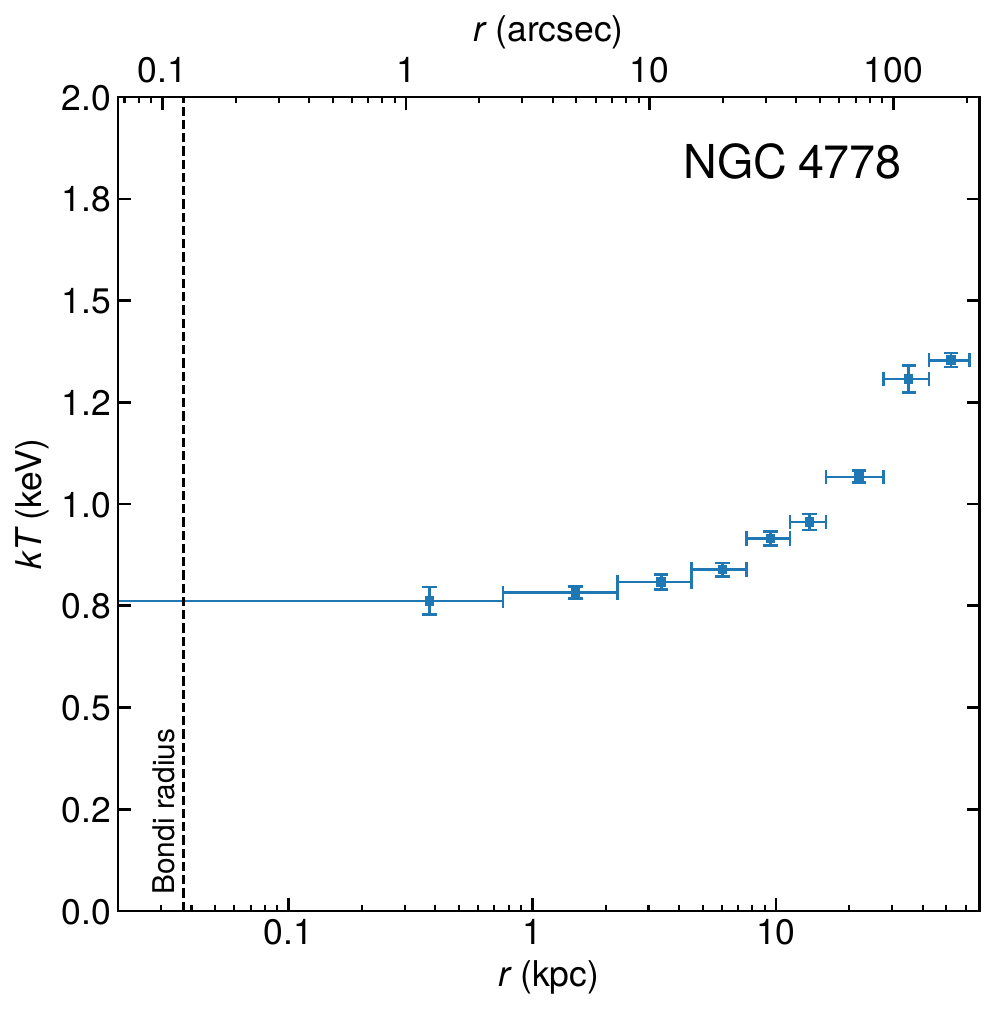}};
\draw (\figxj, \figyj) node {\includegraphics[scale=\figscale]{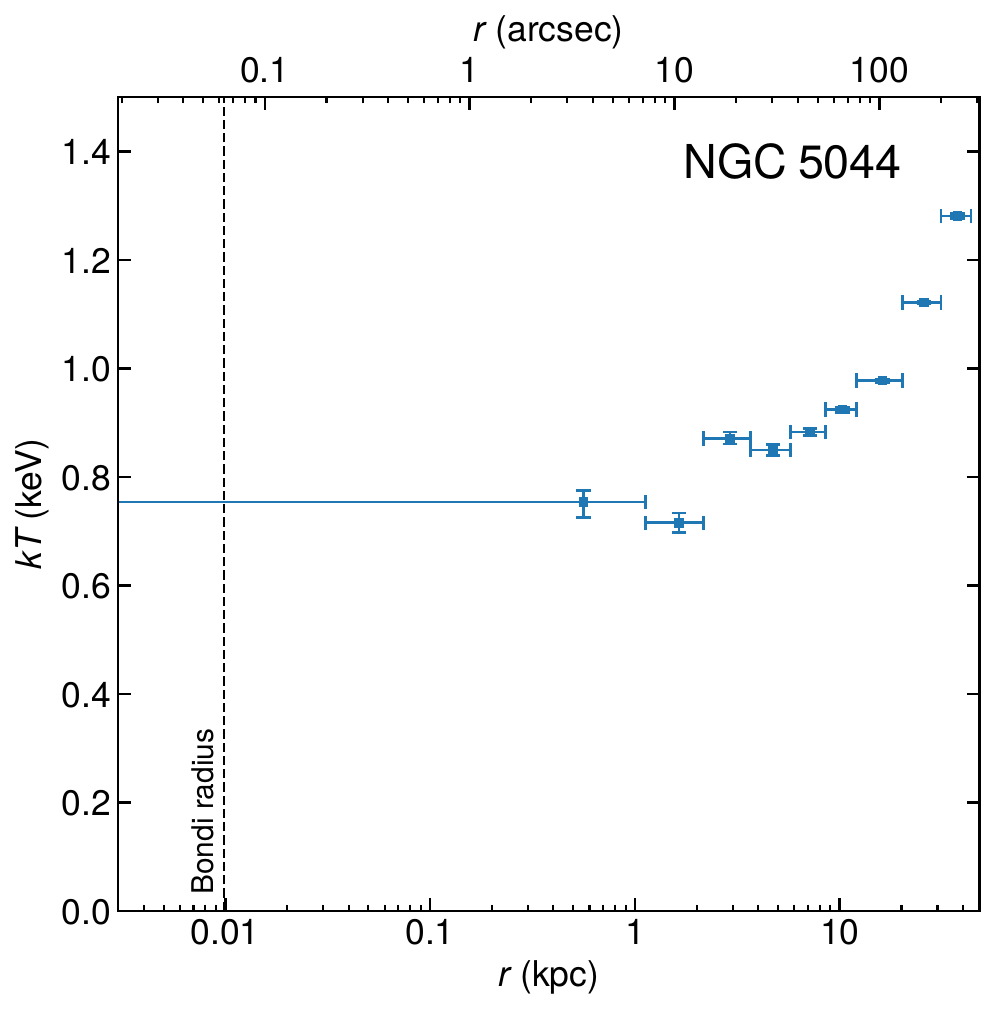}};
\draw (\figxk, \figyj) node {\includegraphics[scale=\figscale]{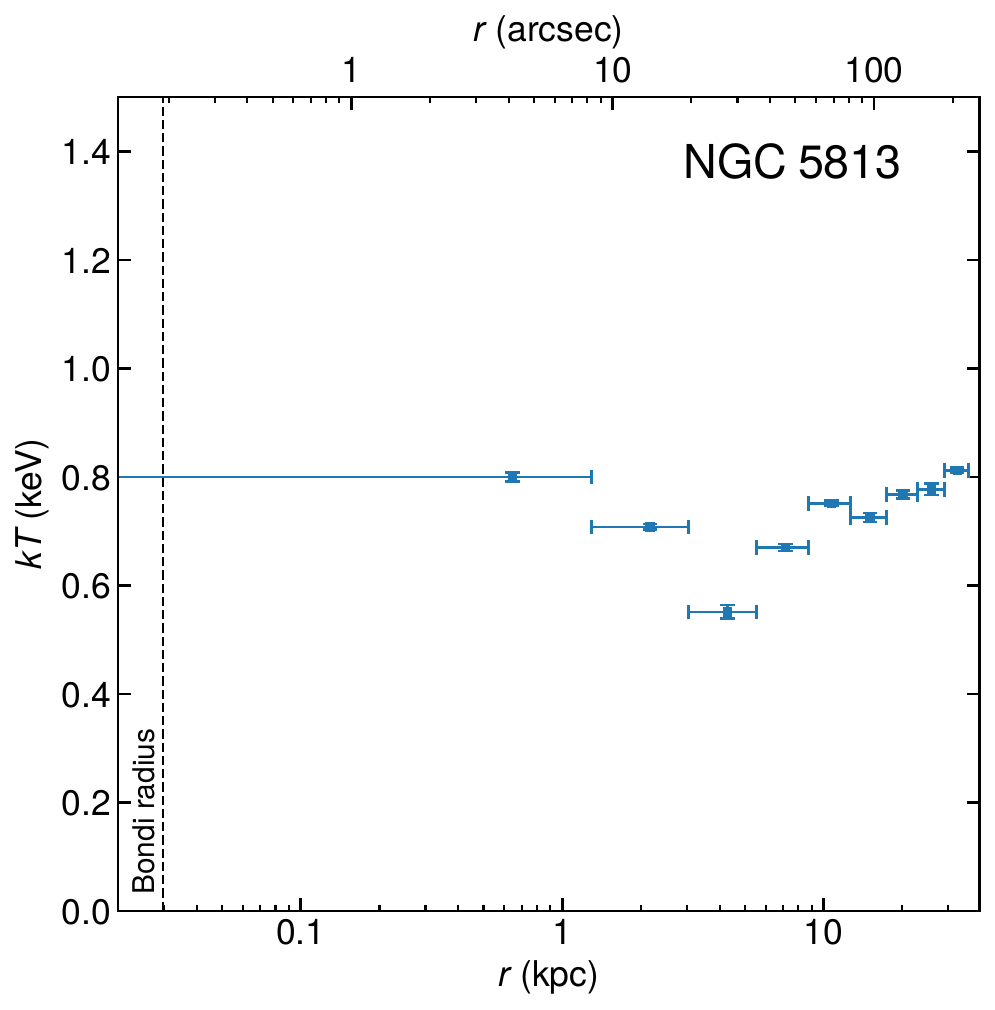}};
\draw (\figxi, \figyk) node {\includegraphics[scale=\figscale]{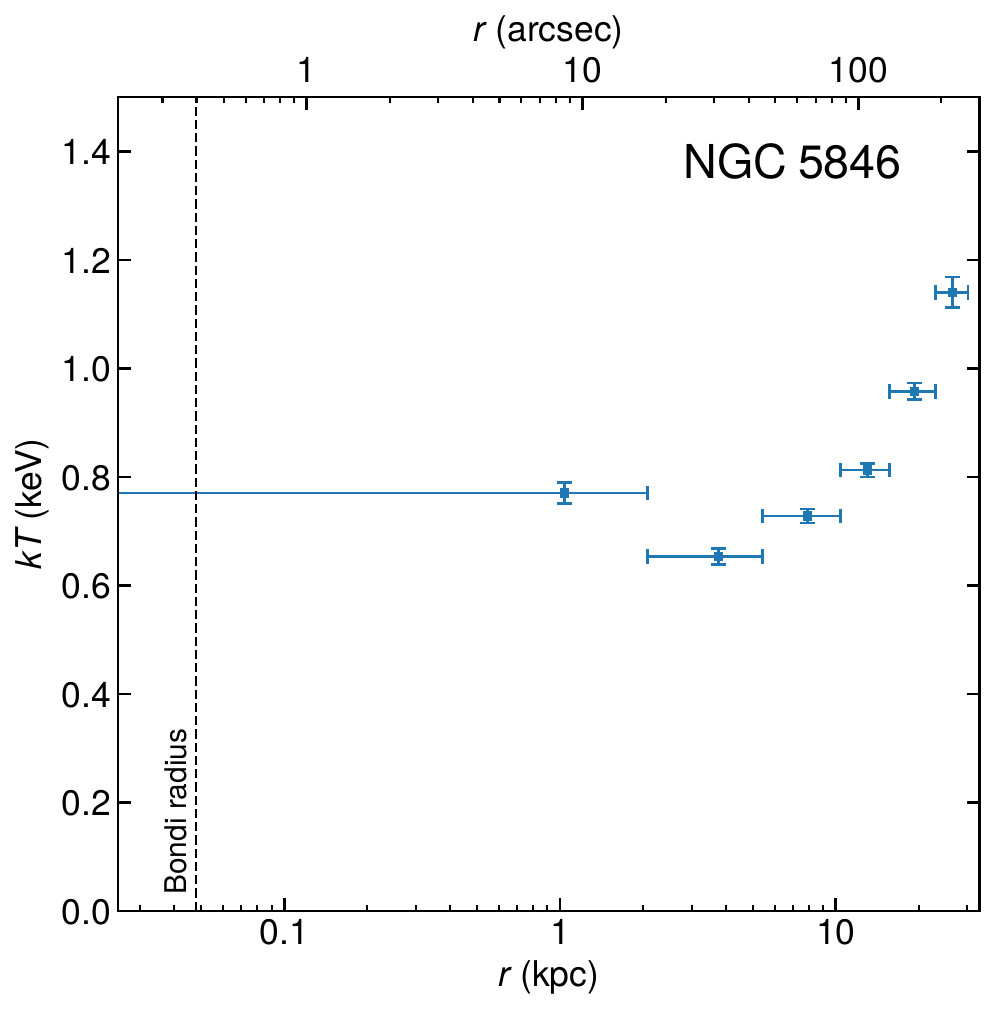}};
\draw (\figxj, \figyk) node {\includegraphics[scale=\figscale]{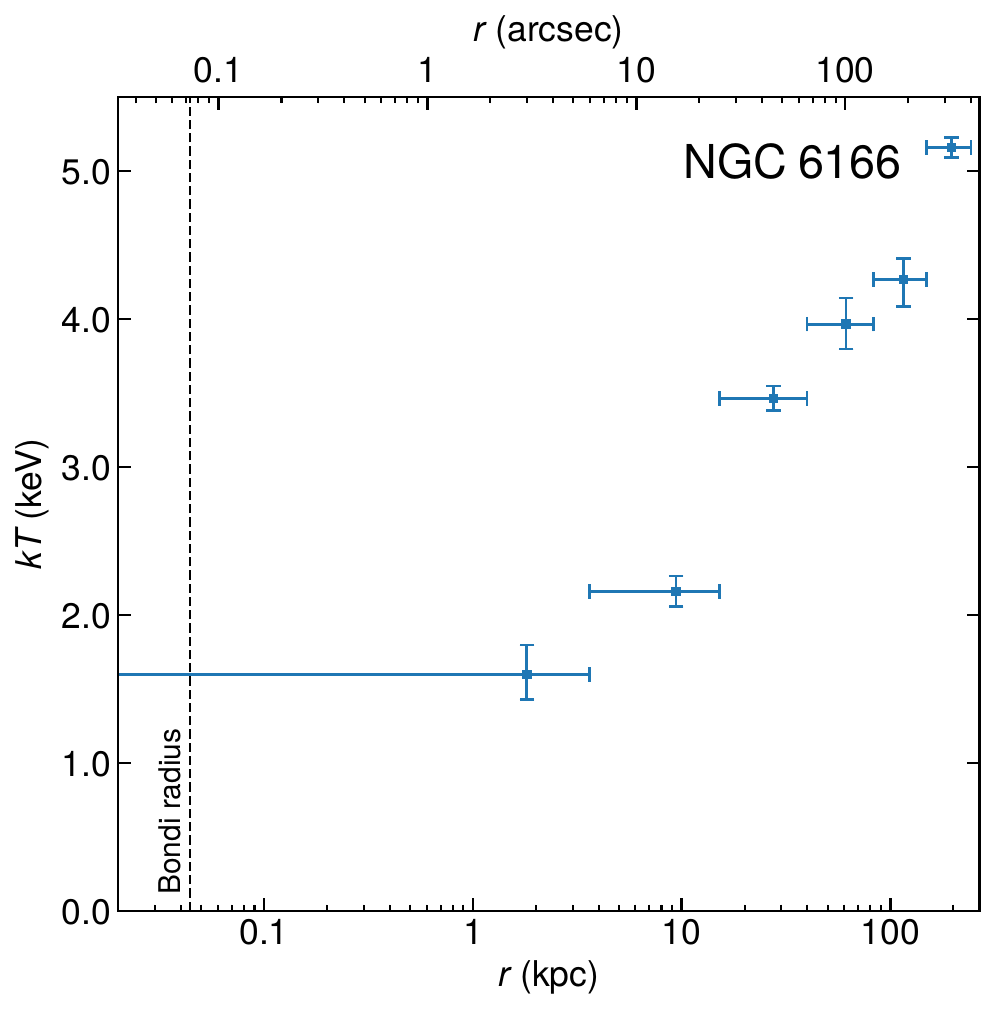}};
\end{tikzpicture}
\caption{Azimuthally averaged radial profiles of temperature $kT$. The vertical dashed line represents the Bondi radius $r_{\text{Bondi}}$. For most galaxies, the temperatures of two or more neighbouring radial bins were tied together. The temperature of the gas inside the Bondi radius was assumed to be the same as the temperature of the innermost radial bin.}
\label{fig:temperature}
\end{figure*}


\begin{figure}
\begin{tikzpicture}
    \draw (\figxi, \figyi) node {\includegraphics[scale=\figscale]{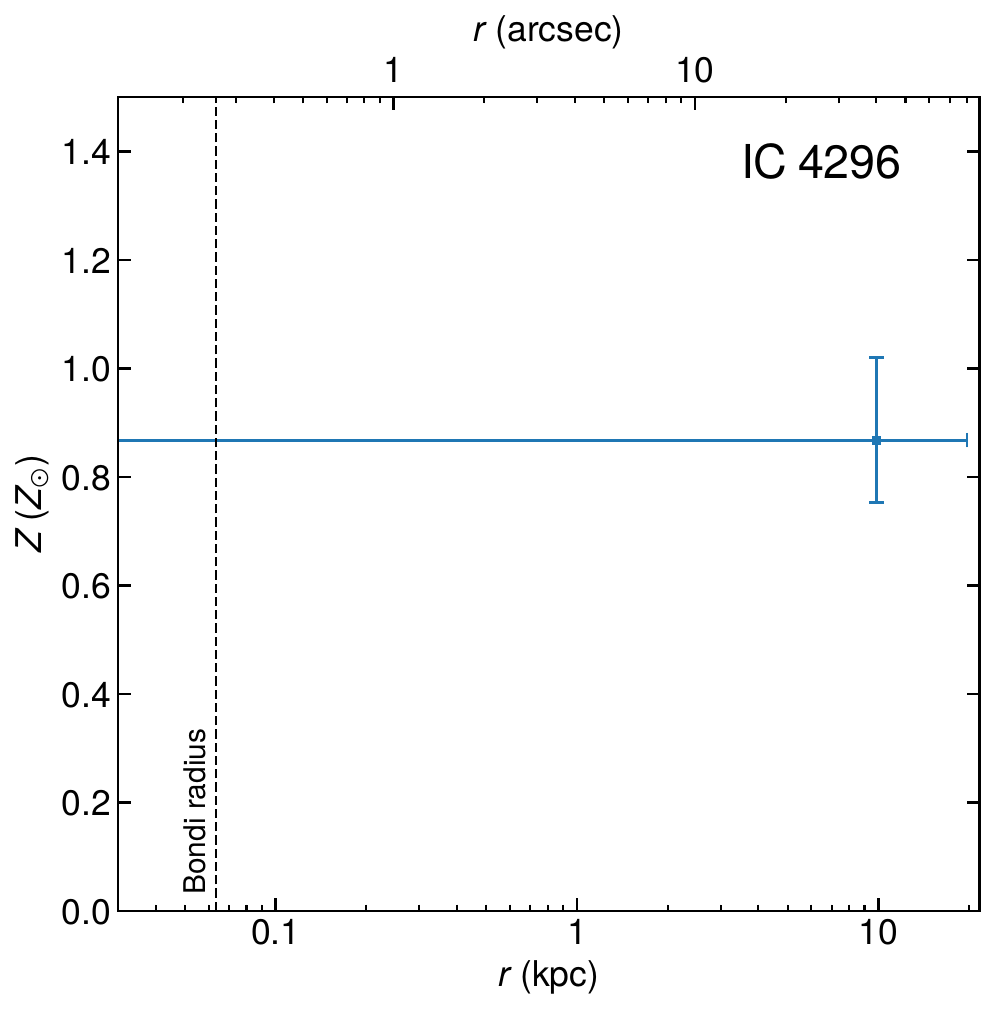}};
    \draw (\figxj, \figyi) node {\includegraphics[scale=\figscale]{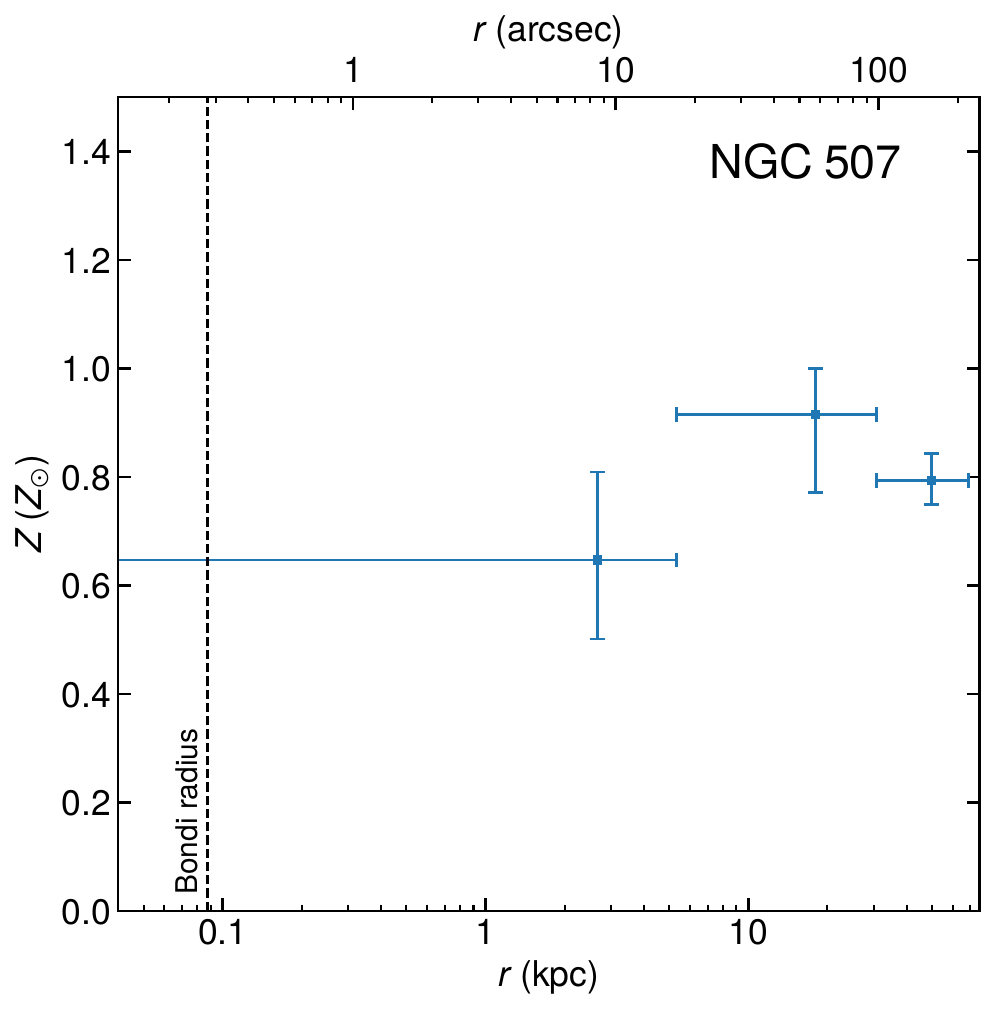}};
    \draw (\figxk, \figyi) node {\includegraphics[scale=\figscale]{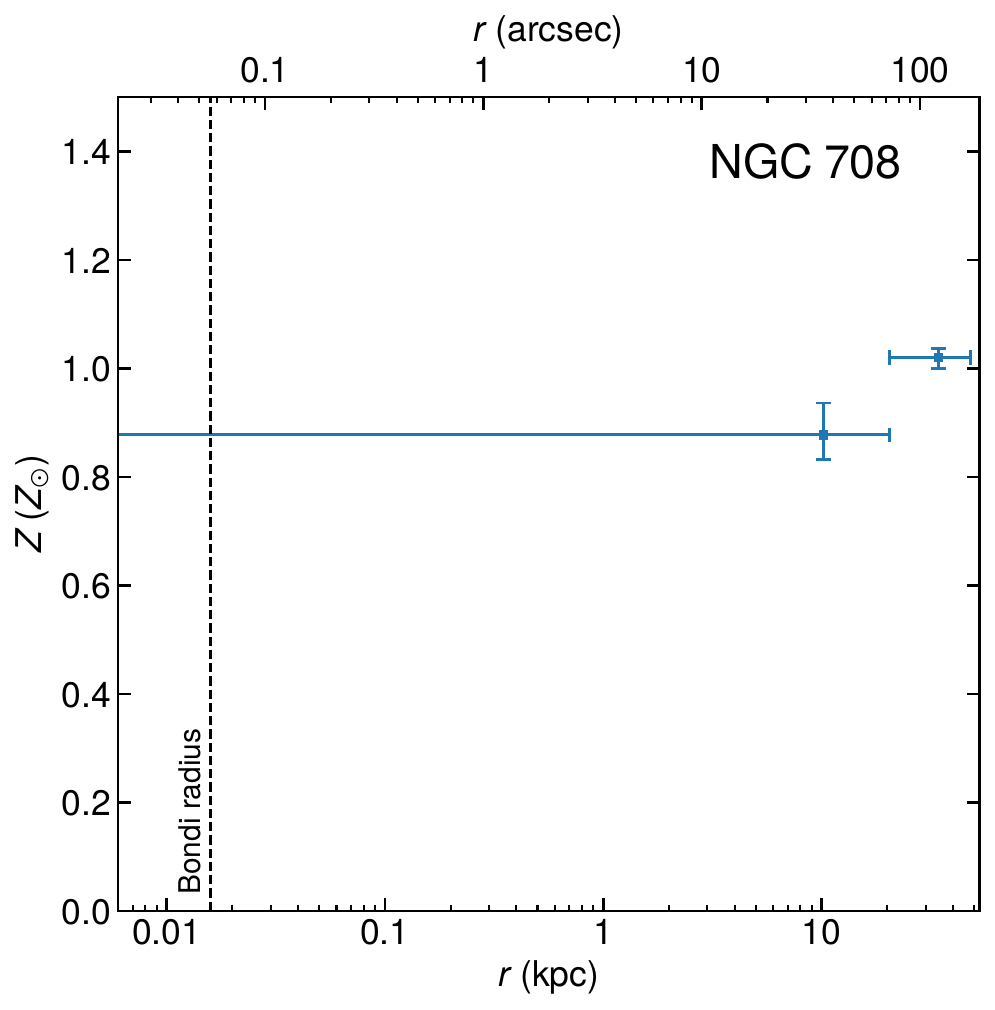}};
    \draw (\figxi, \figyj) node {\includegraphics[scale=\figscale]{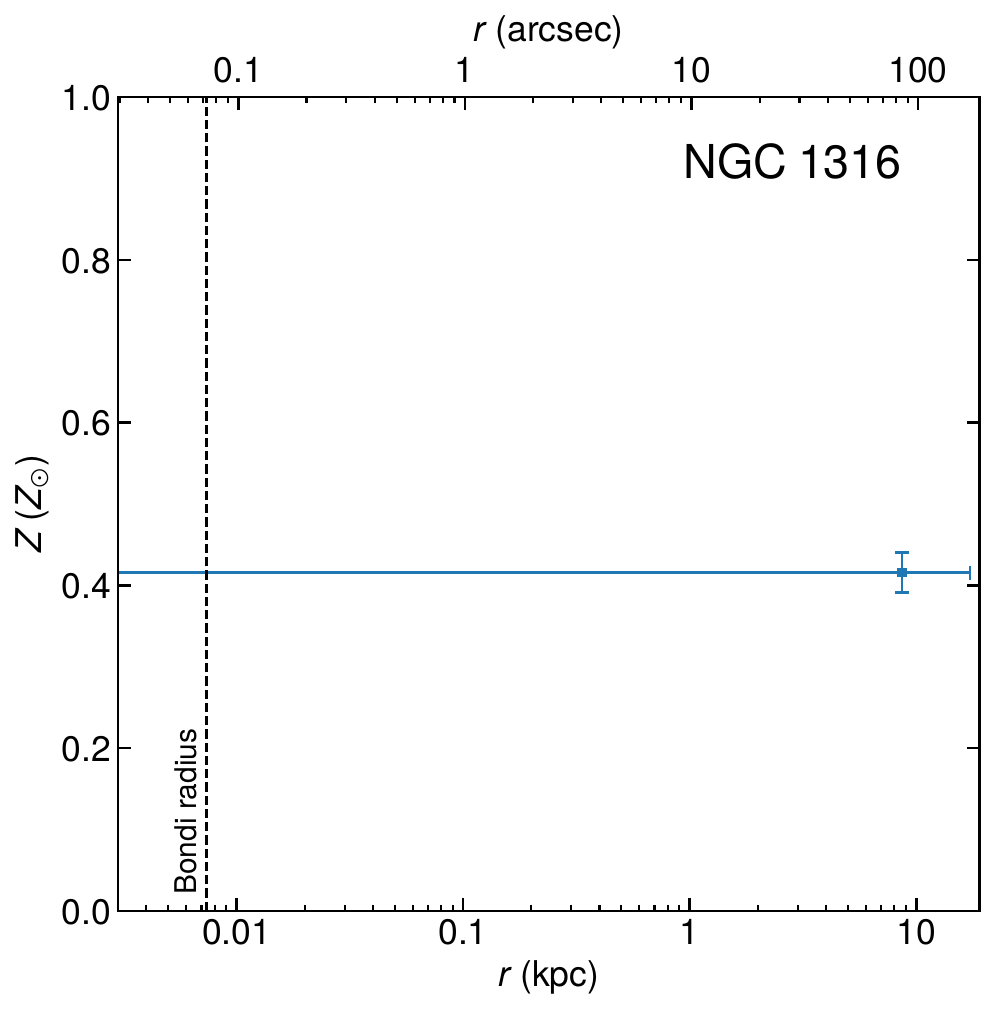}};
    \draw (\figxj, \figyj) node {\includegraphics[scale=\figscale]{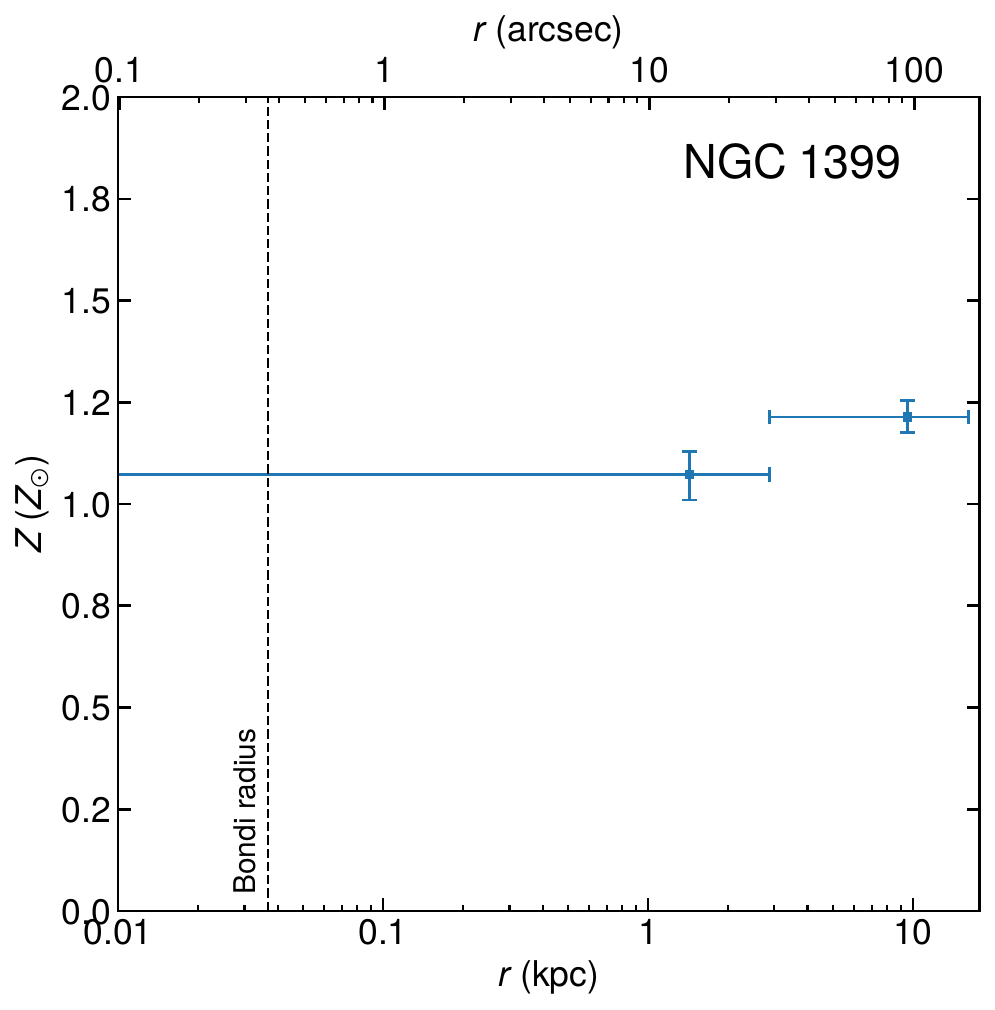}};
    \draw (\figxk, \figyj) node {\includegraphics[scale=\figscale]{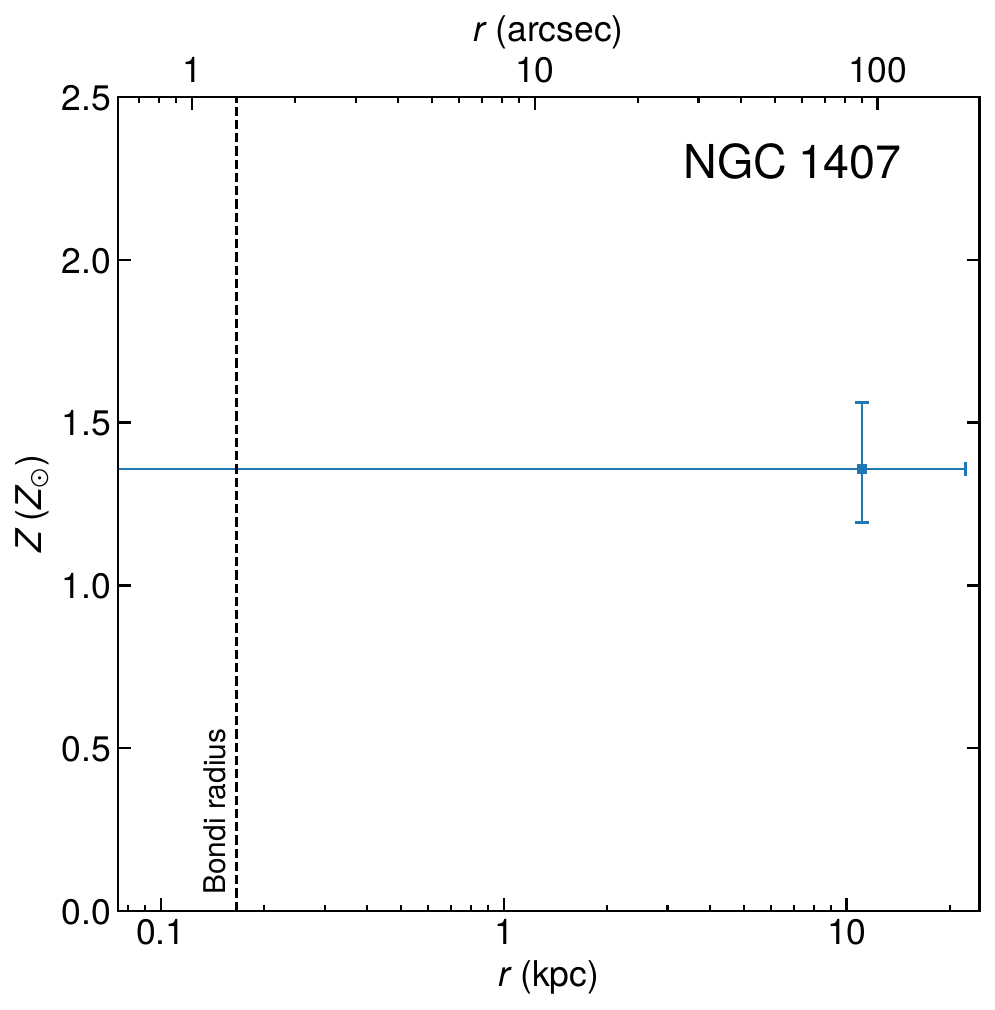}};
    \draw (\figxi, \figyk) node {\includegraphics[scale=\figscale]{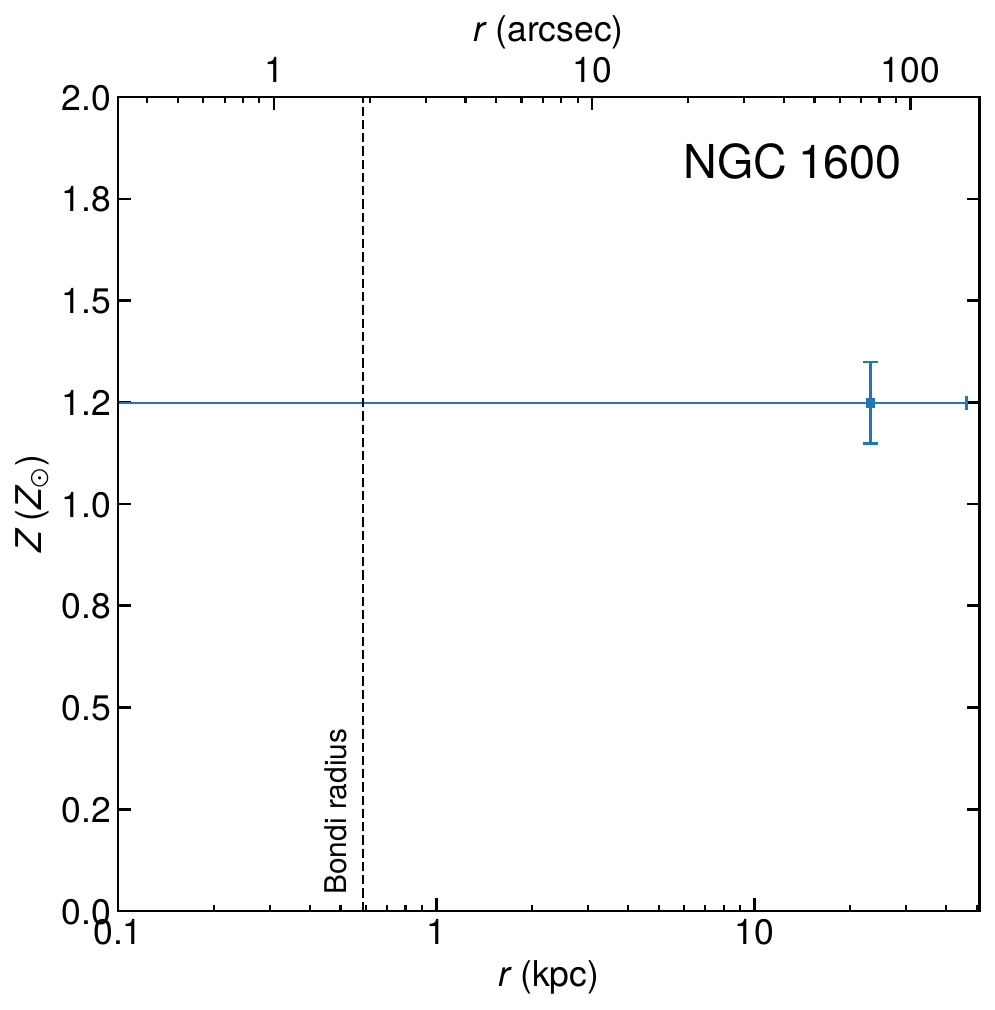}};
    \draw (\figxj, \figyk) node {\includegraphics[scale=\figscale]{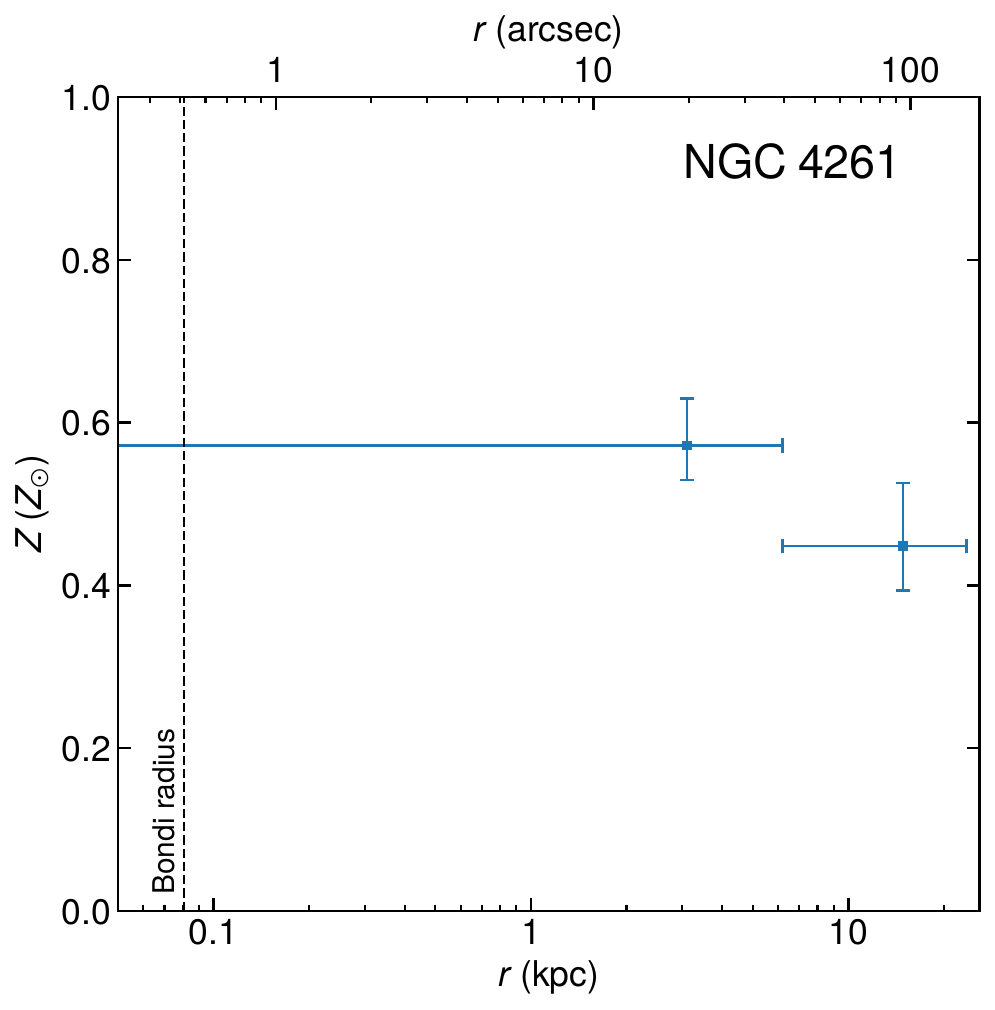}};
    \draw (\figxk, \figyk) node {\includegraphics[scale=\figscale]{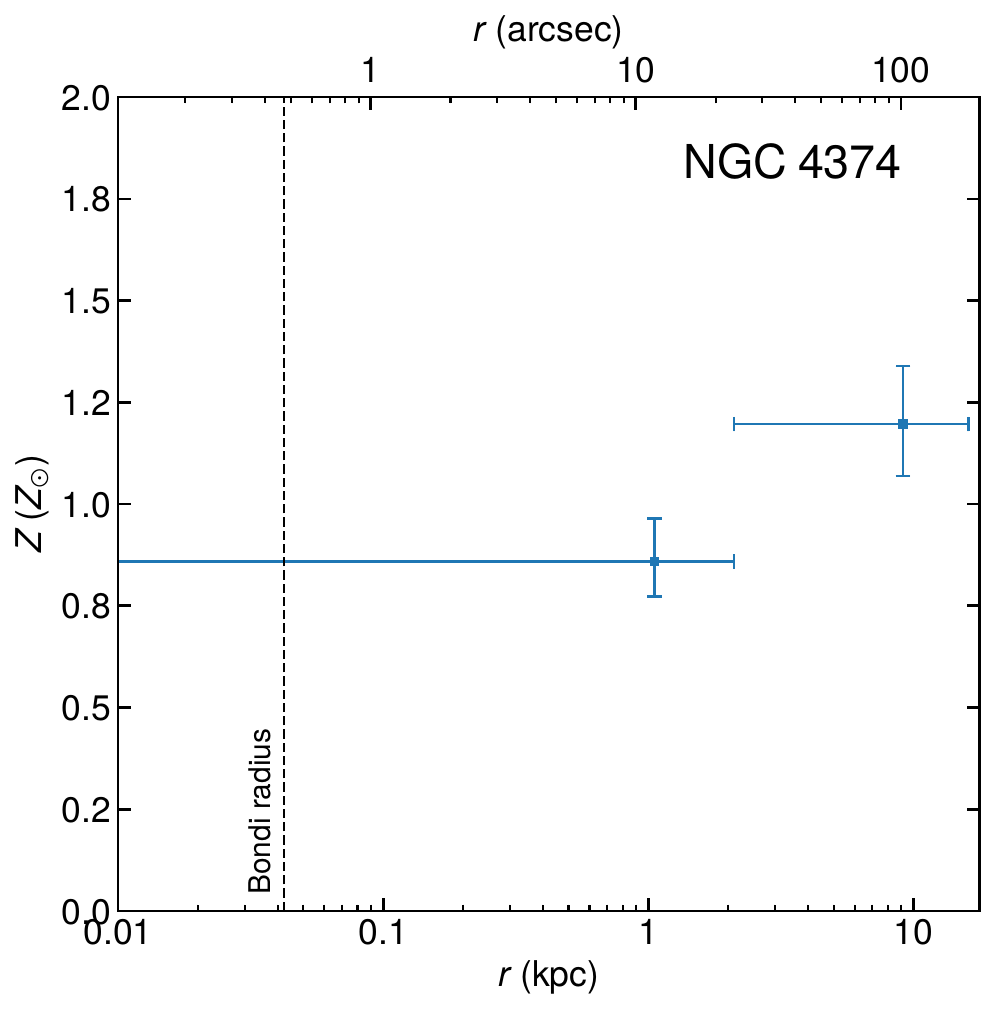}};
    \draw (\figxi, \figyl) node {\includegraphics[scale=\figscale]{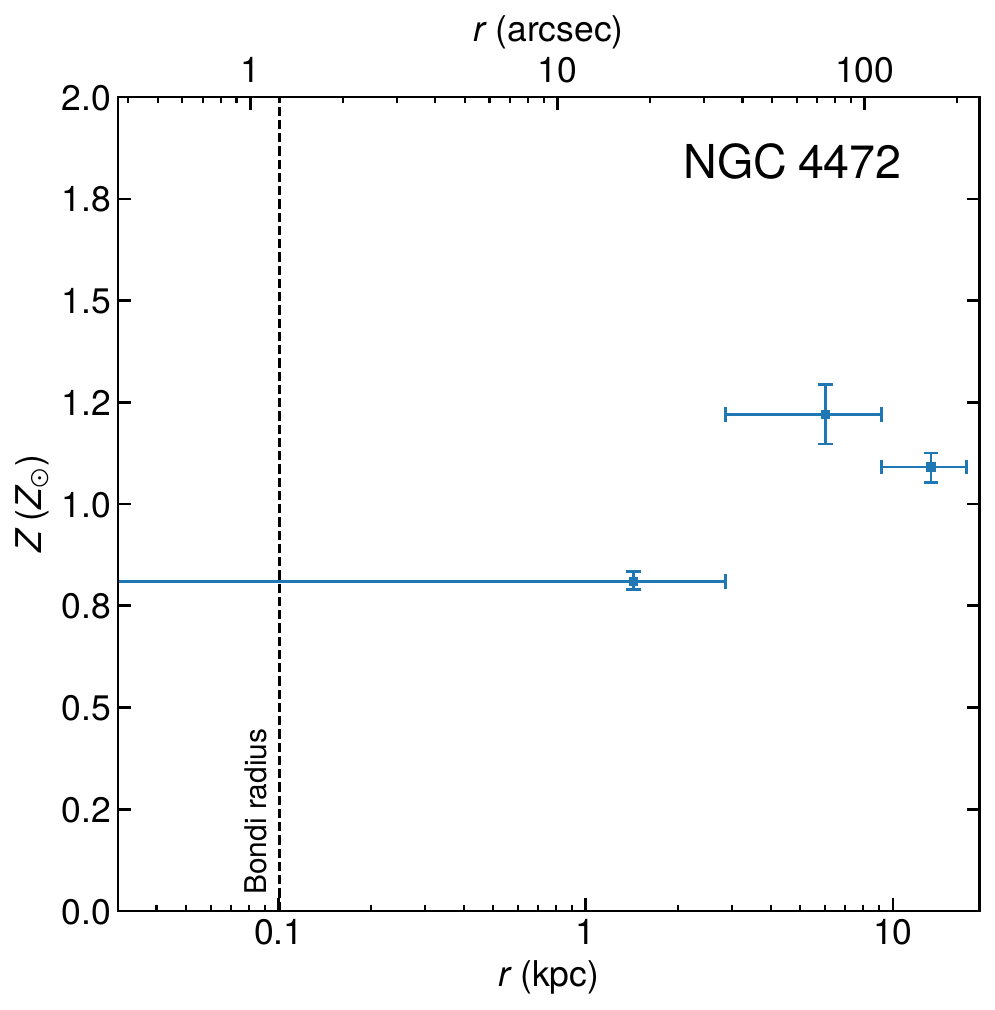}};
    \draw (\figxj, \figyl) node {\includegraphics[scale=\figscale]{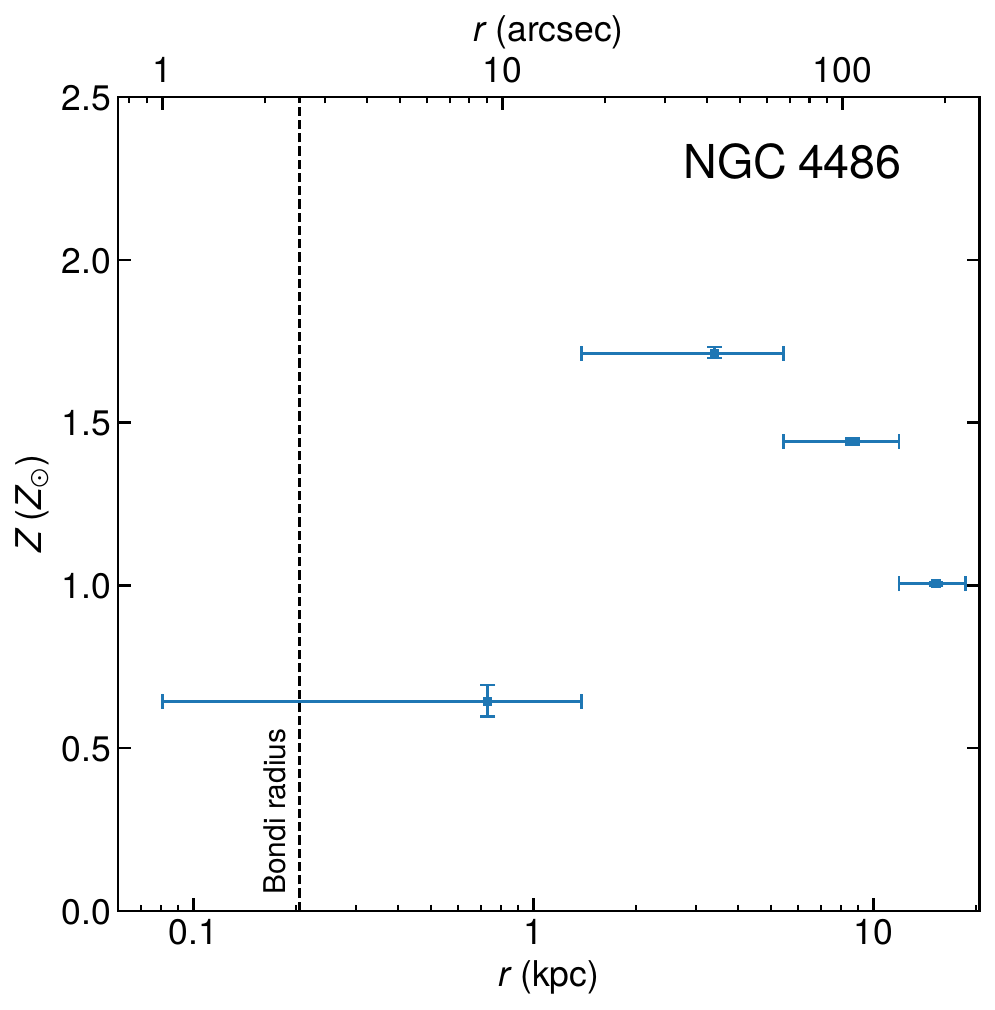}};
    \draw (\figxk, \figyl) node {\includegraphics[scale=\figscale]{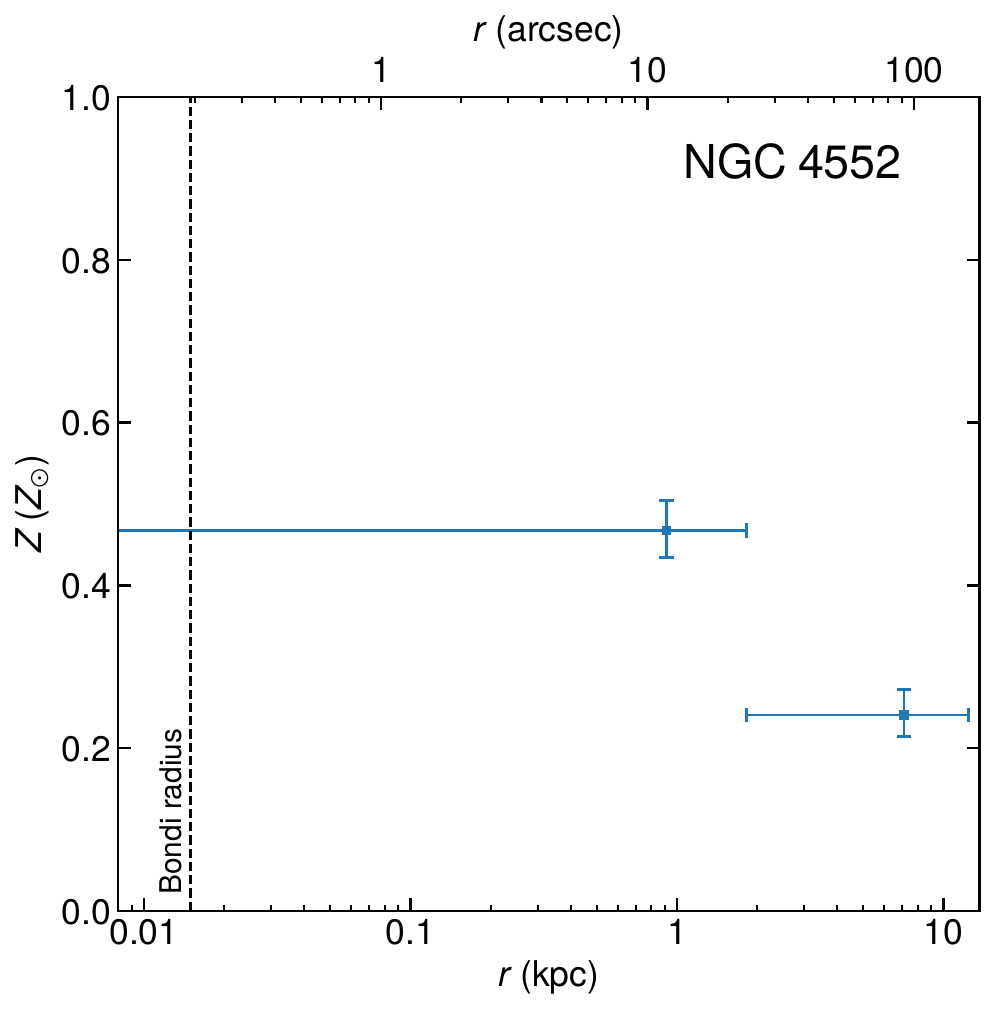}};
\end{tikzpicture}
\end{figure}

\begin{figure*}
\begin{tikzpicture}
    \draw (\figxi, \figyi) node {\includegraphics[scale=\figscale]{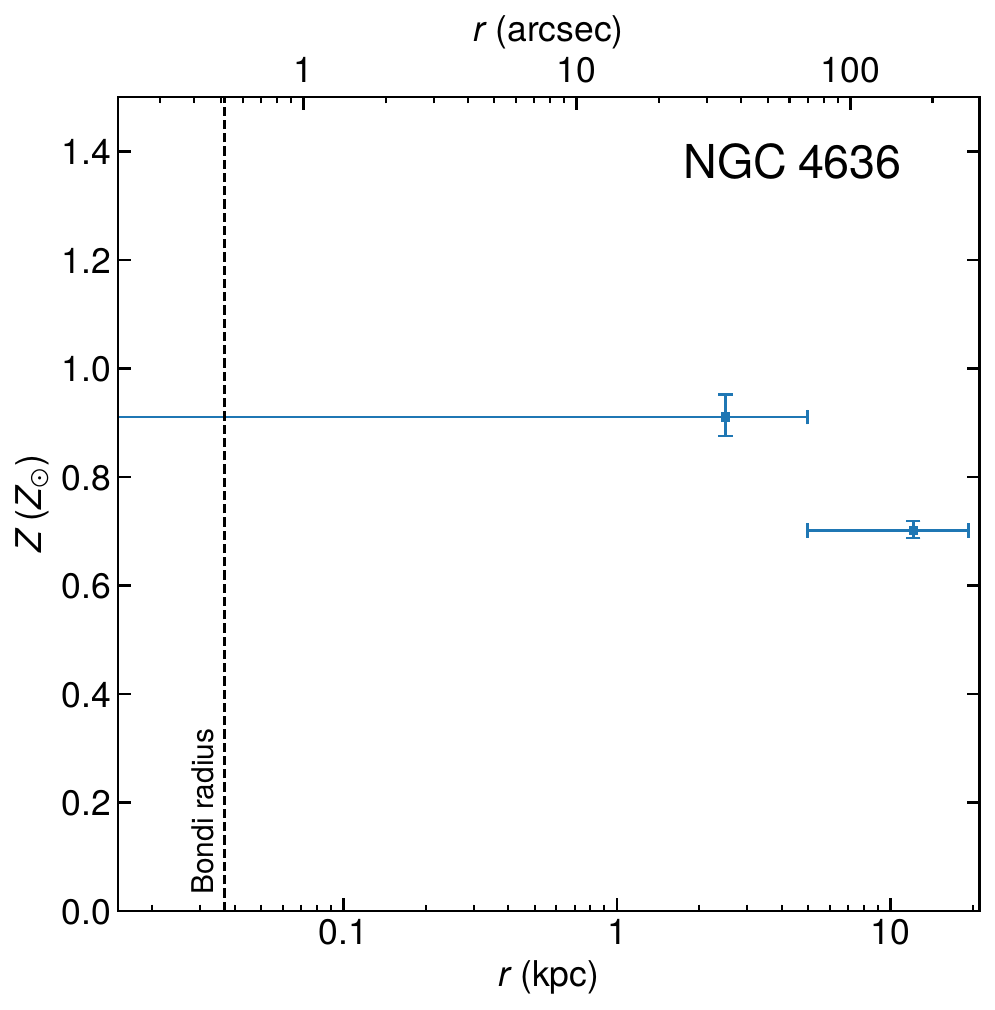}};
    \draw (\figxj, \figyi) node {\includegraphics[scale=\figscale]{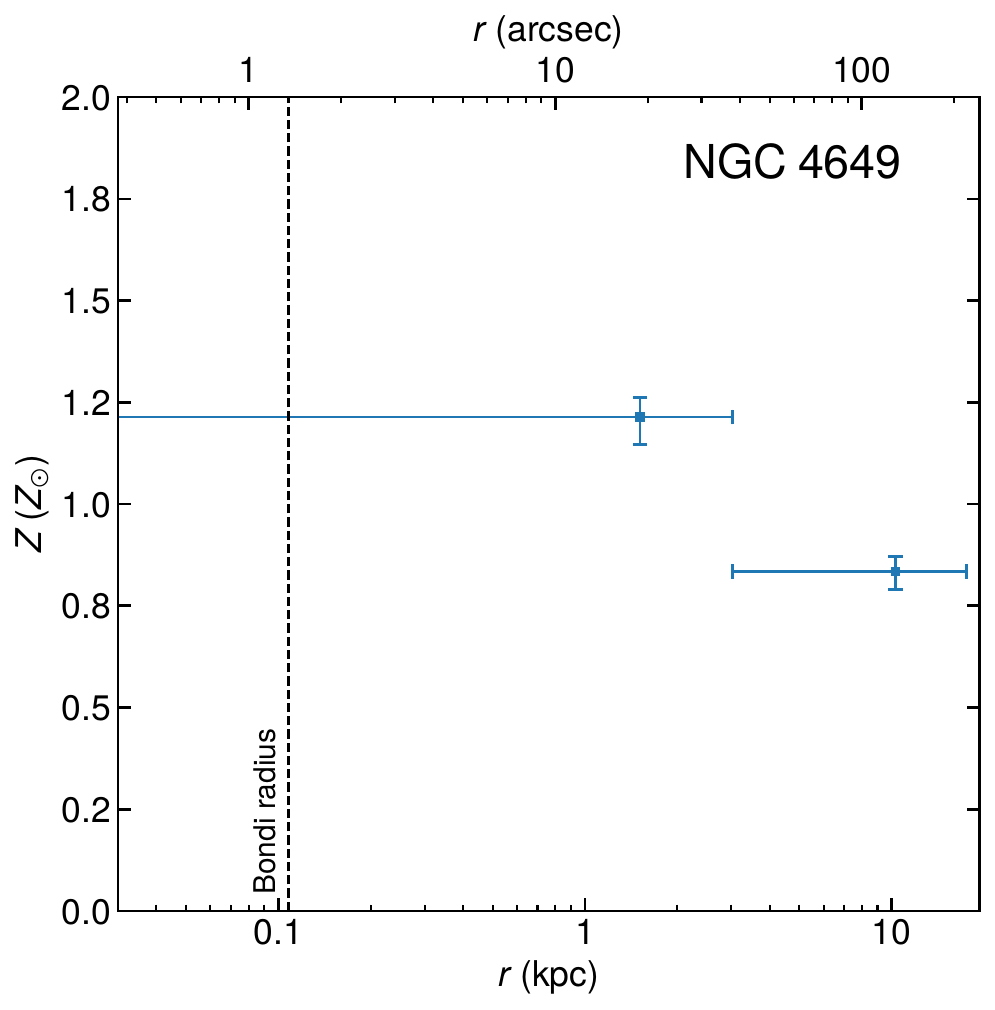}};
    \draw (\figxk, \figyi) node {\includegraphics[scale=\figscale]{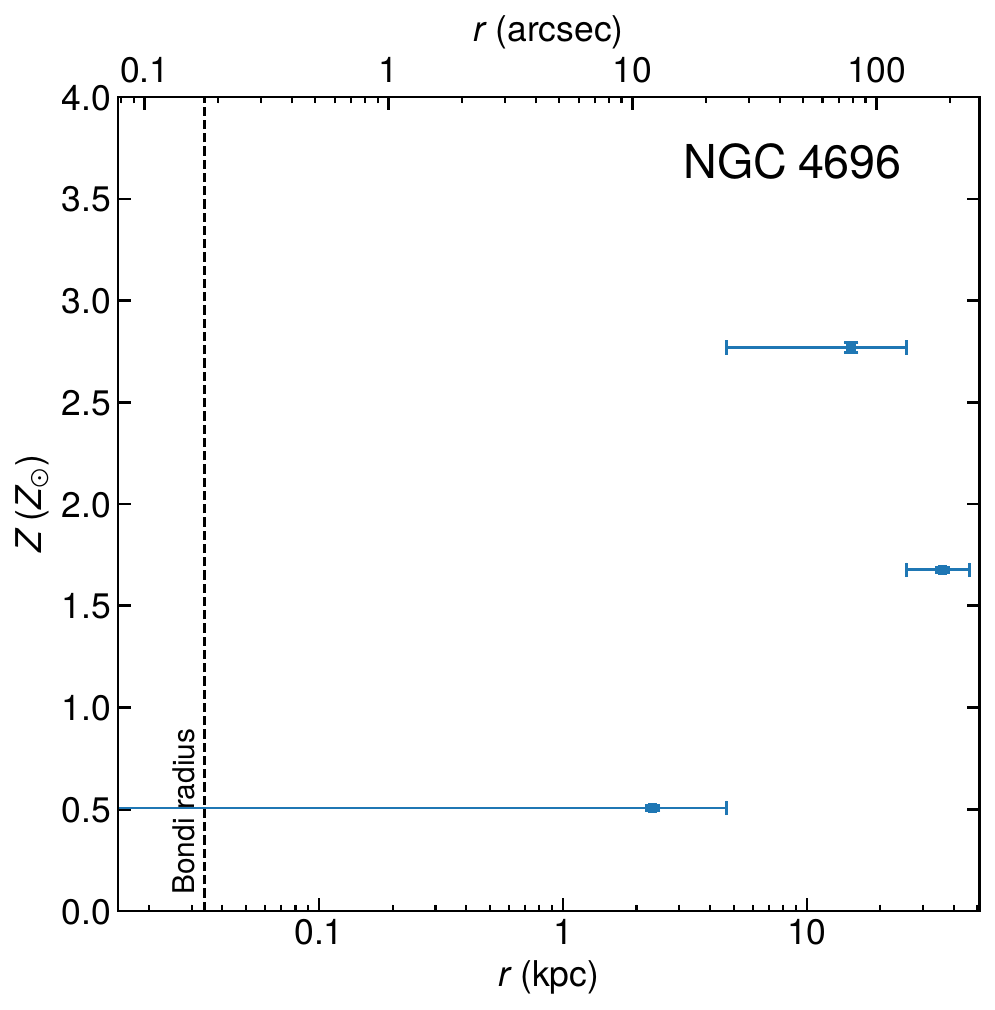}};
    \draw (\figxi, \figyj) node {\includegraphics[scale=\figscale]{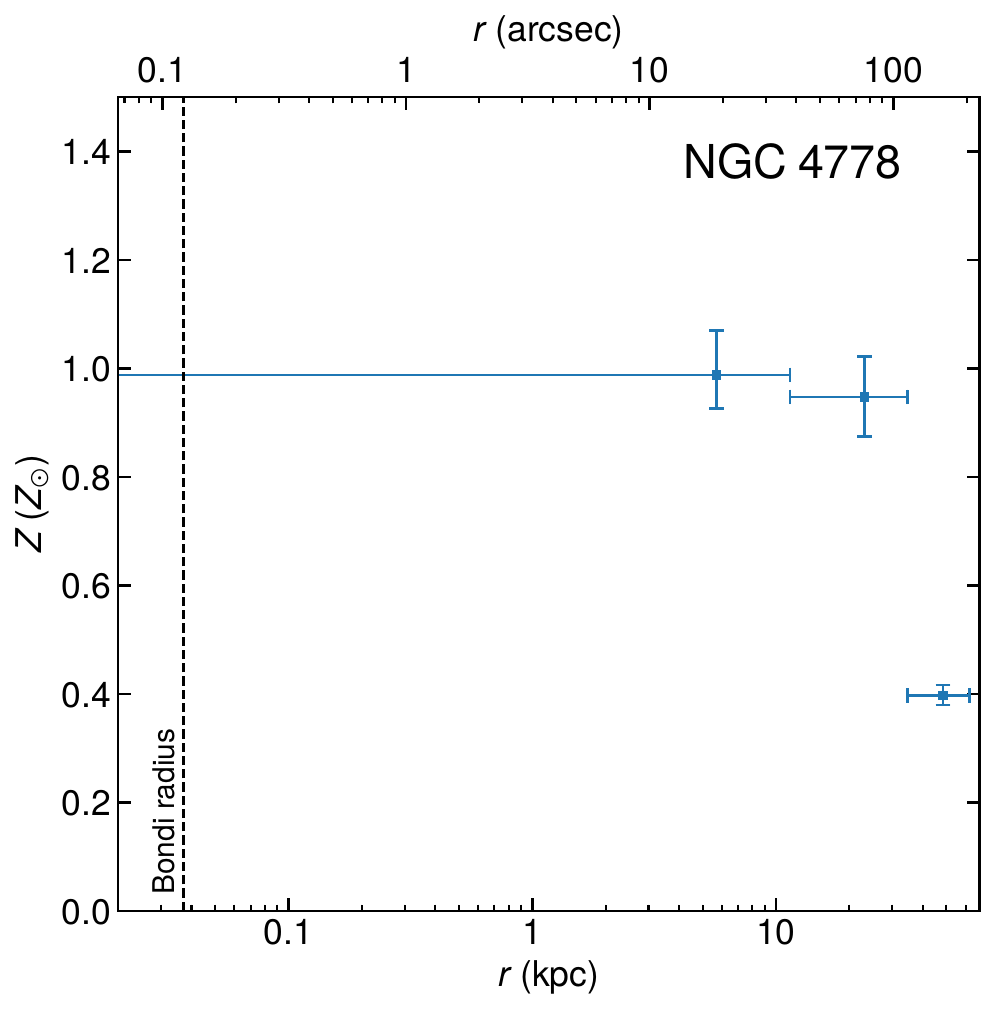}};
    \draw (\figxj, \figyj) node {\includegraphics[scale=\figscale]{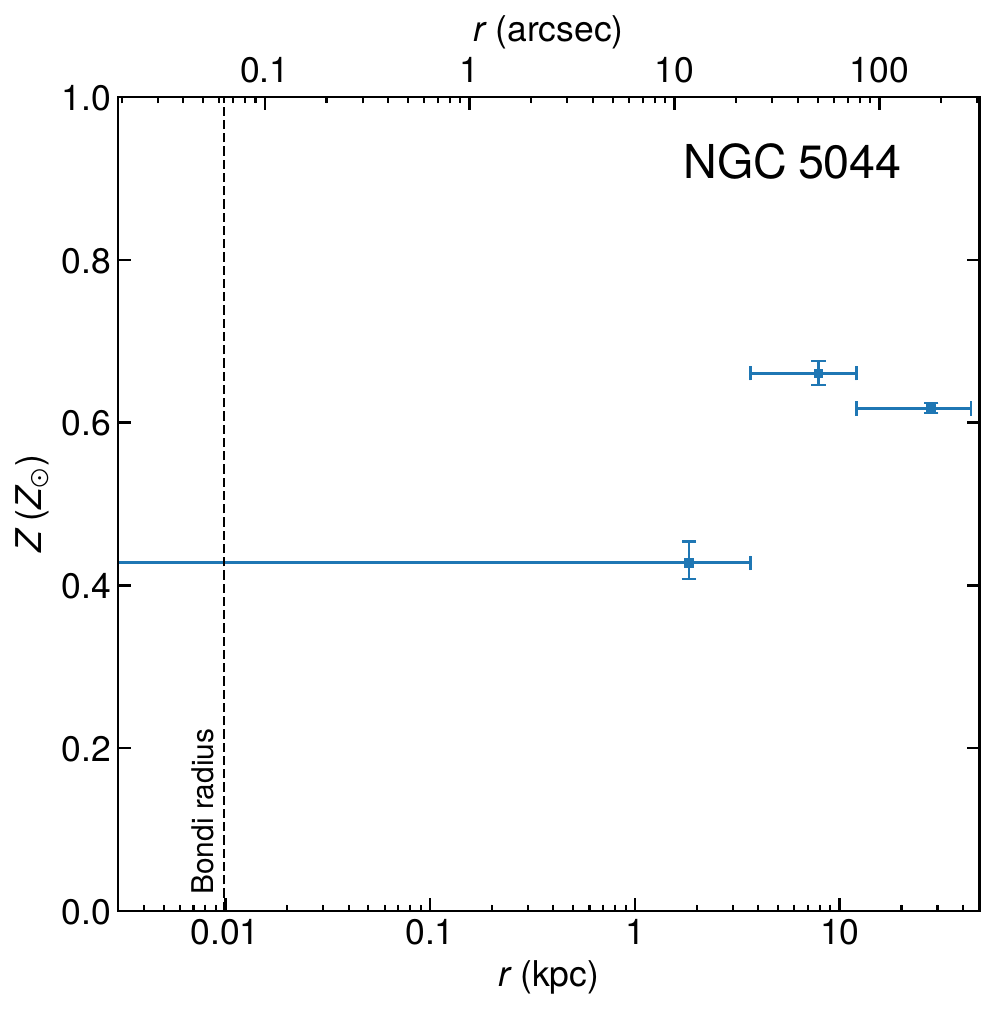}};
    \draw (\figxk, \figyj) node {\includegraphics[scale=\figscale]{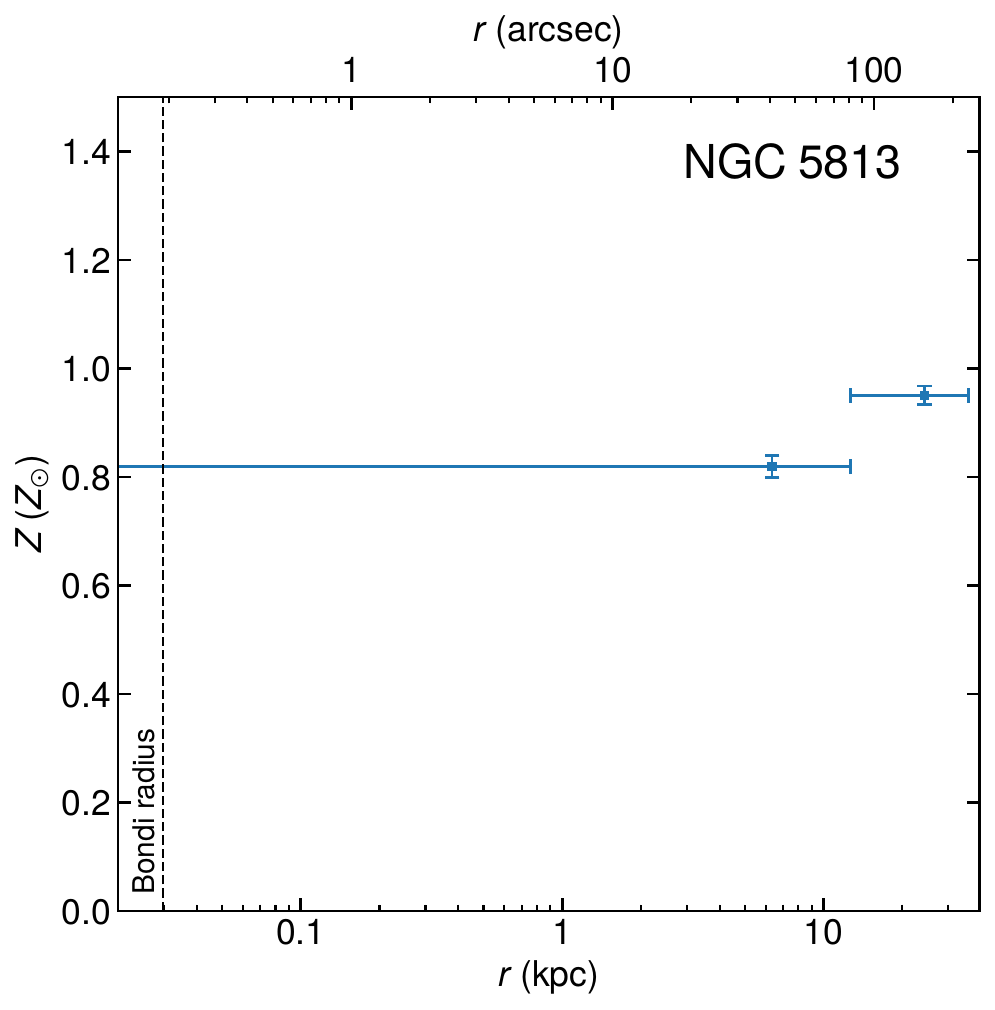}};
    \draw (\figxi, \figyk) node {\includegraphics[scale=\figscale]{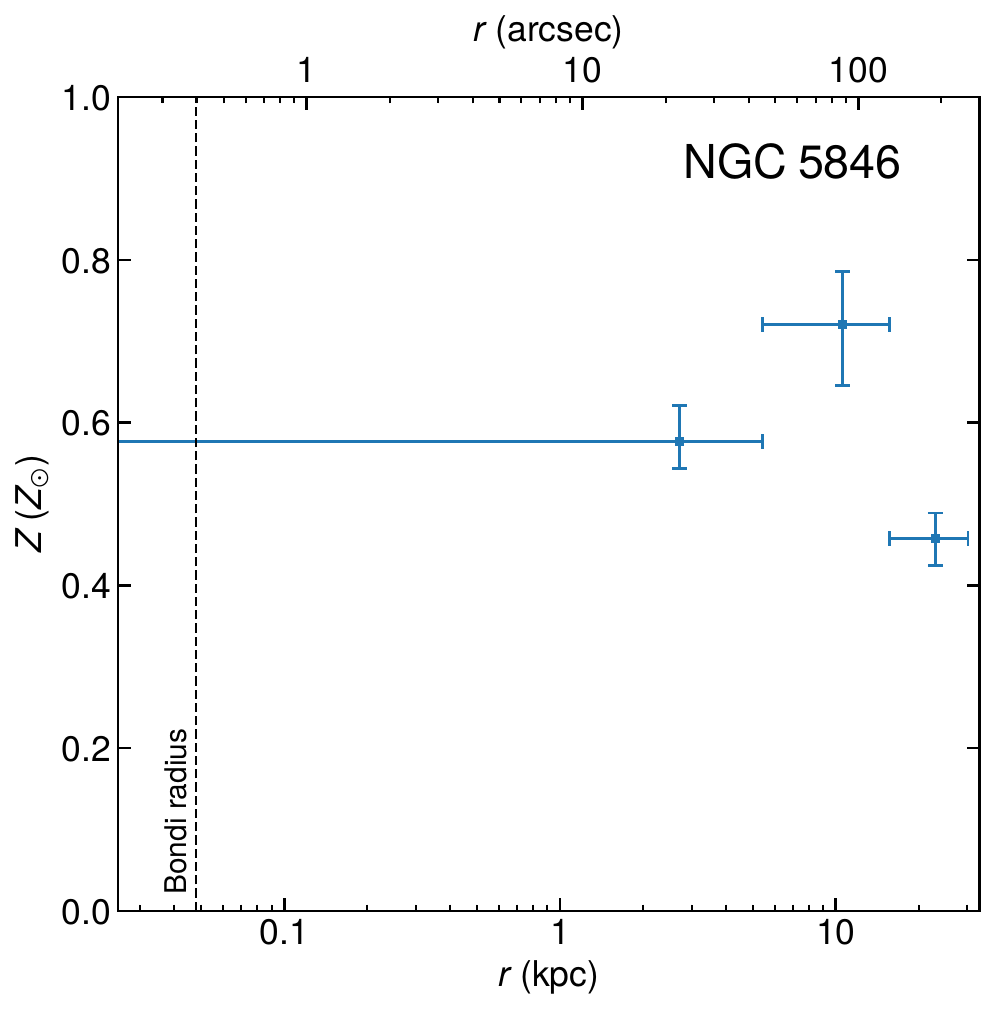}};
    \draw (\figxj, \figyk) node {\includegraphics[scale=\figscale]{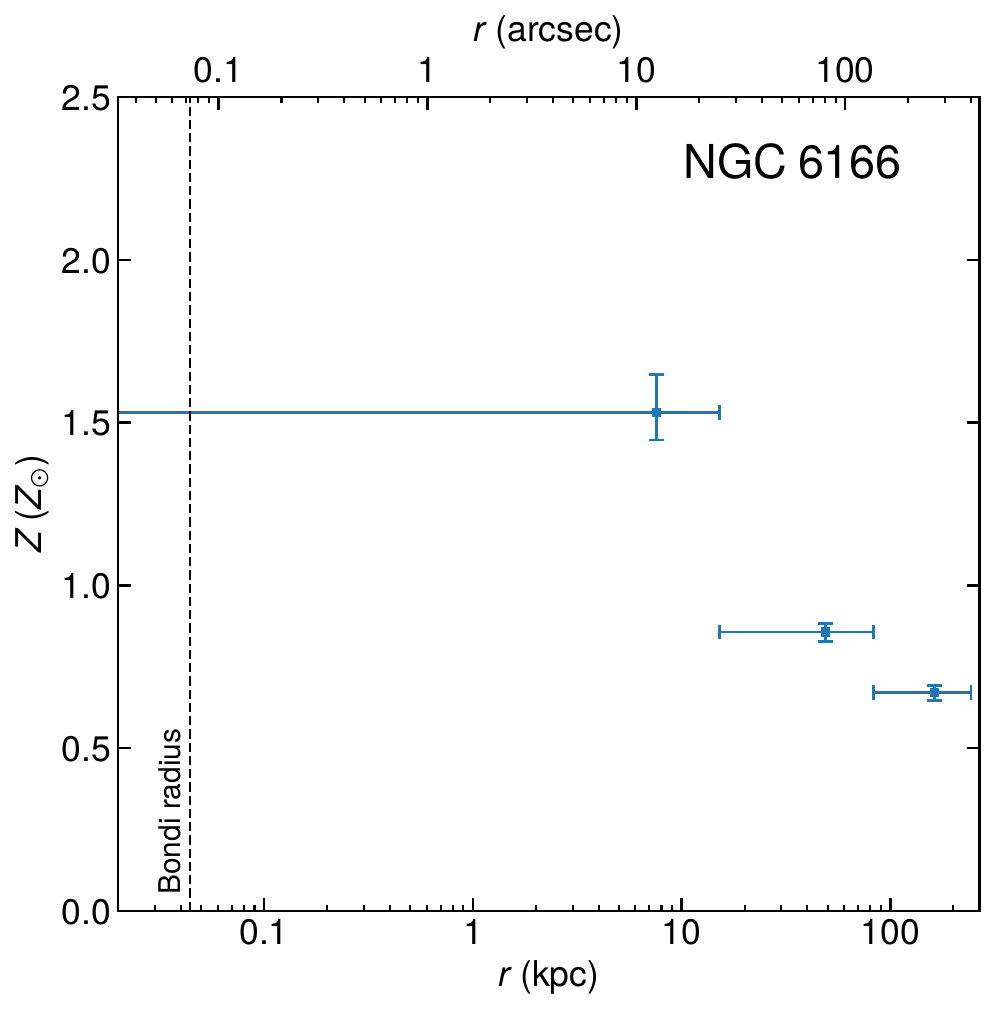}};
\end{tikzpicture}
\caption{Azimuthally averaged radial profiles of the abundance of heavier elements $Z$. The vertical dashed line represents the Bondi radius $r_{\text{Bondi}}$. The abundances are expressed relatively with respect to solar abundance measurements reported by \protect\cite{Lodders2003}. For most galaxies, the abundances of two or more neighbouring radial bins were tied together. The abundance inside the Bondi radius was assumed to be the same as that of the innermost radial bin.}
\label{fig:abund}
\end{figure*}


\begin{figure}
\begin{tikzpicture}
\draw (\figxi, \figyi) node {\includegraphics[scale=\figscale]{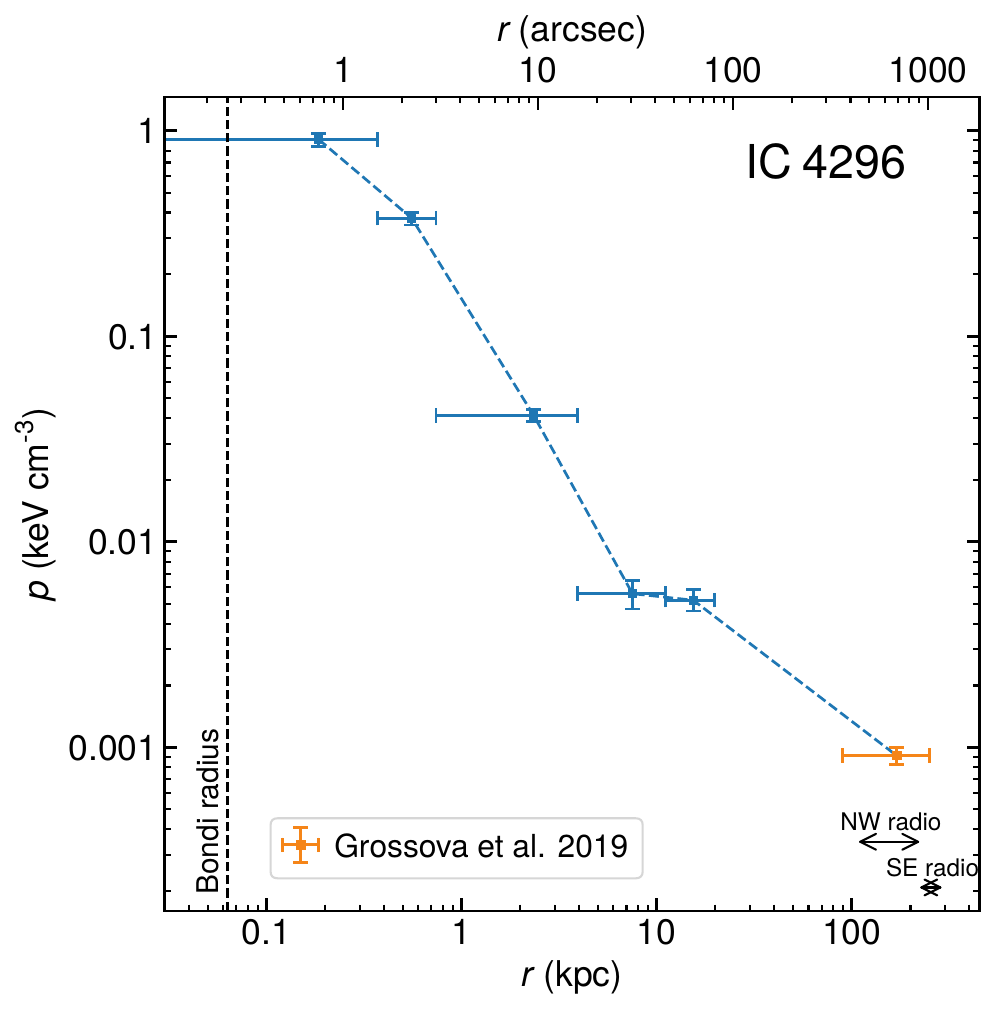}};
\draw (\figxj, \figyi) node {\includegraphics[scale=\figscale]{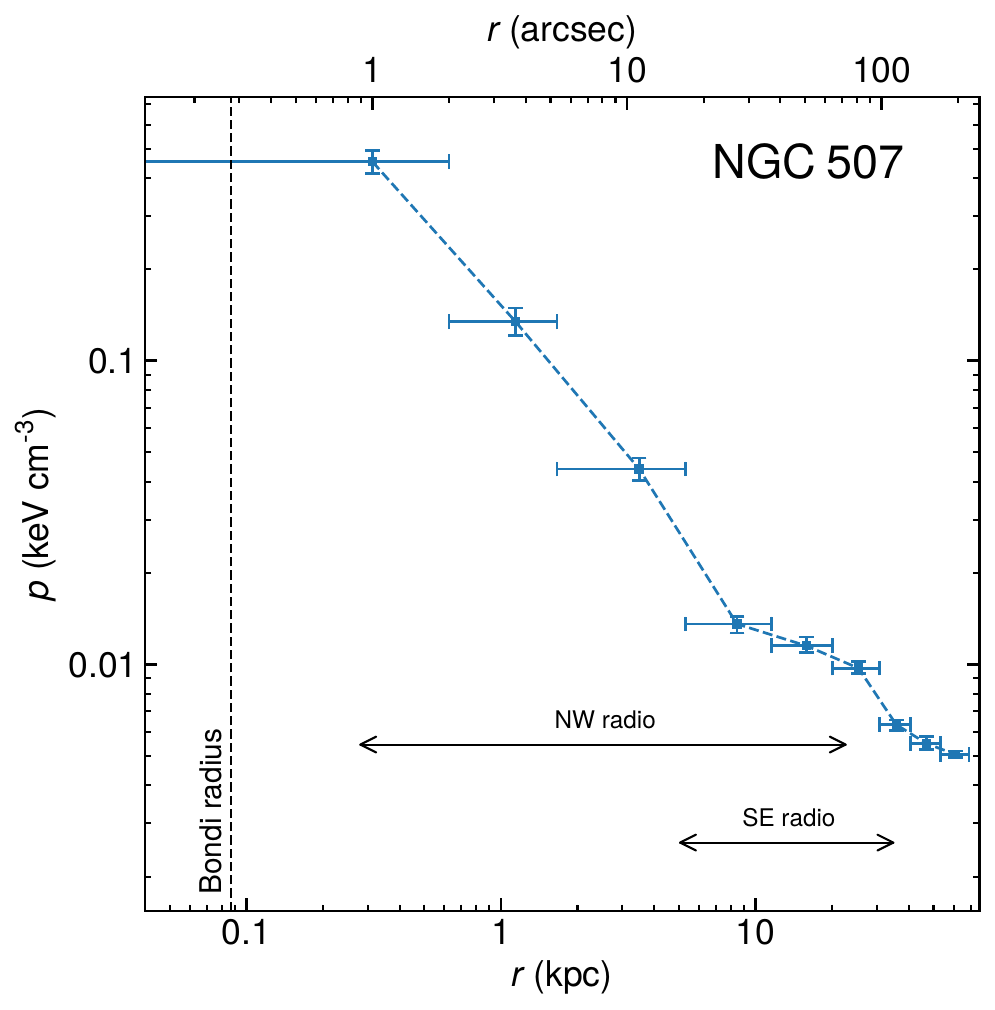}};
\draw (\figxk, \figyi) node {\includegraphics[scale=\figscale]{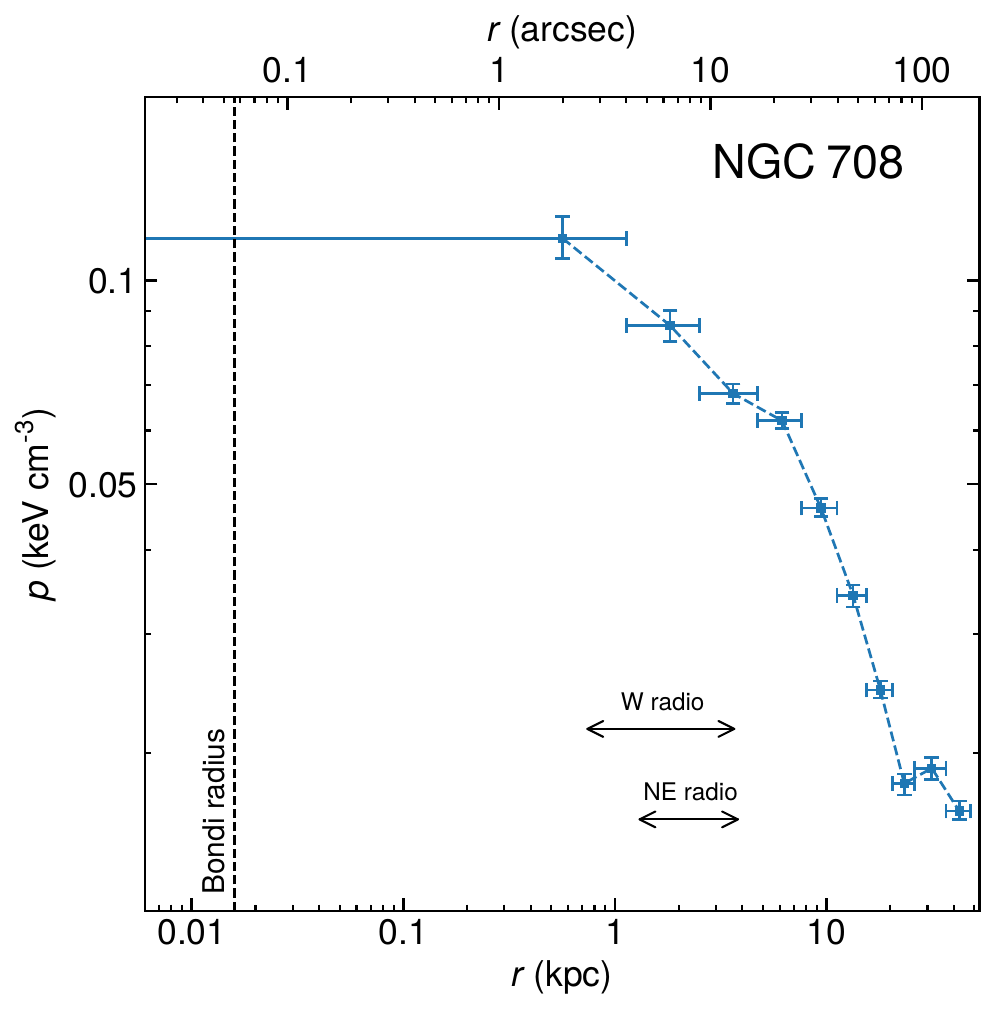}};
\draw (\figxi, \figyj) node {\includegraphics[scale=\figscale]{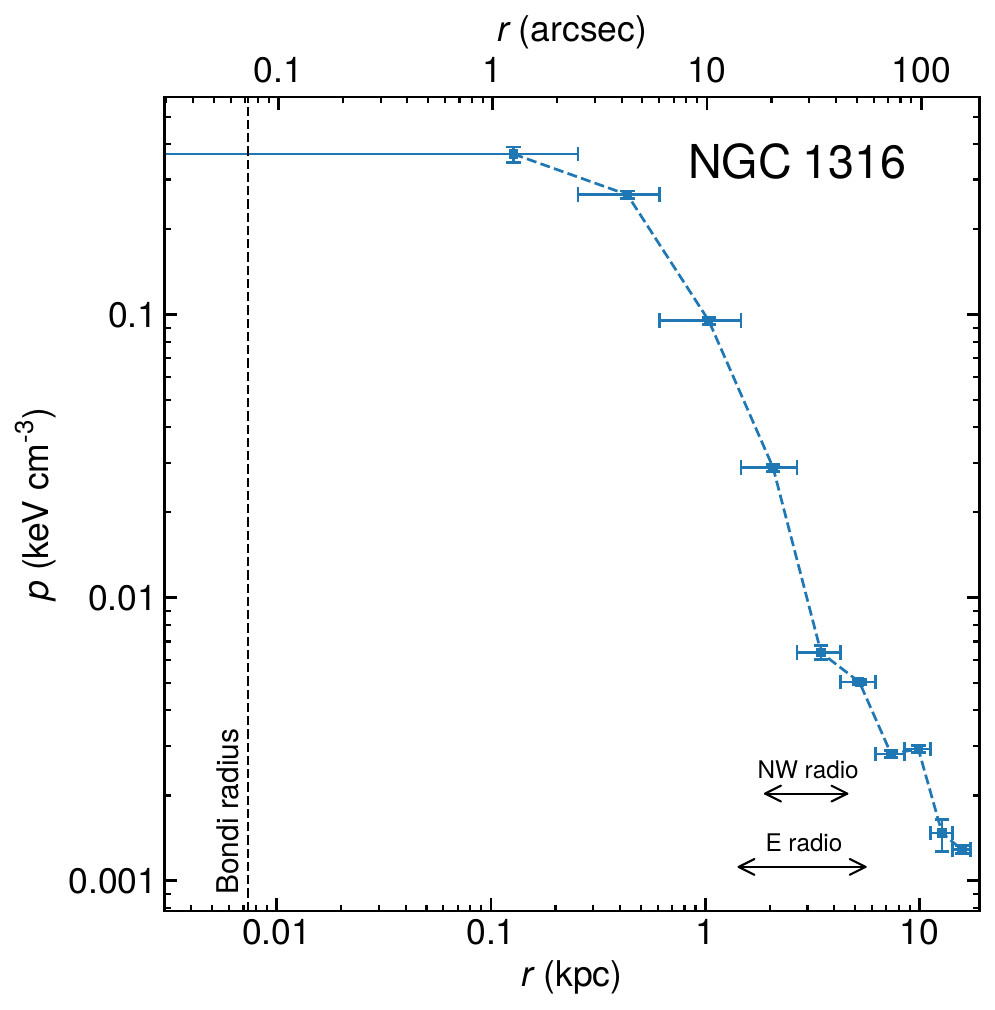}};
\draw (\figxj, \figyj) node {\includegraphics[scale=\figscale]{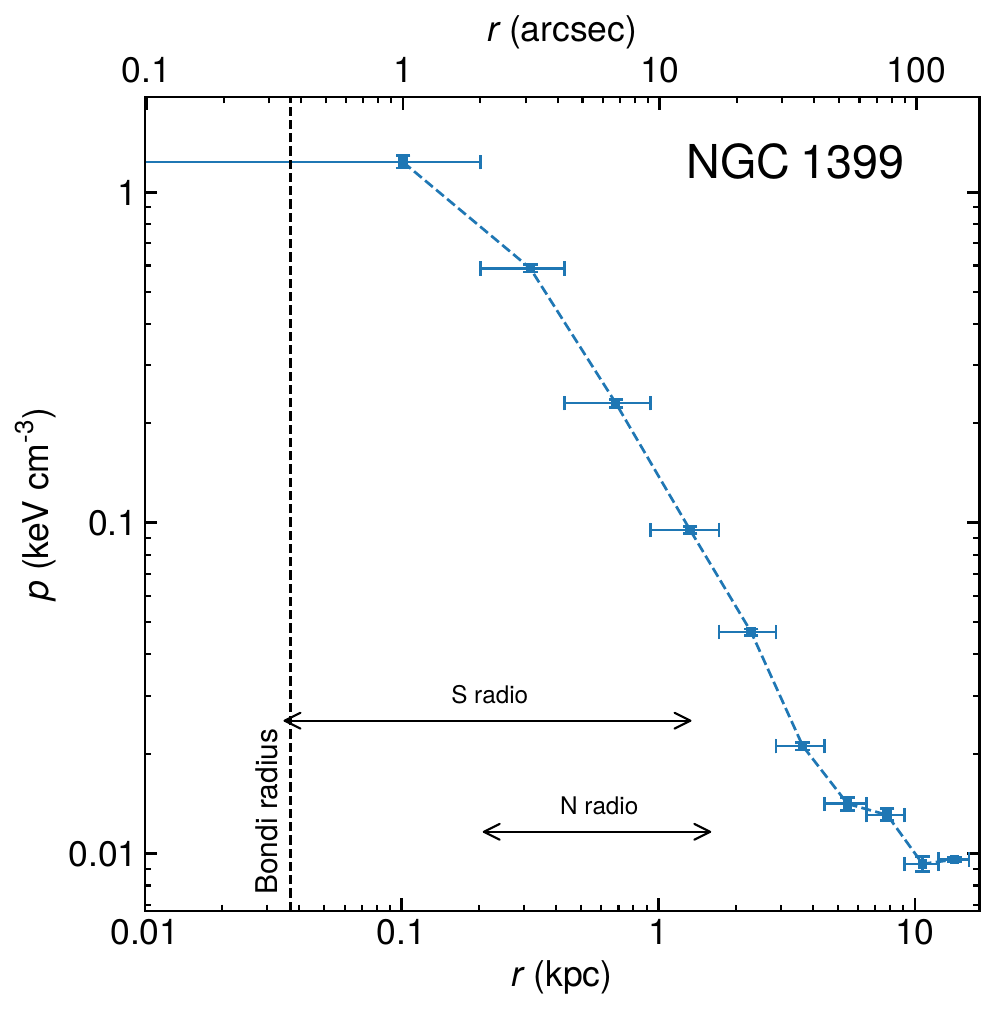}};
\draw (\figxk, \figyj) node {\includegraphics[scale=\figscale]{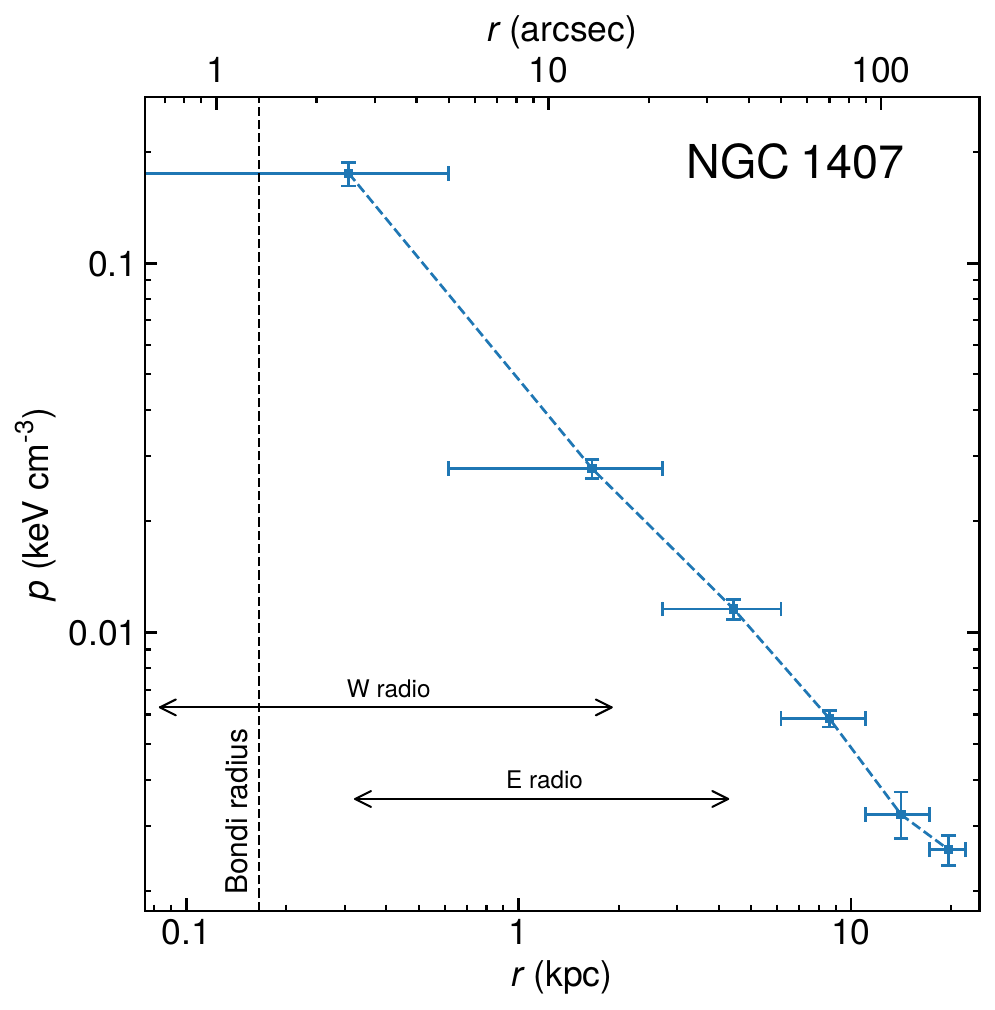}};
\draw (\figxi, \figyk) node {\includegraphics[scale=\figscale]{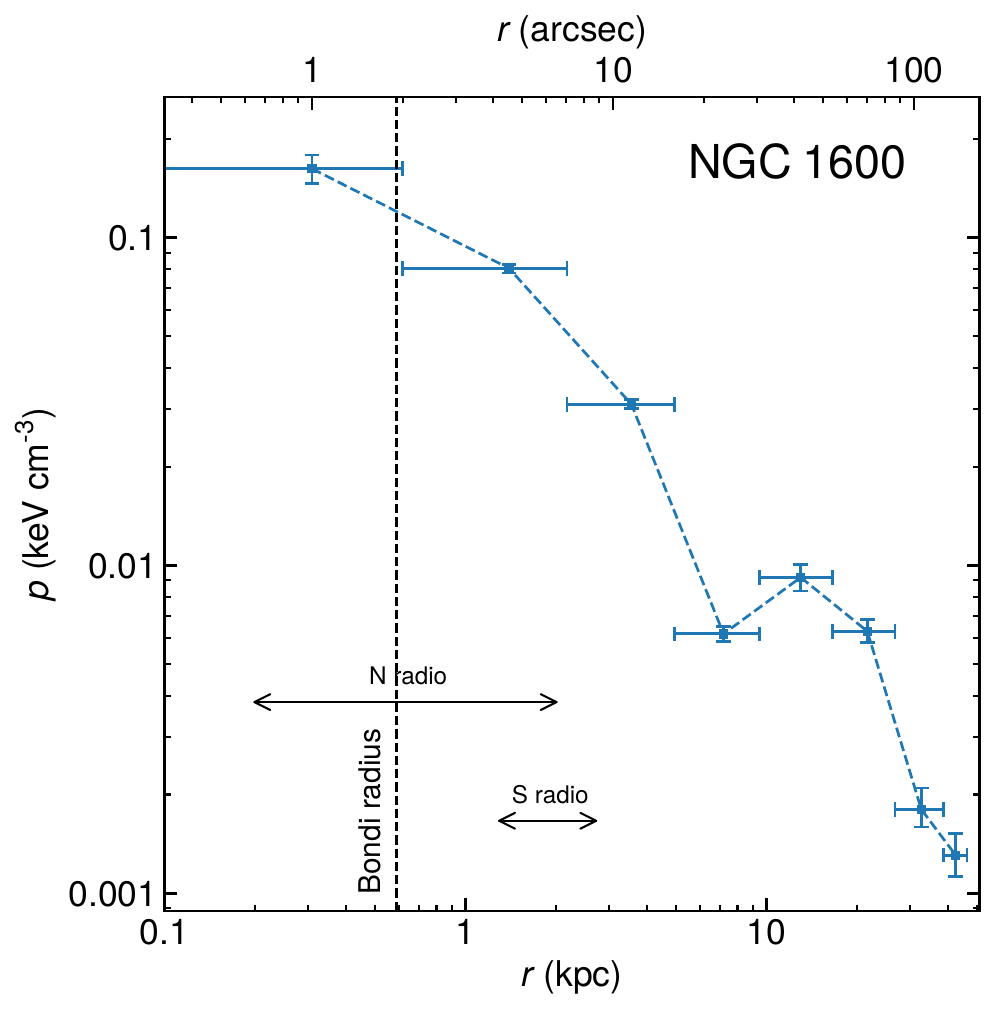}};
\draw (\figxj, \figyk) node {\includegraphics[scale=\figscale]{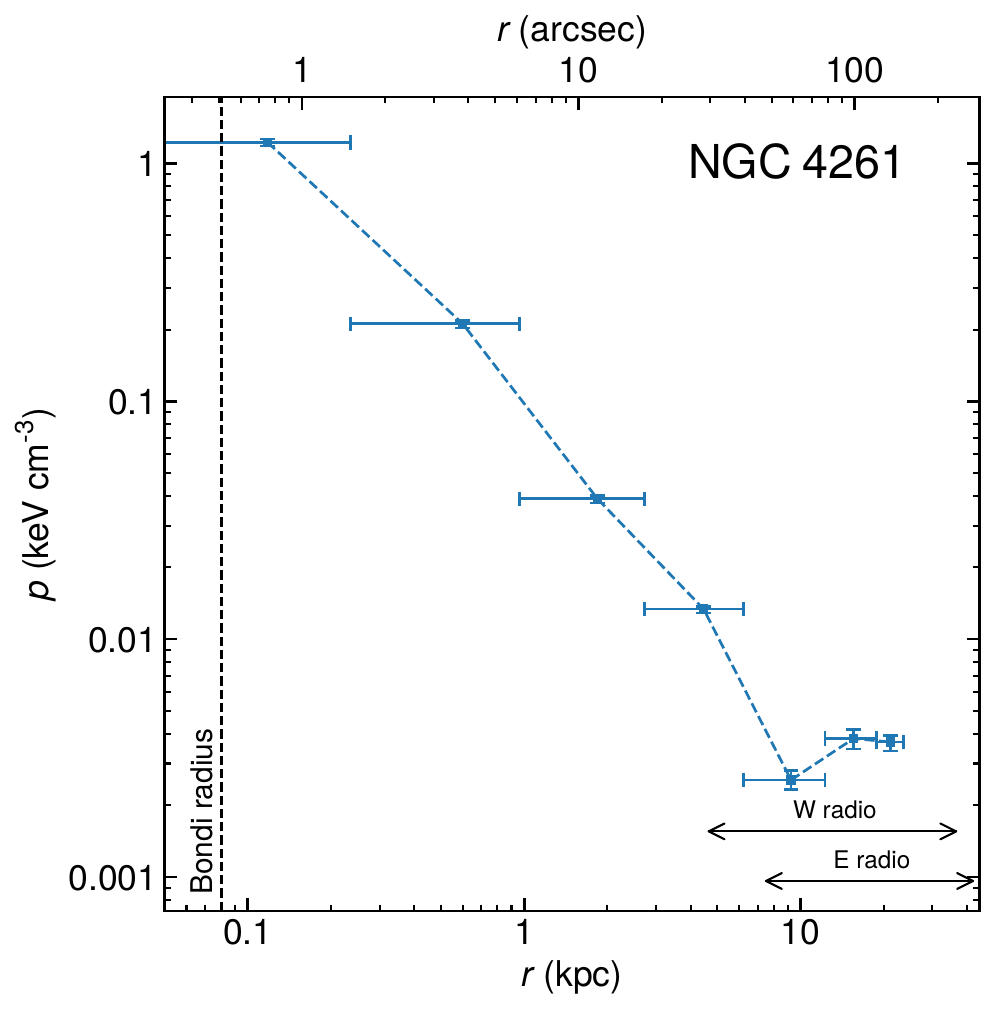}};
\draw (\figxk, \figyk) node {\includegraphics[scale=\figscale]{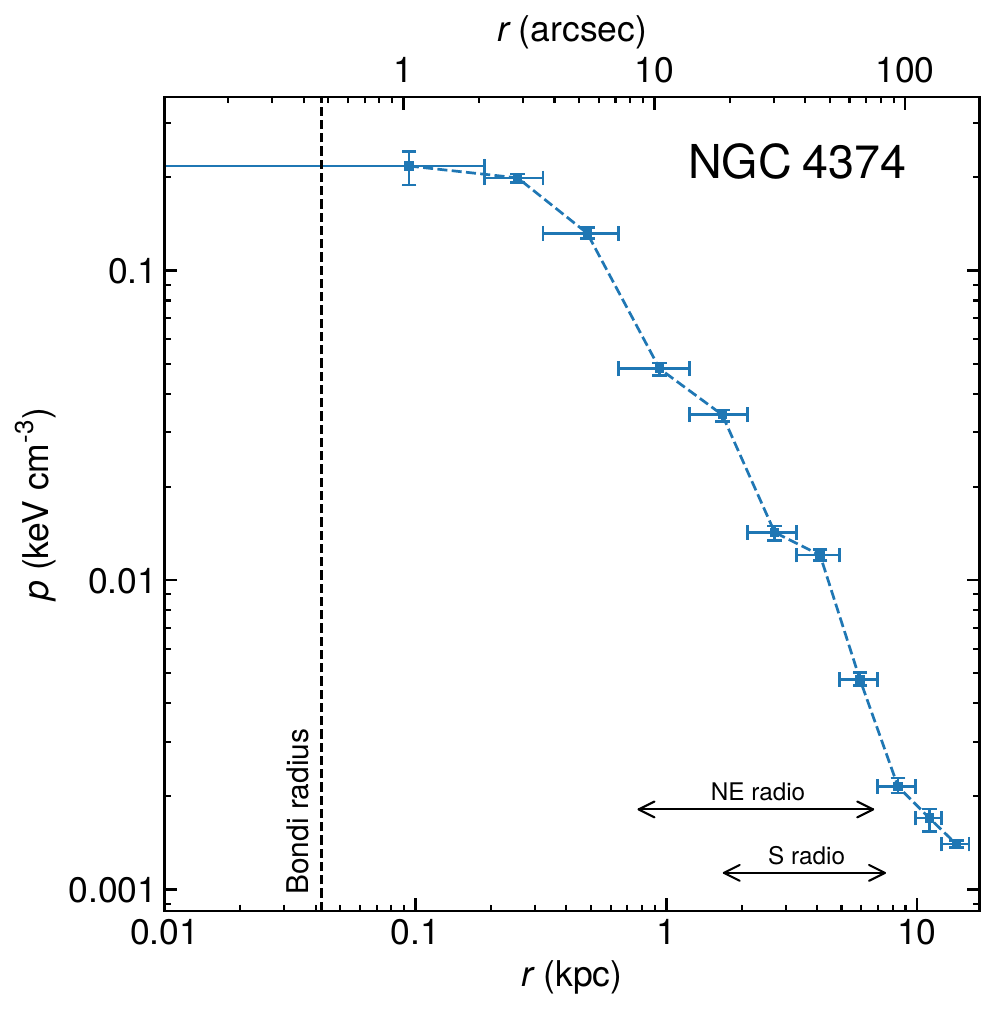}};
\draw (\figxi, \figyl) node {\includegraphics[scale=\figscale]{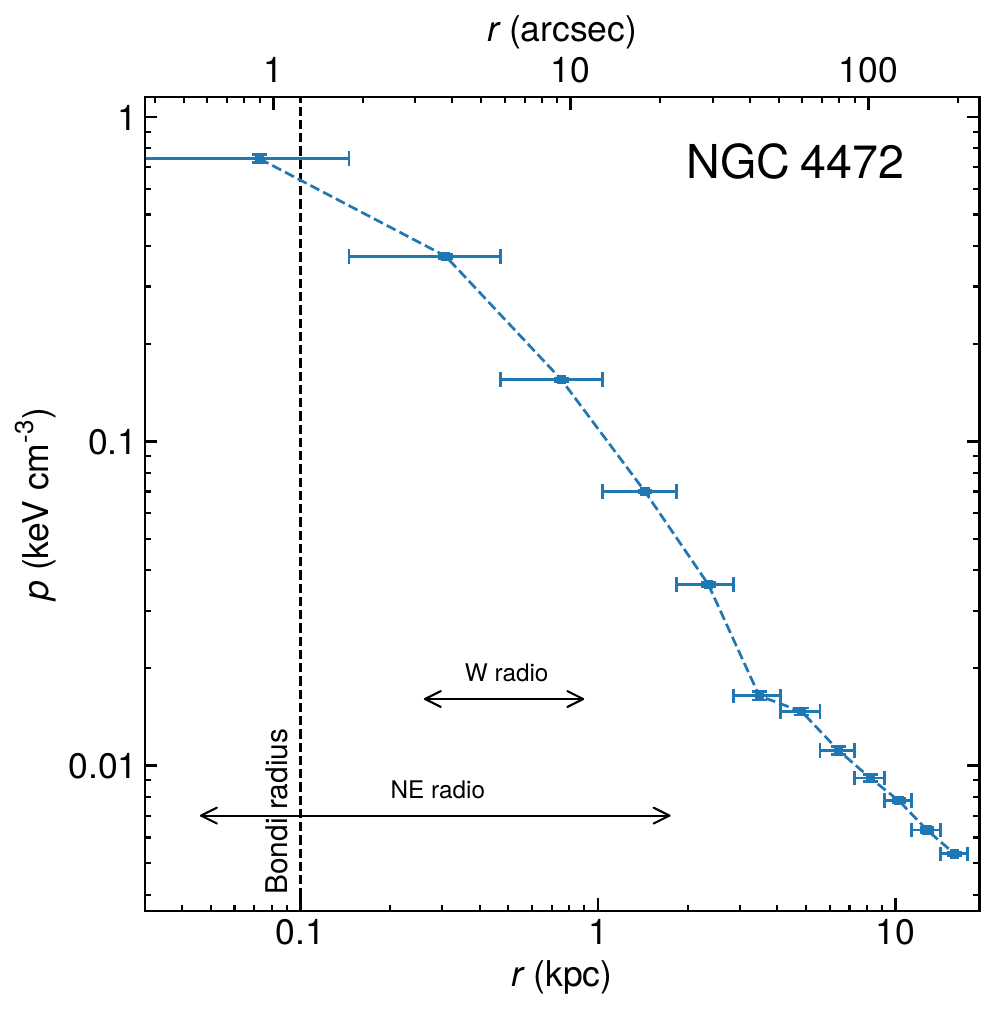}};
\draw (\figxj, \figyl) node {\includegraphics[scale=\figscale]{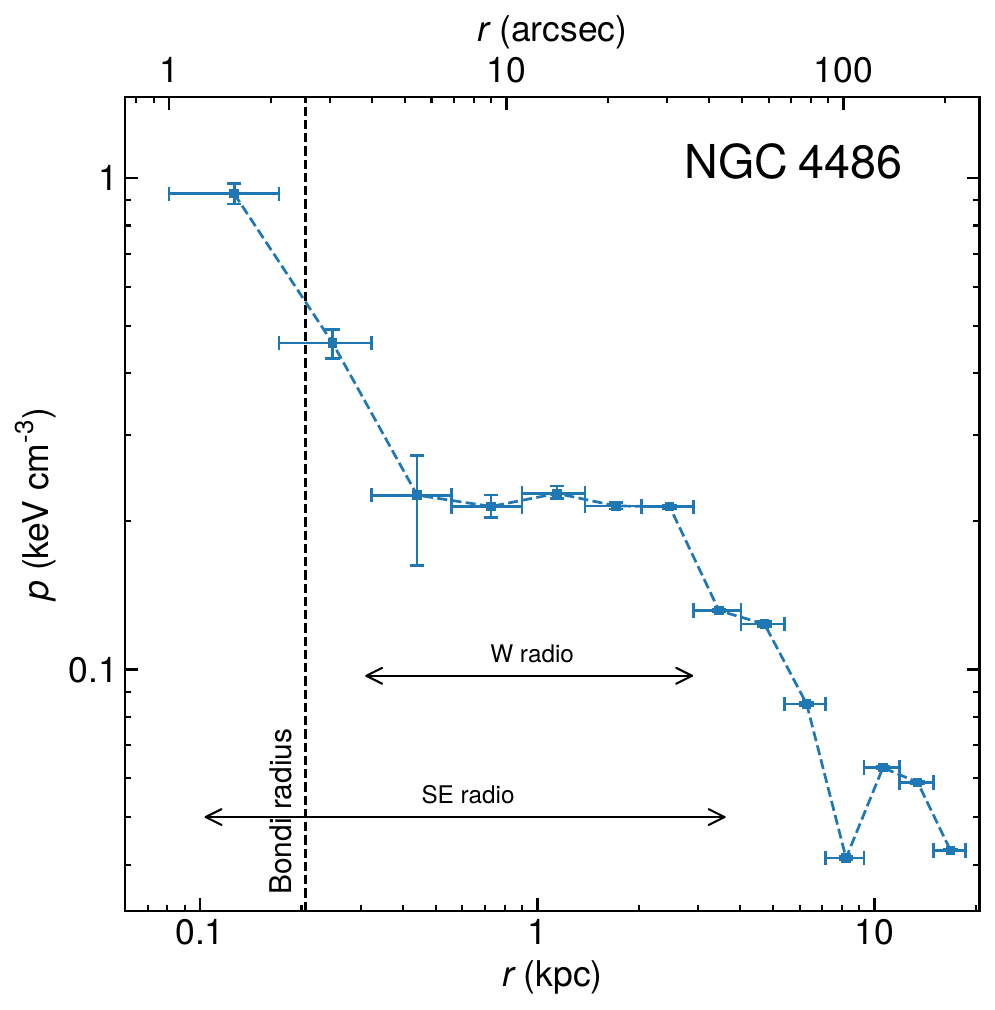}};
\draw (\figxk, \figyl) node {\includegraphics[scale=\figscale]{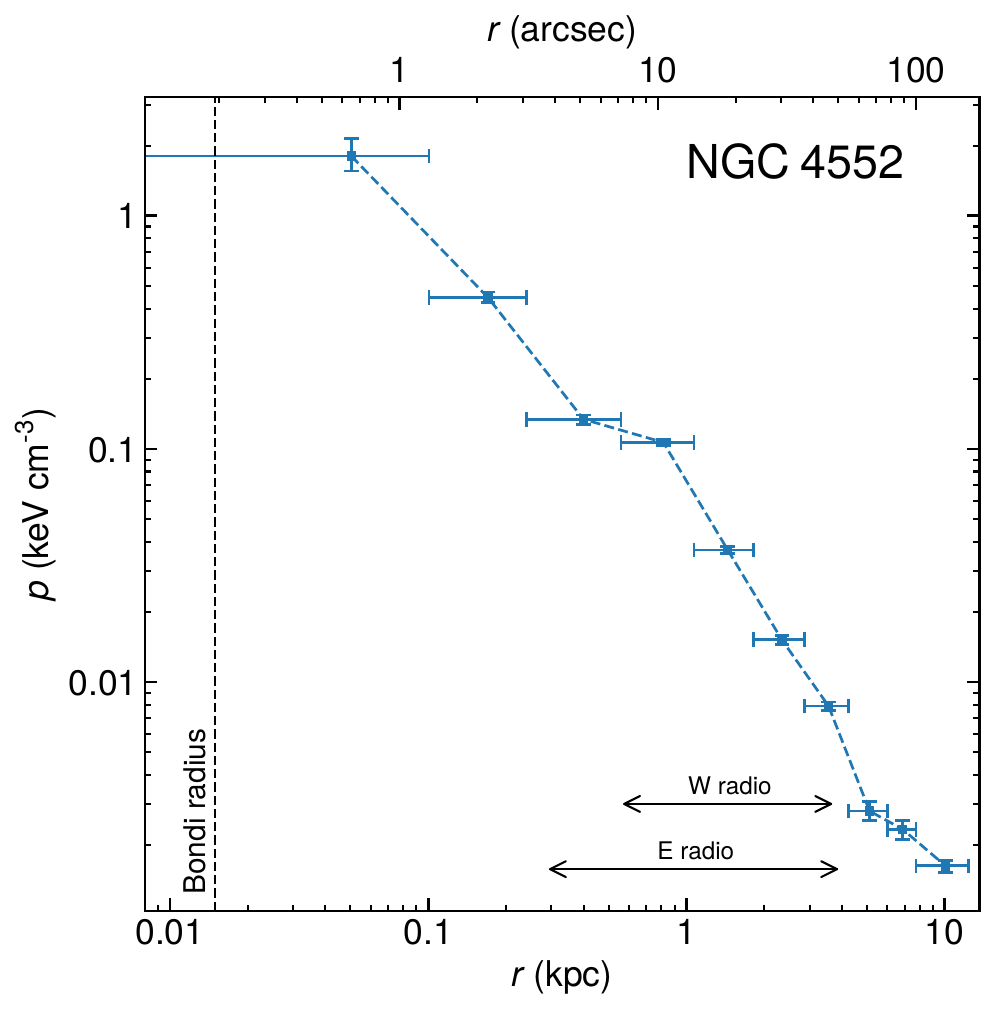}};
\end{tikzpicture}
\end{figure}

\begin{figure*}
\begin{tikzpicture}
\draw (\figxi, \figyi) node {\includegraphics[scale=\figscale]{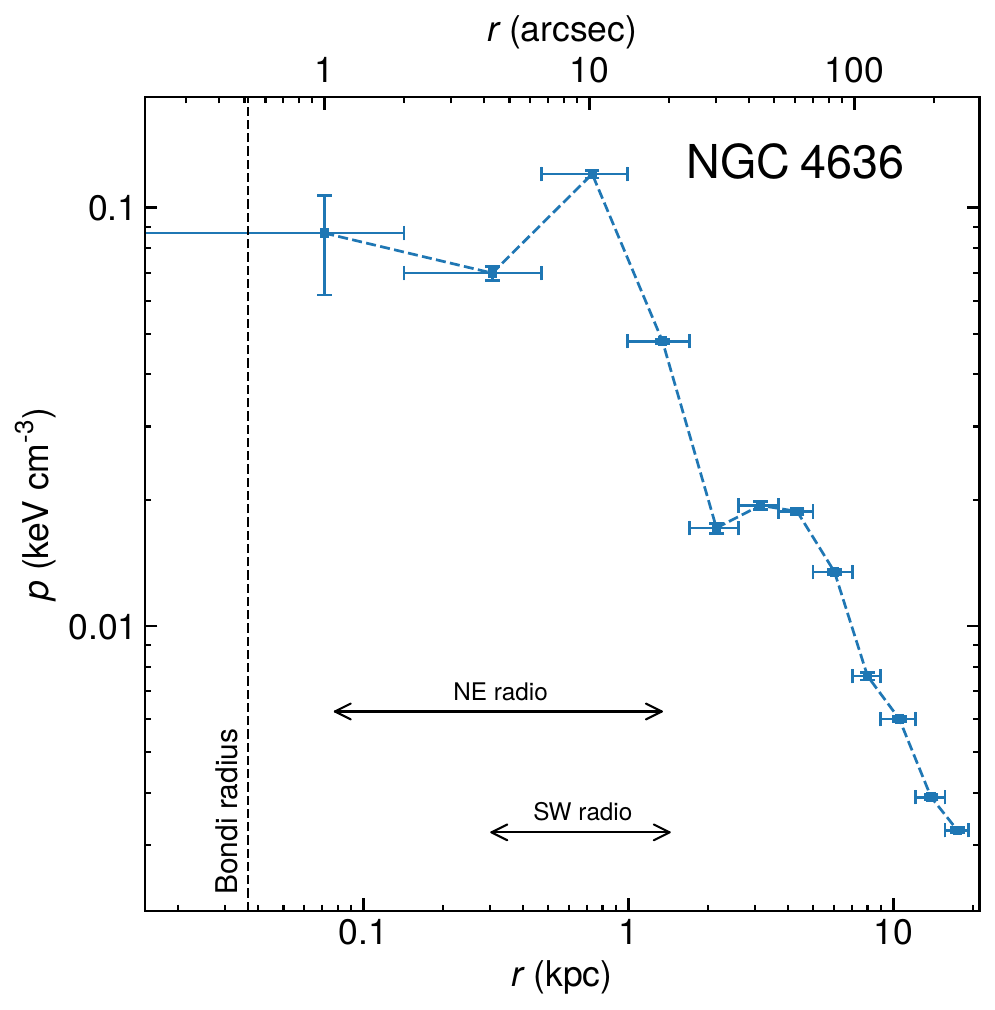}};
\draw (\figxj, \figyi) node {\includegraphics[scale=\figscale]{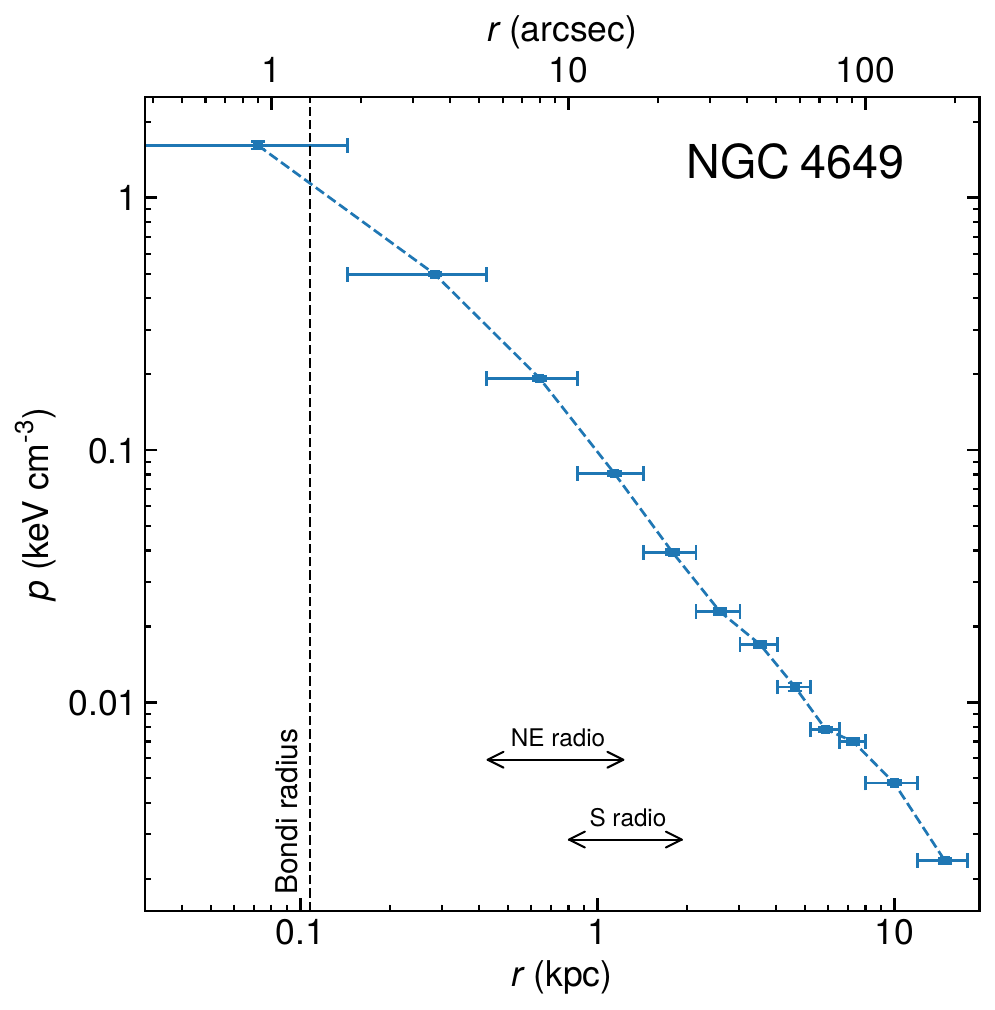}};
\draw (\figxk, \figyi) node {\includegraphics[scale=\figscale]{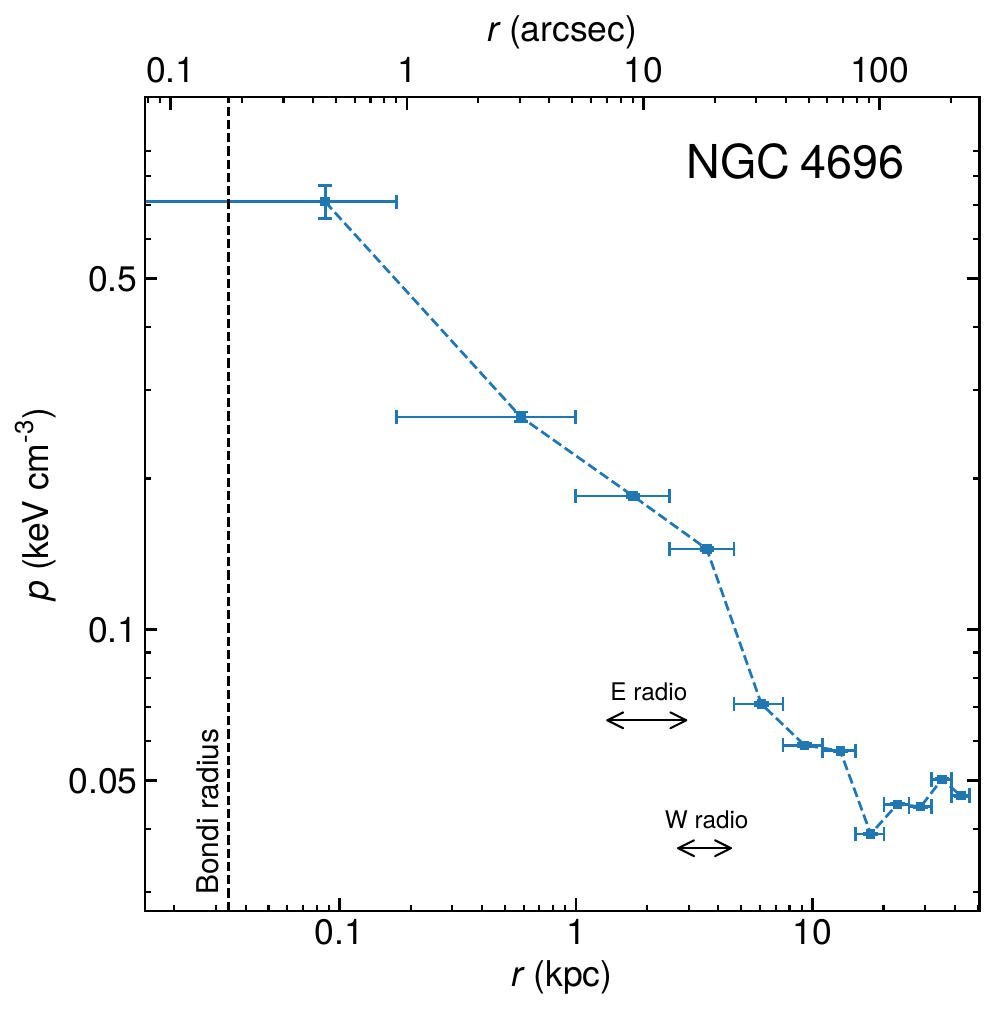}};
\draw (\figxi, \figyj) node {\includegraphics[scale=\figscale]{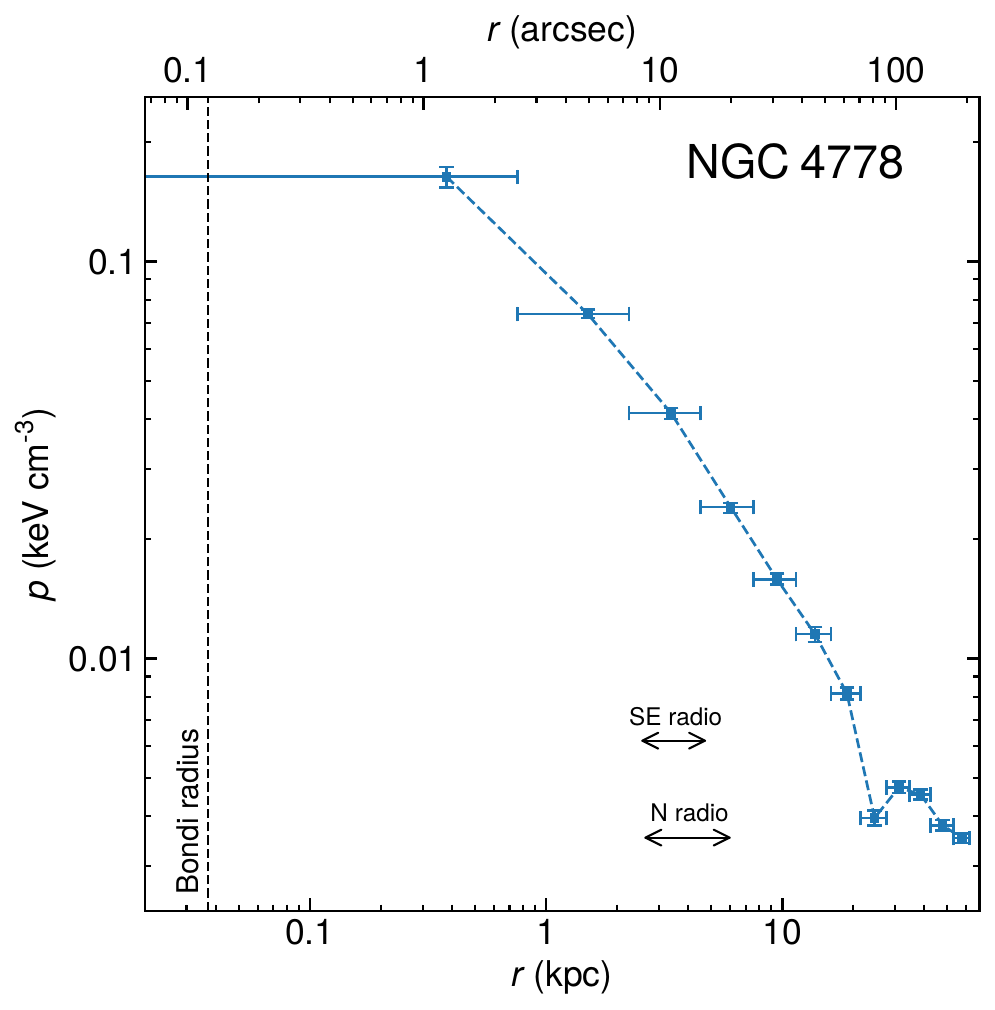}};
\draw (\figxj, \figyj) node {\includegraphics[scale=\figscale]{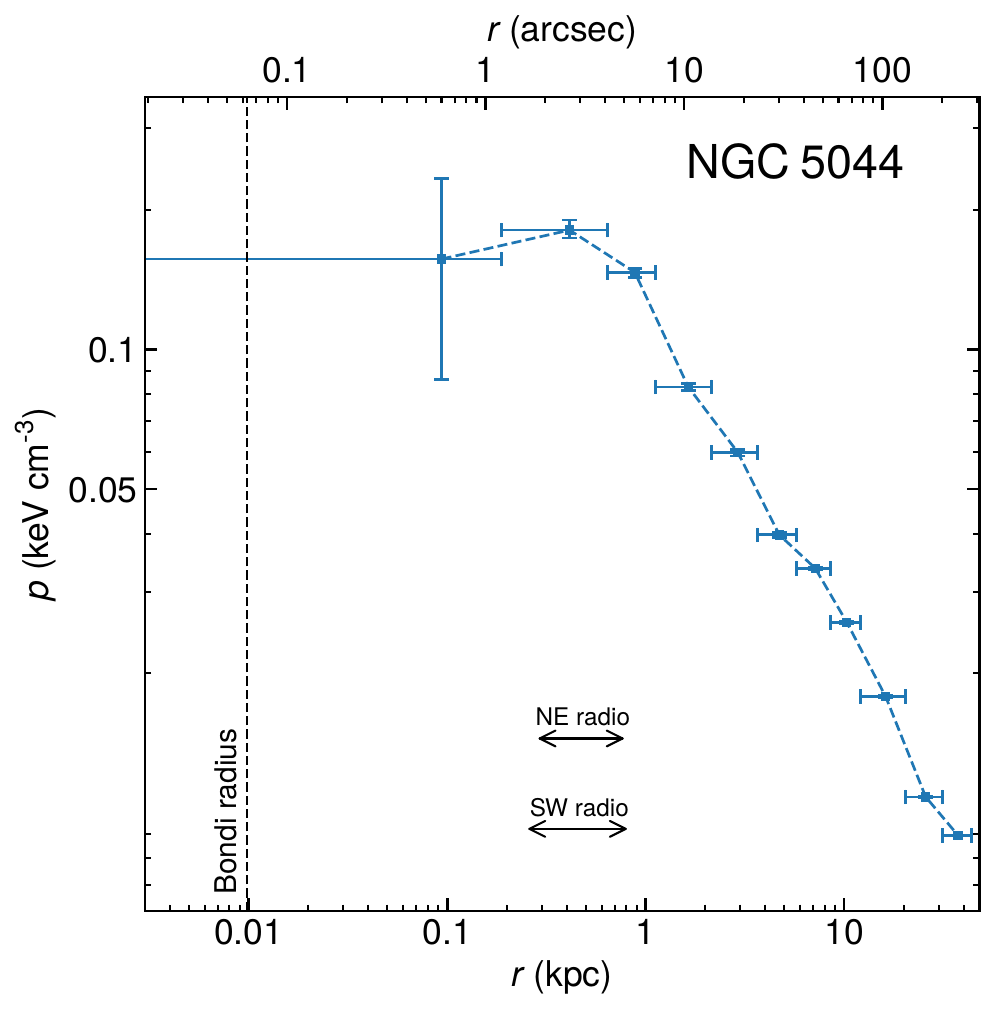}};
\draw (\figxk, \figyj) node {\includegraphics[scale=\figscale]{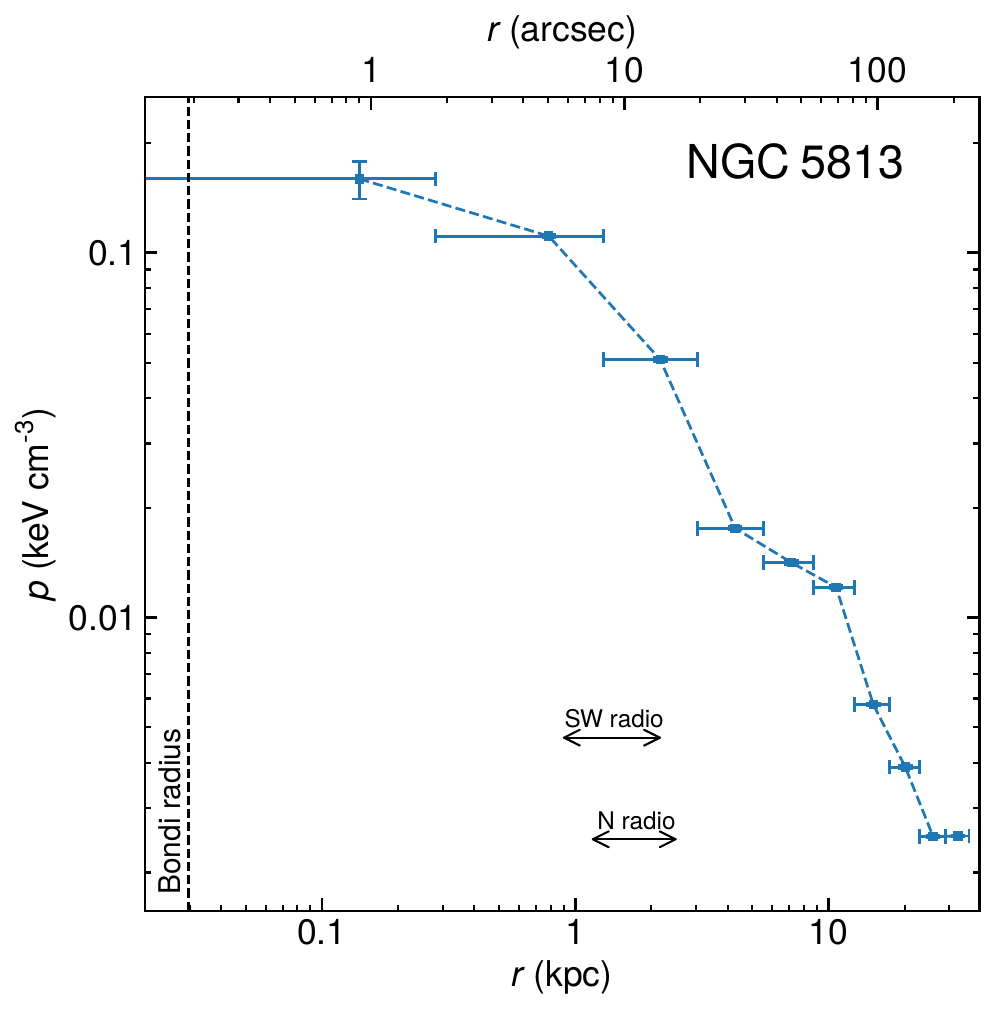}};
\draw (\figxi, \figyk) node {\includegraphics[scale=\figscale]{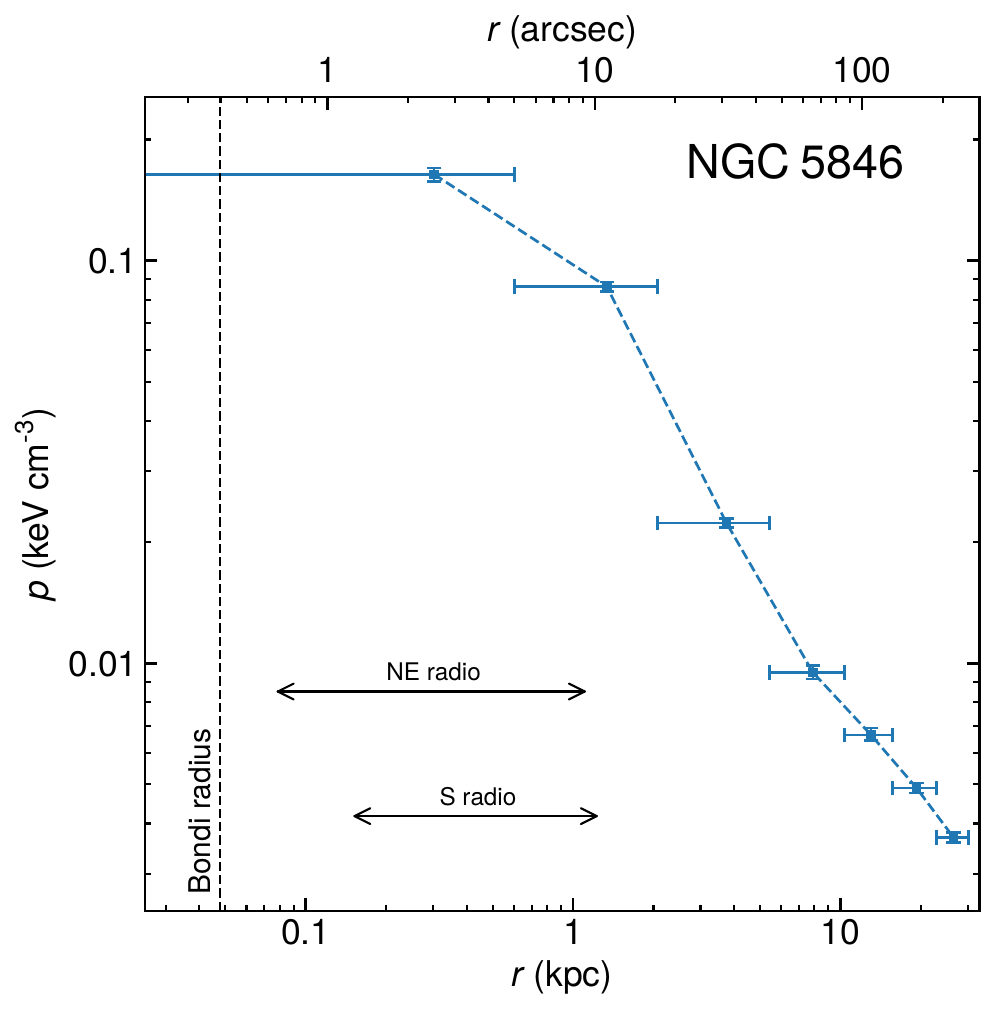}};
\draw (\figxj, \figyk) node {\includegraphics[scale=\figscale]{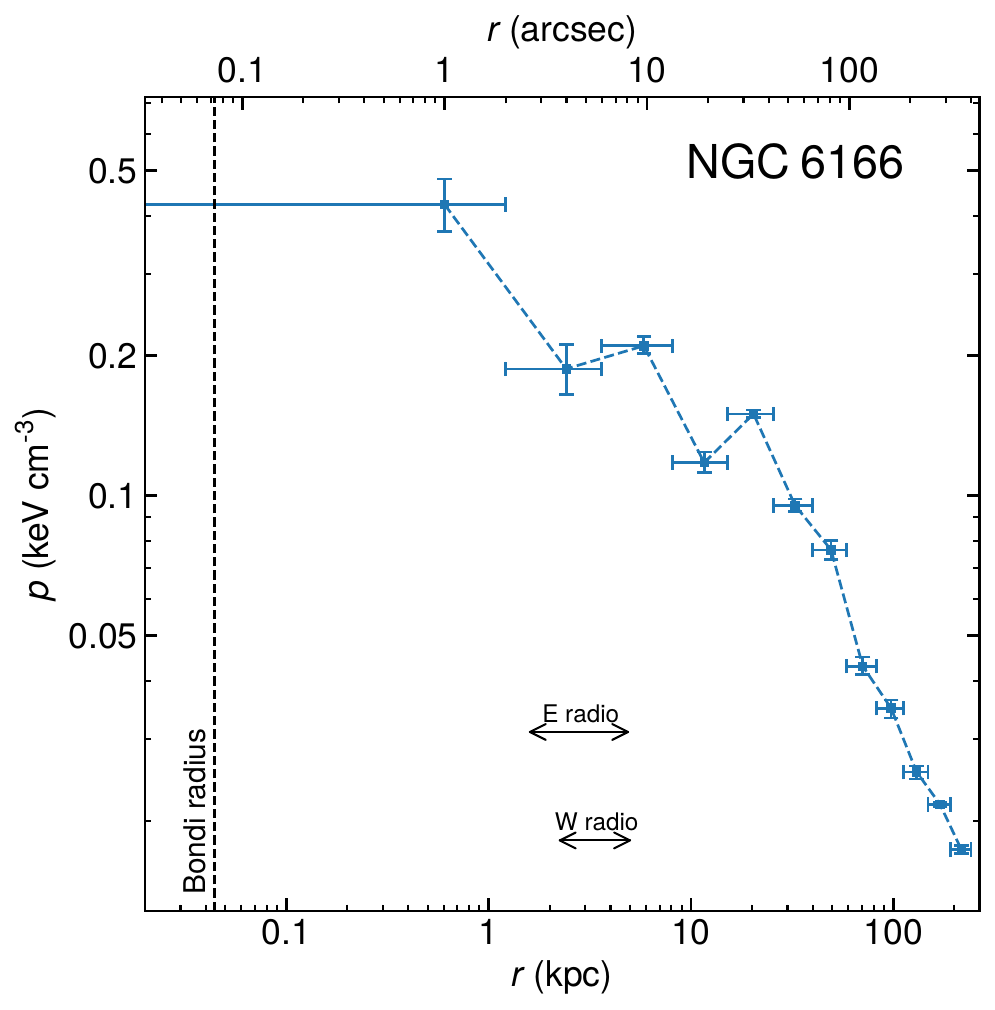}};
\end{tikzpicture}
\caption{Azimuthally averaged radial profiles of thermal pressure $p = n \, kT$. The vertical dashed line represents the Bondi radius $r_{\text{Bondi}}$. The double-sided arrows represent radial ranges of individual radio lobes and the labels express the corresponding position from the galactic centre.}
\label{fig:pressure}
\end{figure*}


\begin{figure}
\begin{tikzpicture}
\draw (\figxi, \figyi) node {\includegraphics[scale=\figscale]{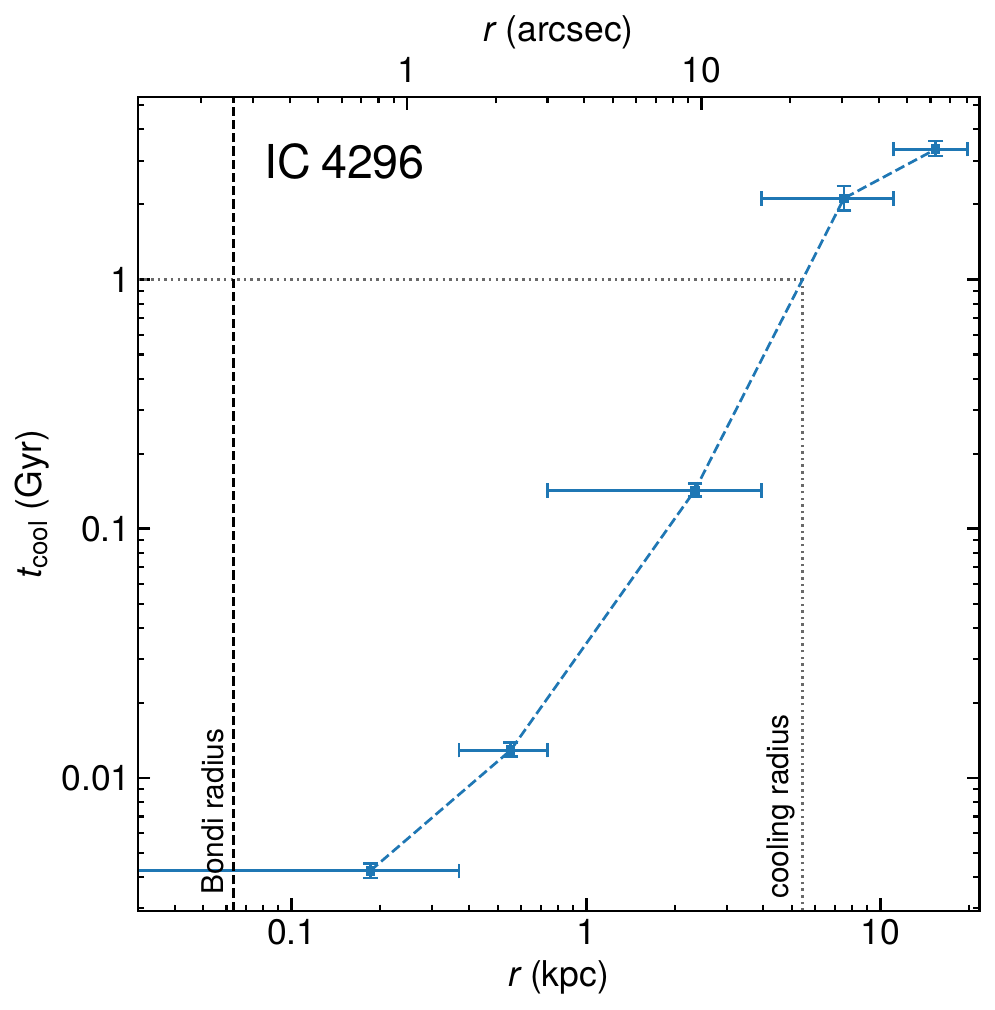}};
\draw (\figxj, \figyi) node {\includegraphics[scale=\figscale]{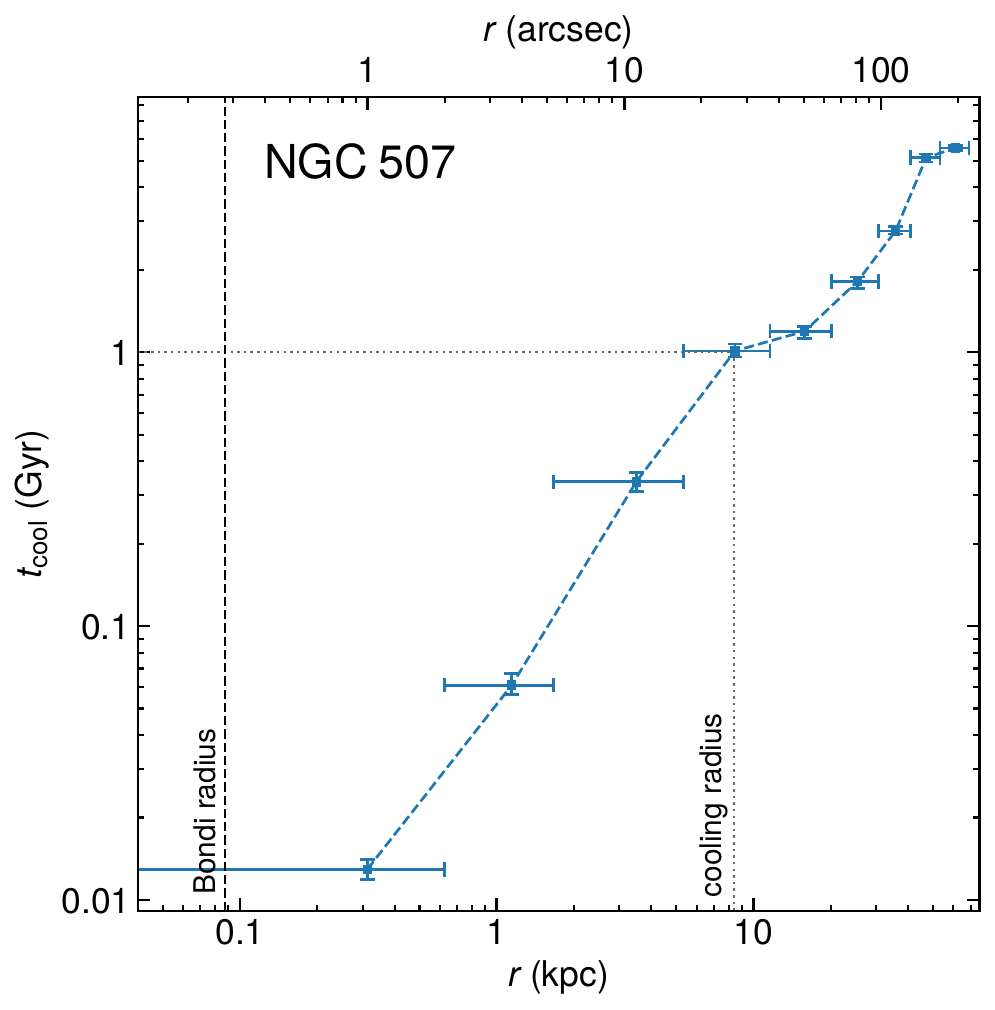}};
\draw (\figxk, \figyi) node {\includegraphics[scale=\figscale]{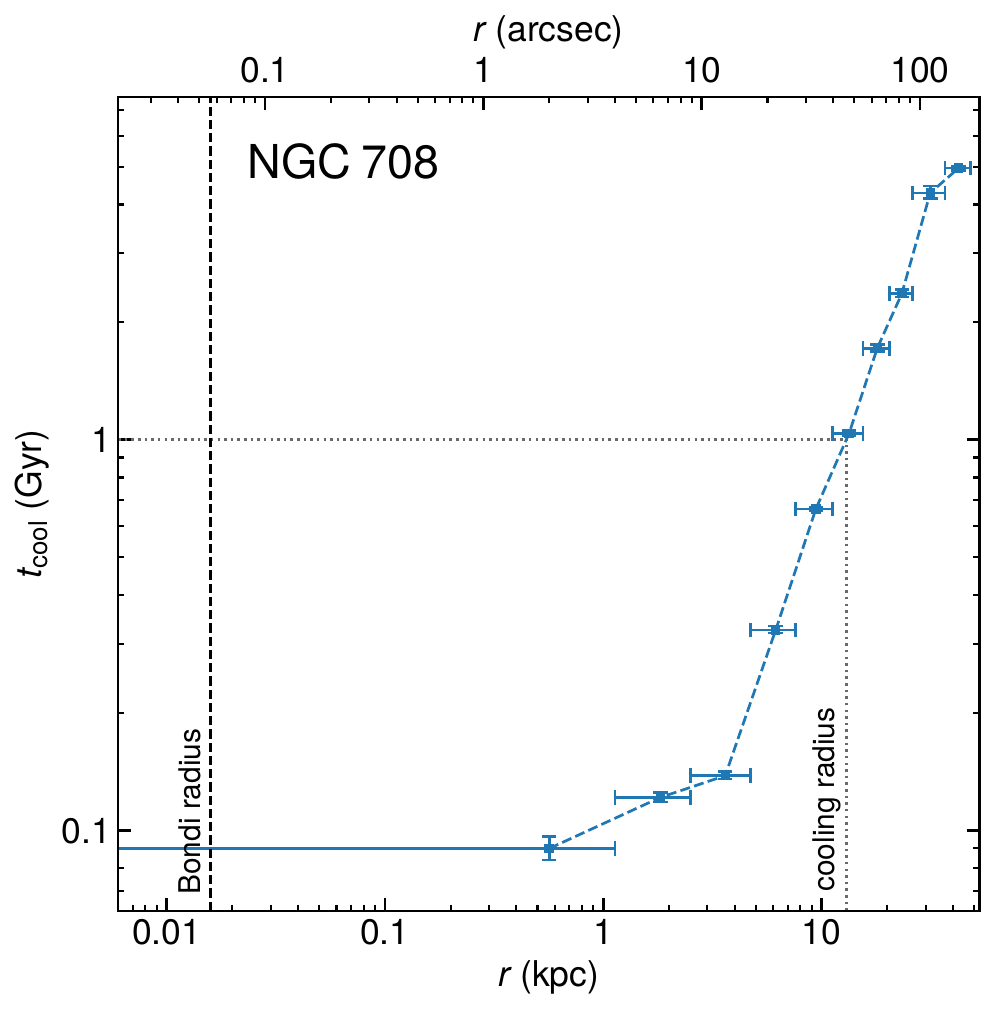}};
\draw (\figxi, \figyj) node {\includegraphics[scale=\figscale]{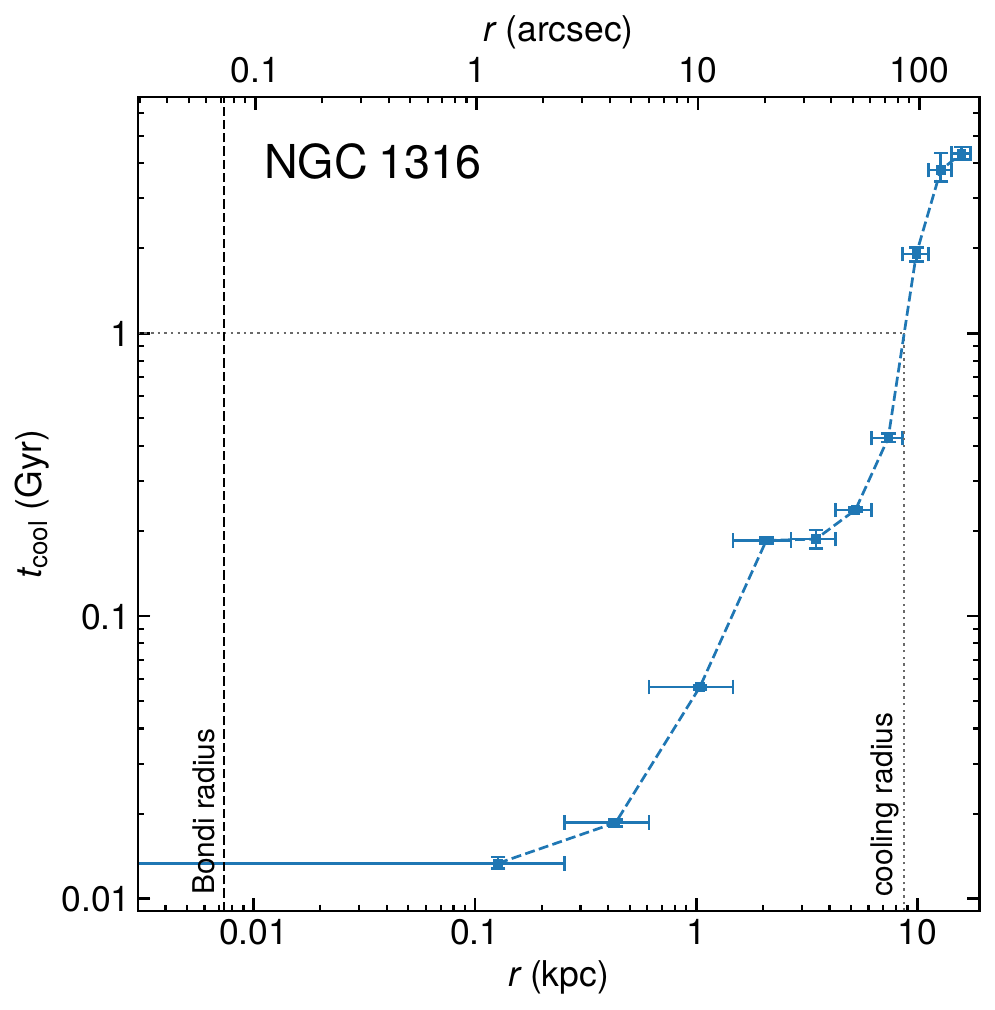}};
\draw (\figxj, \figyj) node {\includegraphics[scale=\figscale]{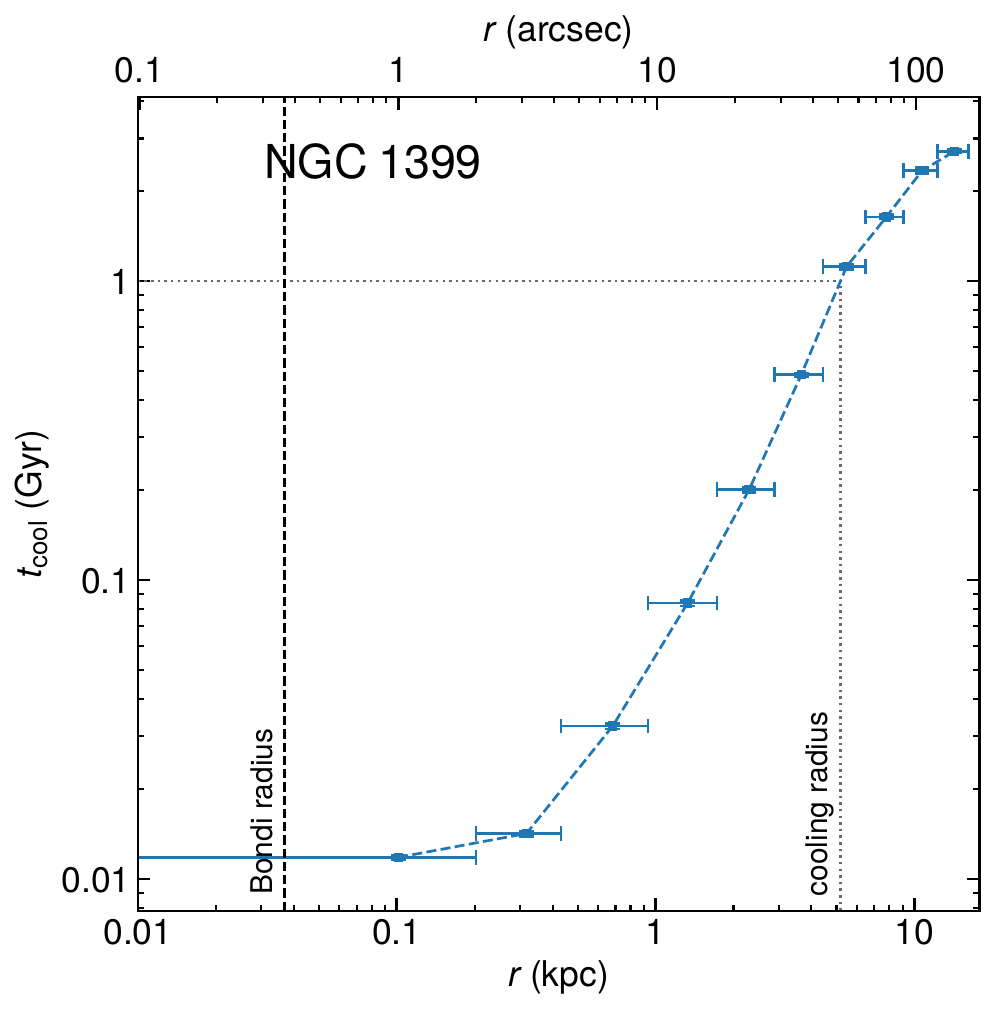}};
\draw (\figxk, \figyj) node {\includegraphics[scale=\figscale]{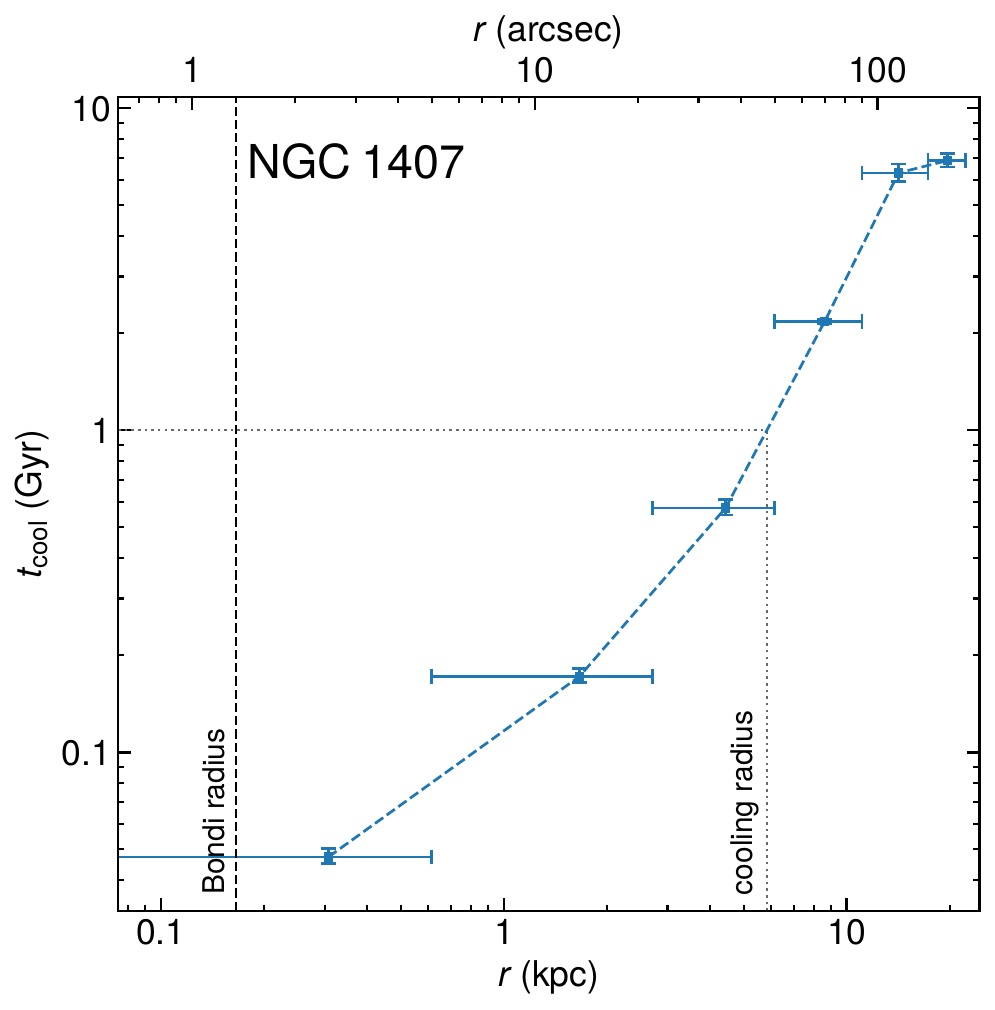}};
\draw (\figxi, \figyk) node {\includegraphics[scale=\figscale]{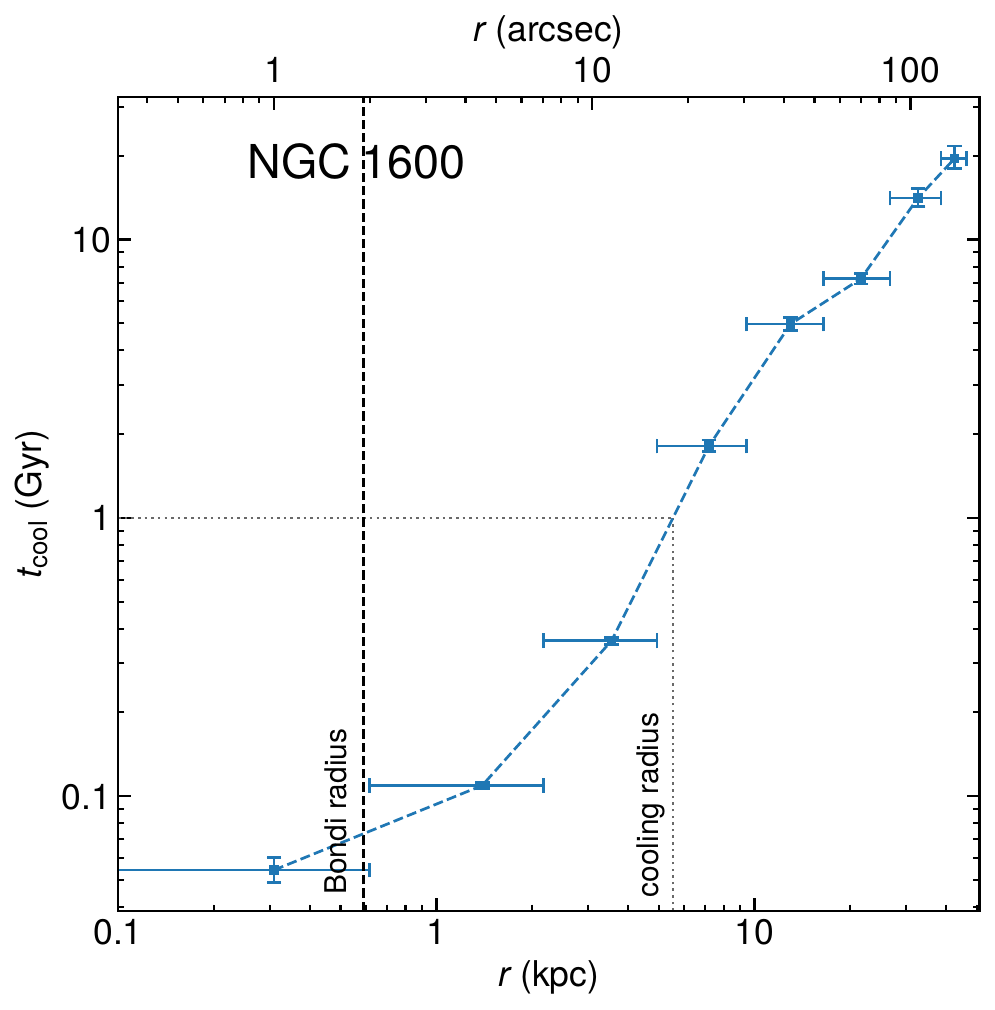}};
\draw (\figxj, \figyk) node {\includegraphics[scale=\figscale]{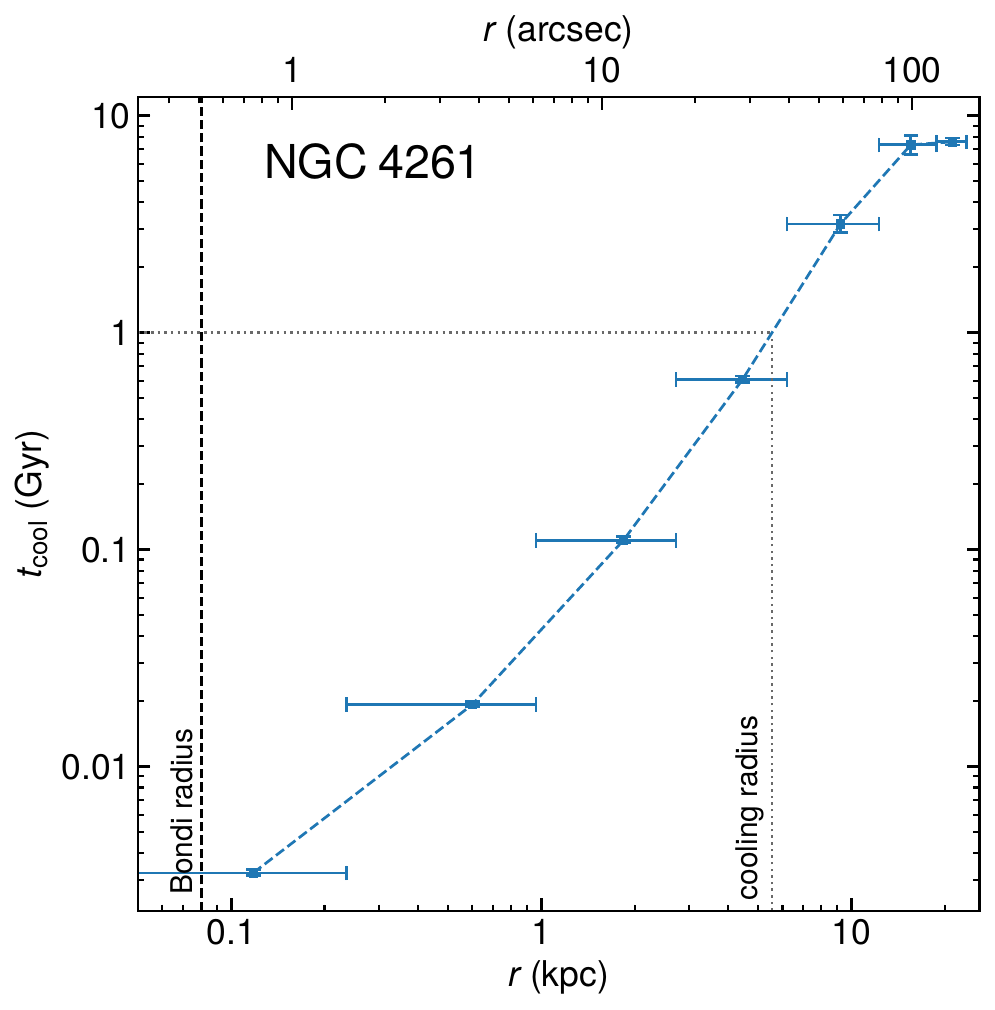}};
\draw (\figxk, \figyk) node {\includegraphics[scale=\figscale]{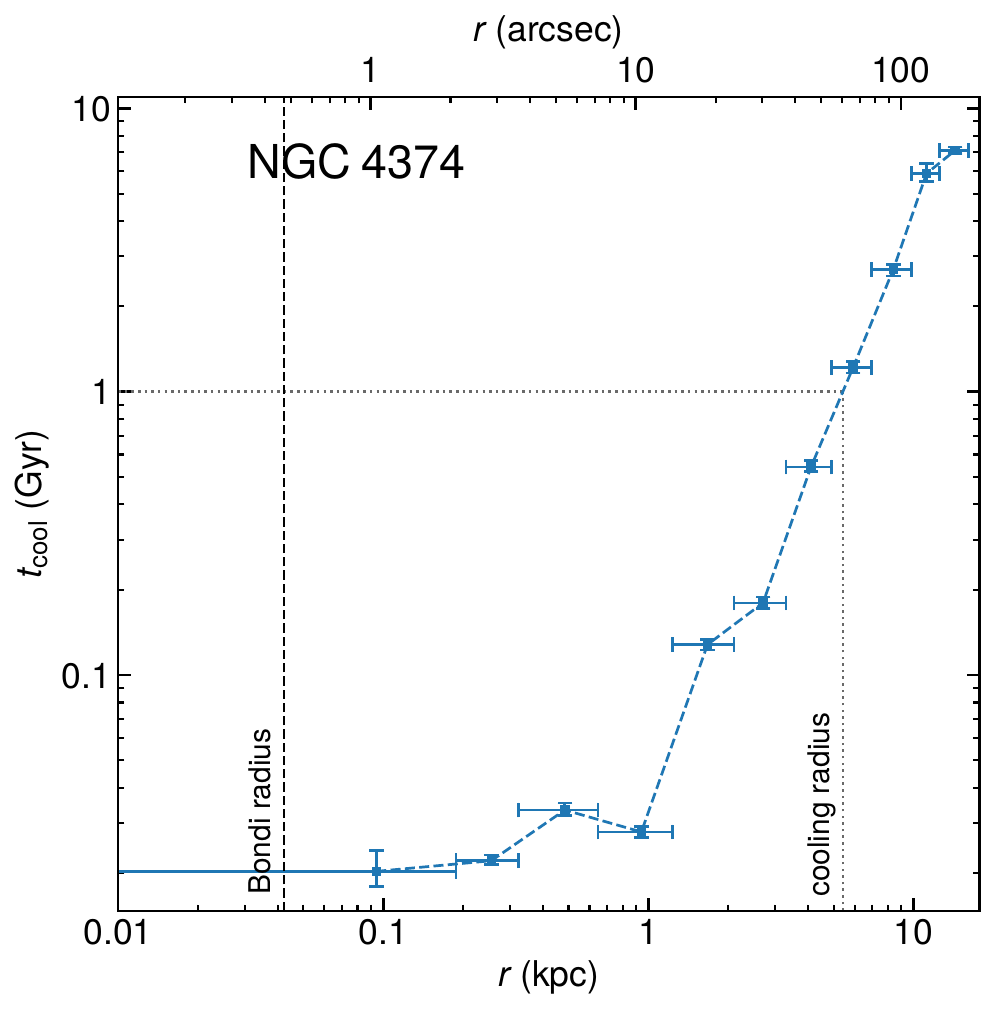}};
\draw (\figxi, \figyl) node {\includegraphics[scale=\figscale]{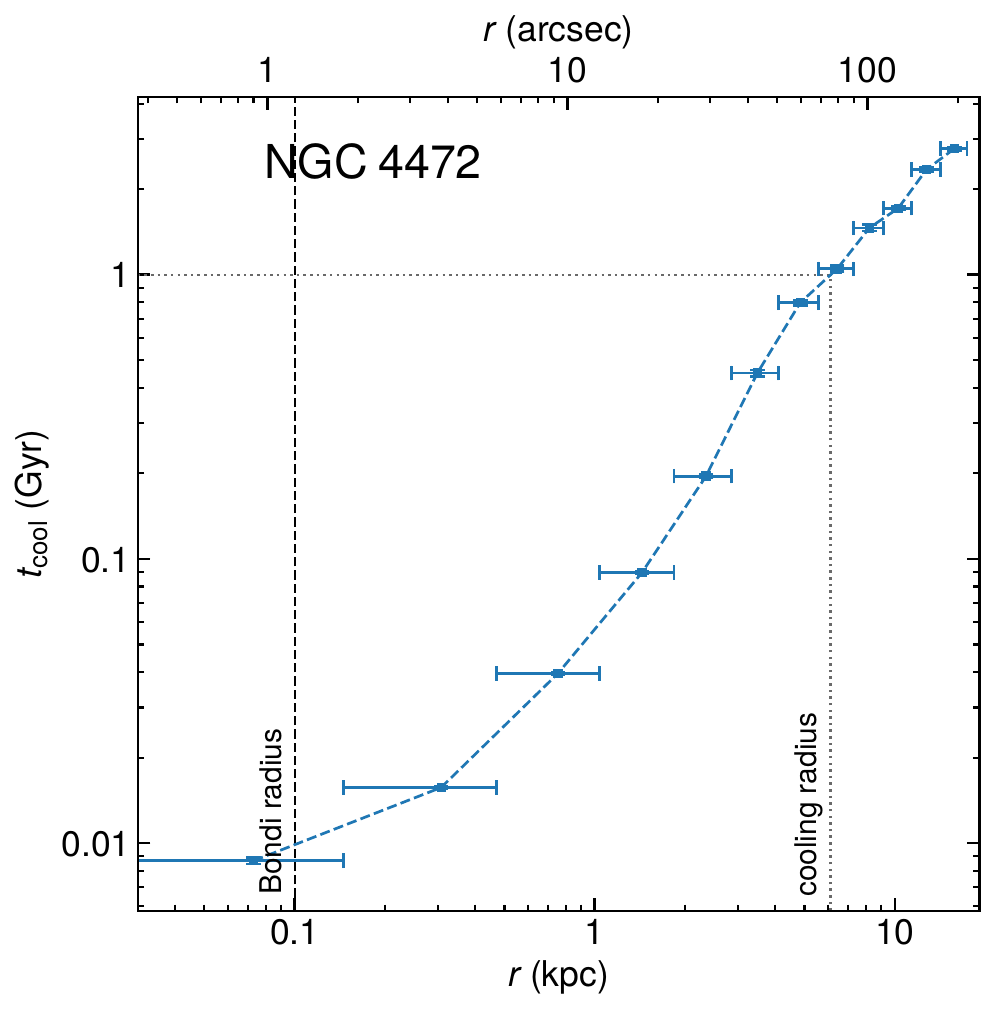}};
\draw (\figxj, \figyl) node {\includegraphics[scale=\figscale]{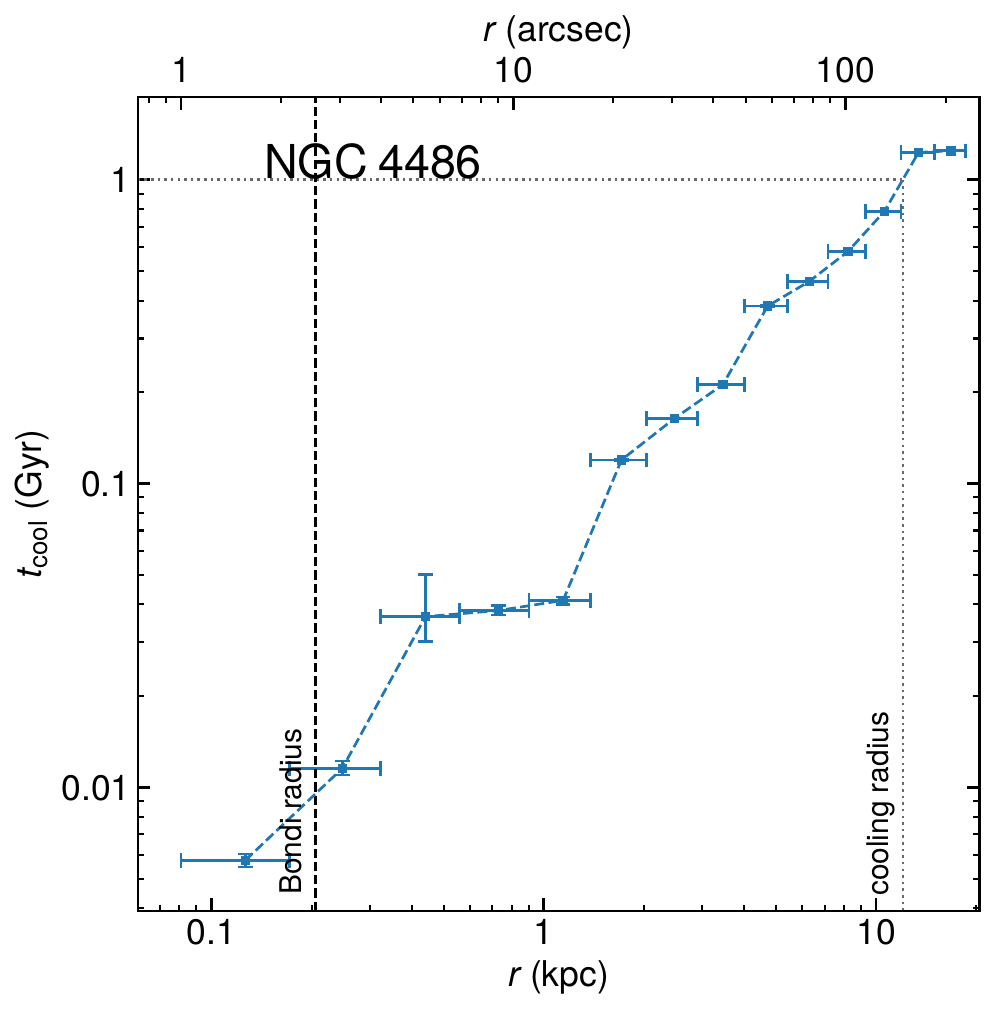}};
\draw (\figxk, \figyl) node {\includegraphics[scale=\figscale]{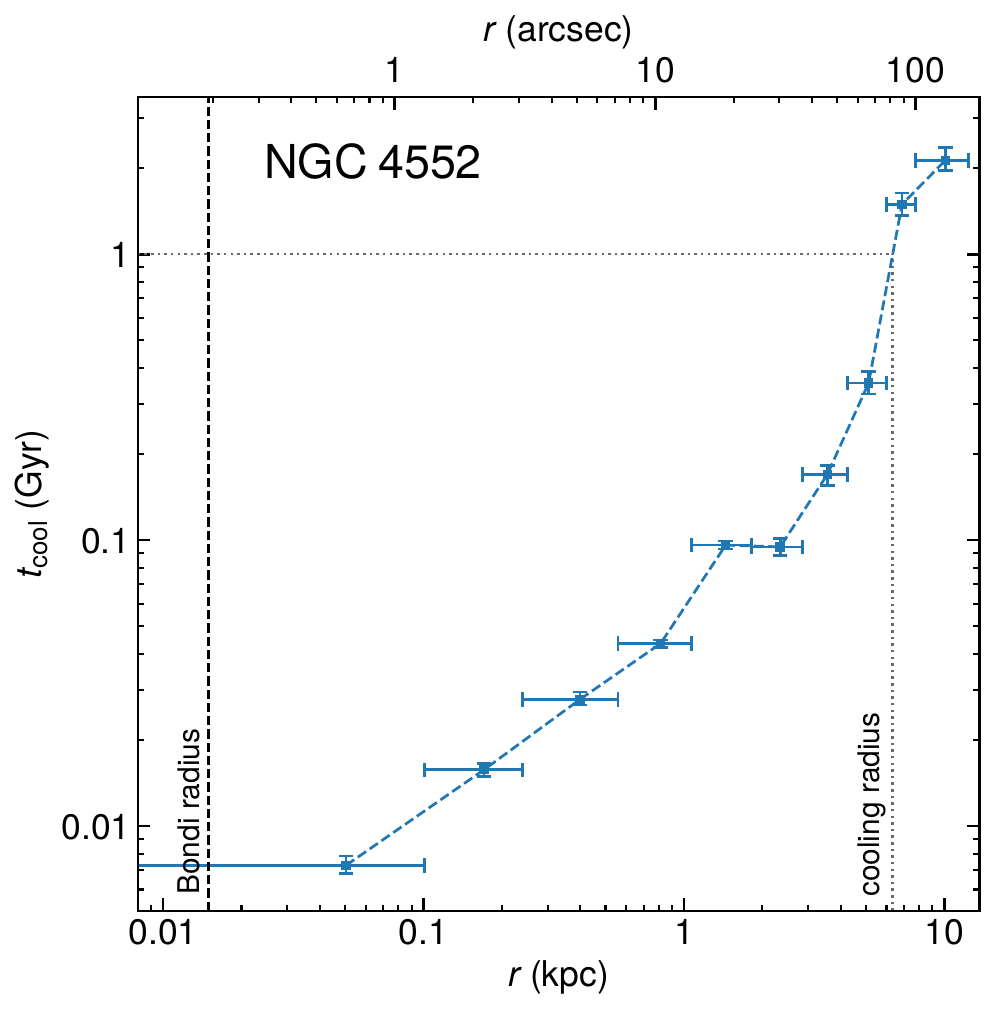}};
\end{tikzpicture}
\end{figure}

\begin{figure*}
\begin{tikzpicture}
\draw (\figxi, \figyi) node {\includegraphics[scale=\figscale]{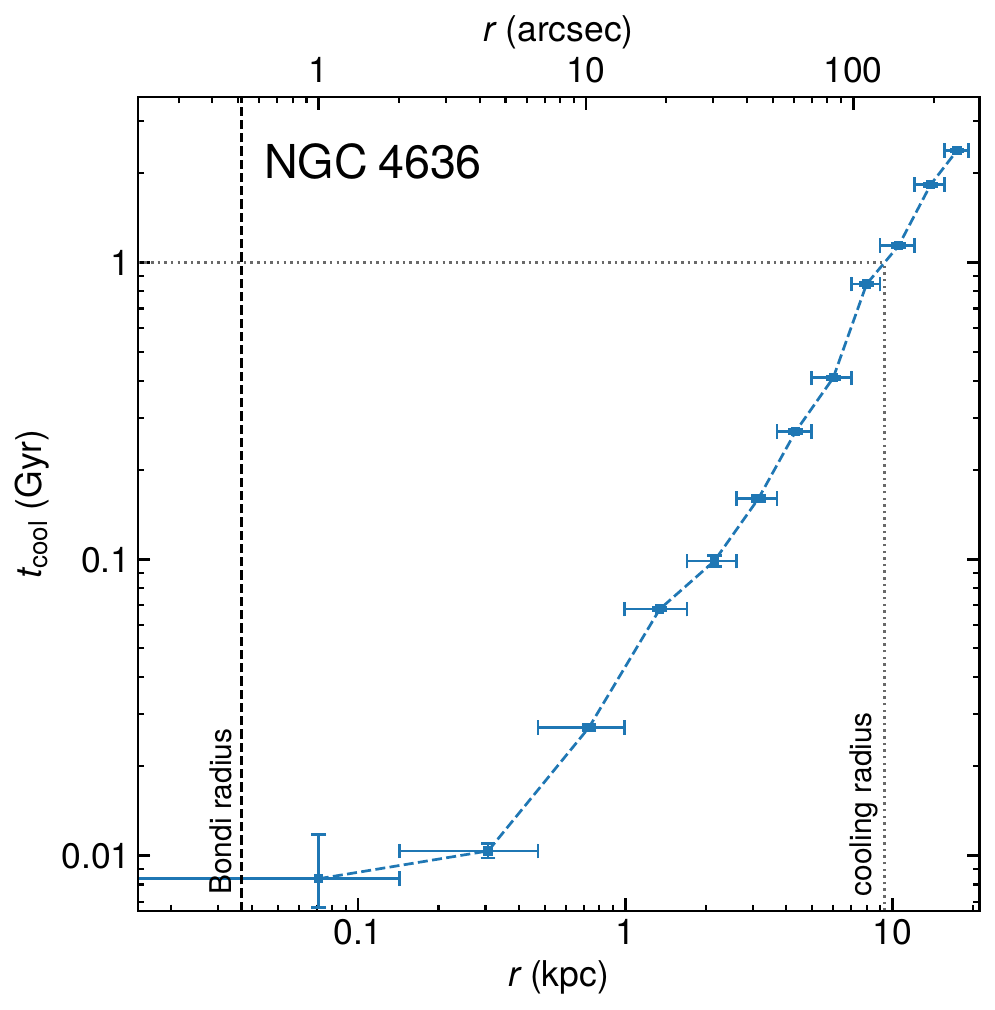}};
\draw (\figxj, \figyi) node {\includegraphics[scale=\figscale]{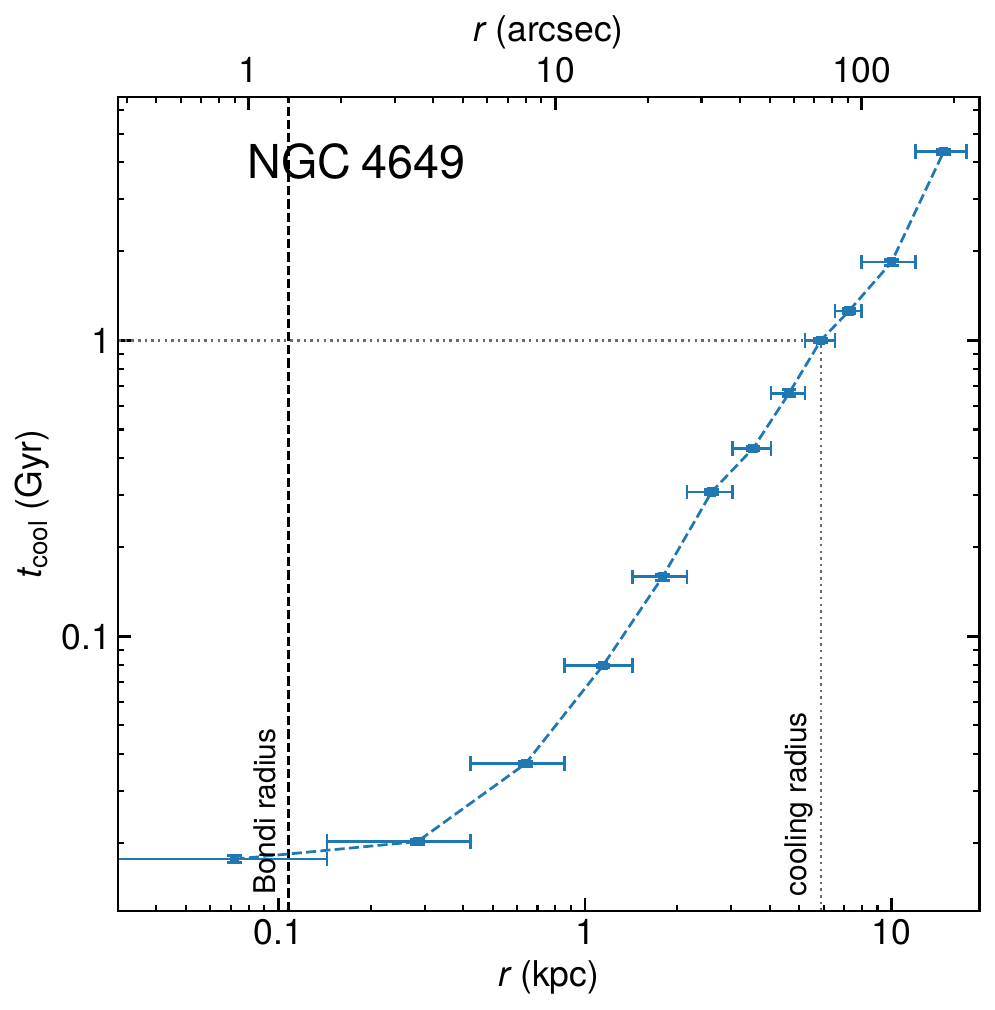}};
\draw (\figxk, \figyi) node {\includegraphics[scale=\figscale]{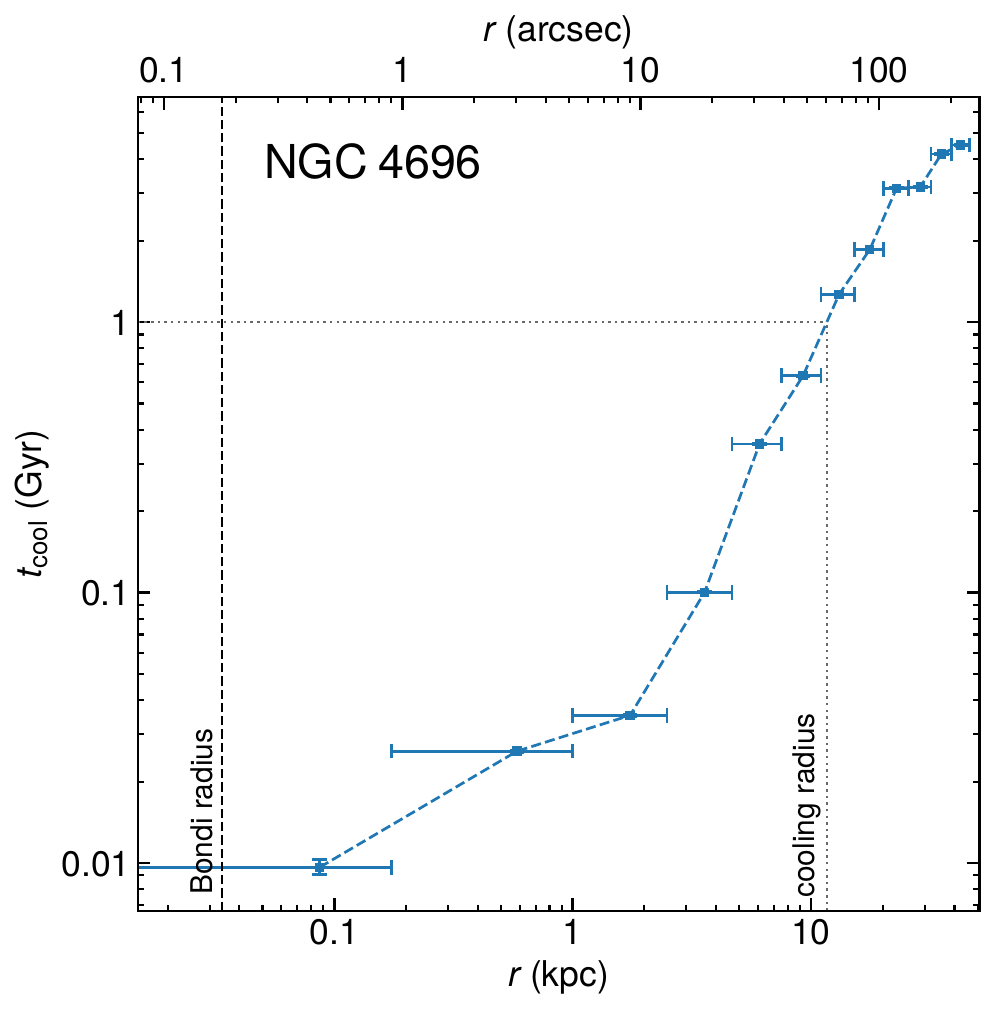}};
\draw (\figxi, \figyj) node {\includegraphics[scale=\figscale]{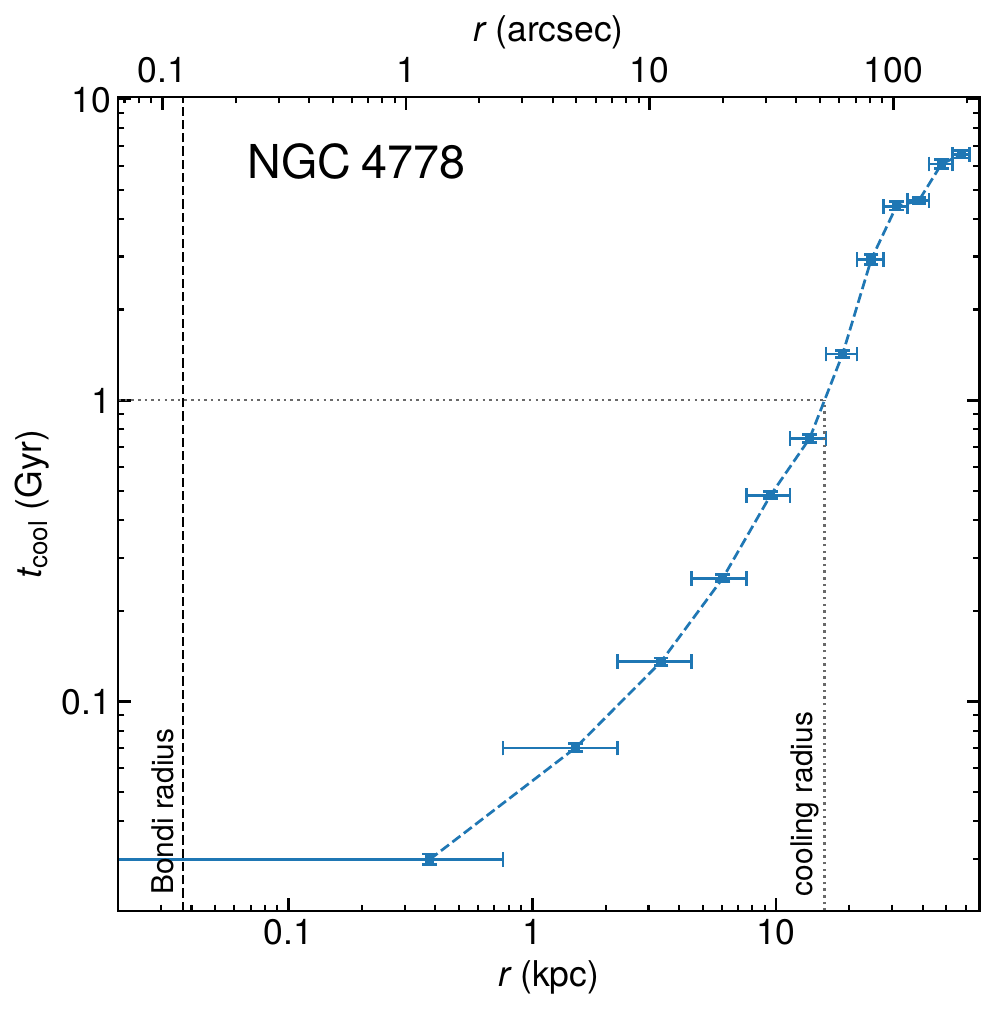}};
\draw (\figxj, \figyj) node {\includegraphics[scale=\figscale]{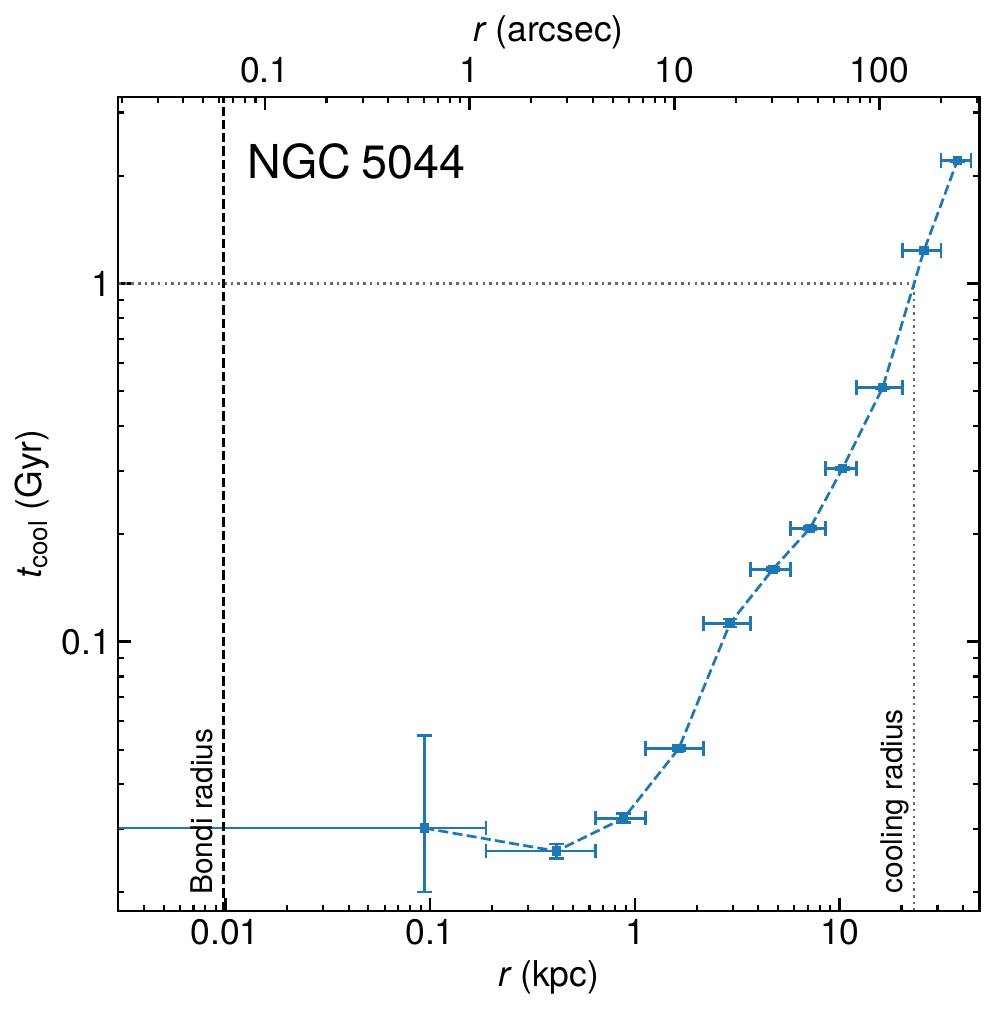}};
\draw (\figxk, \figyj) node {\includegraphics[scale=\figscale]{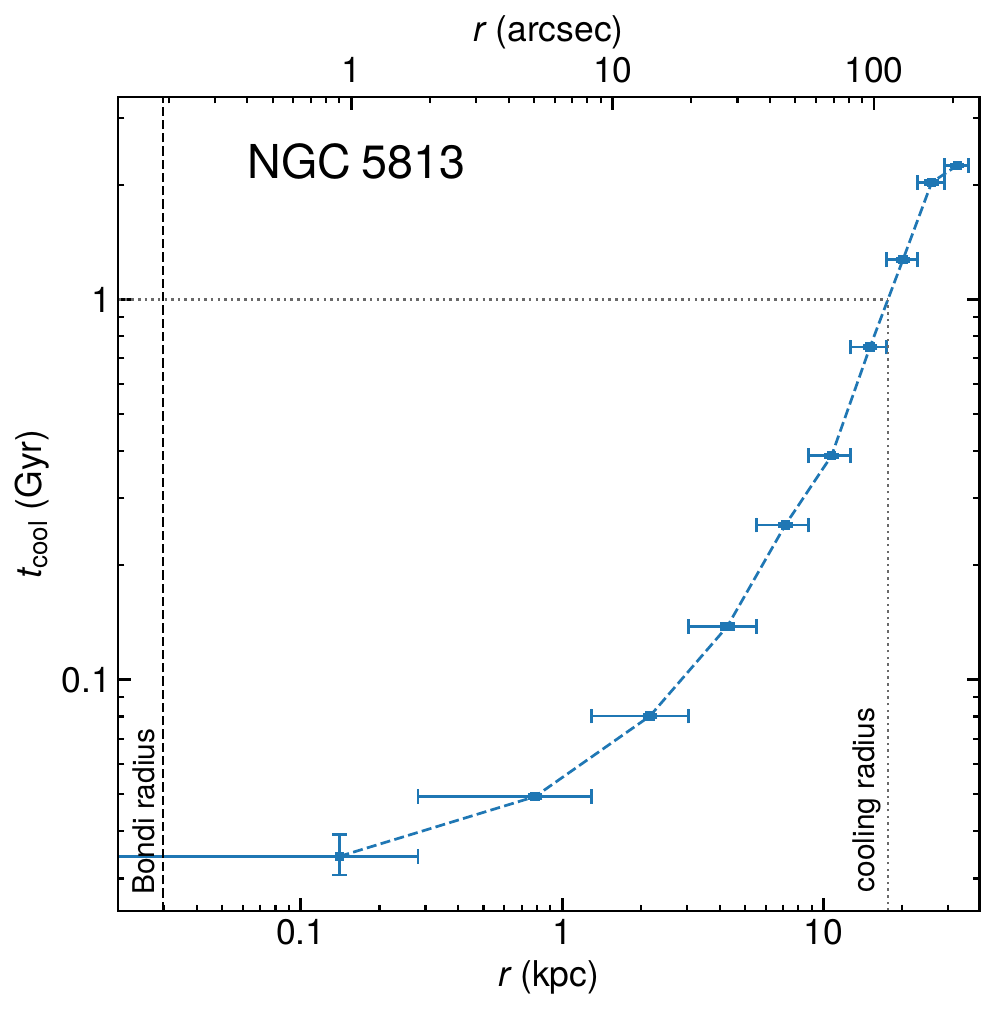}};
\draw (\figxi, \figyk) node {\includegraphics[scale=\figscale]{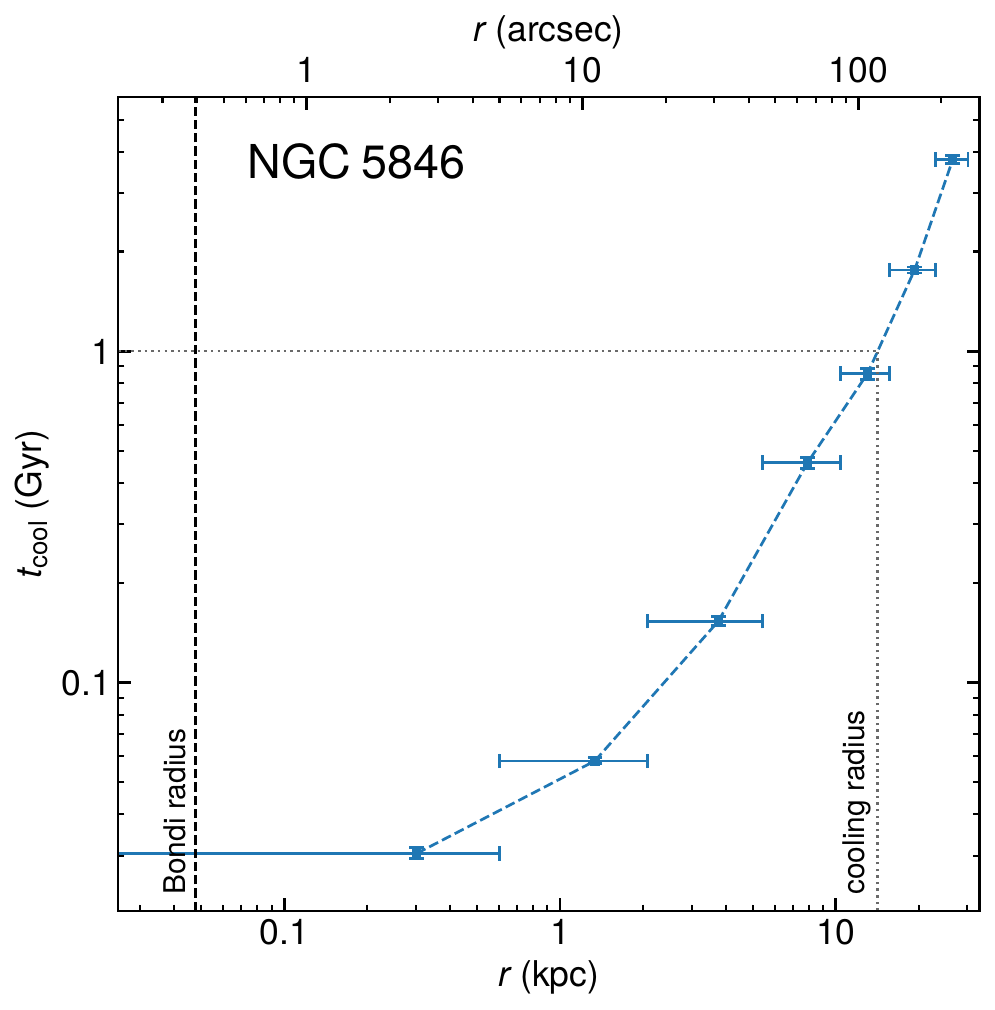}};
\draw (\figxj, \figyk) node {\includegraphics[scale=\figscale]{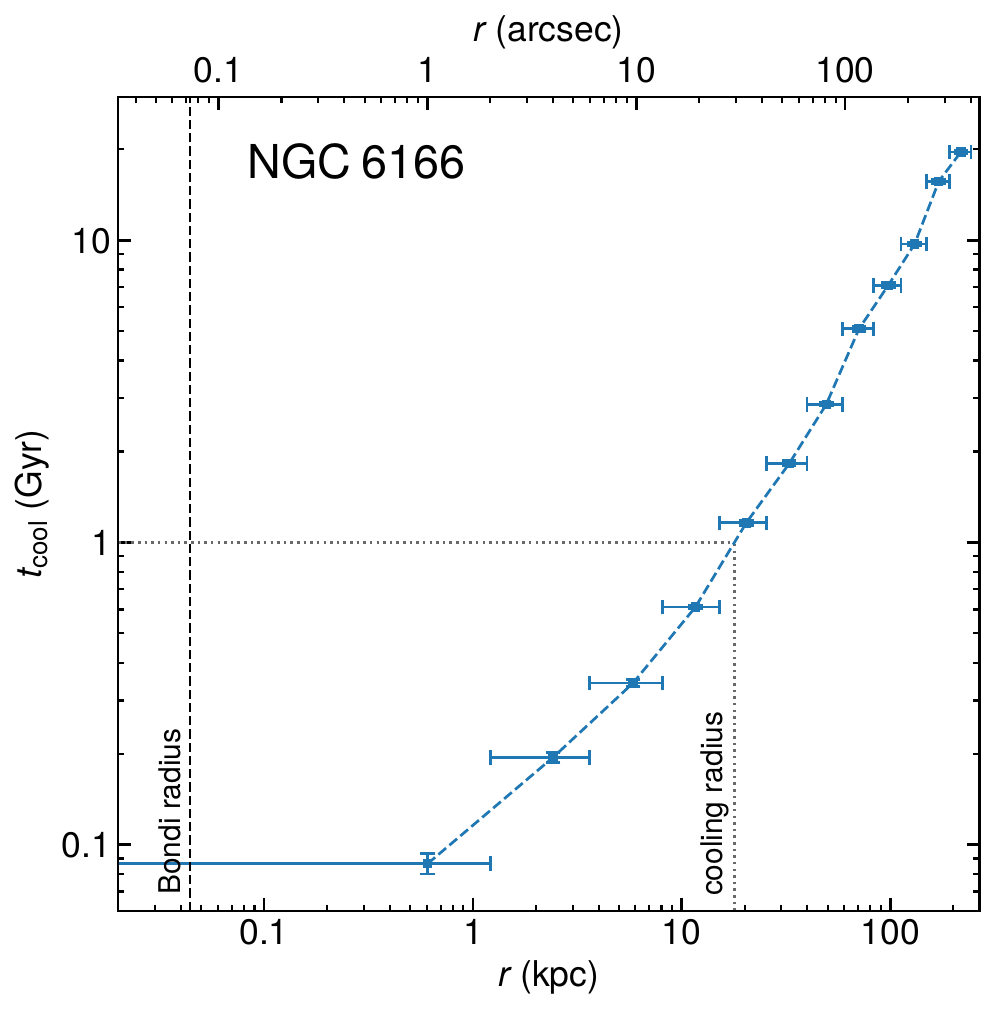}};
\end{tikzpicture}
\caption{Azimuthally averaged radial profiles of cooling time $t_{\text{cool}}$. The vertical dashed line represents the Bondi radius $r_{\text{Bondi}}$ while the dotted lines show the 1 Gyr limit and the corresponding cooling radius.}
\label{fig:tcool}
\end{figure*}


\begin{figure}
\begin{tikzpicture}
\draw (\figxi, \figyi) node {\includegraphics[scale=\figscale]{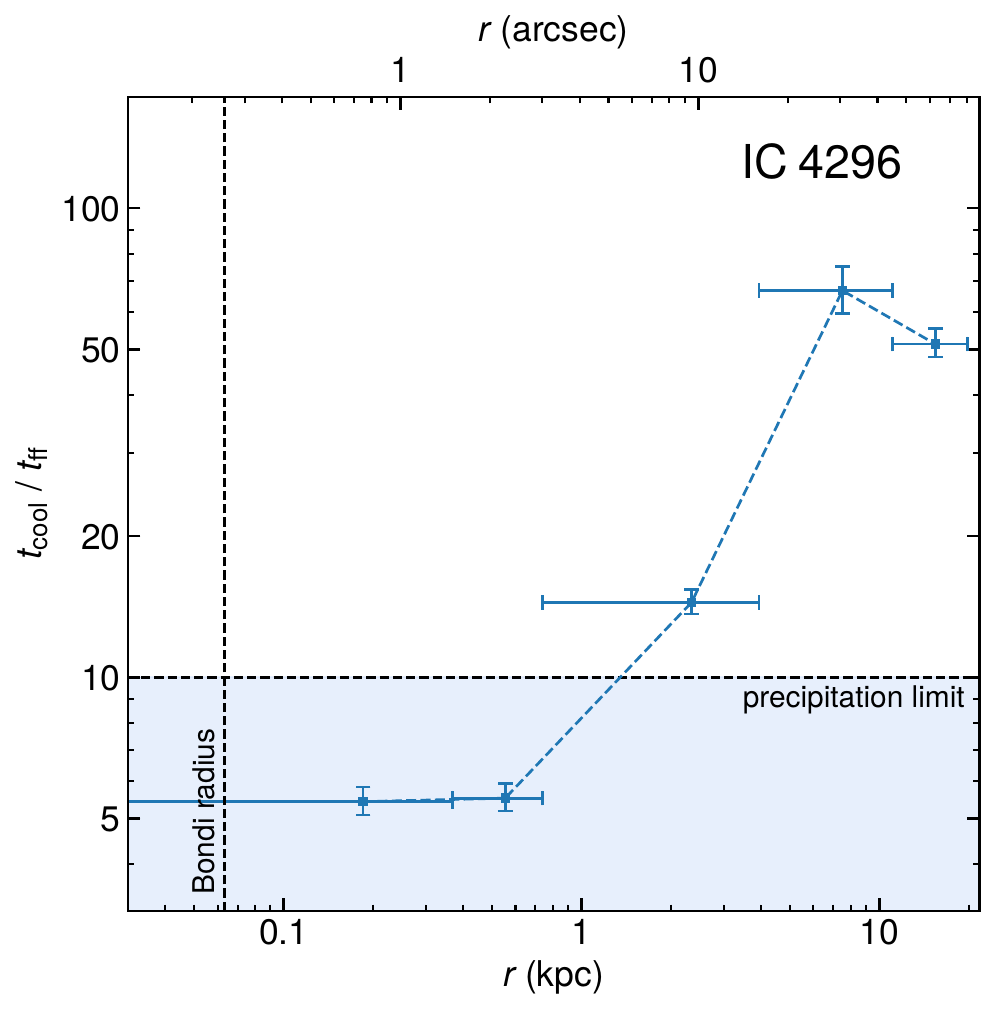}};
\draw (\figxj, \figyi) node {\includegraphics[scale=\figscale]{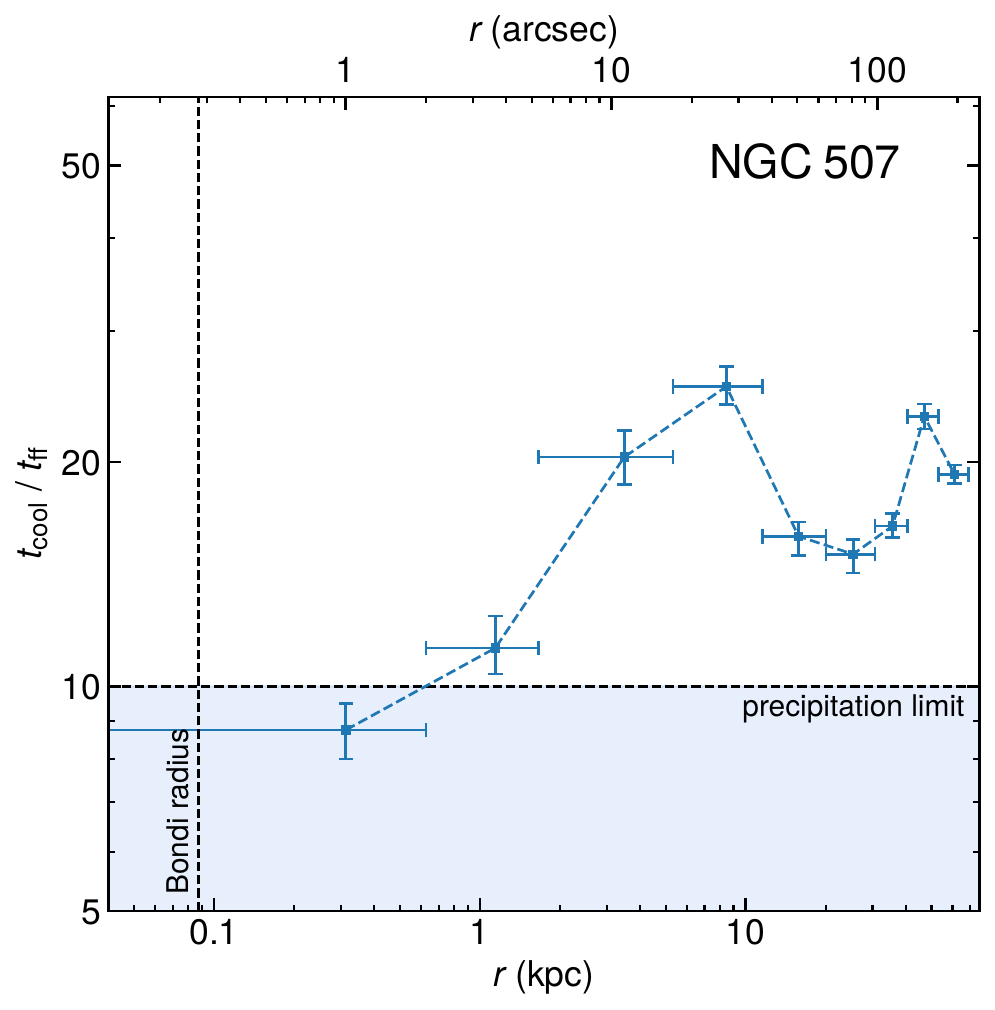}};
\draw (\figxk, \figyi) node {\includegraphics[scale=\figscale]{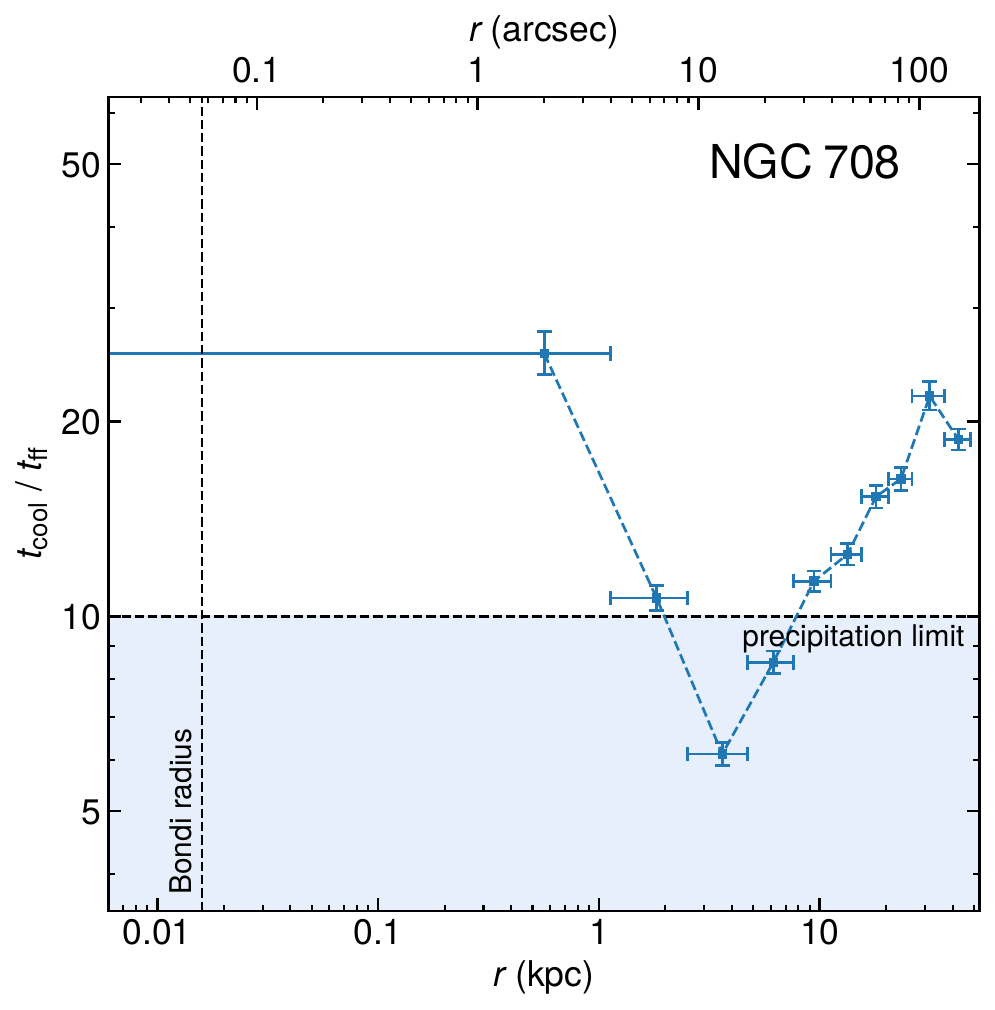}};
\draw (\figxi, \figyj) node {\includegraphics[scale=\figscale]{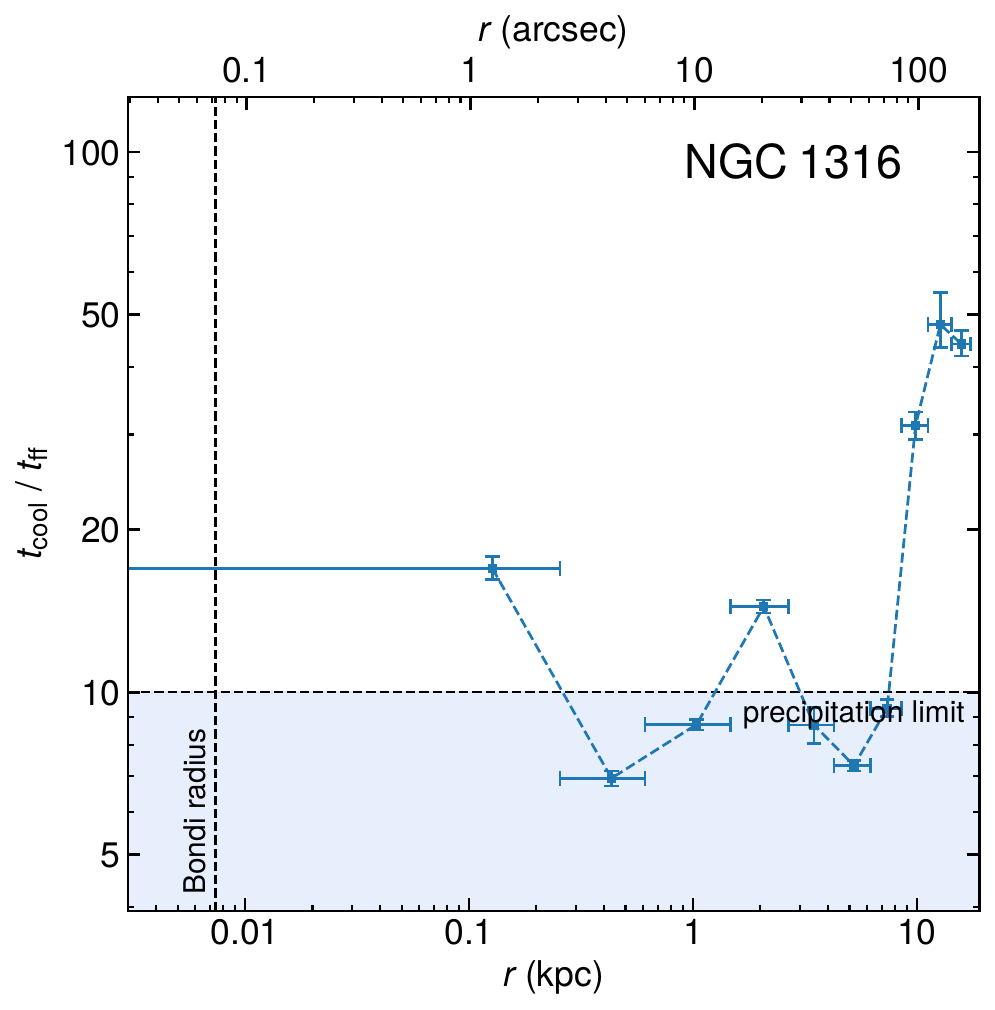}};
\draw (\figxj, \figyj) node {\includegraphics[scale=\figscale]{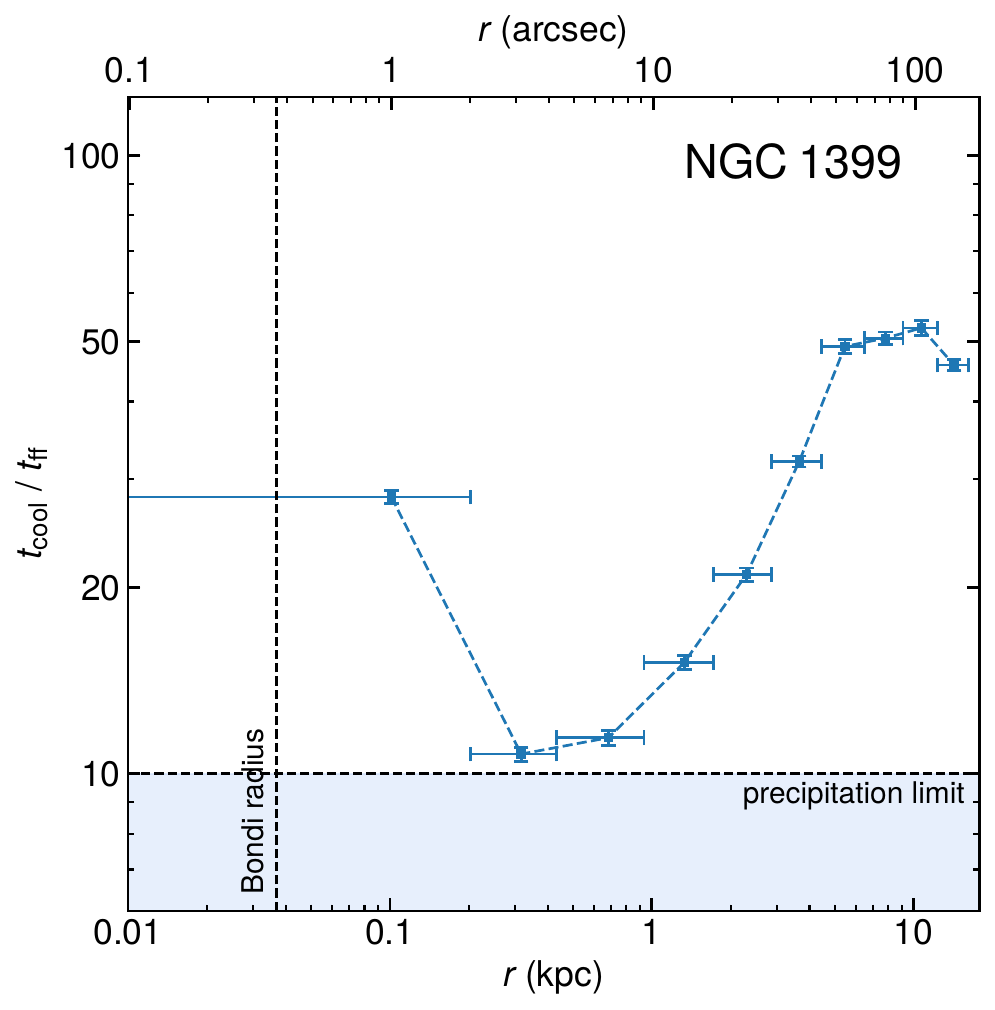}};
\draw (\figxk, \figyj) node {\includegraphics[scale=\figscale]{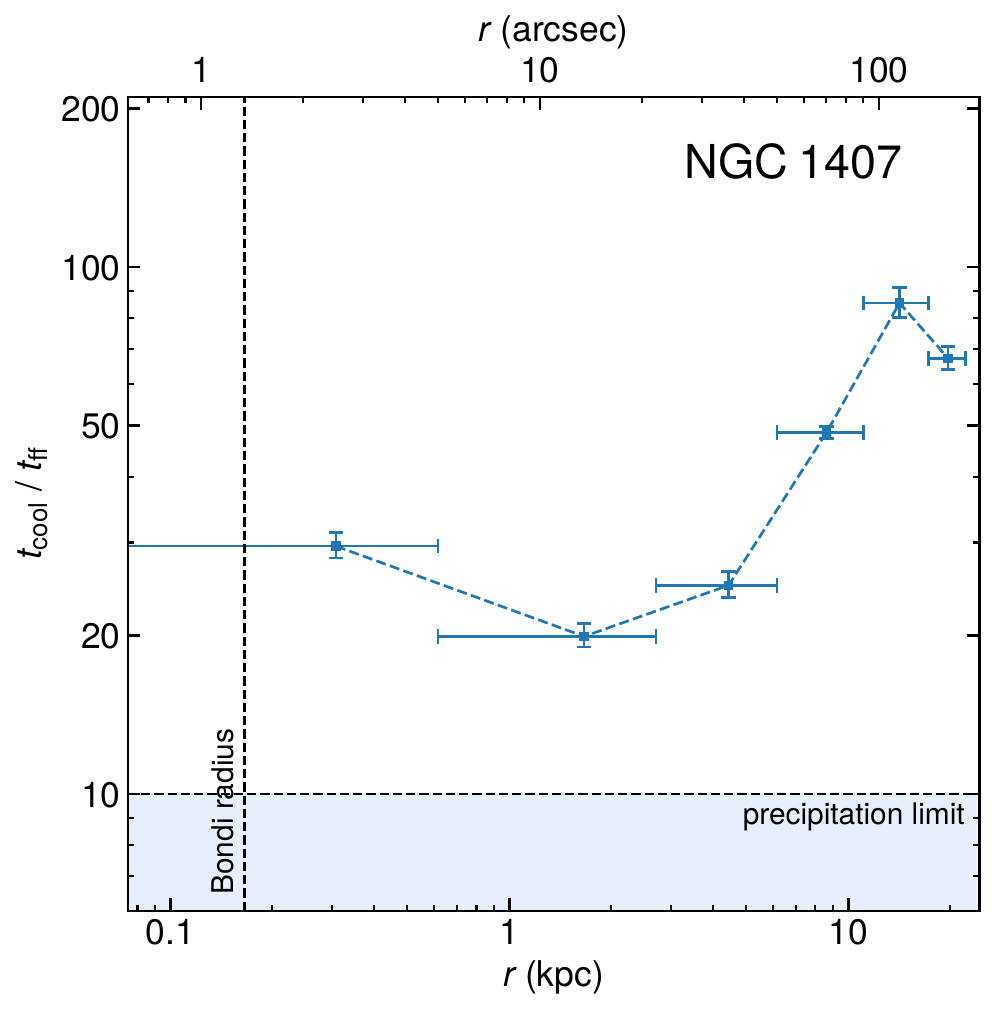}};
\draw (\figxi, \figyk) node {\includegraphics[scale=\figscale]{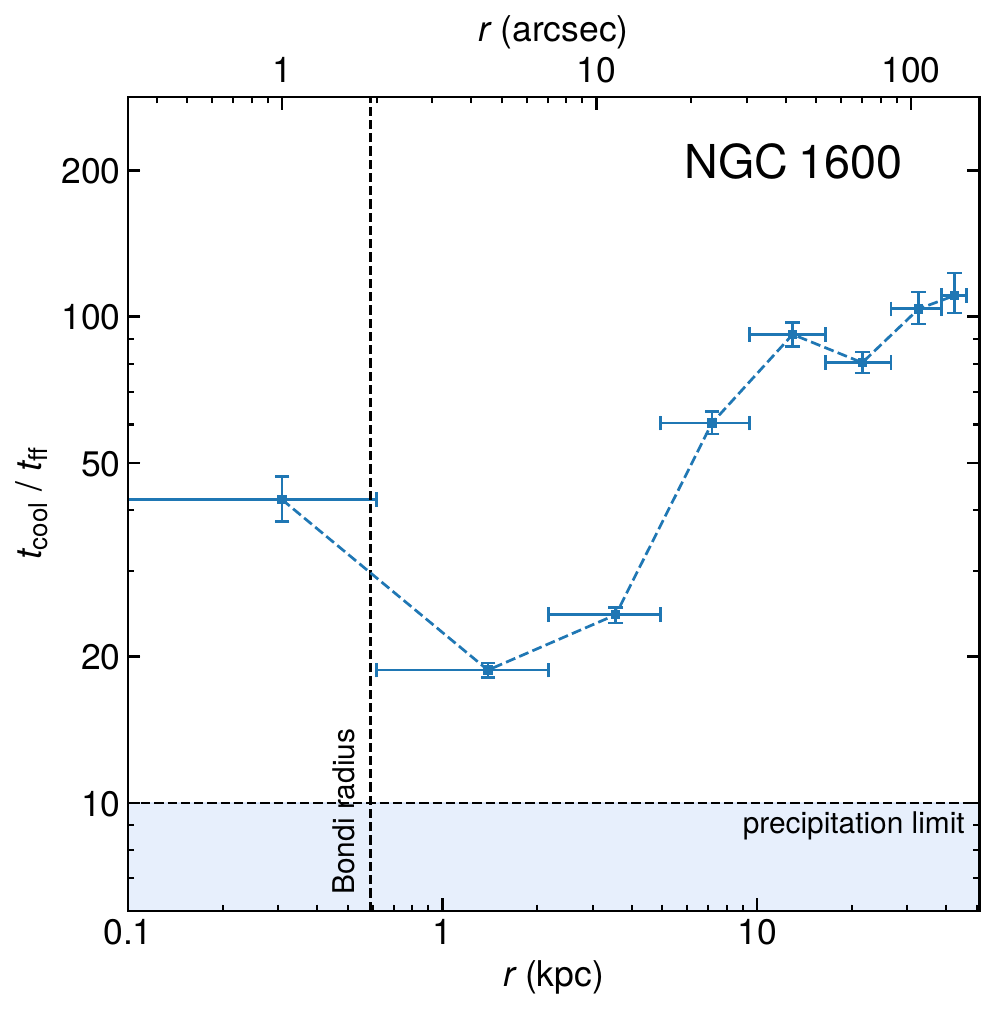}};
\draw (\figxj, \figyk) node {\includegraphics[scale=\figscale]{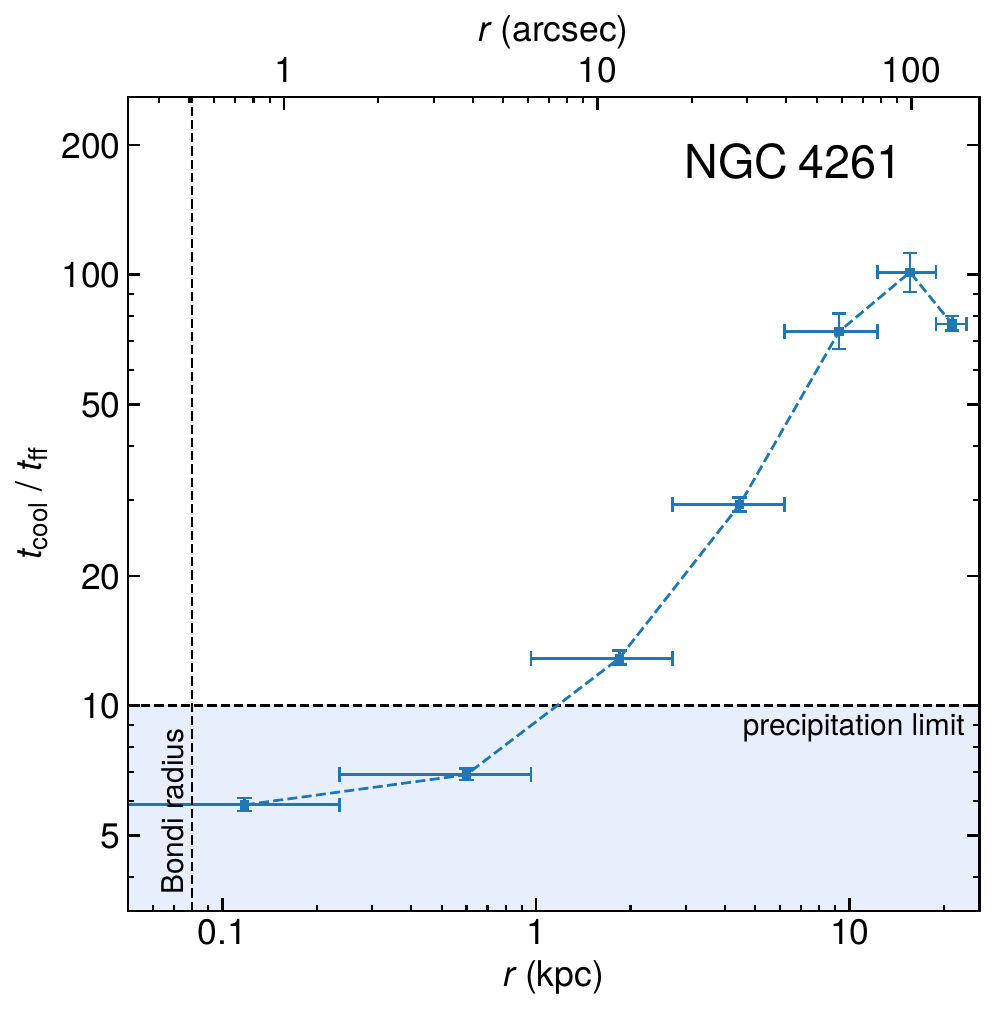}};
\draw (\figxk, \figyk) node {\includegraphics[scale=\figscale]{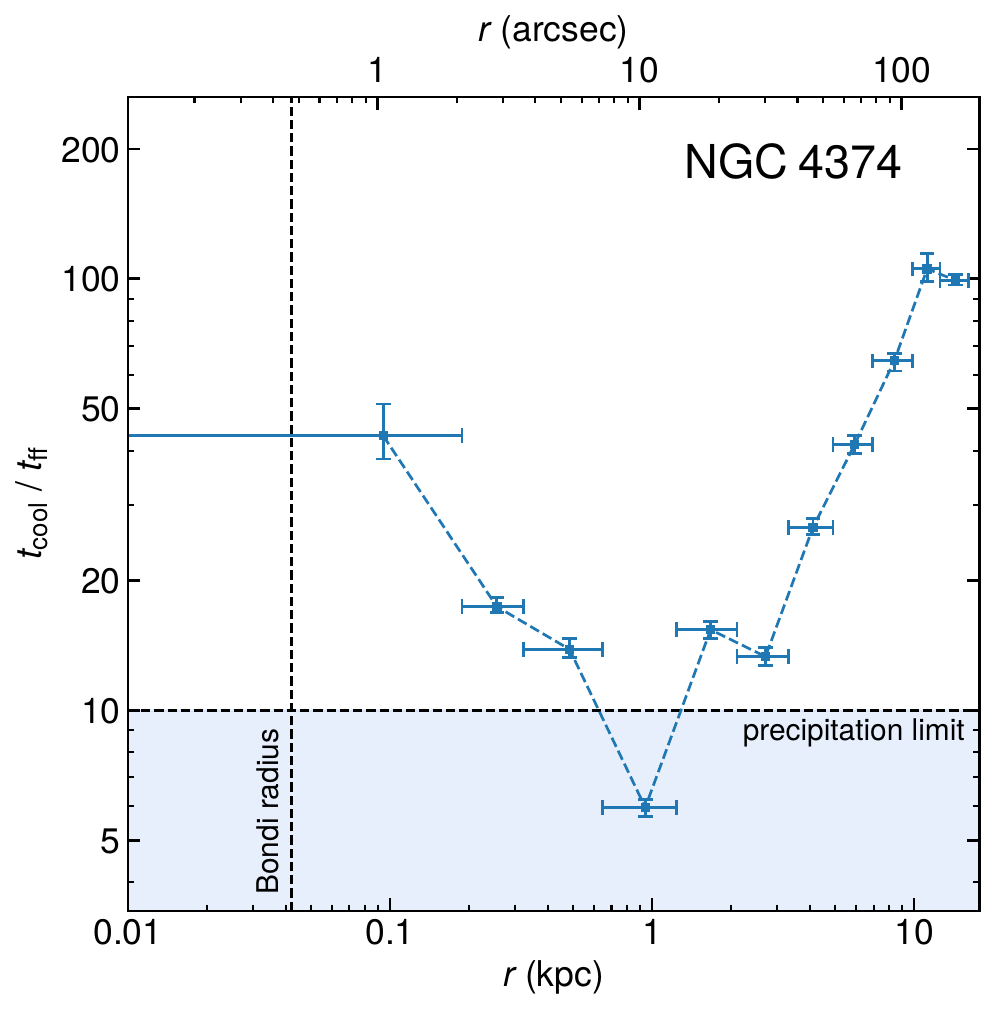}};
\draw (\figxi, \figyl) node {\includegraphics[scale=\figscale]{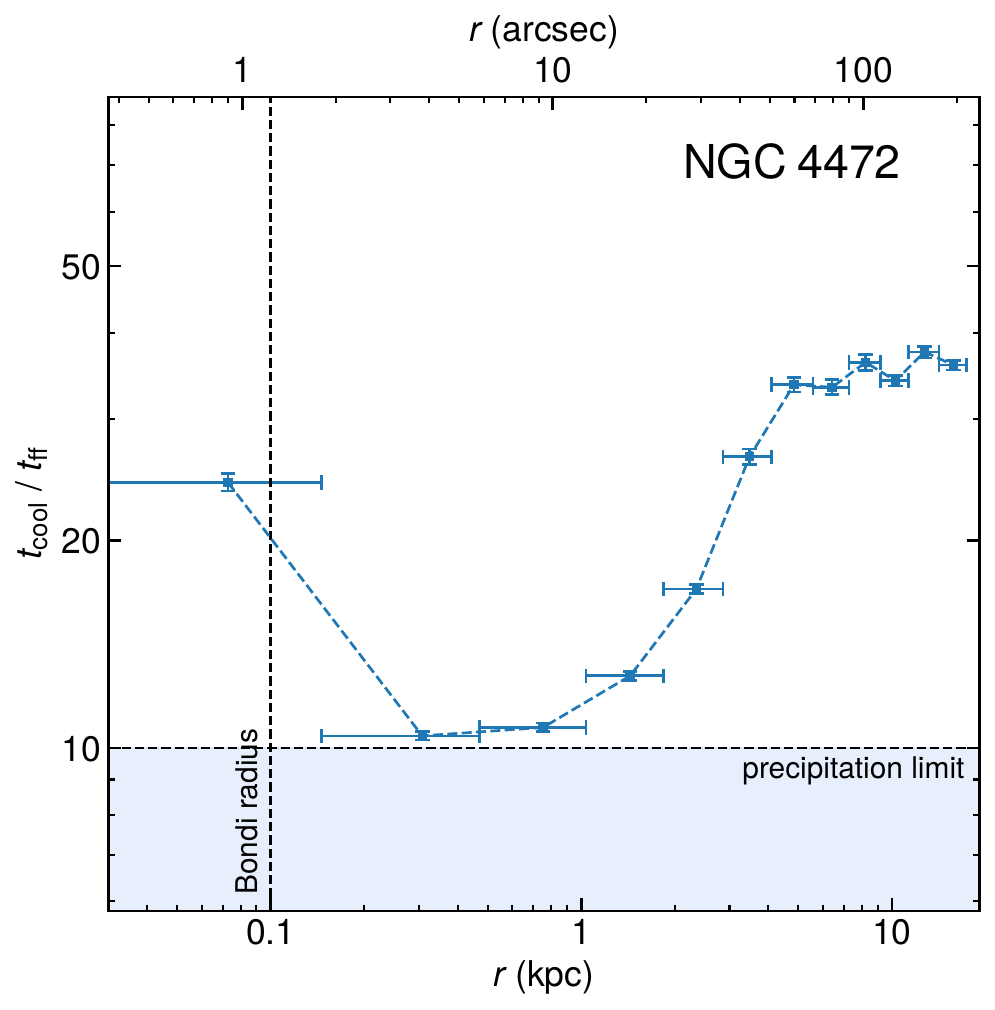}};
\draw (\figxj, \figyl) node {\includegraphics[scale=\figscale]{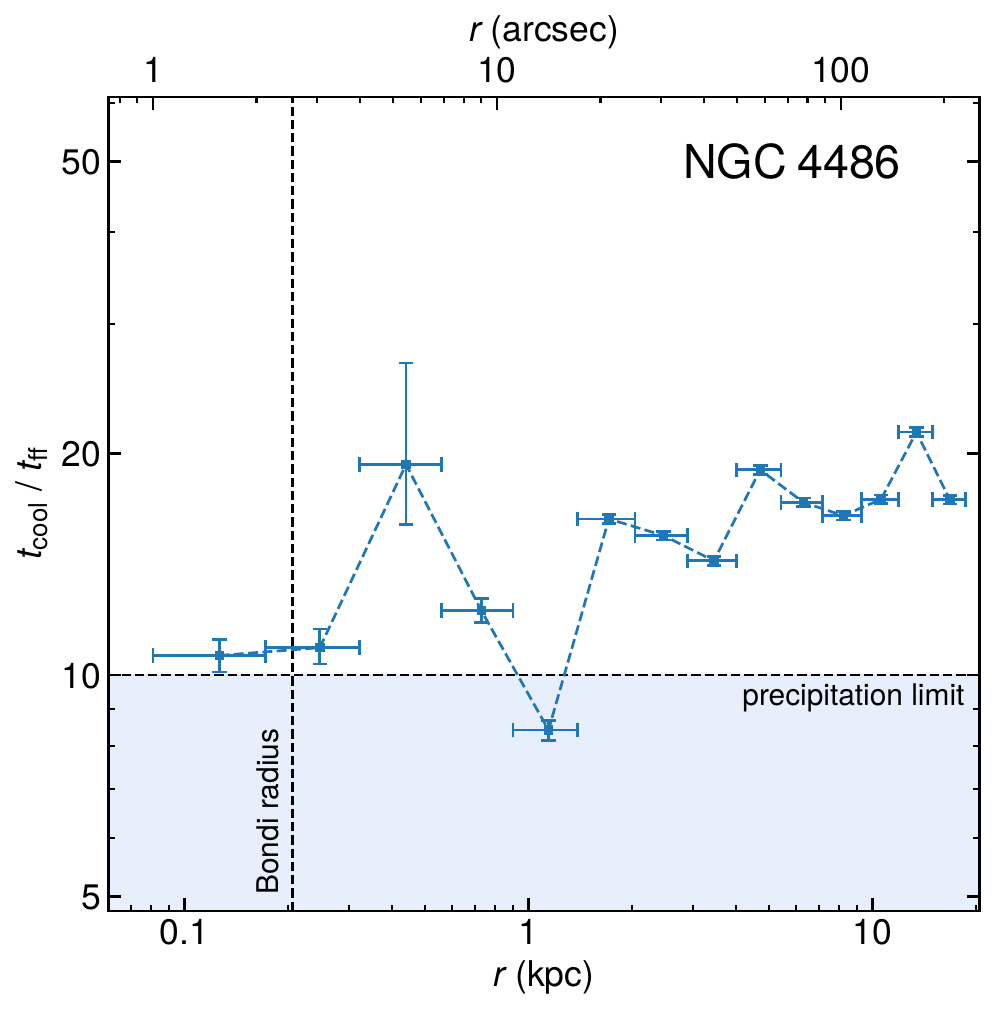}};
\draw (\figxk, \figyl) node {\includegraphics[scale=\figscale]{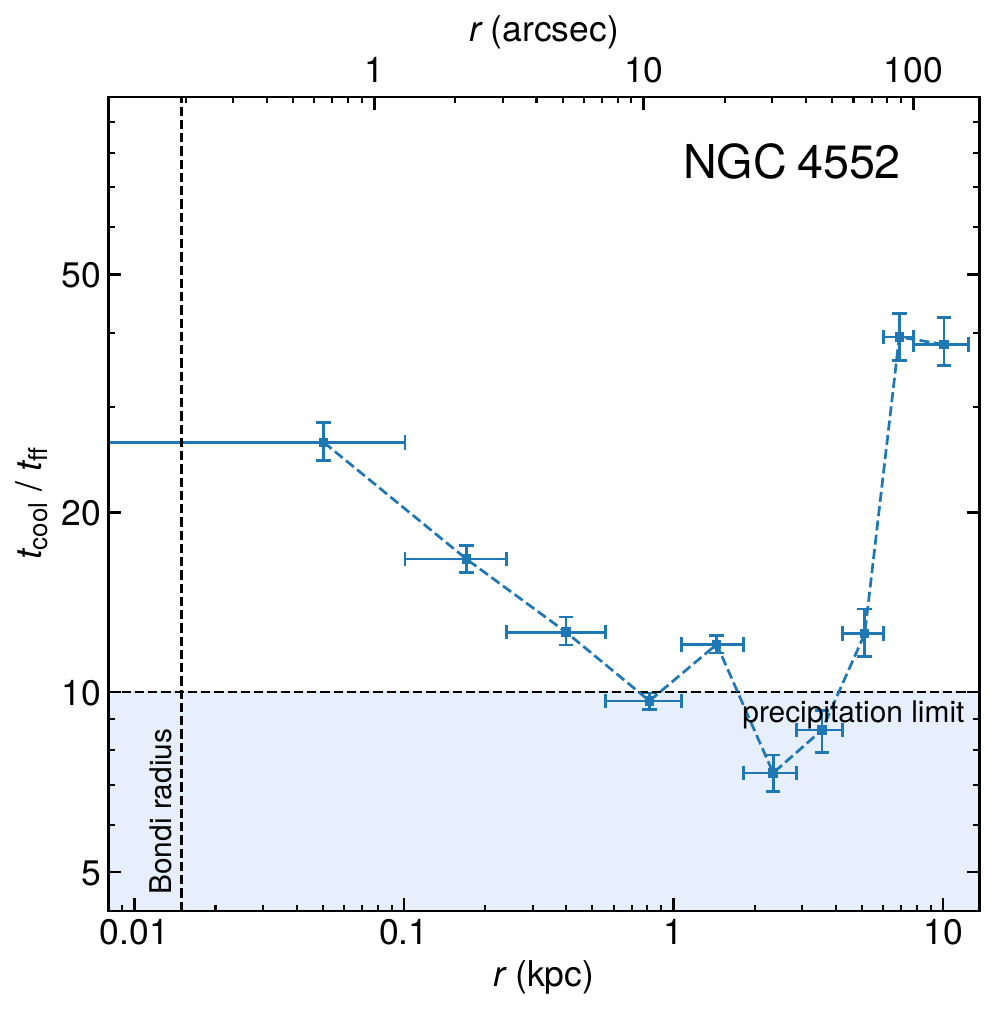}};
\end{tikzpicture}
\end{figure}

\begin{figure*}
\begin{tikzpicture}
\draw (\figxi, \figyi) node {\includegraphics[scale=\figscale]{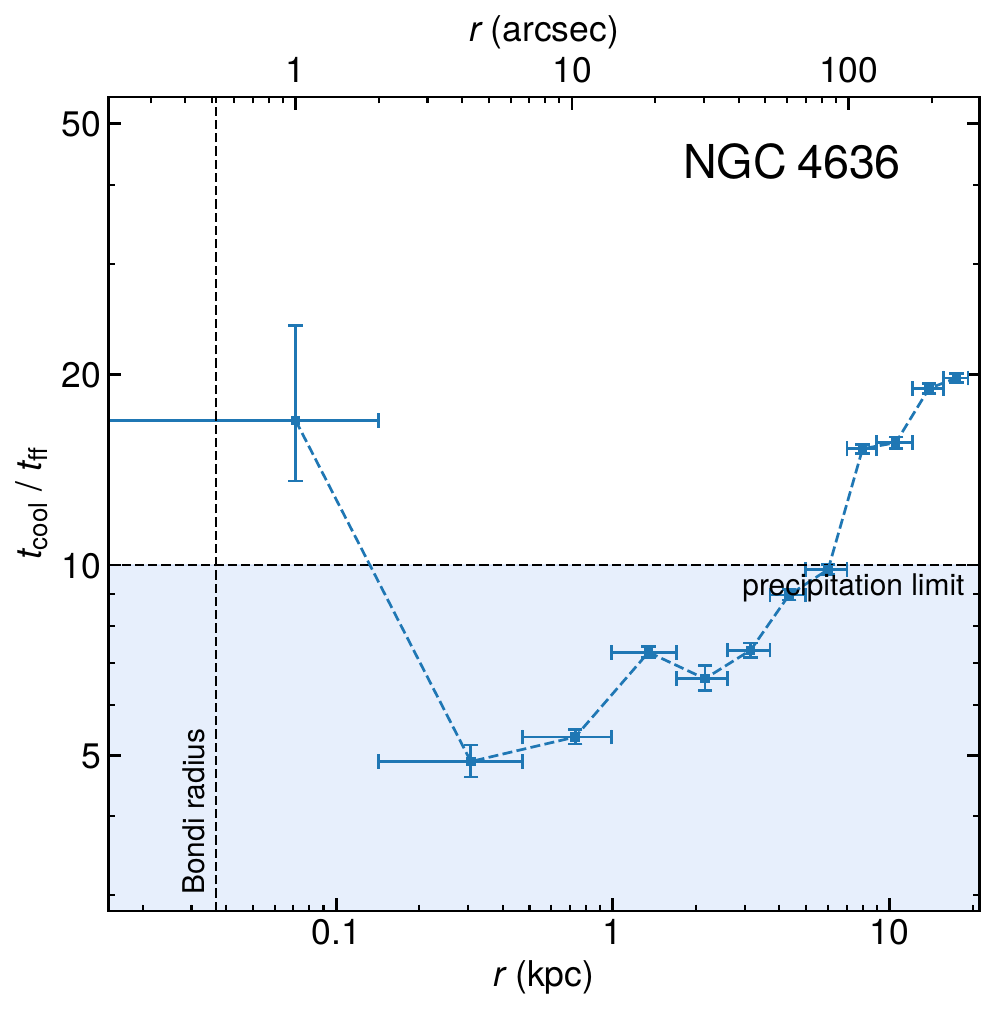}};
\draw (\figxj, \figyi) node {\includegraphics[scale=\figscale]{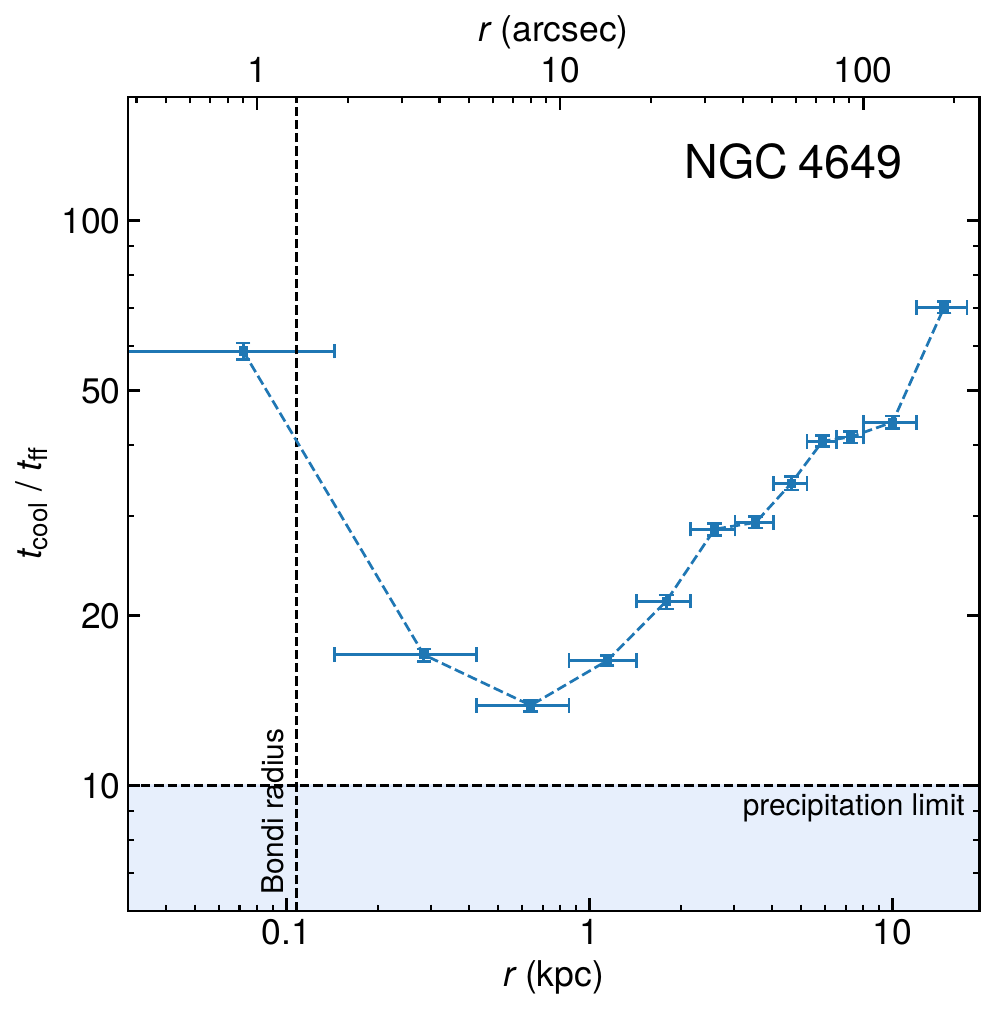}};
\draw (\figxk, \figyi) node {\includegraphics[scale=\figscale]{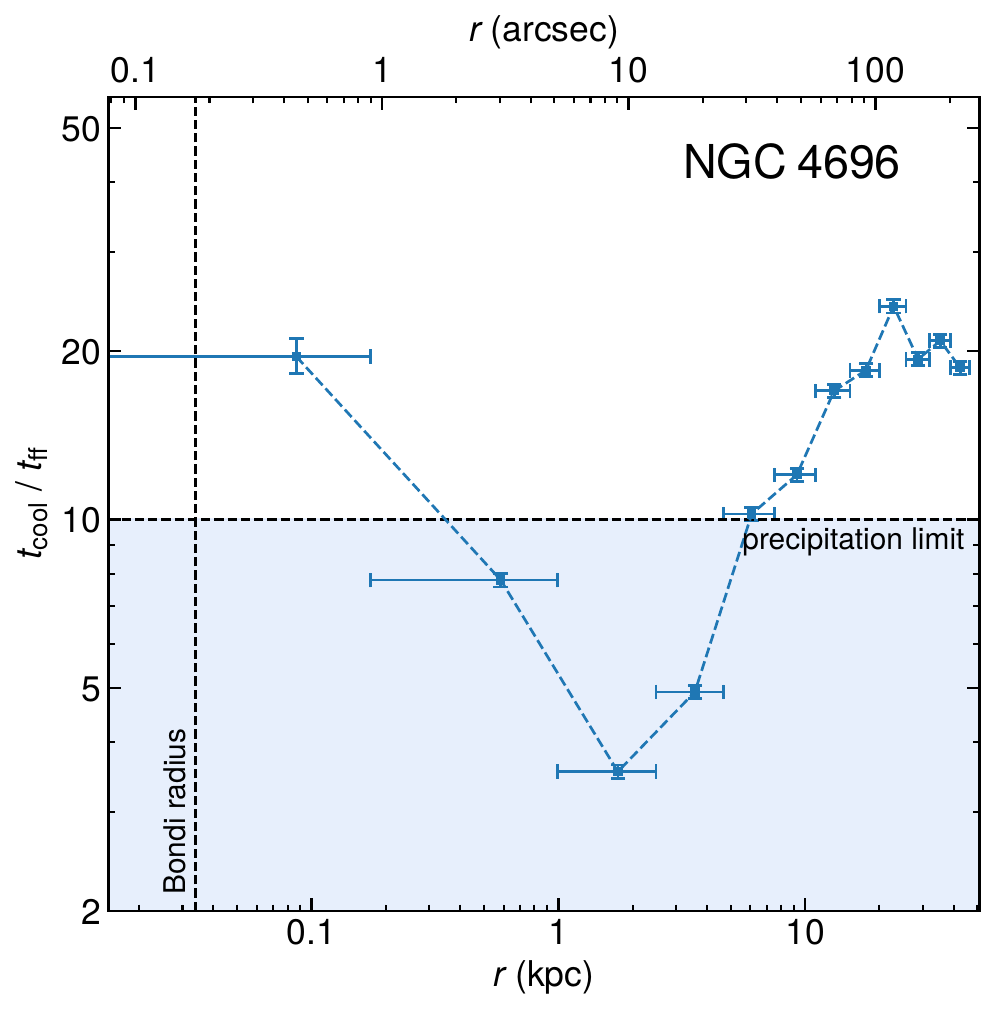}};
\draw (\figxi, \figyj) node {\includegraphics[scale=\figscale]{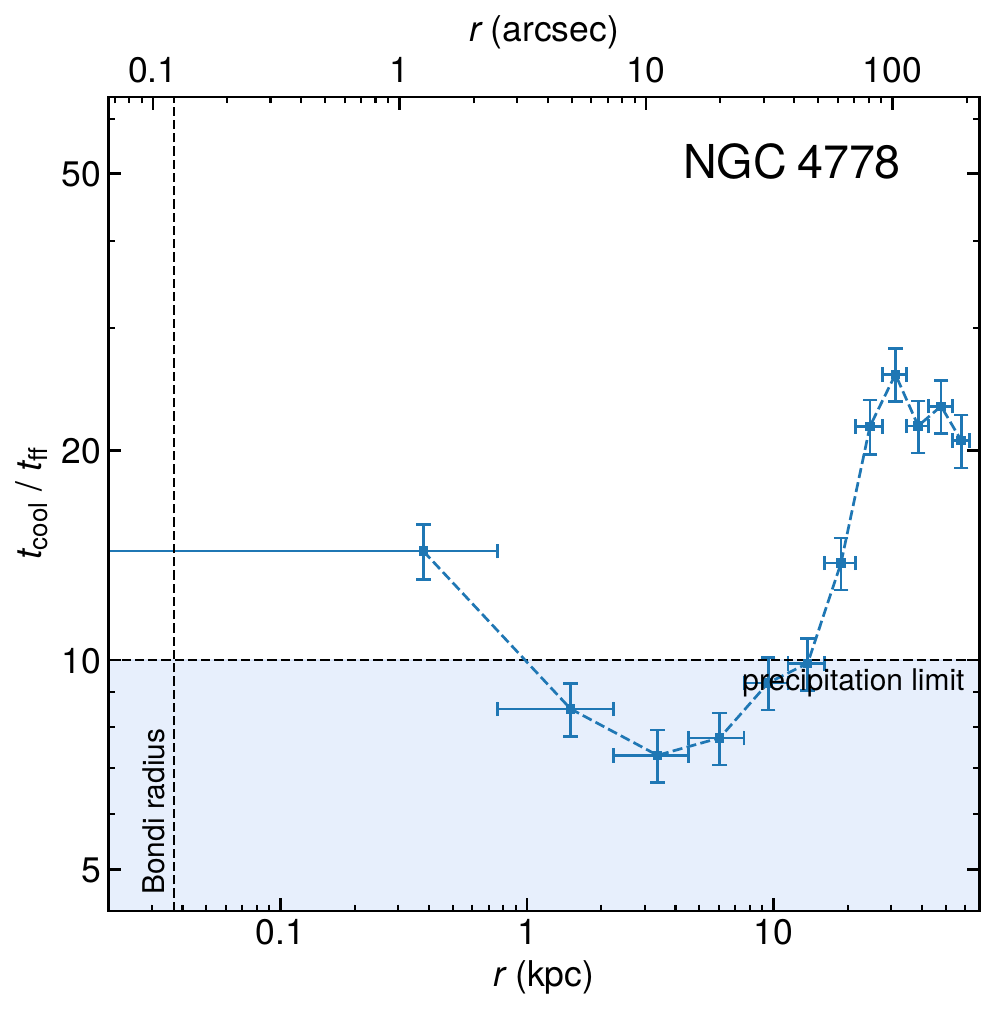}};
\draw (\figxj, \figyj) node {\includegraphics[scale=\figscale]{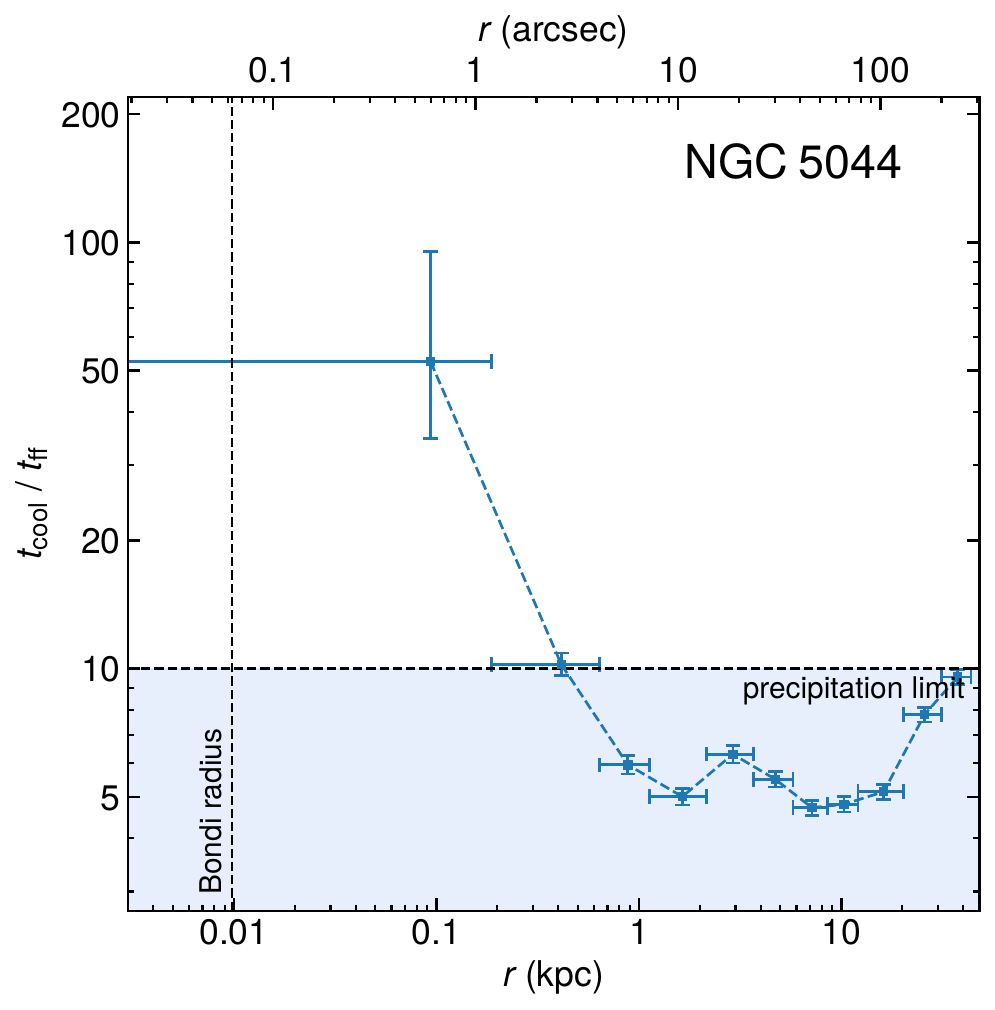}};
\draw (\figxk, \figyj) node {\includegraphics[scale=\figscale]{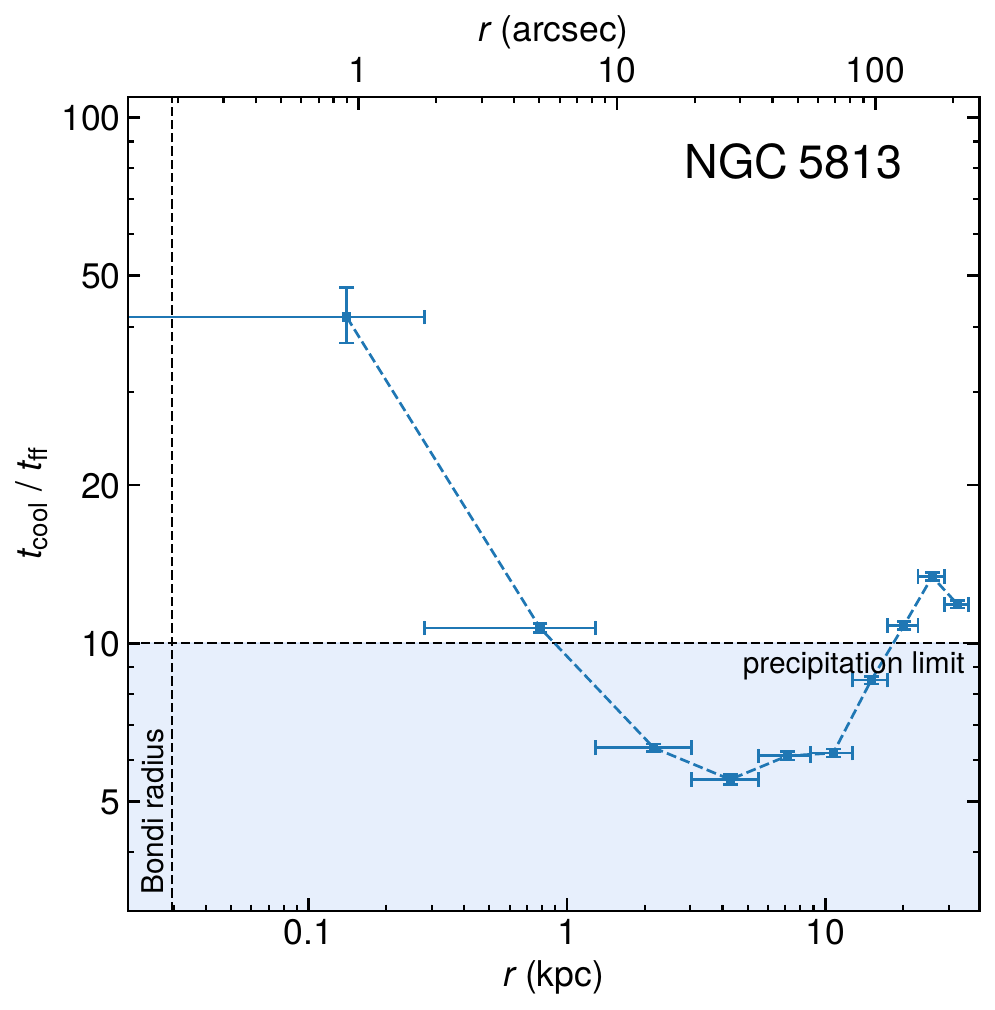}};
\draw (\figxi, \figyk) node {\includegraphics[scale=\figscale]{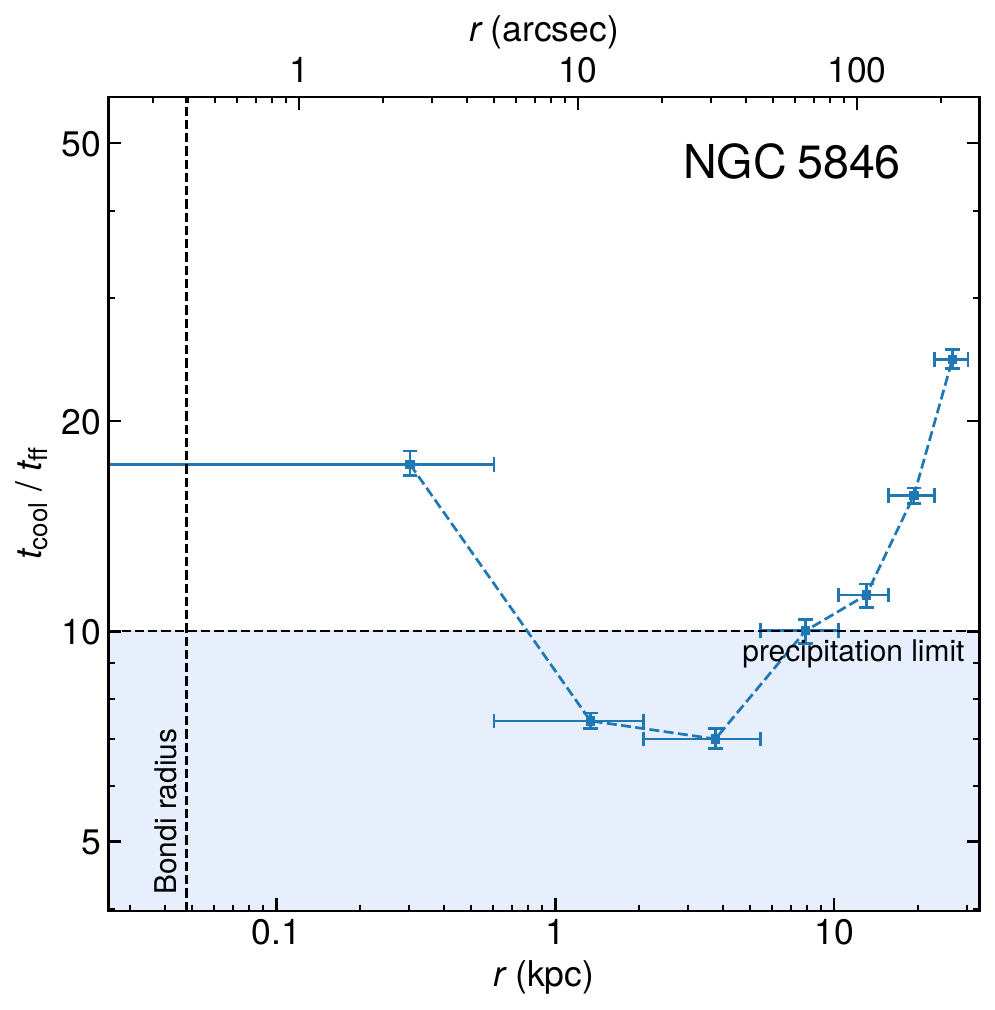}};
\draw (\figxj, \figyk) node {\includegraphics[scale=\figscale]{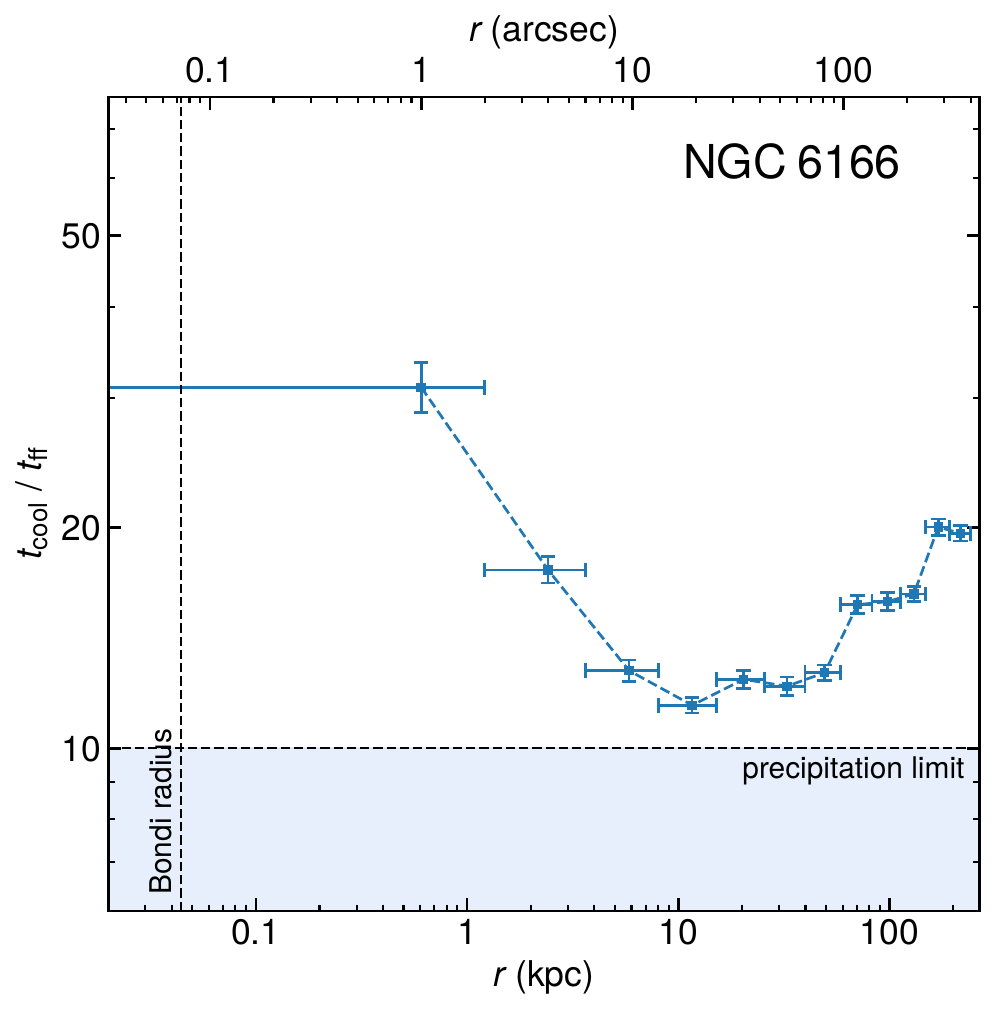}};
\end{tikzpicture}
\caption{Azimuthally averaged radial profiles of cooling time to free fall time ratio $t_{\mathrm{cool}} / t_{\mathrm{ff}}$ estimated from the assumption of an isothermal sphere. The vertical dashed line represents the Bondi radius $r_{\text{Bondi}}$, while the horizontal dashed line is the precipitation limit $t_{\mathrm{cool}} / t_{\mathrm{ff}} \approx 10$.}
\label{fig:tcooltff}
\end{figure*}

\clearpage


\section{Radio lobes}
\label{appendix:lobes}

\begin{figure}
\begin{tikzpicture}[overlay]
\draw (\figxi, \figyii) node {\includegraphics[height=\imgheight]{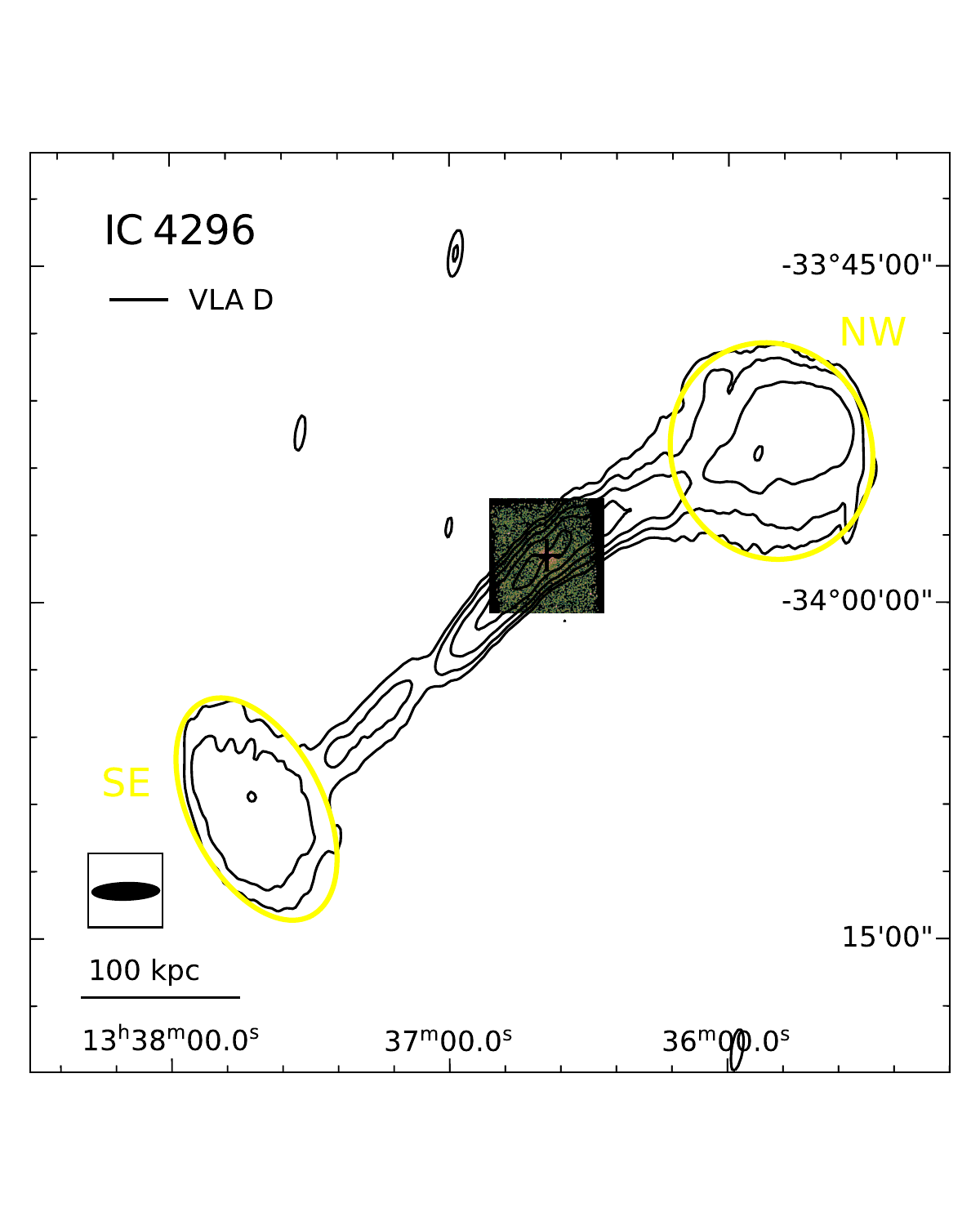}};
\draw (\figxj, \figyii) node {\includegraphics[height=\imgheight]{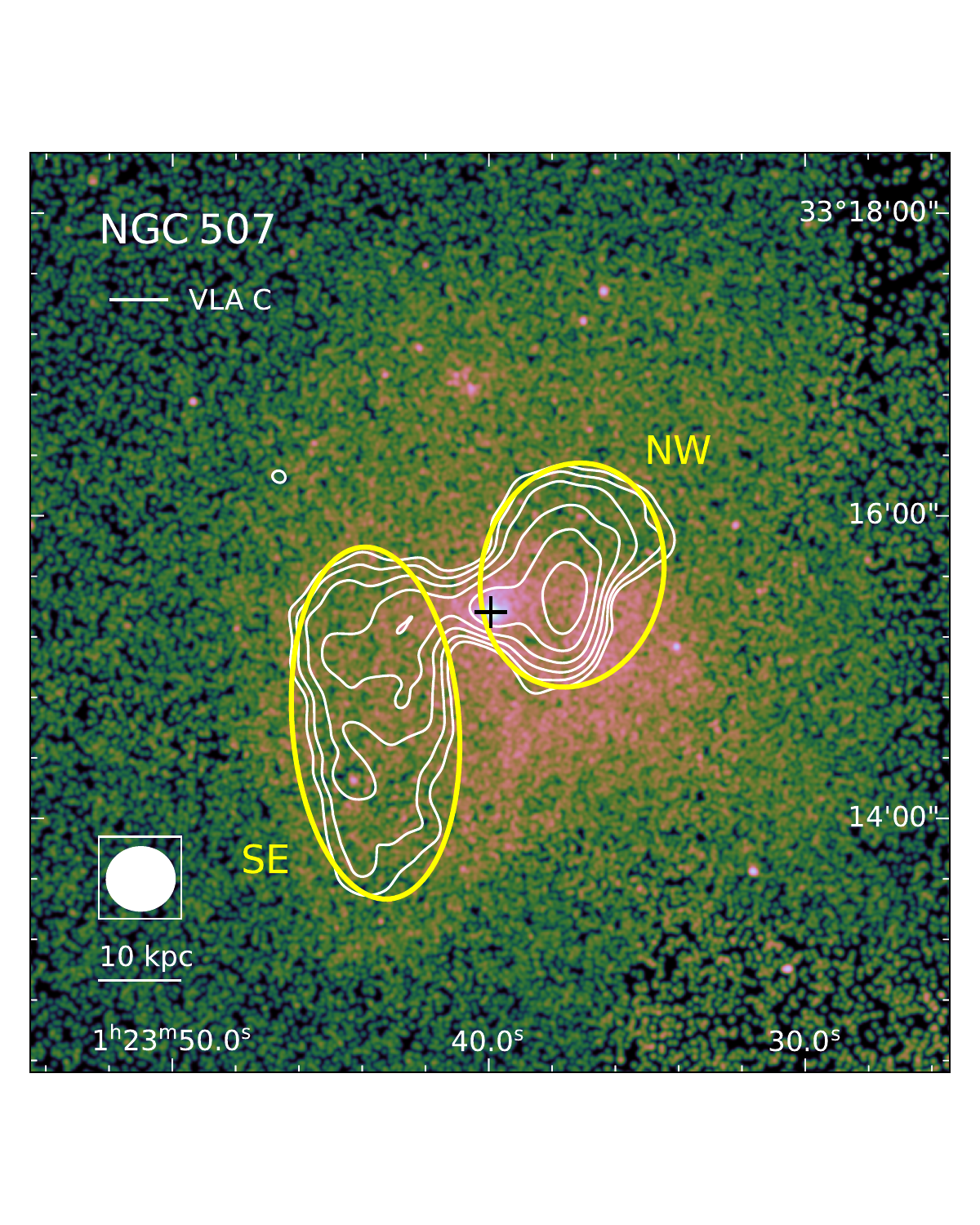}};
\draw (\figxk, \figyii) node {\includegraphics[height=\imgheight]{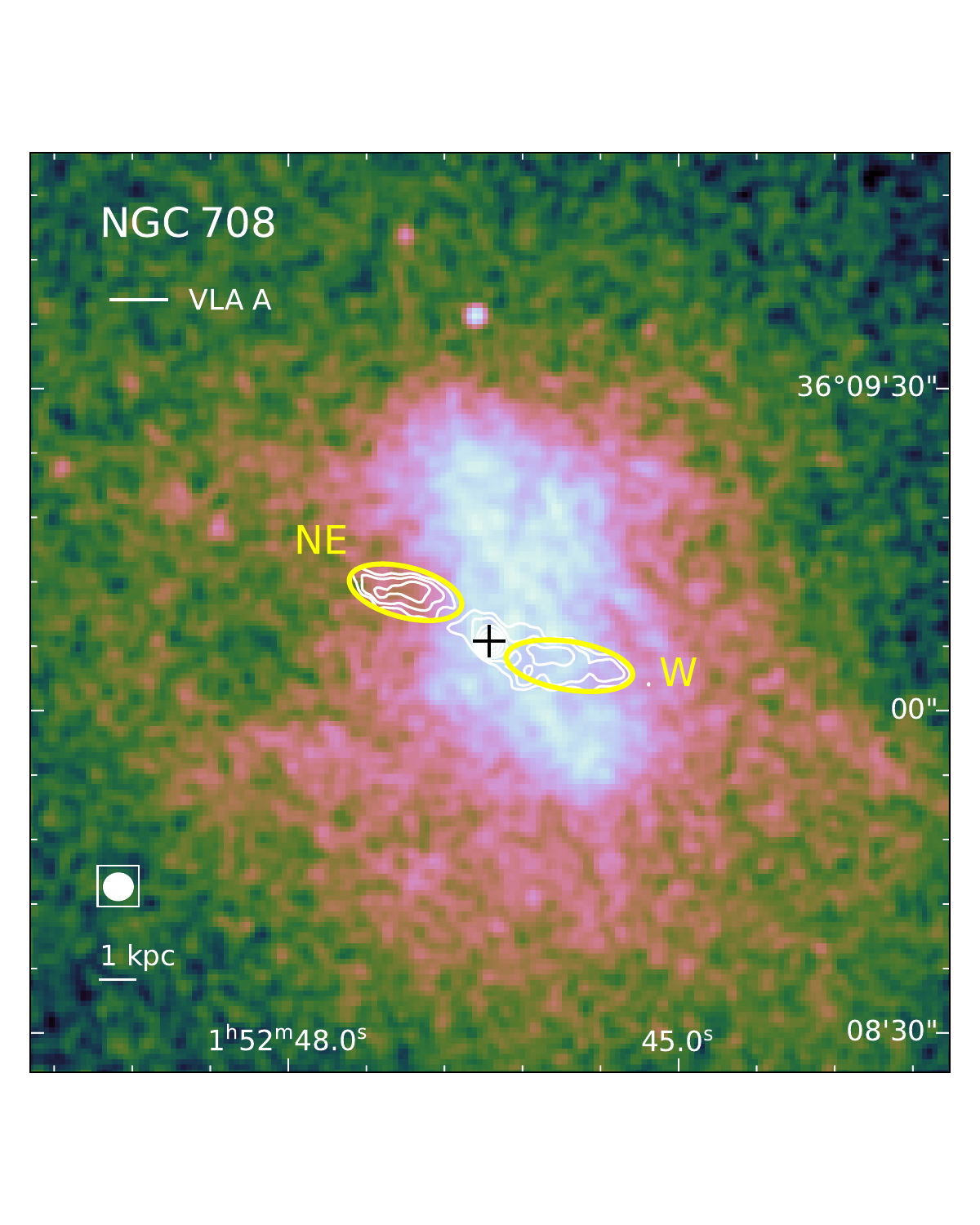}};
\draw (\figxi, \figyjj) node {\includegraphics[height=\imgheight]{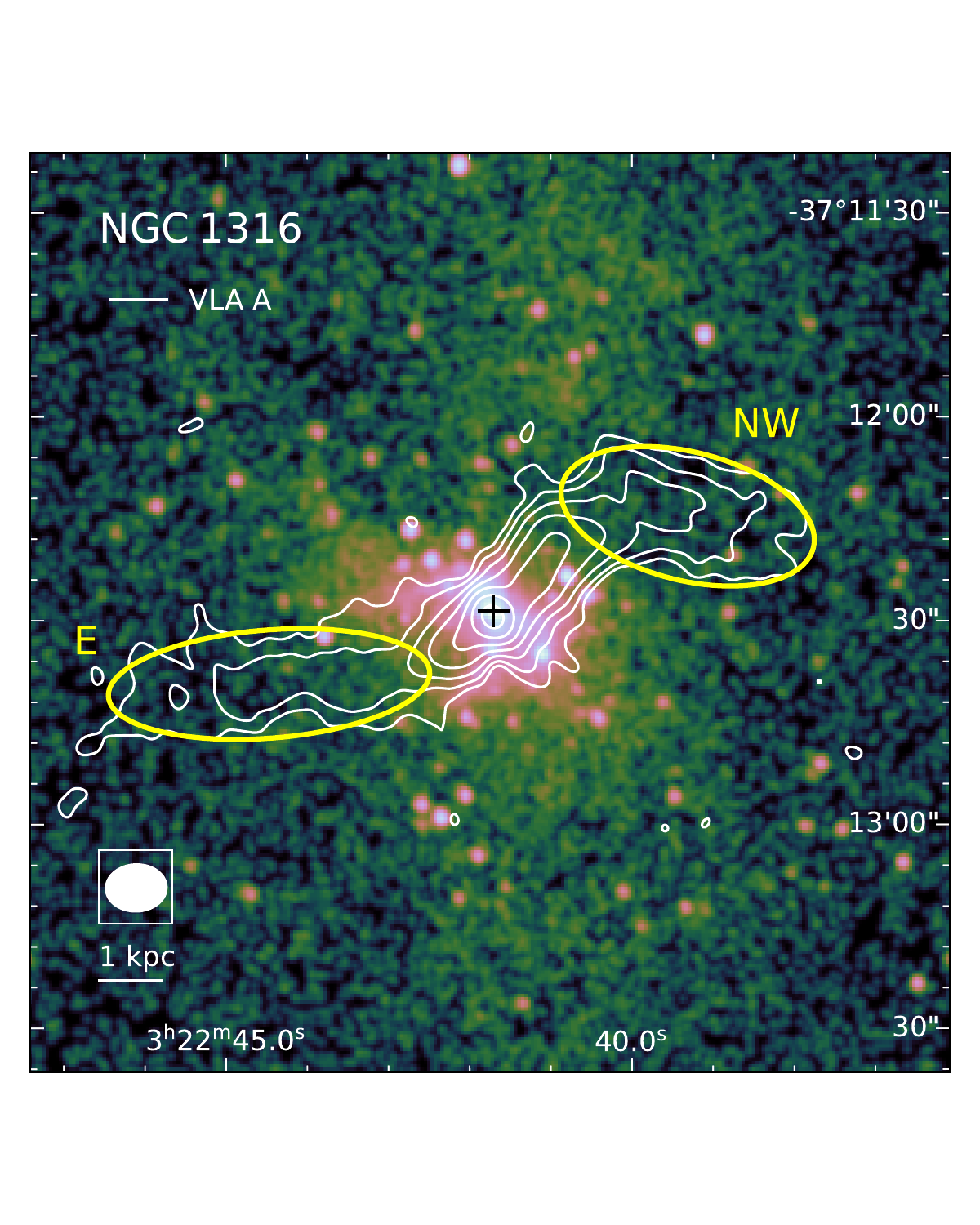}};
\draw (\figxj, \figyjj) node {\includegraphics[height=\imgheight]{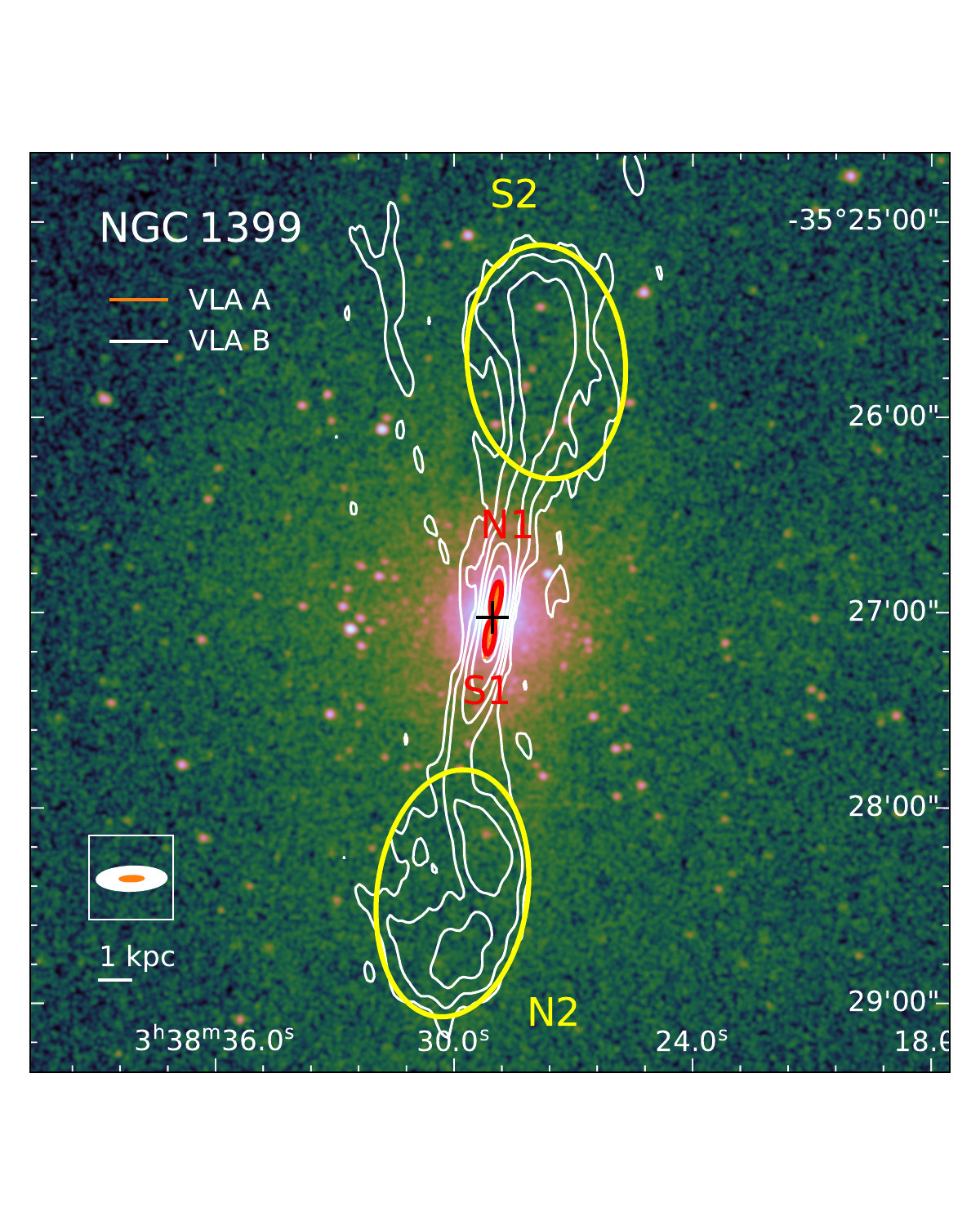}};
\draw (\figxk, \figyjj) node {\includegraphics[height=\imgheight]{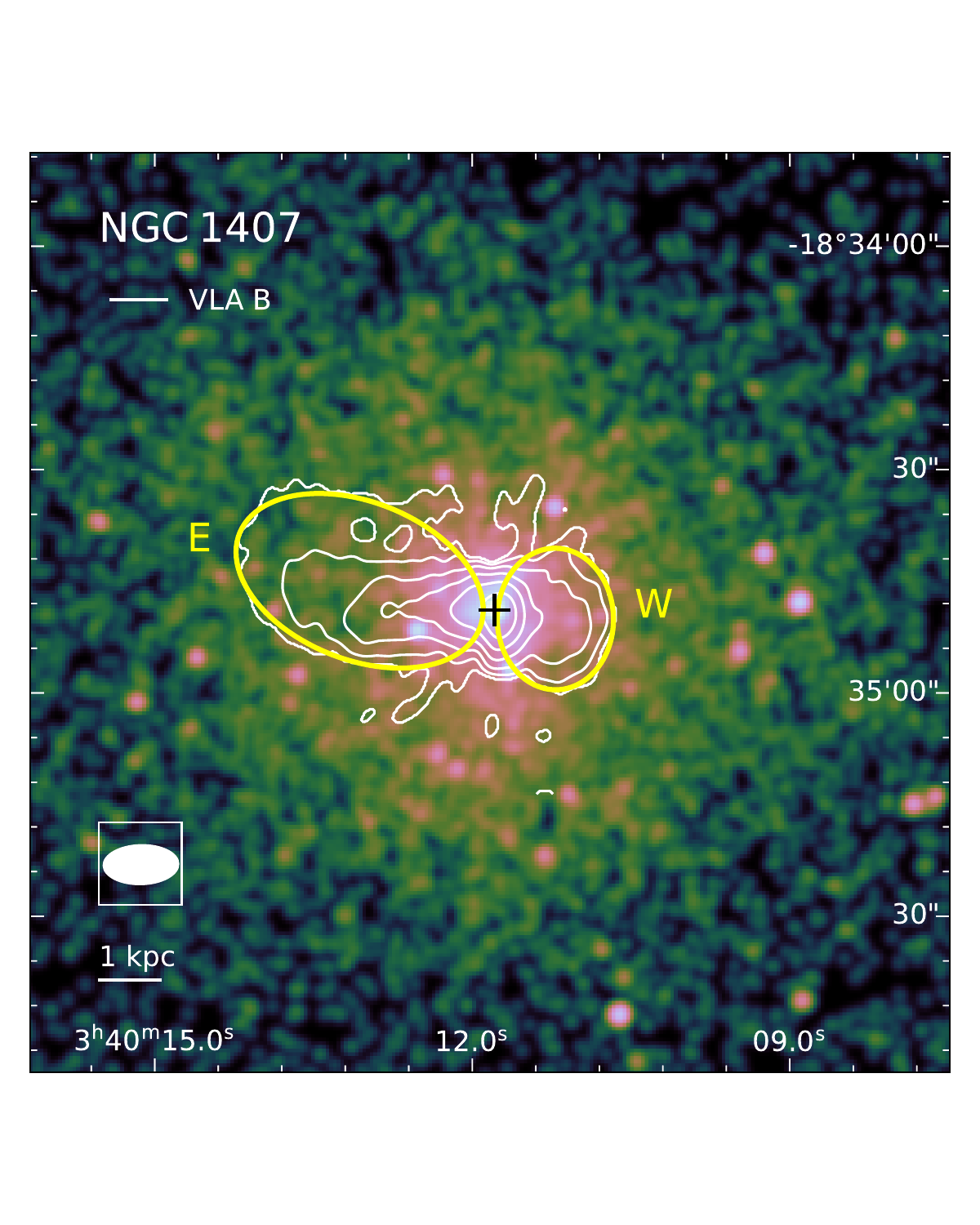}};
\draw (\figxi, \figykk) node {\includegraphics[height=\imgheight]{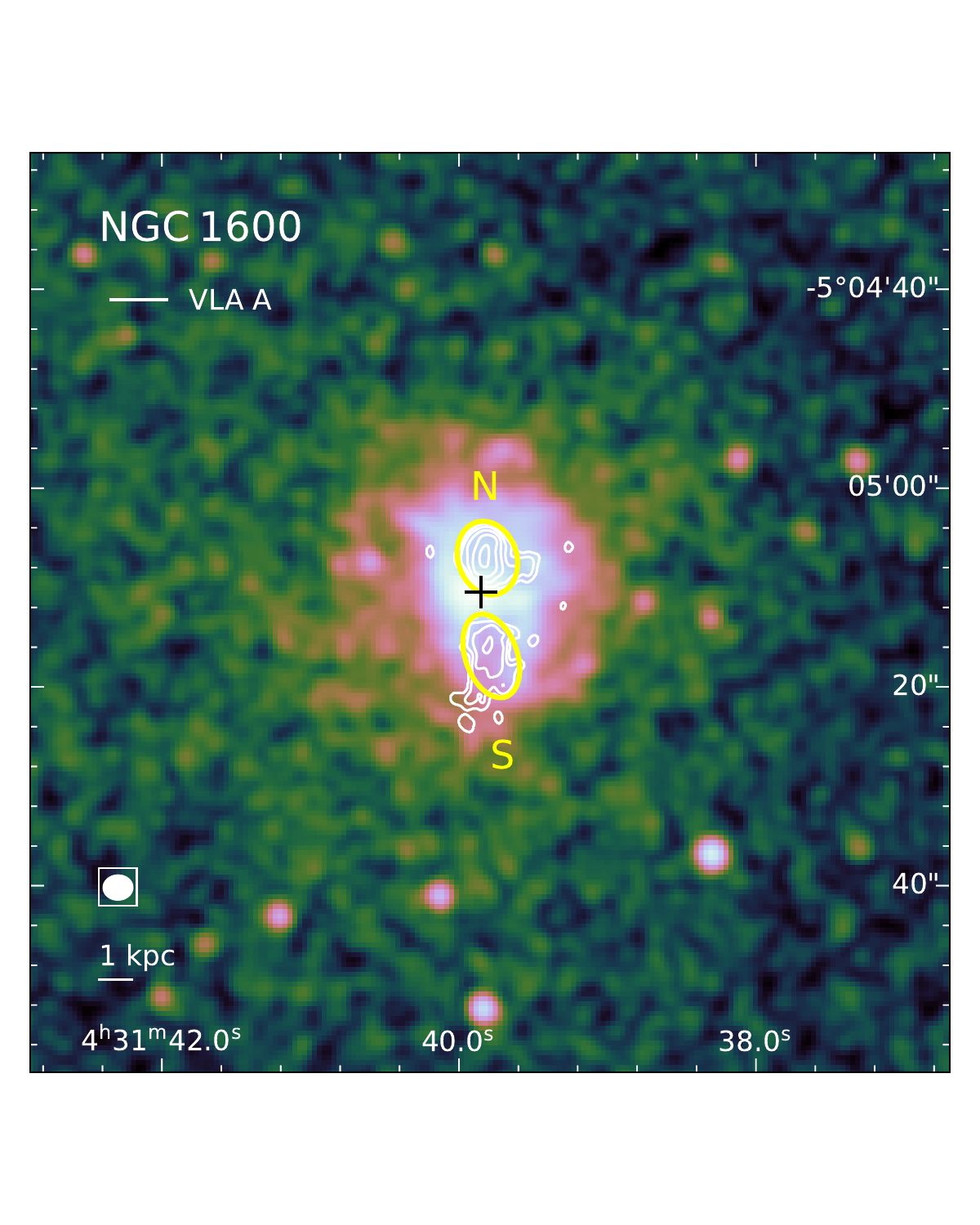}};
\draw (\figxj, \figykk) node {\includegraphics[height=\imgheight]{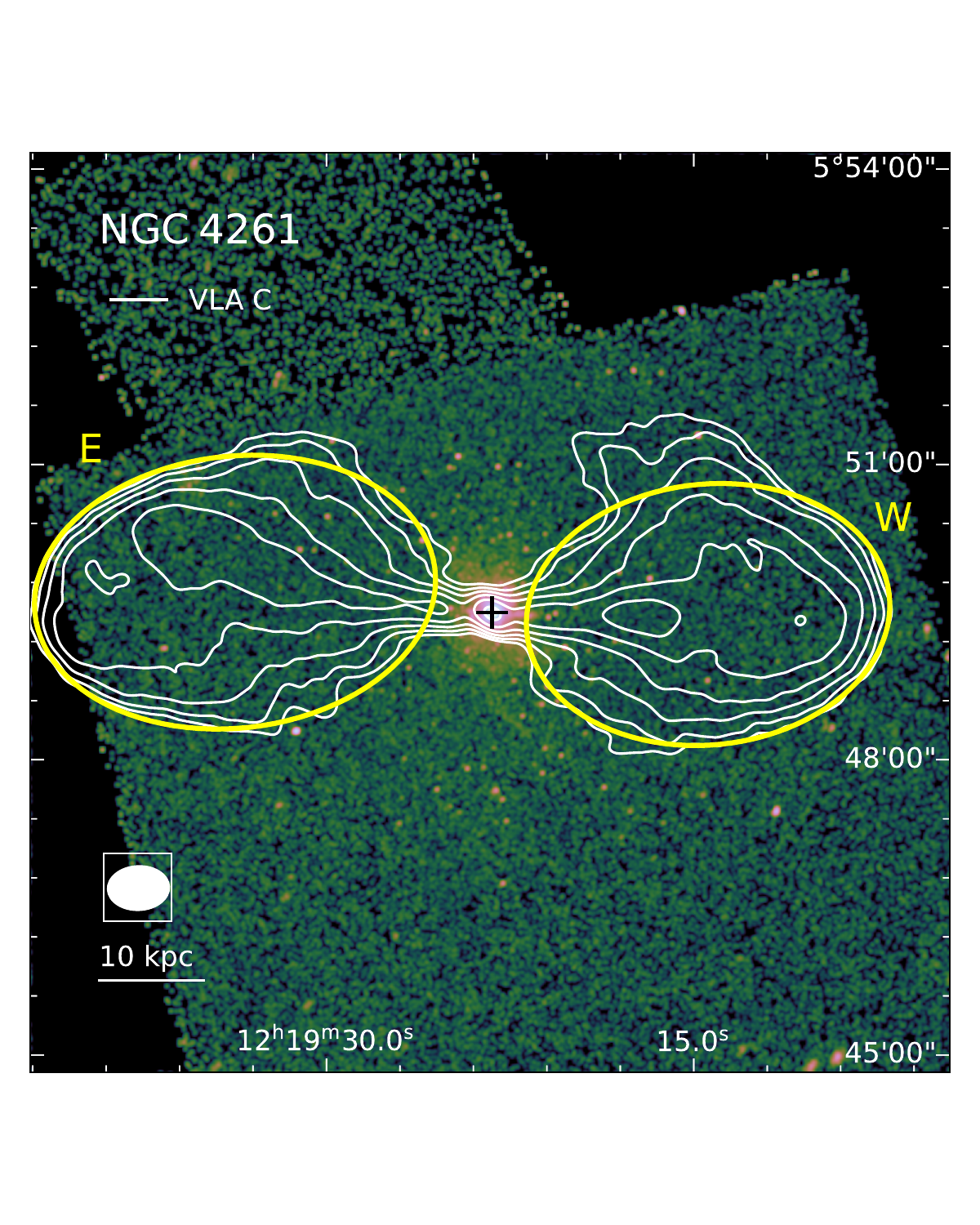}};
\draw (\figxk, \figykk) node {\includegraphics[height=\imgheight]{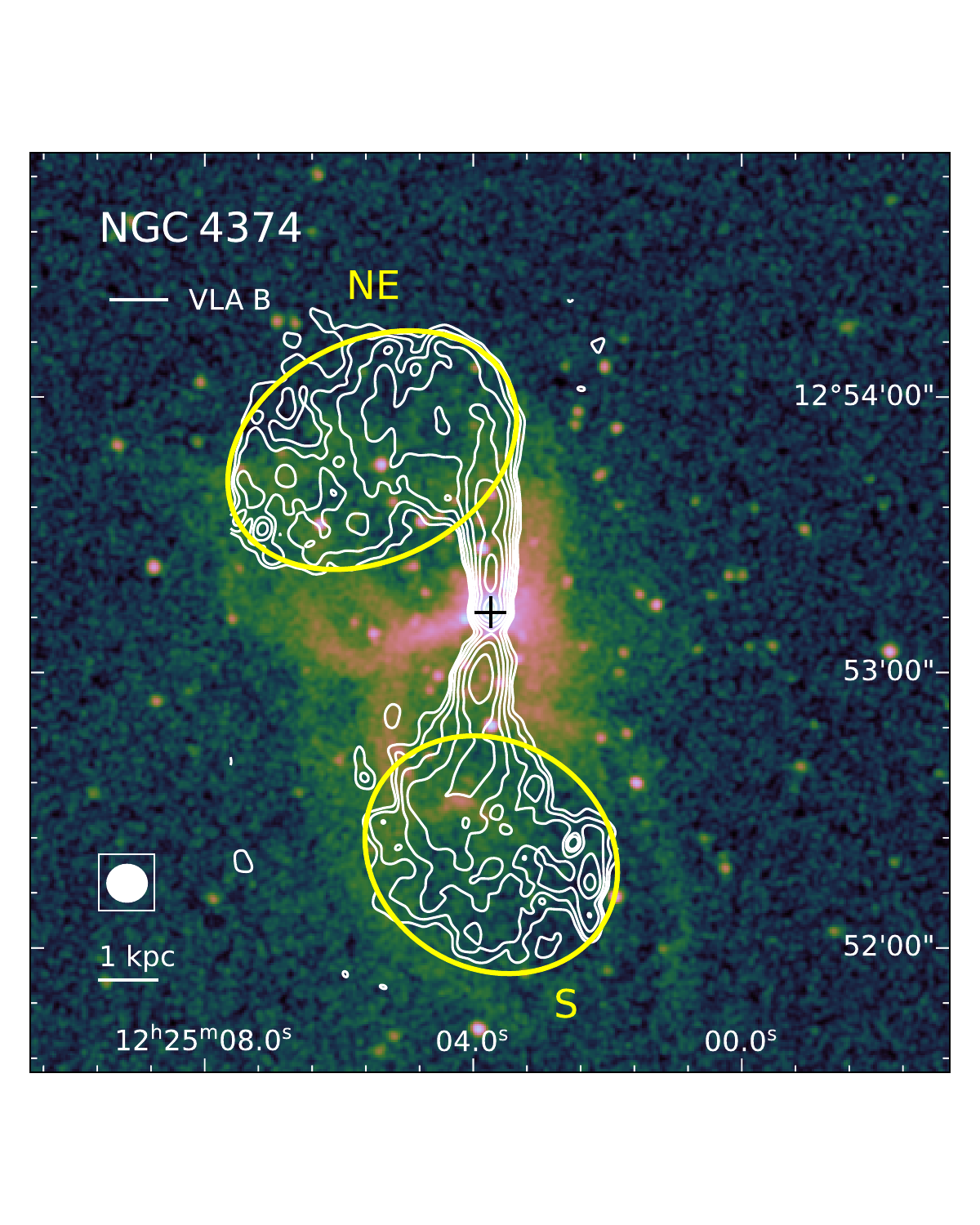}};
\draw (\figxi, \figyll) node {\includegraphics[height=\imgheight]{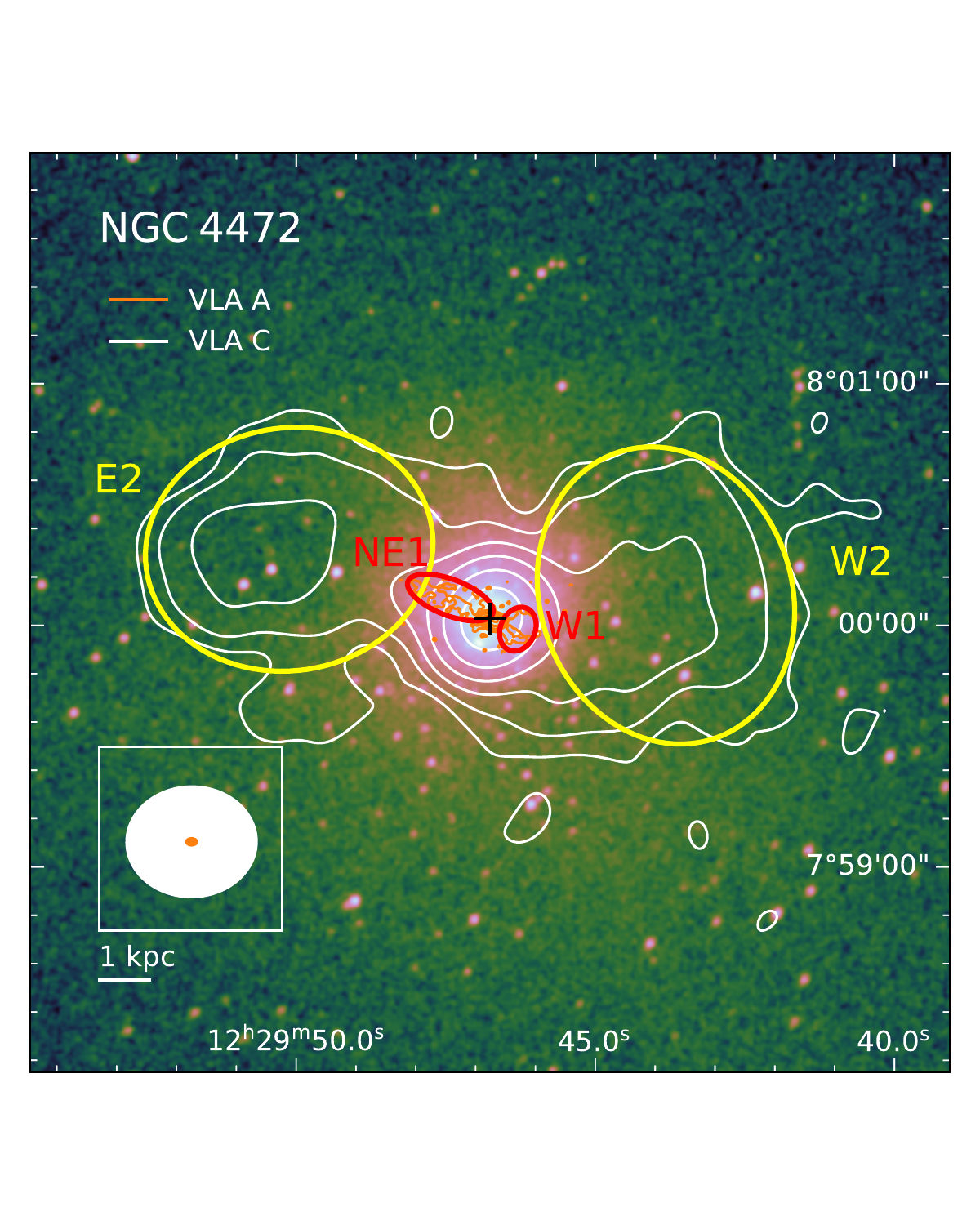}};
\draw (\figxj, \figyll) node {\includegraphics[height=\imgheight]{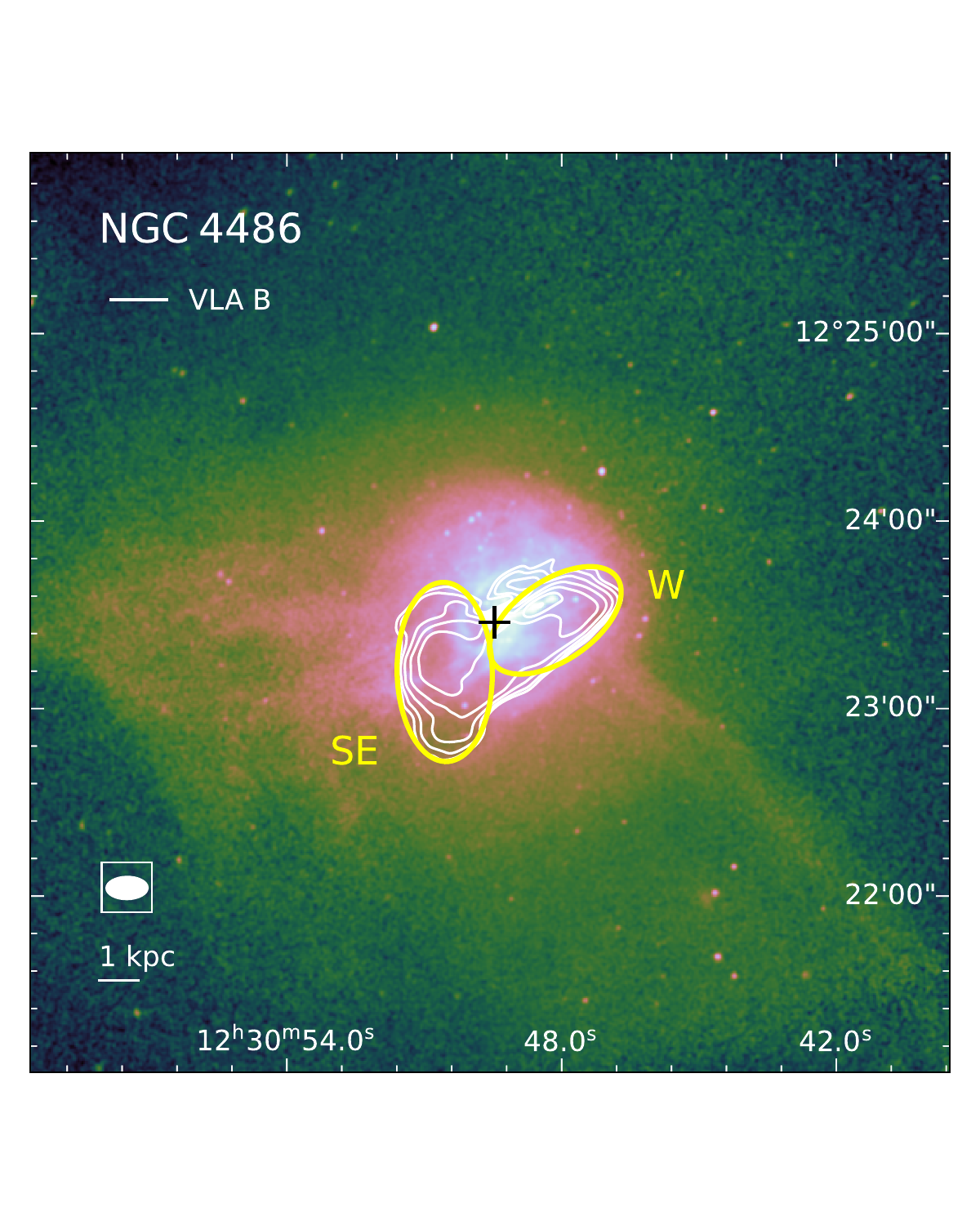}};
\draw (\figxk, \figyll) node {\includegraphics[height=\imgheight]{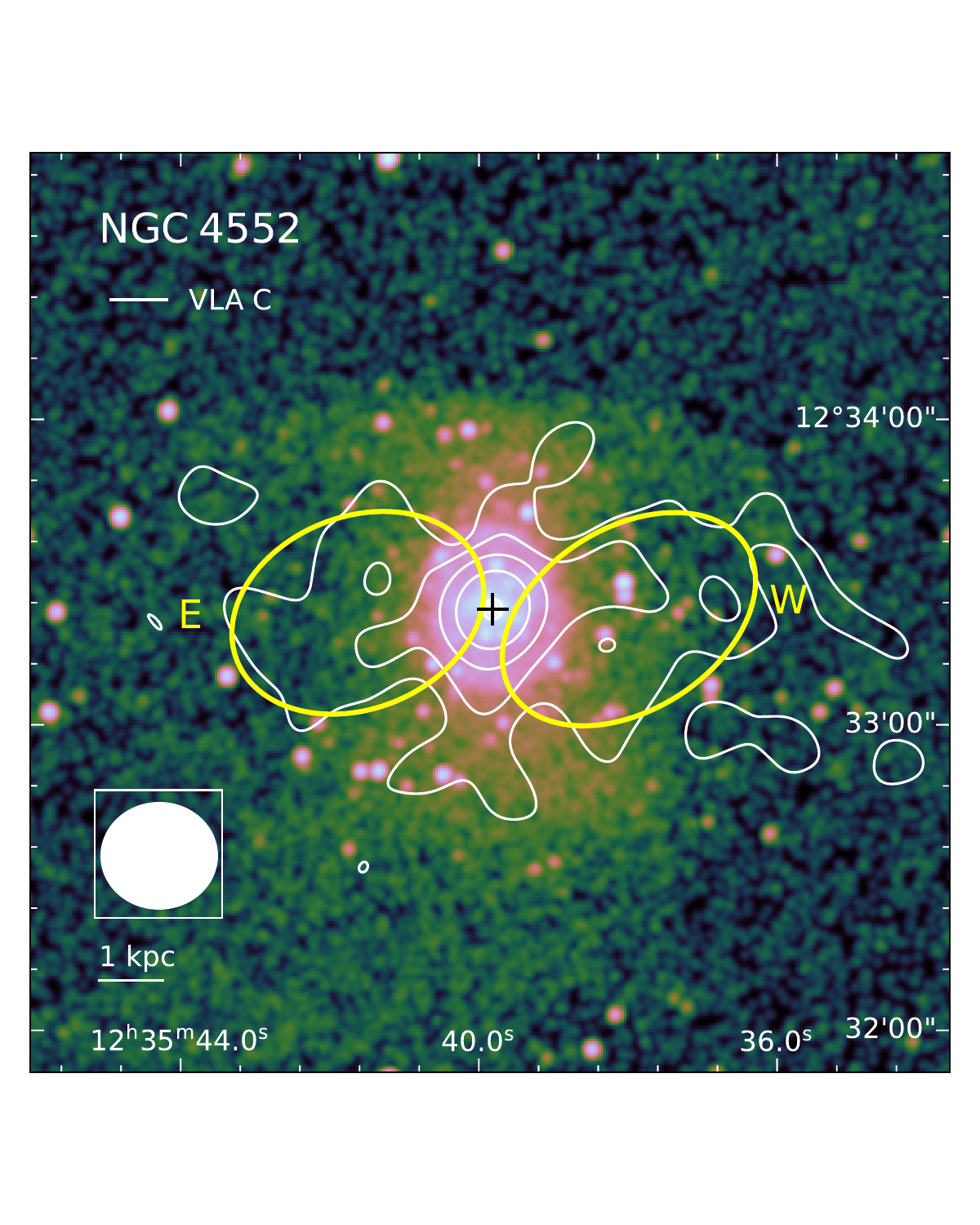}};
\end{tikzpicture}
\end{figure}

\begin{figure*}
\begin{tikzpicture}
\draw (\figxi, \figyi) node {\includegraphics[height=\imgheight]{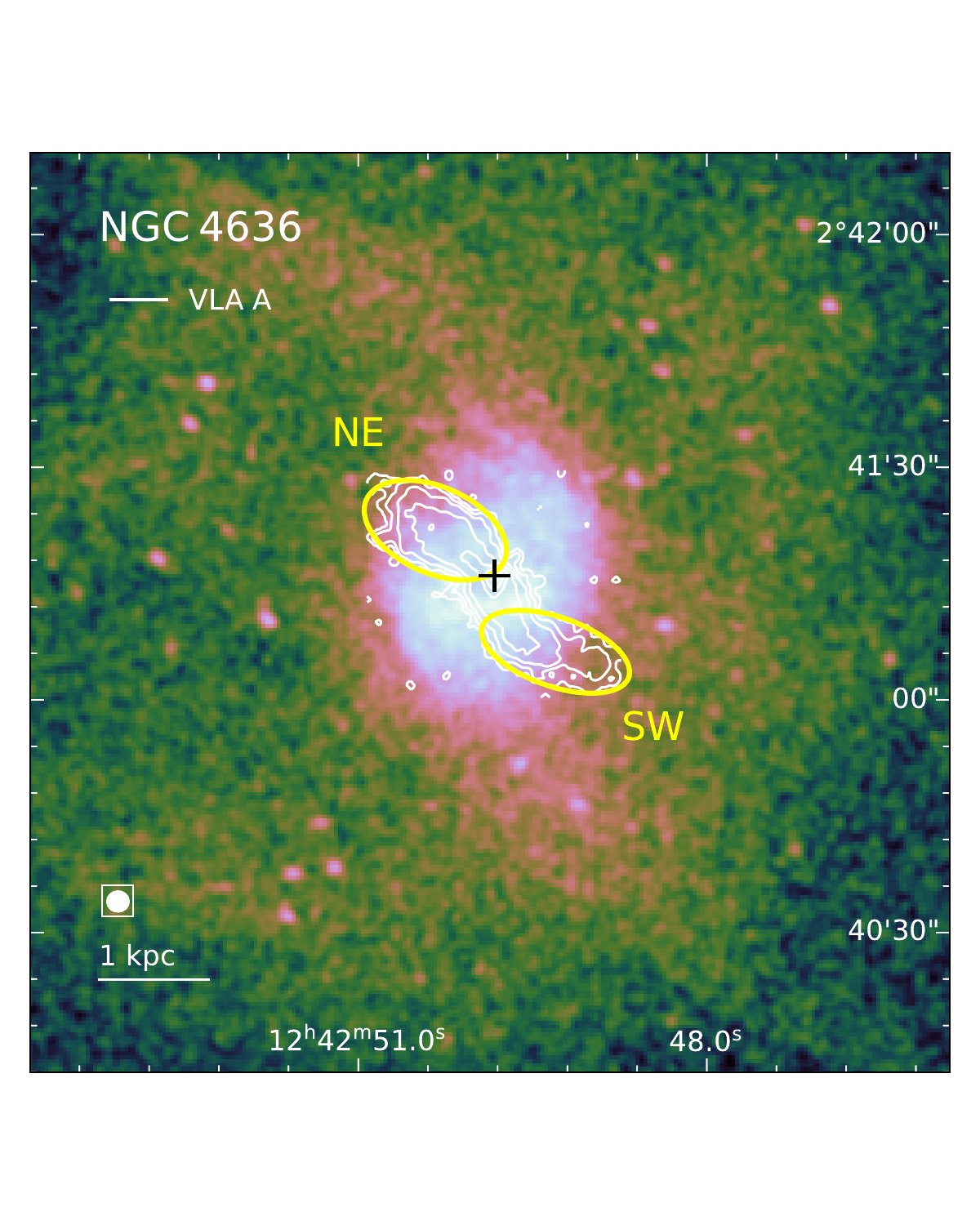}};
\draw (\figxj, \figyi) node {\includegraphics[height=\imgheight]{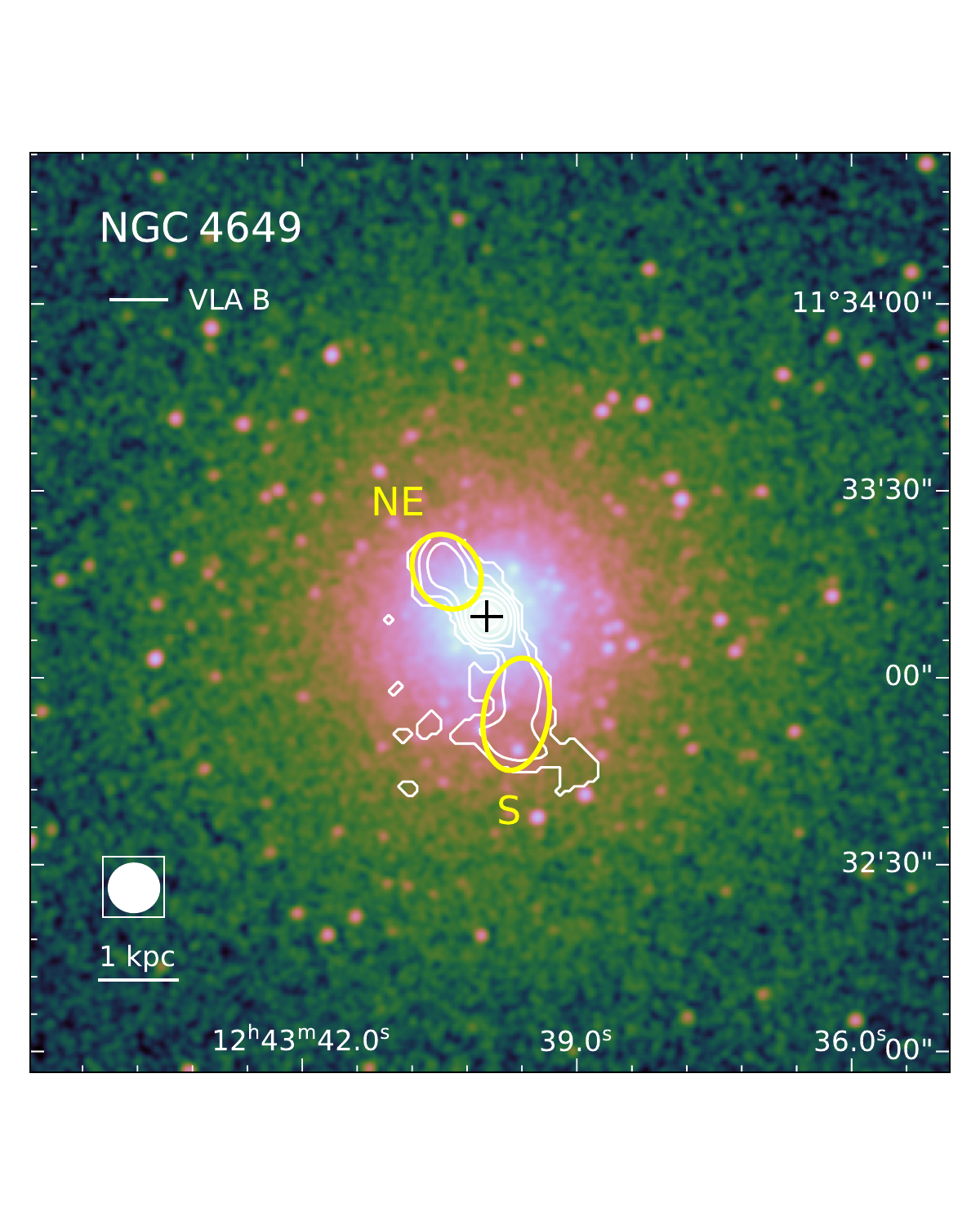}};
\draw (\figxk, \figyi) node {\includegraphics[height=\imgheight]{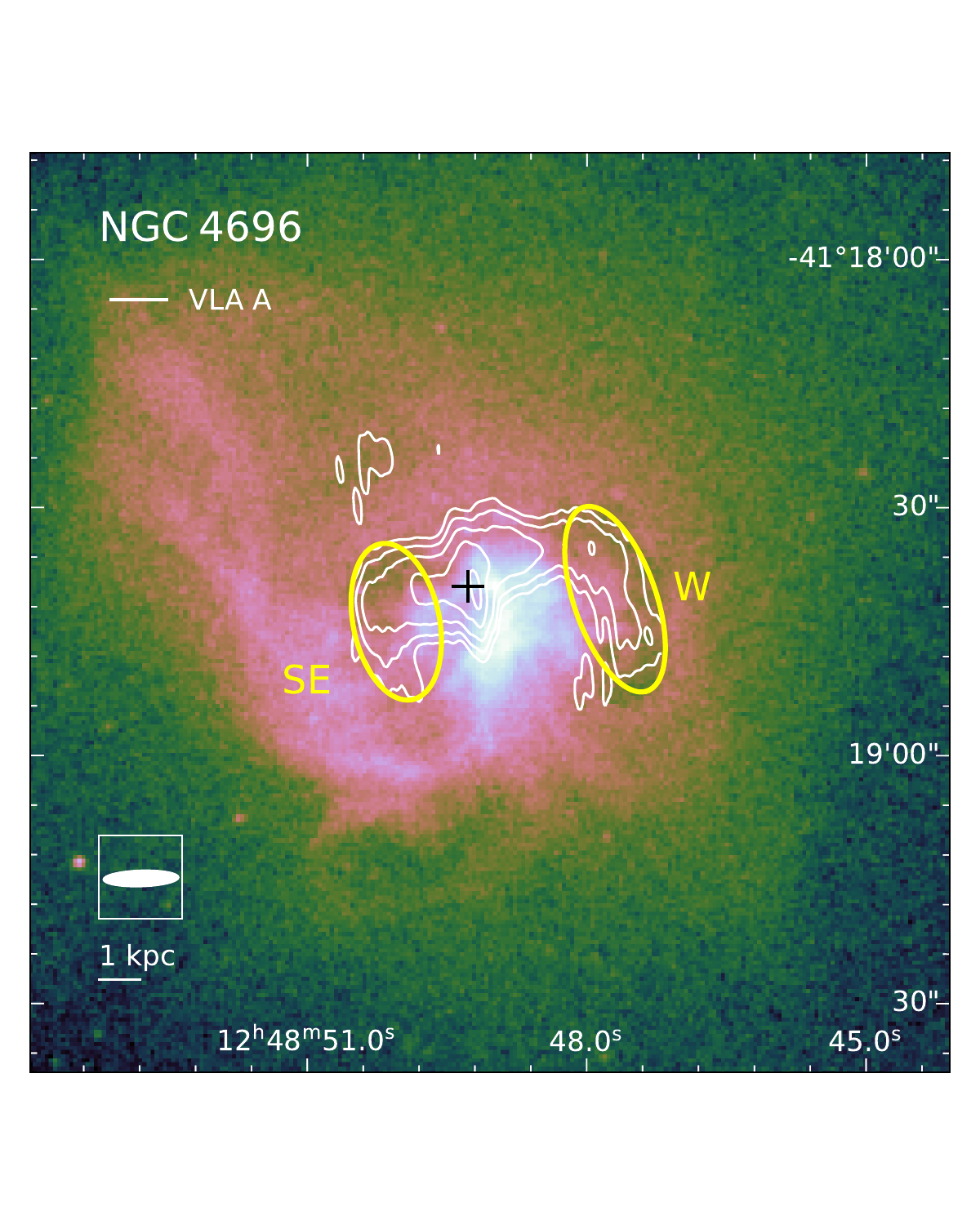}};
\draw (\figxi, \figyj) node {\includegraphics[height=\imgheight]{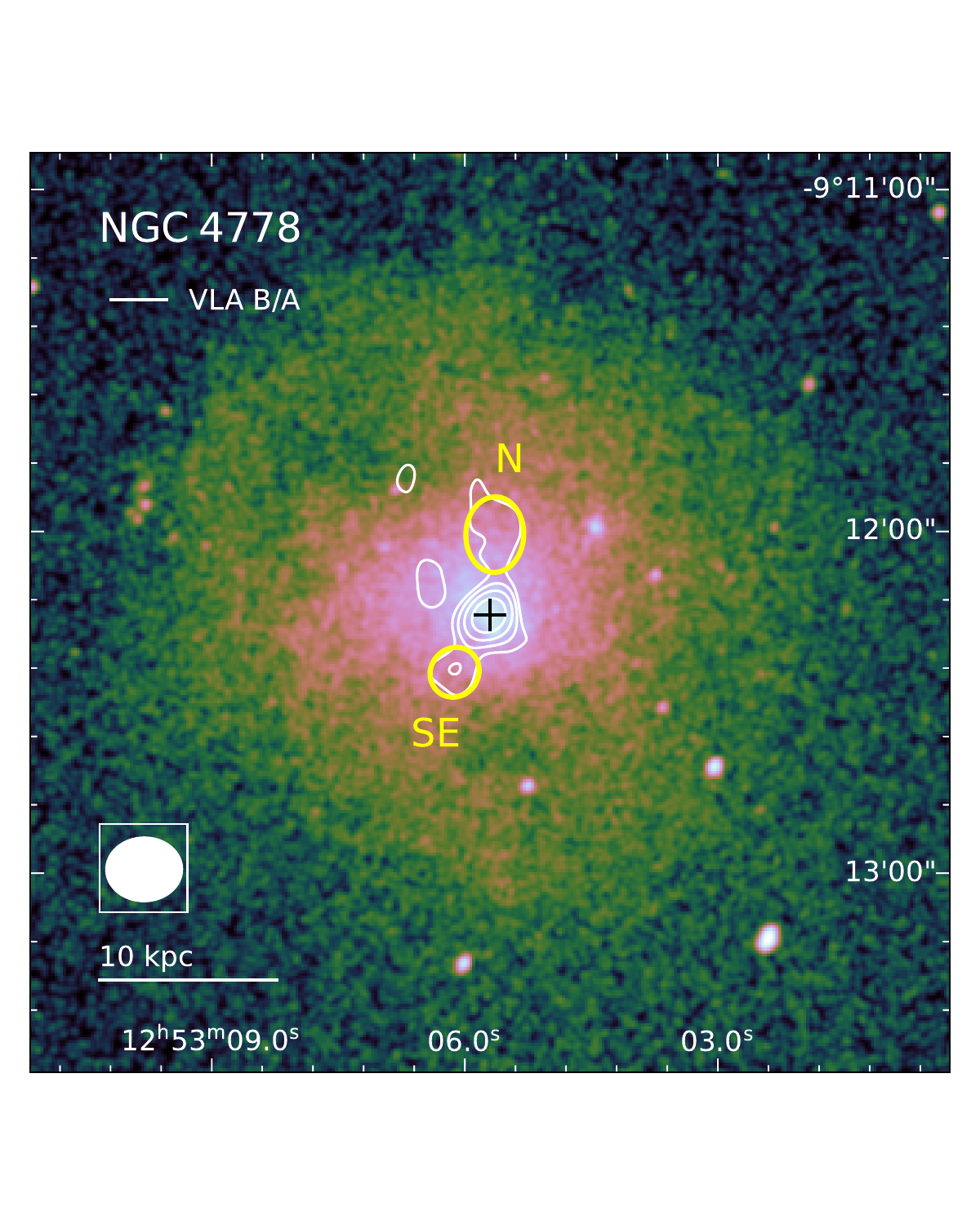}};
\draw (\figxj, \figyj) node {\includegraphics[height=\imgheight]{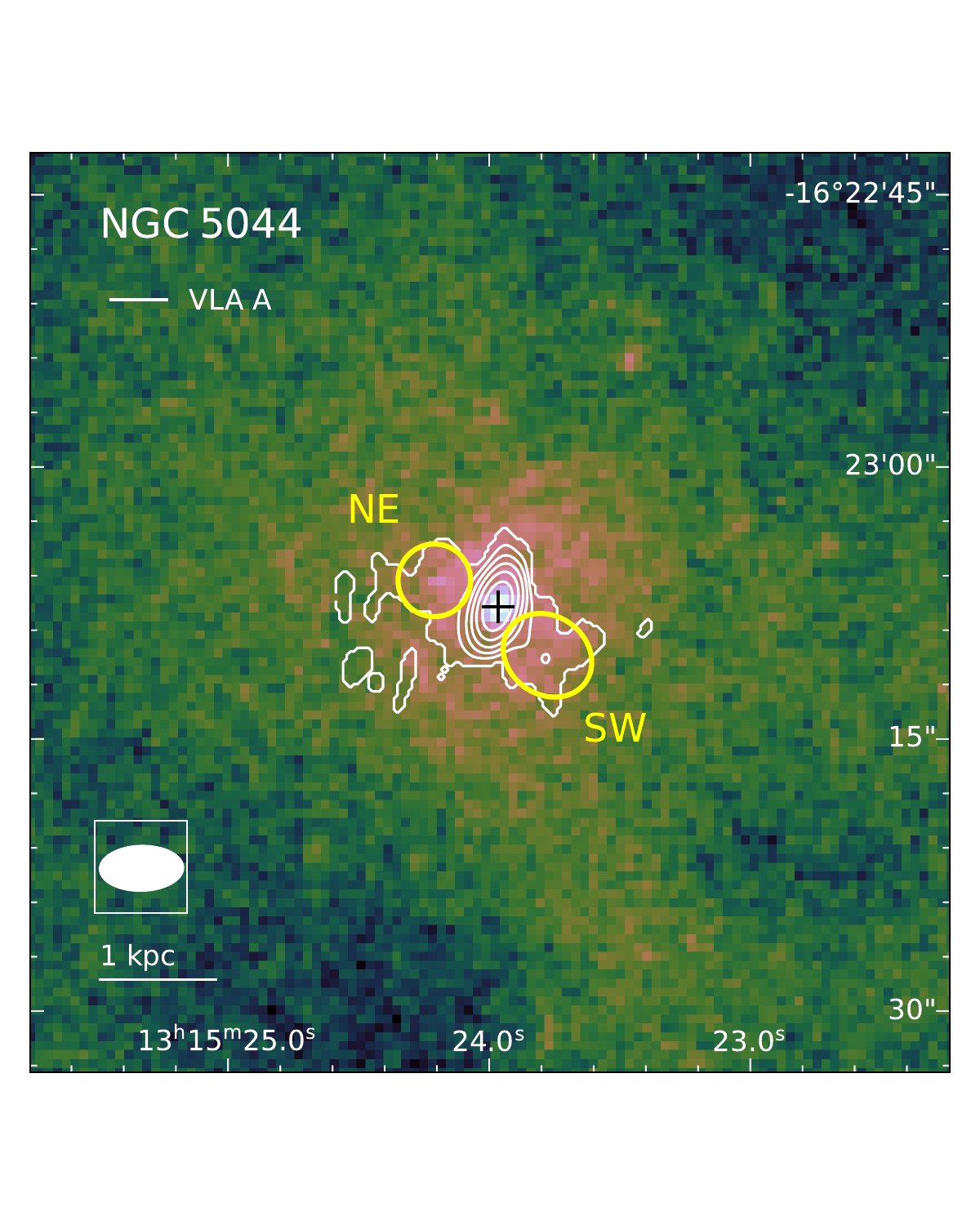}};
\draw (\figxk, \figyj) node {\includegraphics[height=\imgheight]{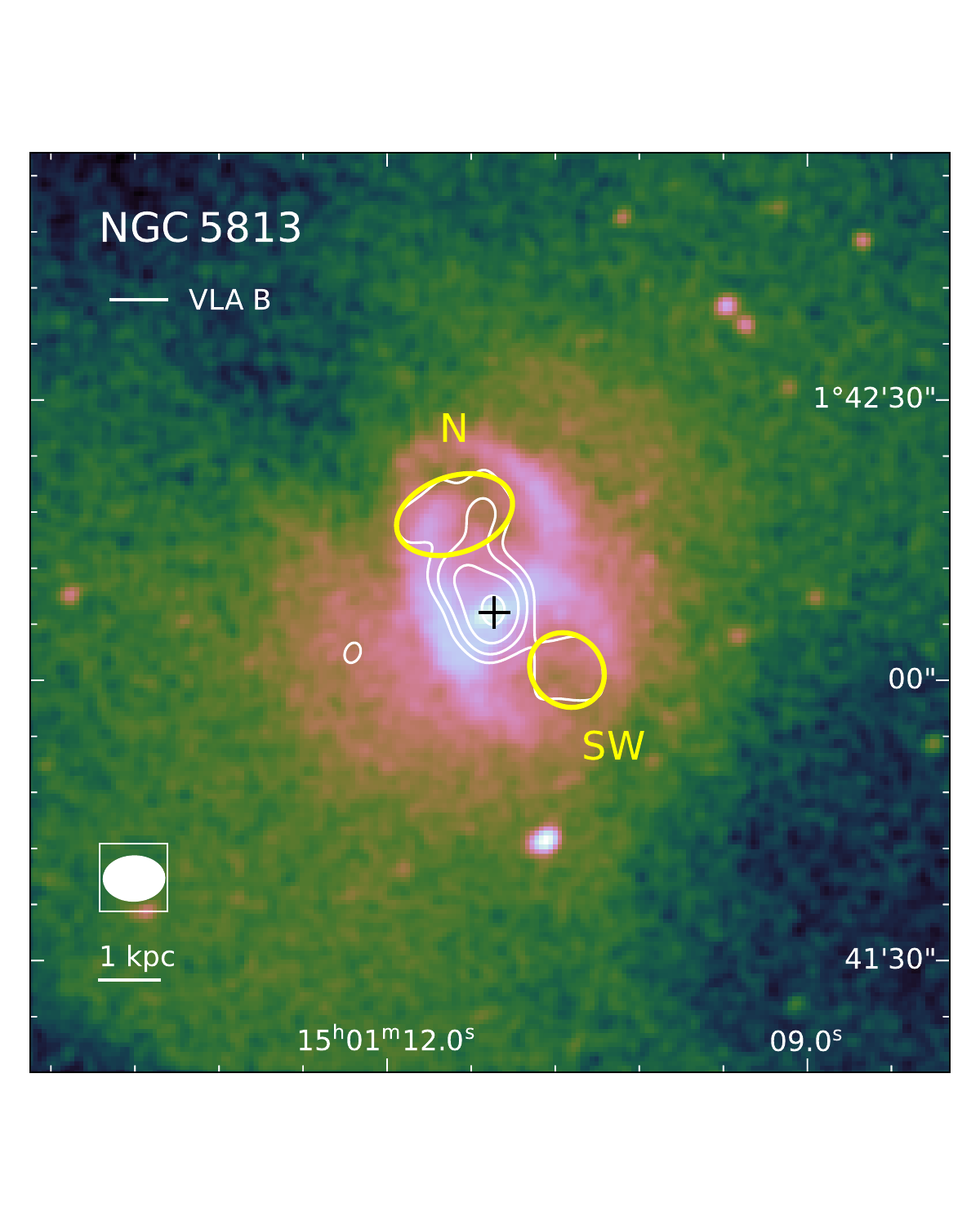}};
\draw (\figxi, \figyk) node {\includegraphics[height=\imgheight]{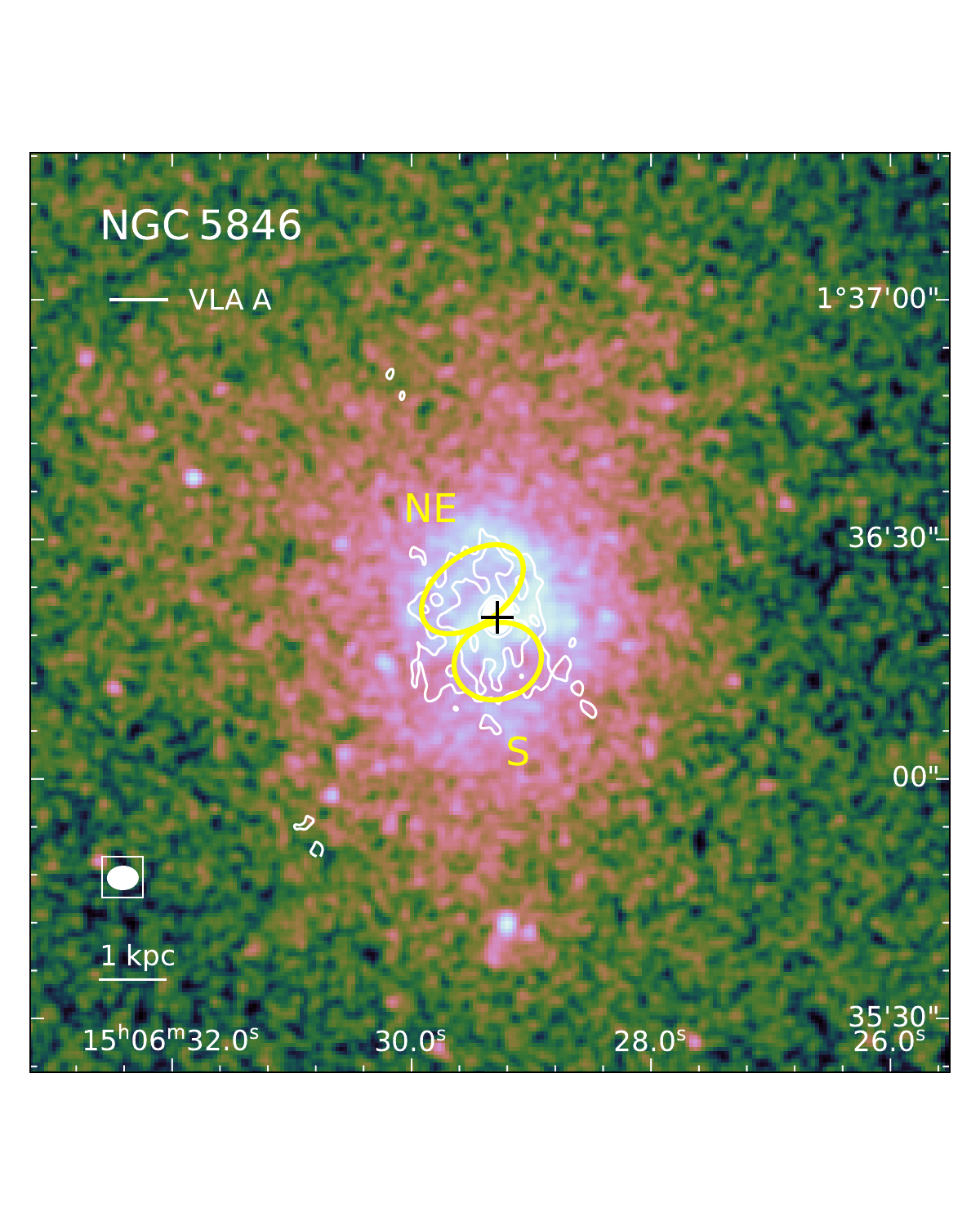}};
\draw (\figxj, \figyk) node {\includegraphics[height=\imgheight]{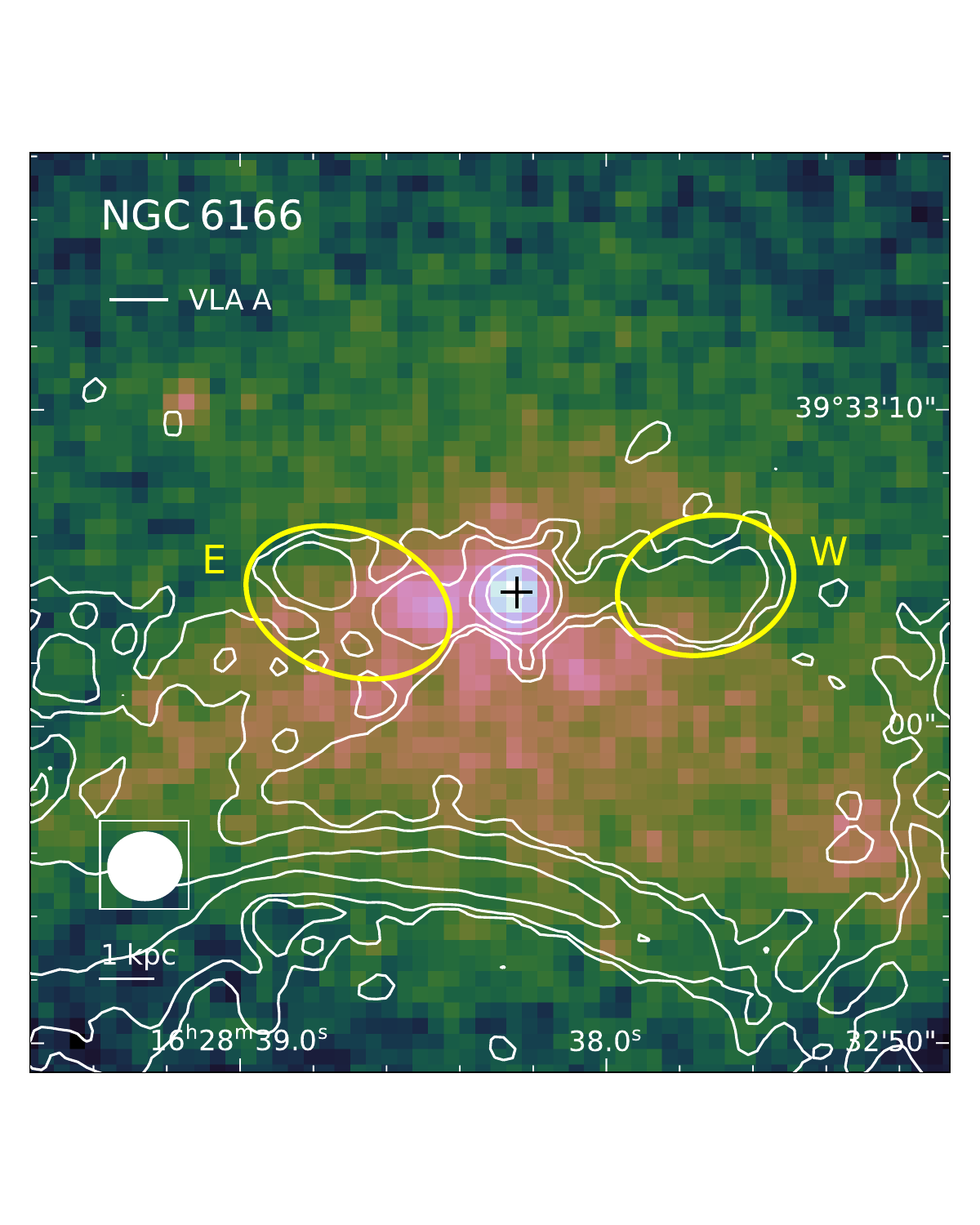}};
\end{tikzpicture}
\caption{Exposure-corrected \textit{Chandra} ($0.5-7.0\;$keV) images overlaid by VLA 1.4 GHz radio contours (\textit{white}) and corresponding ellipse regions (\textit{red}) which were used for size-estimation of radio lobes. VLA array configurations for individual galaxies are stated in the upper left corner of each plot while the corresponding beam sizes are shown in the lower-left corner together with physical scales (1 or 10 kpc). The central black cross represents the exact centre of the galaxy (position of the SMBH).}
\label{fig:cavities}
\end{figure*}

\clearpage

\section{Cavity powers}

\setlength{\tabcolsep}{7.0pt}
\renewcommand{\arraystretch}{1.37}
\begin{table*}
\centering
\caption{Properties of individual radio lobes: cavity position with respect to galactic centre, semi-axis along the jet direction (length) $r_{\text{l}}$, semi-axis perpendicular to the jet direction (width) $r_{\text{w}}$, cavity volume $V$, galactocentric distance $R$, age of the cavity assuming inflation at the speed of sound $t_{\text{age}}$, total enthalpy of the cavity $E$, and estimated mechanical power required for its inflation $P_{\text{cav}}$.}
\begin{tabular}{l c c c c c c c c}
    \toprule \vspace{-0.8mm}
    \multirow{2}{3.2em}{Galaxy} & \multirow{2}{1.8em}{Side} & $r_{\text{w}}$ & $r_{\text{l}}$ & $V$ & $R$ & $t_{\text{age}}$ & $E$ & $P_{\text{cav}}$\\
    &  & (kpc) & (kpc) & (m$^3$) & (kpc) & (Myr) & (erg) & (erg$\,$s$^{-1}$)\\
    \bottomrule \vspace{-3.5mm}
    \\ 
IC4296&SE&78.0&49.0&$3.6 \times 10^{64}$&250.0&$529.03_{-10.08}^{+11.54}$&$2.1_{-1.0}^{+2.1} \times 10^{59}$&$1.3_{-0.6}^{+1.2} \times 10^{43}$\\  
 &N&78.0&70.0&$5.2 \times 10^{64}$&160.0&$343.9_{-6.55}^{+7.5}$&$3.0_{-1.5}^{+3.0} \times 10^{59}$&$2.9_{-1.4}^{+2.8} \times 10^{43}$\\ 
NGC507&NW&14.0&11.0&$2.7 \times 10^{62}$&11.0&$21.25_{-0.2}^{+0.18}$&$2.3_{-1.1}^{+2.3} \times 10^{58}$&$3.5_{-1.7}^{+3.4} \times 10^{43}$\\  
 &SE&22.0&10.0&$6.1 \times 10^{62}$&20.0&$37.57_{-0.3}^{+0.26}$&$3.7_{-1.8}^{+3.7} \times 10^{58}$&$3.2_{-1.5}^{+3.1} \times 10^{43}$\\ 
NGC708&NE&0.66&1.5&$8.2 \times 10^{58}$&2.5&$4.83\pm0.1$&$4.0_{-2.0}^{+4.0} \times 10^{55}$&$2.7_{-1.3}^{+2.6} \times 10^{41}$\\  
 &W&0.64&1.7&$8.3 \times 10^{58}$&2.2&$4.22\pm0.09$&$4.3_{-2.1}^{+4.3} \times 10^{55}$&$3.3_{-1.6}^{+3.2} \times 10^{41}$\\ 
NGC1316&E&0.81&2.4&$2.0 \times 10^{59}$&3.5&$7.72\pm0.05$&$1.3_{-0.7}^{+1.3} \times 10^{55}$&$5.5_{-2.6}^{+5.3} \times 10^{40}$\\  
 &NW&0.96&1.9&$2.2 \times 10^{59}$&3.2&$7.06\pm0.05$&$1.4_{-0.7}^{+1.4} \times 10^{55}$&$6.3_{-3.0}^{+6.0} \times 10^{40}$\\ 
NGC1399&S1&0.22&0.71&$4.2 \times 10^{57}$&0.67&$1.26_{-0.02}^{+0.01}$&$8.4_{-4.2}^{+8.4} \times 10^{54}$&$2.2_{-1.0}^{+2.1} \times 10^{41}$\\  
 &N1&0.26&0.76&$6.3 \times 10^{57}$&0.88&$1.7\pm0.01$&$8.0_{-4.0}^{+8.0} \times 10^{54}$&$1.5_{-0.7}^{+1.5} \times 10^{41}$\\ 
 &S2&3.2&6.8&$8.7 \times 10^{60}$&8.5&$15.86\pm0.06$&$6.9_{-3.4}^{+6.9} \times 10^{56}$&$1.4_{-0.7}^{+1.4} \times 10^{42}$\\  
 &N2&3.3&6.4&$8.4 \times 10^{60}$&7.9&$15.04_{-0.05}^{+0.06}$&$7.0_{-3.5}^{+7.0} \times 10^{56}$&$1.5_{-0.7}^{+1.5} \times 10^{42}$\\ 
NGC1407&E&1.3&2.1&$4.5 \times 10^{59}$&2.3&$4.53_{-0.08}^{+0.09}$&$6.7_{-3.3}^{+6.7} \times 10^{55}$&$4.8_{-2.3}^{+4.6} \times 10^{41}$\\  
 &W&1.2&0.96&$1.6 \times 10^{59}$&1.0&$2.02\pm0.04$&$4.5_{-2.2}^{+4.5} \times 10^{55}$&$7.3_{-3.5}^{+6.9} \times 10^{41}$\\ 
NGC1600&N&0.91&1.2&$1.2 \times 10^{59}$&1.1&$2.12_{-0.03}^{+0.02}$&$6.3_{-3.2}^{+6.3} \times 10^{55}$&$9.8_{-4.6}^{+9.3} \times 10^{41}$\\  
 &S&0.8&1.3&$1.1 \times 10^{59}$&2.0&$3.93\pm0.05$&$3.6_{-1.8}^{+3.6} \times 10^{55}$&$3.0_{-1.4}^{+2.9} \times 10^{41}$\\ 
NGC4261&E&13.0&19.0&$4.0 \times 10^{62}$&25.0&$49.89_{-0.34}^{+0.43}$&$9.1_{-4.6}^{+9.1} \times 10^{57}$&$6.0_{-2.9}^{+5.7} \times 10^{42}$\\  
 &W&12.0&17.0&$3.3 \times 10^{62}$&21.0&$41.78_{-0.28}^{+0.36}$&$7.7_{-3.8}^{+7.7} \times 10^{57}$&$6.0_{-2.9}^{+5.7} \times 10^{42}$\\ 
NGC4374&NE&4.0&3.1&$6.2 \times 10^{60}$&3.7&$8.89\pm0.08$&$3.9_{-1.9}^{+3.9} \times 10^{56}$&$1.4_{-0.7}^{+1.4} \times 10^{42}$\\  
 &S&2.9&3.9&$4.0 \times 10^{60}$&4.6&$10.64\pm0.09$&$2.0_{-1.0}^{+2.0} \times 10^{56}$&$6.2_{-3.0}^{+6.0} \times 10^{41}$\\ 
NGC4472&W1&0.46&0.35&$8.9 \times 10^{57}$&0.6&$1.27\pm0.01$&$1.0_{-0.5}^{+1.0} \times 10^{55}$&$2.6_{-1.2}^{+2.5} \times 10^{41}$\\  
 &NE1&0.36&0.91&$1.5 \times 10^{58}$&0.9&$1.89\pm0.01$&$1.4_{-0.7}^{+1.4} \times 10^{55}$&$2.4_{-1.2}^{+2.3} \times 10^{41}$\\ 
 &E2&2.4&2.8&$2.0 \times 10^{60}$&4.3&$8.87\pm0.03$&$2.0_{-1.0}^{+2.0} \times 10^{56}$&$7.3_{-3.5}^{+7.0} \times 10^{41}$\\  
 &W2&3.0&2.5&$2.9 \times 10^{60}$&3.2&$6.71_{-0.03}^{+0.02}$&$4.1_{-2.0}^{+4.1} \times 10^{56}$&$2.0_{-0.9}^{+1.9} \times 10^{42}$\\ 
NGC4486&SE&2.3&1.2&$8.2 \times 10^{59}$&1.8&$3.55\pm0.03$&$1.0_{-0.5}^{+1.0} \times 10^{57}$&$9.3_{-4.4}^{+9.0} \times 10^{42}$\\  
 &W&1.0&1.9&$2.6 \times 10^{59}$&1.6&$3.22_{-0.04}^{+0.03}$&$3.5_{-1.7}^{+3.5} \times 10^{56}$&$3.5_{-1.7}^{+3.4} \times 10^{42}$\\ 
NGC4552&E&1.5&2.0&$5.2 \times 10^{59}$&2.1&$4.38_{-0.08}^{+0.06}$&$8.0_{-4.0}^{+8.0} \times 10^{55}$&$6.0_{-2.8}^{+5.6} \times 10^{41}$\\  
 &W&1.4&2.1&$5.0 \times 10^{59}$&2.1&$4.43_{-0.08}^{+0.07}$&$6.5_{-3.2}^{+6.5} \times 10^{55}$&$4.8_{-2.3}^{+4.6} \times 10^{41}$\\ 
NGC4636&NE&0.4&0.69&$1.3 \times 10^{58}$&0.69&$2.41_{-0.08}^{+0.09}$&$7.5_{-3.8}^{+7.5} \times 10^{54}$&$1.0_{-0.5}^{+1.0} \times 10^{41}$\\  
 &SW&0.32&0.71&$8.8 \times 10^{57}$&0.89&$2.68_{-0.04}^{+0.05}$&$4.6_{-2.3}^{+4.6} \times 10^{54}$&$5.6_{-2.7}^{+5.4} \times 10^{40}$\\ 
NGC4649&NE&0.69&0.38&$2.2 \times 10^{58}$&1.0&$1.81\pm0.02$&$1.3_{-0.7}^{+1.3} \times 10^{55}$&$2.4_{-1.1}^{+2.3} \times 10^{41}$\\  
 &S&0.46&0.75&$2.0 \times 10^{58}$&1.3&$2.4\pm0.02$&$8.1_{-4.1}^{+8.1} \times 10^{54}$&$1.1_{-0.5}^{+1.1} \times 10^{41}$\\ 
NGC4696&E&1.8&0.84&$3.2 \times 10^{59}$&2.1&$4.3\pm0.03$&$3.4_{-1.7}^{+3.4} \times 10^{56}$&$2.6_{-1.2}^{+2.5} \times 10^{42}$\\  
 &W&2.0&0.82&$4.1 \times 10^{59}$&3.5&$7.26\pm0.05$&$3.4_{-1.7}^{+3.4} \times 10^{56}$&$1.5_{-0.7}^{+1.5} \times 10^{42}$\\ 
NGC4778&SE&1.3&1.4&$2.7 \times 10^{59}$&3.6&$7.86\pm0.08$&$6.5_{-3.3}^{+6.5} \times 10^{55}$&$2.7_{-1.3}^{+2.6} \times 10^{41}$\\  
 &N&1.5&2.0&$5.9 \times 10^{59}$&4.3&$9.34\pm0.07$&$1.2_{-0.6}^{+1.2} \times 10^{56}$&$4.3_{-2.0}^{+4.1} \times 10^{41}$\\ 
NGC5044&NE&0.28&0.27&$2.7 \times 10^{57}$&0.52&$1.16\pm0.02$&$2.9_{-1.4}^{+2.9} \times 10^{54}$&$8.1_{-3.9}^{+7.8} \times 10^{40}$\\  
 &SW&0.27&0.31&$2.8 \times 10^{57}$&0.53&$1.18\pm0.02$&$2.9_{-1.5}^{+2.9} \times 10^{54}$&$8.0_{-3.8}^{+7.8} \times 10^{40}$\\ 
NGC5813&SW&0.79&0.74&$5.6 \times 10^{58}$&1.5&$3.27_{-0.01}^{+0.02}$&$2.3_{-1.2}^{+2.3} \times 10^{55}$&$2.3_{-1.1}^{+2.3} \times 10^{41}$\\  
 &N&1.2&0.77&$1.4 \times 10^{59}$&1.8&$3.81\pm0.02$&$4.8_{-2.4}^{+4.8} \times 10^{55}$&$4.1_{-1.9}^{+4.0} \times 10^{41}$\\ 
NGC5846&NE&0.88&0.52&$5.0 \times 10^{58}$&0.57&$1.24\pm0.02$&$3.5_{-1.7}^{+3.5} \times 10^{55}$&$9.2_{-4.4}^{+8.8} \times 10^{41}$\\  
 &S&0.67&0.57&$3.2 \times 10^{58}$&0.66&$1.46\pm0.02$&$2.2_{-1.1}^{+2.2} \times 10^{55}$&$5.0_{-2.4}^{+4.8} \times 10^{41}$\\ 
NGC6166&W&1.3&1.7&$3.6 \times 10^{59}$&3.6&$5.51\pm0.31$&$4.5_{-2.2}^{+4.5} \times 10^{56}$&$2.7_{-1.3}^{+2.5} \times 10^{42}$\\  
 &E&1.4&2.0&$4.7 \times 10^{59}$&3.2&$4.92\pm0.28$&$5.9_{-2.9}^{+5.9} \times 10^{56}$&$3.9_{-1.9}^{+3.7} \times 10^{42}$\\ 
    \bottomrule
\end{tabular}
\label{tab:cavities}
\end{table*}

\clearpage

\section{List of OBSIDs}

\vspace{3.5mm}

\small
\setlength{\tabcolsep}{5.0pt}
\renewcommand{\arraystretch}{1.068}
\tablefirsthead{\multicolumn{5}{l}{\small \hspace{-4mm} \textbf{Table D1.} List of individual \textit{Chandra} ACIS observations. \vspace{2mm}}\\
\toprule Galaxy & OBSID & Instrument & Date & Exptime (ks) \\ \toprule}
\tablehead{\multicolumn{2}{l}{{\small \hspace{-4mm} \textbf{Table D1.} Continued. \vspace{2mm}}} \\ \toprule
\tabletail{\bottomrule}
\tablelasttail{\bottomrule}
Galaxy & OBSID & Instrument & Date & Exptime (ks)\\ \toprule}
\begin{supertabular}{c c c c c}
IC$\,$4926 & 3394 & ACIS-S & 2001-12-15 & 18.6\\ 
\hline  
NGC$\,$507 & 317 & ACIS-S & 2000-10-11 & 26.9\\
 & 2882 & ACIS-I & 2002-01-08 & 43.6\\ 
\hline  
NGC$\,$708 & 2215 & ACIS-S & 2001-08-03 & 28.7\\ 
 & 7921 & ACIS-S & 2006-11-20 & 110.7\\ 
\hline  
NGC$\,$1316 & 2022 & ACIS-S & 2001-04-17 & 28.4\\  
 & 20340 & ACIS-S & 2019-04-16 & 45.0\\  
 & 20341 & ACIS-S & 2019-04-22 & 51.4\\  
 & 22179 & ACIS-S & 2019-04-17 & 39.0\\  
 & 22180 & ACIS-S & 2019-04-20 & 13.6\\ 
 & 22187 & ACIS-S & 2019-04-25 & 53.2\\ 
\hline  
NGC$\,$1399 & 319 & ACIS-S & 2000-01-18 & 56.0\\  
 & 9530 & ACIS-S & 2008-06-08 & 59.4\\  
 & 14527 & ACIS-S & 2013-07-01 & 27.8\\  
 & 14529 & ACIS-S & 2015-11-06 & 31.6\\
 & 16639 & ACIS-S & 2014-10-12 & 29.7\\
\hline
NGC$\,$1407 & 791 & ACIS-S & 2000-08-16 & 44.5\\
\hline  
NGC$\,$1600 & 4283 & ACIS-S & 2002-09-18 & 22.7\\  
 & 4371 & ACIS-S & 2002-09-20 & 26.8\\  
 & 21374 & ACIS-S & 2018-12-03 & 25.7\\  
 & 21375 & ACIS-S & 2019-11-28 & 42.2\\  
 & 21998 & ACIS-S & 2018-12-03 & 13.9\\  
 & 22878 & ACIS-S & 2019-11-25 & 45.0\\  
 & 22911 & ACIS-S & 2019-11-01 & 31.0\\ 
 & 22912 & ACIS-S & 2019-11-02 & 35.6\\ 
\hline  
NGC$\,$4261 & 834 & ACIS-S & 2000-05-06 & 30.92\\ 
 & 9569 & ACIS-S & 2008-02-12 & 100.9\\ 
\hline  
NGC$\,$4374 & 20539 & ACIS-S & 2019-04-05 & 39.5\\  
 & 20540 & ACIS-S & 2019-02-26 & 30.2\\  
 & 20541 & ACIS-S & 2019-04-10 & 11.3\\  
 & 20542 & ACIS-S & 2019-03-18 & 34.6\\  
 & 20543 & ACIS-S & 2019-04-27 & 54.3\\  
 & 21845 & ACIS-S & 2019-03-28 & 27.7\\  
 & 21852 & ACIS-S & 2019-02-18 & 15.6\\  
 & 21867 & ACIS-S & 2019-03-13 & 23.6\\  
 & 22113 & ACIS-S & 2019-02-20 & 21.8\\  
 & 22126 & ACIS-S & 2019-02-28 & 35.1\\  
 & 22127 & ACIS-S & 2019-03-02 & 22.8\\  
 & 22128 & ACIS-S & 2019-03-03 & 23.8\\  
 & 22142 & ACIS-S & 2019-03-14 & 20.8\\  
 & 22143 & ACIS-S & 2019-03-16 & 22.8\\  
 & 22144 & ACIS-S & 2019-03-15 & 31.8\\  
 & 22153 & ACIS-S & 2019-03-23 & 21.1\\  
 & 22163 & ACIS-S & 2019-03-29 & 35.6\\  
 & 22164 & ACIS-S & 2019-03-31 & 32.6\\  
 & 22166 & ACIS-S & 2019-04-06 & 38.6\\  
 & 22174 & ACIS-S & 2019-04-11 & 49.4\\  
 & 22175 & ACIS-S & 2019-04-12 & 27.2\\  
 & 22176 & ACIS-S & 2019-04-13 & 51.4\\  
 & 22177 & ACIS-S & 2019-04-14 & 36.6\\  
 & 22195 & ACIS-S & 2019-04-28 & 38.1\\ 
 & 22196 & ACIS-S & 2019-05-07 & 20.6\\
\bottomrule
\\
NGC$\,$4472 & 321 & ACIS-S & 2000-06-12 & 34.5\\  
 & 11274 & ACIS-S & 2010-02-27 & 39.7\\  
 & 12888 & ACIS-S & 2011-02-21 & 159.3\\
 & 12889 & ACIS-S & 2011-02-14 & 133.5\\
\hline
NGC$\,$4486 & 352 & ACIS-S & 2000-07-29 & 37.7\\  
 & 2707 & ACIS-S & 2002-07-06 & 98.7\\  
 & 18232 & ACIS-S & 2016-04-27 & 18.2\\  
 & 18233 & ACIS-S & 2016-02-23 & 37.2\\  
 & 18781 & ACIS-S & 2016-02-24 & 39.5\\  
 & 18782 & ACIS-S & 2016-02-26 & 34.1\\  
 & 18783 & ACIS-S & 2016-04-20 & 36.1\\  
 & 18836 & ACIS-S & 2016-04-28 & 38.9\\  
 & 18837 & ACIS-S & 2016-04-30 & 13.7\\  
 & 18838 & ACIS-S & 2016-05-28 & 56.3\\  
 & 18856 & ACIS-S & 2016-06-12 & 25.5\\  
 & 20034 & ACIS-S & 2017-04-11 & 13.1\\  
 & 20035 & ACIS-S & 2017-04-14 & 13.1\\  
 & 21075 & ACIS-S & 2018-04-22 & 9.1\\  
 & 21076 & ACIS-S & 2018-04-24 & 9.0\\  
 & 21457 & ACIS-S & 2019-03-27 & 14.1\\
 & 21458 & ACIS-S & 2019-03-28 & 12.8\\ 
\hline  
NGC$\,$4552 & 2072 & ACIS-S & 2001-04-22 & 54.4\\  
 & 13985 & ACIS-S & 2012-04-22 & 49.4\\  
 & 14358 & ACIS-S & 2012-08-10 & 49.4\\ 
 & 14359 & ACIS-S & 2012-04-23 & 48.1\\ 
\hline
NGC$\,$4636 & 323 & ACIS-S & 2000-01-26 & 45.1\\  
 & 3926 & ACIS-I & 2003-02-14 & 74.7\\ 
 & 4415 & ACIS-I & 2003-02-15 & 74.4\\ 
\hline
NGC$\,$4649 & 785 & ACIS-S & 2000-04-20 & 26.9\\  
 & 8182 & ACIS-S & 2007-01-30 & 49.5\\  
 & 8507 & ACIS-S & 2007-02-01 & 17.5\\  
 & 12975 & ACIS-S & 2011-08-08 & 84.9\\ 
 & 12976 & ACIS-S & 2011-02-24 & 101.0\\ 
 & 14328 & ACIS-S & 2011-08-12 & 14.0\\ 
\hline
NGC$\,$4696 & 504 & ACIS-S & 2000-05-22 & 31.8\\  
 & 505 & ACIS-S & 2000-06-08 & 10.0\\  
 & 4954 & ACIS-S & 2004-04-01 & 89.0\\  
 & 4955 & ACIS-S & 2004-04-02 & 44.7\\  
 & 5310 & ACIS-S & 2004-04-04 & 49.3\\  
 & 16223 & ACIS-S & 2014-05-26 & 179.0\\  
 & 16224 & ACIS-S & 2014-04-09 & 42.3\\  
 & 16225 & ACIS-S & 2014-04-26 & 30.1\\  
 & 16534 & ACIS-S & 2014-06-05 & 55.4\\  
 & 16607 & ACIS-S & 2014-04-12 & 45.7\\  
 & 16608 & ACIS-S & 2014-04-07 & 34.1\\  
 & 16609 & ACIS-S & 2014-05-04 & 82.3\\ 
 & 16610 & ACIS-S & 2014-04-27 & 17.3\\ 
\hline
NGC$\,$4778 & 921 & ACIS-S & 2000-01-25 & 48.5\\  
 & 10462 & ACIS-S & 2009-03-02 & 67.2\\ 
 & 10874 & ACIS-S & 2009-03-03 & 51.4\\
\hline
NGC$\,$5044 & 798 & ACIS-S & 2000-03-19 & 20.5\\
 & 9399 & ACIS-S & 2008-03-07 & 82.7\\
 & 17195 & ACIS-S & 2015-06-06 & 78.0\\
 & 17196 & ACIS-S & 2015-05-11 & 88.9\\
 & 17653 & ACIS-S & 2015-05-07 & 35.5\\
 & 17654 & ACIS-S & 2015-05-10 & 25.0\\
 & 17666 & ACIS-S & 2015-08-23 & 88.5\\
NGC$\,$5813 & 5907 & ACIS-S & 2005-04-02 & 48.4\\
 & 9517 & ACIS-S & 2008-06-05 & 98.8\\
 & 12951 & ACIS-S & 2011-03-28 & 74.0\\
 & 12952 & ACIS-S & 2011-04-05 & 143.1\\
 & 12953 & ACIS-S & 2011-04-07 & 31.8\\
 & 13246 & ACIS-S & 2011-03-30 & 45.0\\
 & 13247 & ACIS-S & 2011-03-31 & 35.8\\
 & 13253 & ACIS-S & 2011-04-08 & 118.0\\
 & 13255 & ACIS-S & 2011-04-10 & 43.4\\
\hline
NGC$\,$5846 & 788 & ACIS-S & 2000-05-24 & 23.4\\ 
 & 7923 & ACIS-I & 2007-06-12 & 90.0\\ 
\hline
NGC$\,$6166 & 497 & ACIS-S & 2000-05-13 & 19.5\\  
 & 498 & ACIS-S & 1999-12-11 & 17.9\\  
 & 10748 & ACIS-I & 2009-11-19 & 40.6\\  
 & 10803 & ACIS-I & 2009-11-17 & 30.2\\  
 & 10804 & ACIS-I & 2009-06-23 & 18.8\\ 
 & 10805 & ACIS-I & 2009-11-23 & 30.3\\
\end{supertabular}
\label{tab:obsids}


\bsp	
\label{lastpage}
\end{document}